\documentclass[a4paper,11pt,DIV=9,BCOR=2cm]{scrbook} 
\usepackage{makeidx}
\usepackage{xcolor}
\usepackage[utf8]{inputenc} 
\usepackage[T1]{fontenc}
\usepackage[ngerman,english]{babel}

\usepackage[toc,page]{appendix}
\usepackage{breakurl}
\usepackage{graphicx}
\usepackage{wrapfig}
\usepackage{color}
\usepackage{eufrak}
\usepackage{amsthm}
\usepackage{manyfoot}

\theoremstyle{definition}

\theoremstyle{remarka}

\theoremstyle{remarkb}

\usepackage[backend=biber,authordate,useprefix=true]{biblatex-chicago}
\usepackage{hyperref}
\DeclareNewFootnote{A}[arabic]
\DeclareNewFootnote{B}[alph]
\usepackage{alphalph}

\usepackage[bottom]{footmisc}

\begin{document}
\title{Von Neumann's 1927 Trilogy on the Foundations of Quantum Mechanics. \\ Annotated Translations}
\author{Anthony Duncan}
\date{\today}

\maketitle

\tableofcontents

\newpage


  \nocite{Einstein:1987,WheelerZurek:1983,Ludwig:1968,VanderWaerden:1968,Schroedinger:1927,Schroedinger:1982}

\chapter{Introduction}

In his witty and insightful biography of John von Neumann, 
Norman Macrae \autocite*{Macrae:1992} suggests that von Neumann had realized ``nearly all his
achievements while he was mainly engaged in something else.'' Perhaps the best illustration of this paradox can
be found in the year 1927, when von Neumann, while completing the requirements for his habilitation (basically,
approval to lecture officially at a German University) at the Friedrich-Wilhelm University in Berlin by submitting
\emph{two} separate theses, one on mathematical logic (``The axiomatization of set theory''), the other in
functional analysis (``General eigenvalue theory of symmetric functional operators''), also managed to produce three
papers laying the foundations for a mathematically rigorous and conceptually coherent formulation of quantum theory \autocite{VonNeumann:1927a,VonNeumann:1927b,VonNeumann:1927c}.
The object of the present article is to present  new, fully annotated translations of these remarkable papers.\footnote{The article will elaborate on the treatment of these papers in Duncan and Janssen \autocite*[Vol.\ 2, Ch.\ 17]{DuncanJanssen:2013,DuncanJanssen:2019-2023}.} 
We begin
with some brief biographical information on von Neumann; this will be followed by a short discussion of the scientific context
of von Neumann's incursion into physics; and, finally, a (very condensed!) description of the contents of each paper of the trilogy.   

\section{Early life: from J\'anos to Johann}
  Neumann J\'anos Lajos 
  (in Hungarian the family name comes first) was born in Budapest on December 28, 1903. His father, Neumann Miksa (Max),
  was a prominent and successful banker with a law degree; on his mother's (Kann Margit) side, the family had access to the considerable resources of
  the Kann-Heller hardware firm, whose sales offices occupied the ground floor of an imposing building, while the Neumanns lived in the 18 room
  apartment on the top floor. The Neumann's were, like many other middle class Hungarian Jews, secular and completely acculturated to the ``k.\ u.\ k.''
  (\emph{kaiserlich und k\"oniglich}, ``imperial and royal'') society of turn of the century Austria-Hungary.\footnote{For a nontechnical biography of von Neumann,
  see Bhattacharya \autocite*{Bhattacharya:2021}. Also nontechnical, but with somewhat more detail on the conceptual advances initiated by von Neumann, Israel and Gasca \autocite*{IsraelGasca:2000}.}
 
    The term ``child prodigy'' seems almost inadequate to describe the young J\'anos, who \ possessed
    remarkable powers of retention: he apparently memorized huge tracts of Oncken's multivolume world history, and received tutoring 
    (on the insistence of his father, for the practical reason that it would allow his children freedom of mobility in a still very fractured
    European continent \autocite{Bhattacharya:2021}) in 
    several modern European languages (English, German, French, and Italian) as well as the obligatory Latin and (Ancient) Greek. His
    future mathematical aptitude \autocite[see][]{Halmos:1973} first appeared in his capacity (age 6) for arithmetical mental gymnastics (of the ``divide two eight digits numbers'' type),
    but soon matured into a mastery of calculus (age 8), and of contemporary real and complex analysis (age 12), as presented in Borel's \autocite*{Borel:1898} treatise.  
    
    In 1913 Max Neumann was ennobled for his services to the Austro-Hungarian state, and the family name became Margittai Neumann (Hungarian for ``Neumann of
    Margitta'', the latter being a place name chosen by Max in honor of his wife), later, following German practice, replaced by ``von Neumann'' (omitting the object of the ``von'').
     J\'anos entered secondary school a year later at the Lutheran gymnasium (of which most of the students were Jewish), and fortunately found in L\'aszl\'o R\'atz, the mathematics teacher, a mentor who immediately recognized
     the extraordinary aptitude of his new pupil. R\'atz arranged for J\'anos to obtain additional tutoring directly from instructors of mathematics at the University of Budapest: first, from
     G\'abor Szeg\"o, and later, from Lip\'ot Fej\'er (one of the foremost Hungarian mathematicians at the time), and Fej\'er's student Mih\'aly Fekete. Other than in mathematics, J\'anos'
     secondary school education followed the standard Lutheran gymnasium syllabus. While still at the gymnasium, J\'anos coauthored a paper with Fekete on the zeroes of Chebyshef polynomials. He 
     also began to think about the set-theoretical  foundations of mathematics---specifically, how to avoid the infamous paradoxes discovered by Bertrand Russell, and the development of a 
     completely rigorous foundation for the natural numbers (first by defining ordinal numbers, and from them, the associated cardinals). These all too abstract preoccupations worried his father
     Max sufficiently that he insisted that his eldest son transfer his attention to a more practical area of study: chemical engineering, for which the \emph{Eidgen\"ossische Technische Hochschule} (ETH) in Z\"urich was the obvious destination.
     At this point however the aspiring engineer's grasp of the prerequisite chemistry was inadequate for admission to the prestigious Swiss school, and it was decided that J\'anos should spend some
     time first at the University of Berlin to acquire the required background---while simultaneously working on a doctorate in the axiomatization of set theory at the University of Budapest, a concession
     no doubt extracted by J\'anos from his father in exchange for his consent to chemical studies. J\'anos arrived in Berlin in September 1921---and adopted for his matriculation the German equivalent---Johann---for his Hungarian given
     name.
     
     \section{Berlin-Z\"urich-G\"ottingen}  
     
      For the next eight years von Neumann moved back and forth between Berlin, Z\"urich,\footnote{As pointed out by Macrae  \autocite*[Ch.\ 4]{Macrae:1992}, Z\"urich was probably a practical necessity, given the
      rampant inflation in Germany: in Switzerland money was stable and one could avoid the financial vertigo of hyperinflation.
 Von Neumann came from a very well-to-do family in Budapest, and had
      no intention of enduring the Bohemian poverty of many German students at the time.} and G\"ottingen.
 While fully completing the chemical engineering prerequisites for admission to the ETH (1923), he mainly
      occupied himself with mathematical matters. In particular, he interacted frequently with the mathematician  Erhard Schmidt at the
(Friedrich Wilhelms) University of Berlin: Schmidt was a student of Hilbert who had contributed 
      substantially to the early development of functional analysis.
 It was no doubt through the offices of Schmidt that Hilbert (in G\"ottingen) was kept informed about the activities of
       the new Hungarian prodigy, who was already making profound contributions
      to the foundations of mathematics, a central area of interest for Hilbert.

  In von Neumann's Berlin student years, chemistry clearly played a secondary role to his mathematical studies: with Erhard Schmidt, Issai Schur,
Richard von Mises, and Ludwig Bieberbach in senior positions the University attracted a large number of talented and engaged students
in mathematics  at all levels, and the
mathematics seminar was typically fully attended, with lively discussions between the students following the lectures. Von Neumann's fellow students at this
time have testified to his preternatural speed in mental computation. He clearly impressed Schmidt---an expert in function theory, and in fact the person who
had first given a precise description of the class of infinite dimensional linear spaces which were becoming known (somewhat unfairly to Schmidt) as \emph{Hilbert}
spaces. In particular, Schmidt was interested in the foundational issues in the formal axiomatization of mathematics (particularly set theory, and the Cantorian
theory of transfinite numbers), which von Neumann was in the process of revolutionizing. On the side, von Neumann acquired the knowledge of chemistry needed
to matriculate in the ETH for a course in chemical engineering.

   The move to Z\"urich for his course at the ETH (beginning in the Fall of 1923) did not distract von Neumann from mathematics: as a mentor he simply replaced
Schmidt with Hermann Weyl (another student of Hilbert), who, in addition to an interest in axiomatics, was deeply interested in a number of areas of mathematical
physics (most prominently, relativity theory). By the end of summer 1926, von Neumann had passed his doctoral exam at the University of Budapest (with a thesis
on the axiomatization of set theory), as well as graduating from the ETH with the long-planned degree in chemical engineering.

  In G\"ottingen, David Hilbert, and his assistant Richard Courant, had by this point  (starting in January 1926) begun the process of applying to the
International Education Board (established by J. D. Rockefeller as a source of fellowship grants to promising young scholars trying to survive the miserable
economic climate of the early 1920s in Europe) for a grant to bring von Neumann to G\"ottingen. By the Fall of 1926 we find von Neumann installed 
in G\"ottingen, where a new topic of interest---the quantum mechanical theories of Heisenberg and Schr\"odinger, so different in form, but apparently
identical in physical content---had, at least temporarily, displaced axiomatics in Hilbert's focus of activities. 

  Hilbert had long displayed a keen interest in developments in theoretical physics. In 1915/16 he was involved in a somewhat acrimonious priority dispute 
with Einstein on the discovery of generally covariant gravitational field equations (a dispute now universally acknowledged to have been resolved in Einstein's
favor). He also followed closely developments in quantum theory, giving lectures on the old quantum theory in the Winter semester of 1922/23, which were 
written up by Lothar Nordheim, a doctoral student of Max Born \autocite[see][]{SauerMajer:2009}. Nordheim had been explicitly recruited by Hilbert from
Sommerfeld's group in Munich as a sort of liaison for Hilbert with the theoretical physics community. 

After the dust had at least partially settled following the explosive developments
in matrix mechanics (Fall of 1925) and wave mechanics (Winter/Spring 1926), Hilbert  enlisted Nordheim's help in addressing the mathematical formulation
of the new theory (Winter semester, 1926/27). In particular, in tune with Hilbert's preference for axiomatic formulations, the recent article of Pascual Jordan,
entitled ``On a new foundation of quantum mechanics'' \autocite[\emph{\"Uber eine neue Begr\"undung der Quantenmechanik,''}][]{Jordan:1927b}, which purported
to provide a single axiomatic framework encompassing the various apparently distinct versions of the new theory, became the central focus of Hilbert's
activity in this area at just the point when the recently arrived (now, Johann) von Neumann joined the fray.  Nordheim recalled later (AIP interview
of Lothar Nordheim by John Heilbron, July 30, 1962),
when asked by Heilbron if von Neumann had been concerned with quantum mechanics prior to arriving in G\"ottingen,
\begin{quote}
 I think that he got his start in this field in G\"ottingen, and being so tremendously fast, he got hold of it very soon. Yes, he had one of the
fastest minds I have ever met. His brains always worked on roller bearings [!].
\end{quote}
   
\section{Von Neumann becomes a physicist}

  When von Neumann arrived in G\"ottingen, the broad outlines of what would before long become a consistent, conceptually sound theory
of atomic phenomena were already visible.\footnote{For a detailed account of the development of quantum theory from about 1900 to 1927, see Duncan and Janssen \autocite*{DuncanJanssen:2019-2023}.} 
Werner Heisenberg and his collaborators Max Born and Pascual Jordan had developed the 
former's proposal to replace classical mechanical quantities with complex matrices constrained by certain nonlinear conditions (the commutation
relation of classically conjugate variables, such as position and momentum) into a well-defined theory of energy levels and quantum transitions,
at least for systems with a discrete spectrum (such as bound electrons in atoms). In a completely separate development, Erwin Schr\"odinger,
in Z\"urich, had developed an approach to quantization in which classical quantities were replaced by linear operations (including
derivatives) on a complex function
$\psi(q)$ (where $q$ represents all the coordinate variables of the particle(s) described by this ``wave function''). Schr\"odinger had also
been able to show that his theory gave identical results to the G\"ottingen matrix mechanics for systems with a purely discrete spectrum. However,
his theory clearly encompassed non-discrete systems (such as free, or unbound, electrons) with ease, which could not be said for the matrix
approach.

   At the point where von Neumann arrived in G\"ottingen (Fall, 1926), there were two outstanding issues of great conceptual importance which the new theories
posed, to wit:
\begin{enumerate}
\item The essentially stochastic nature of the emerging quantum theory was only understood in an incomplete and sometimes contradictory way.
\item The precise mathematical structure underlying the two (superficially completely different) versions (matrix vs.\ wave) of the theory was not yet visible.
\end{enumerate}
  The remarkable achievement of von Neumann was to provide a completely rigorous mathematical formulation of the theory in which its probabilistic
aspects---specifically, the Born rule for probabilities---would emerge naturally once a few basic assumptions on the properties of measurements of
quantum systems were made. 

  Hilbert's desire to keep abreast of the latest developments in quantum theory provided an ideal gateway for von Neumann's entry into the field.
Presumably, Nordheim, Born's doctoral student, was the natural conduit  to von Neumann of the activities of Born, Heisenberg, and, increasingly of relevance to our
story here, Pascual Jordan.
Jordan's attempt to axiomatize quantum theory,\footnote{For detailed discussion of Jordan's approach, see Duncan and Janssen \autocite*[Vol.\ 2, Sec.\ 16.2]{DuncanJanssen:2013,DuncanJanssen:2019-2023}.} 
and thereby to subsume its disparate versions under a single theoretical roof, would clearly have appealed to von Neumann, a disciple of Hilbert for whom  an exact formal foundation was the ultimate desideratum of any mathematically based science.

  In the summer of 1926, Max Born, in his analysis of scattering processes in quantum mechanics, proposed an explicitly probabilistic interpretation of 
the Schr\"odinger wave function: its absolute square would provide an indication of the probability of finding the scattered particle in a given region (more
exactly, traveling in a given angular direction) after the scattering:
\begin{quote}
  One obtains \emph{no} answer to the question ``what is the state [of the scattered particle] after the collision'', but rather, to the question ``how probable
is any prescribed effect of the collision?'' \autocite[p.\ 866]{Born:1926a}
\end{quote}
   
   In the Fall of 1926, Pascual Jordan, stimulated by numerous communications with Wolfgang Pauli, attempted to construct an axiomatic formulation of
quantum theory in which (a) the peculiarities of the new theory would be subsumed in postulates of a probabilistic nature, but in which \emph{probability
amplitudes} (of which the Schr\"odinger wave function was the best known special case) would obey combination laws which mimic those of classical
probability theory---despite the fact that actual probabilities were related nonlinearly (via the absolute square, as emphasized by Born) to the underlying
(in general complex) amplitudes, and (b) these amplitudes would serve as a basis for a completely general axiomatic formulation of the theory, capable
of reproducing all previous versions (matrix mechanics, wave mechanics, Born-Wiener theory, and the Dirac $q$-number theory).

  We will briefly recall here the axiomatic framework proposed by Jordan in the first installment of his ``New Foundation of Quantum Mechanics'' \autocite{Jordan:1927b}.
Jordan's notation is extraordinarily clumsy so we shall abandon it in favor of a cleaner formulation.\footnote{We also eliminate the ``supplementary amplitude''
introduced by Jordan in his first paper to allow for non-hermitian quantities. The treatment throughout is for a system of a single degree of freedom, and for
observables with a continuous spectrum.} The framework erected by Jordan posits for the given quantum mechanical system (say, a particle on a line subject to an
imposed potential energy function) a set of mechanical quantities $A,B,C, \ldots$ which stand in a definite kinematical relation to one another. For example, classically
such quantities would be definite functions $A(Q,P), B(Q,P),C(Q,P)\ldots$ of the position/momentum canonical pair $Q/P$ for the particle. For simplicity, Jordan assumes
that these quantities can only assume continuous values: in particular, the observable $A$ can be found  to have the (real number) value $\alpha$, the observable
$B$ a value $\beta$, and so on. Jordan postulates
\begin{enumerate}
 \item \emph{Postulate I.} If the system is known to have a value $\beta$ of quantity $B$, then the probability that the [measured] value  of quantity $A$ lies 
in the interval $(\alpha,\alpha+d\alpha)$ is $|\varphi_{AB}(\alpha,\beta)|^{2}d\alpha$. Here, the functional dependence of the probability amplitude $\varphi_{AB}i$ on
its two arguments is completely determined by the kinematical relation between the mechanical quantities $A$ and $B$: \emph{it is completely indifferent to
the particular dynamics of the system, for example, the particular form of the Hamiltonian energy function $H(Q,P)$, which for Jordan is just one more of the
accessible mechanical quantities of the system.}
\item \emph{Postulate II.} The probability amplitude connecting two quantities is complex symmetric:
\begin{equation}
     \varphi_{AB}(\alpha,\beta) =\varphi_{BA}(\beta,\alpha)^{*}.
\end{equation}
 \item \emph{Postulate III.} The probability amplitudes (and not their squares!) obey standard composition rules of the probability calculus. If two outcomes [German
\emph{Tatsachen}] $F_1, F_2$, associated with probability amplitudes $\varphi_1,\varphi_2$ are mutually exclusive [\emph{ sich ausschliessen}] then the amplitude for 
``$F_1$ or $F_2$'' is $\varphi_{1}+\varphi_{2}$. If two outcomes  $F_1, F_2$, associated with probability amplitudes $\varphi_1,\varphi_2$ are independent [\emph{unabh\"angig }] then the amplitude for 
``$F_1$ and $F_2$'' is $\varphi_{1}\varphi_{2}$. 
\item \emph{Postulate IV.} For every mechanical quantity $q$ there exists a canonically conjugate momentum $p$.
\end{enumerate}
An immediate consequence of Postulate III is the following composition rule for amplitudes:
\begin{equation} 
\label{interferprob}
   \varphi_{AC}(\alpha,\gamma) = \int \varphi_{AB}(\alpha,\beta)\varphi_{BC}(\beta,\gamma)d\beta.
\end{equation}
As an illustration, imagine that the observable $C$ (with measured value $\gamma$) corresponds to the position of a particle emitted from a diffuse source at time $t_0$, observable $B$
to the (now discrete) location $\beta$ of a hole (or slit) in a screen through which the particle passes at a time $t_1>t_0$, and $A$ to the location $\alpha$ at which the particle stops on a fluorescent
screen at $t_{2}>t_{1}>t_{0}$. The independent passage from source to the perforated screen, then from the perforated screen to the final fluorescent one, requires the multiplication of the
$\varphi_{AB}$ and $\varphi_{BC}$ amplitudes. The mutually exclusive choice of which hole (or slit) to pass through then requires a sum over such holes, which would of course mean that a discrete
sum would replace the integral over $\beta$ in Eq.\ (\ref{interferprob}). In the more general case, in which the intermediate state requires a continuous specification, the integral form given above applies.\footnote{More generally
still, the $B$ observable might contain both a discrete and continuous spectrum---say, the position on a screen perforated with slits as well as opeinings of finite size---in which case the right hand side of
Eq.\ (\ref{interferprob}) would involve both an integral and a sum.}
The familiar two-slit example illustrating the interference phenomena peculiar to quantum mechanics furnishes thus just a very simple special case of Jordan's composition rule. 

  There are however two serious problems with Jordan's postulated axioms for probability amplitudes. The first is a deep conceptual one, the second of a more technical nature. Namely:
\begin{enumerate}
\item   First, the exact operational status of the first line of Postulate I---``If the system is known\ldots then the probability that\ldots''---is never clearly specified in \emph{New Foundation}. Von Neumann would, in
the second paper of his trilogy, lay out the precise meaning of this superficially straightforward assertion by applying a rigorous version of the frequentist definition of probabilities which he had absorbed
from Richard von Mises in his Berlin studies. Values  (or, more precisely, \emph{ranges of values}) of observable quantities were to be assigned to ensembles of physically identical quantum systems
subjected to identical measurement interventions. Conditional probabilities would emerge as the limit of frequencies of results obtained by subsequent measurements on such suitably prepared ensembles.
Once the association is made between the values obtained by measurements and the spectrum of associated symmetric [i.e., hermitian] operators, a few almost self-evident assumptions on the statistical properties
of these ensembles would lead von Neumann to an essentially unique construction of the formalism of quantum mechanics, including the Born rule encapsulated in Postulate I.
\item  The mathematical status of Jordan's amplitudes $\varphi_{AB}(\alpha,\beta)$ was extremely obscure. The problem becomes acute whenever a continuous spectrum is present. Consider again
Eq.\ (\ref{interferprob}), but assume that the observables $A$, $C$ are the same, with $B$ a different observable (all with continuous spectra). Using Postulate II, this becomes
\begin{eqnarray}
    \varphi_{AA}(\alpha,\alpha^{\prime}) &\!\!\!=\!\!\!& \int \varphi_{AB}(\alpha,\beta)\varphi_{AB}^{*}(\alpha^{\prime},\beta)d\beta = 0 \;\;\mathrm{if}\;\alpha \neq \alpha^{\prime} \nonumber \\
\label{deltaamp}
  &\!\!\!=\!\!\!& \infty (?)  \;\;\mathrm{if}\;\alpha = \alpha^{\prime}
\end{eqnarray}
The modern physicist will recognize the peculiar behavior of the $\varphi_{AA}$ amplitude as associated with the ``delta function'' $\delta(\alpha-\alpha^{\prime})$ introduced heuristically by Dirac,
but it must be understood that the appropriate mathematical framework for dealing with an ``improper function'' of this kind lay approximately 25 years in the future, with Laurent Schwartz' distribution
theory. In particular, the appearance of such an improper function was particularly embarrassing given the probability base of the theory, as attempting to square it gives nonsense. As we
shall see, von Neumann was
at pains, especially in the first paper of the trilogy, to indicate the mathematical problems introduced by admitting such improper functions, while emphaiszing that their use \emph{was neither physically sensible,
or mathematically necessary.} He does this by taking the examples of  quintessentially continuous observables of quantum mechanics---position and momentum---and showing how the essence of Postulate I can
be maintained for measurements of position and momentum without the appearance of a single ``improper'' function.
\end{enumerate}

   Hilbert, in keeping with his long-standing desire to keep abreast of the latest developments in fundamental theoretical physics, gave a lecture course on quantum theory in the Winter
semester of 1926/1927 in which he presented matrix mechanics and wave mechanics. This was just the time when Jordan was formulating and submitting for publication his ``New Foundation \ldots'' paper
with the four postulates outlined above. No doubt Nordheim, Hilbert's ``spy'' in the G\"ottingen physics department, was keeping a close eye on these developments, which would naturally appeal to
Hilbert's formalist outlook. The idea of building up a theory from formal elements, initially given no interpretation or ``meaning'', but specified solely by posited interrelations and structural connections---the ``axioms''
of the theory---appealed greatly to Hilbert. Only after the nexus of ``theorems'' describing more complex properties derivable from the axioms had been worked out would one then a specific mathematical
(or physical) interpretation to the basic formal elements. Hilbert had followed precisely this approach in his \emph{Grundlagen der Geometrie} [Foundations of Geometry, 1899], in which abstract concepts such
as ``point, line, plane, etc'' are introduced in a purely axiomatic fashion and their aggregates shown to satisfy a rich set of properties, without ever needing to identify these words with their commonplace
equivalents (in two dimensional Euclidean geometry). Hilbert was enthusiastic about the possibility of doing the same with quantum mechanics: a complete structure could be erected using Jordan's complex
probability amplitudes, axiomatically prescribed to have certain properties. Any analytic realization of these properties, supplemented with a physical interpretation of the amplitudes, would then constitute
an acceptable quantum mechanical theory.

    In the Spring of 1927, Hilbert enlisted both Nordheim, and the recently arrived prodigy Johann von Neumann (whose formalist credentials were hardly in doubt, given his work on the axiomatization of 
set theory) in realizing a reworking of Jordan's \emph{Neue Begr\"undung} which would both clarify the axiomatic structure proposed by Jordan, and then show the consistency of these axioms by
exhibiting an analytic realization of the axioms in terms of mathematically well-defined operations. The resultant paper would not be published for another year, and was by the time of its publication 
completely superseded by the trilogy of von Neumann. Nevertheless, it is a valuable index to von Neumann's state of knowledge at the outset of his career in physics. The second paragraph of the paper
gives an indication of the relative roles of the participants in the writing of \autocite{HilbertvonNeumannNordheim:1928}:
\begin{quote}
  The work presented below arose from a seminar given by D.\ Hilbert in the Winter semester 1926/27 on the recent development of quantum mechanics, the preparation of which profited from the essential
assistance of L.\ Nordheim. Important parts of the mathematical execution are due to J.\ v.\ Neumann. 
\end{quote}

  In the Hilbert-Nordheim-von Neumann paper, the theory is built up from a set of six axioms, three of which (I, II, and IV)
correspond to Postulates I-III of
Jordan (the missing one being Jordan's assertion on the existence of conjugate coordinate/momentum pairs). The ``extra'' three are 
really already explicitly  present in Jordan's \emph{Neue Begr\"undung} paper, if not listed as fundamental postulates: namely (a) (III) the ``sharp''
(i.e., delta function) behavior of the amplitudes for an observable with respect to itself (cf.\ Eq.\ (\ref{deltaamp})), (b) (V) the requirement that
the form of the amplitudes depend only on the kinematical relation between the respective observables, not on the particular dynamics of
the theory (e.g., the Hamiltonian), and (c) (VI) the assertion that the probabilities do not depend on the particular choice of coordinate system
in which we choose to work.   

   Despite the implications of the quote above specifically attributing aspects of the mathematical execution to von Neumann, the analytic apparatus employed
in this paper to concretely realize the Jordan axiomatic framework owes much more to Hilbert than to his young acolyte. In an attempt to find
a unified representation of Jordan's amplitudes, Hilbert resorts to the use of integral kernels, a completely natural choice given his central role in
working out a complete spectral theory of linear integral equations (with bounded kernels) in the two decades preceding this paper. Unfortunately,
this approach immediately necessitates the introduction of the improper ``functions'' (such as the delta function) to which von Neumann was strongly
allergic (as he would emphasize in \autocite{VonNeumann:1927a}). If probability amplitudes $\varphi_{AB}(\alpha,\beta)$ were to play a dual role as
kernels of linear transformation operators on spaces of functions,\footnote{We have slightly altered the notation used by Hilbert et al. to conform to our earlier usage.}
\begin{equation}
\label{phikernel}
   f(\beta) \rightarrow g(\alpha) \equiv T_{AB}f (\alpha) = \int \varphi_{AB}(\alpha,\beta)f(\beta)d\beta,
\end{equation}
then the simplest possible linear operator, the identity operation $f\rightarrow f$, must be represented by an integral kernel which is just the Dirac
delta function:
\begin{equation}
     T_{AA}f(\alpha) = f(\alpha) = \int \delta(\alpha-\beta)f(\beta)d\beta.
\end{equation}
Unfortunately, as von Neumann undoubtedly realized from the outset, could not serve as the basis for developing a rigorous spectral theory. While the
formal manipulations of Hilbert et al., which we recognize today as perfectly justified by the distribution theory developed in the late 40's (and by the use of Gelfand triplets,
the so-called ``rigged'' Hilbert spaces), the rigorous mathematical status of these maneuvers was still a complete mystery in the mid 1920s. So the highly formal
manipulations of the Hilbert-Nordheim-von Neumann paper bear only a superficial similarity to the (completely rigorous!) spectral theory of integral operators
developed by the Hilbert school. This latter theory was in fact explicitly---and correctly---deployed in the culminating ``Three Man Paper'' of matrix mechanics, in a section written
by Born \autocite[for discussion, see][Vol.\ 2, Secs.\ 12.3.3 and 12.3.4]{DuncanJanssen:2019-2023}.
What Born realized at the time was that the Hilbert theory (amplified by Hellinger to include a rigorous theory of operators with a partly continuous spectrum)
only applied to \emph{bounded} operators---those, roughly speaking---whose eigenvalues were limited in absolute magnitude. Almost all the operators appearing
in quantum mechanics (position, momentum, energy, etc) are \emph{unbounded} operators. Born, in a footnote \autocite[p.\ 583]{BornHeisenbergJordan:1926}, expressed the hope that an appropriate extension of
the Hilbert theory would allow treatment of such quantities as well. 

Born did not, however  (in 1925), fully appreciate the structure of the function space underlying  the spectral theories of Hilbert and Hellinger,  for which he can hardly
be blamed, as neither did Hilbert! The critical concept, that of a infinite-dimensional vector space, equipped with an (in general complex valued)
inner product, and with a complete denumerable orthogonal basis set dense in the space, had already been introduced by Erhard Schmidt, whom von Neumann had
come to know very well in Berlin, more than a decade earlier. This space, re-dubbed by von Neumann a ``Hilbert space'', would be placed center stage by
von Neumann in his first paper (as sole author) on the mathematical foundations of quantum mechanics. We turn now to a brief review of the contents of this work.

\section{Paper 1: The Mathematical Foundation of Quantum Mechanics}

  The article of Hilbert, Nordheim and von Neumann, although prepared for publication by April of 1927, was not published until the following year, long past the point
where von Neumann's three seminal papers had completely upended and permanently reframed the formal structure of quantum theory. It has therefore become
something of an historical oddity, which, despite the eminence of its provenance, essentially vanished from the mainstream development of the theory. In his first
solo paper on quantum mechanics, \autocite{VonNeumann:1927a}, von Neumann would essentially reject the methodology of the Jordan axiomatization, replacing 
probability amplitudes as the central building blocks of the theory with vectors (and, more generally, projection operators\footnote{In this introductory article, and in the translated papers below, we
shall use the conventional modern terminology, ``projection operator'', to indicate a self-adjoint idempotent operator. The term used by von Neumann, ``Einzeloperator'' (E.\ Op.\ for short) goes
back to Hilbert's ``Einzelform'', a quadratic form with a discrete spectrum consisting of the single eigenvalue 1 \autocite{Hilbert:1912}, and
translates literally to ``single-operator.''} onto linear subspaces) in an infinite
dimensional vector space of a special sort.    

 The paper begins with an itemized summary of the present status of ``quantum mechanics'', a theory---or better, nexus of ideas [\emph{Begriffsbildungen}]---attributed 
explicitly to Heisenberg, Dirac, Born, Schr\"odinger and Jordan. 
\begin{enumerate}
\item The behavior of atomic systems is connected to a particular eigenvalue problem, with the spectrum of eigenvalues coinciding with the possible measured values [of what
quantity, is not stated---but we may presume that von Neumann is talking about the energy here].
\item The peculiar mixture of continuous and discrete aspects appearing in quantum mechanics is therefore attributable to the simple fact that eigenvalue problems
can easily lead to the coexistence of continuous and discrete spectra.
\item The quantum laws of Nature are in general not strictly causal, although in certain restricted cases the behavior of the system can be reliably predicted; thus the basic laws
give probability distributions.
\item The eigenvalue problem can appear in quite distinct forms: as the diagonalisation problem of an infinite matrix, or of a differential equation. These two formulations
can be related by expanding the ``wave function'' in the differential operator case in a (discrete) infinite series of orthogonal functions, in which case the action of
the differential operator can be rewritten as the action of an infinite matrix on the expansion coefficients of the wave function.
\item Both approaches (matrix/differential equation) have their own difficulties. In the matrix case, the transformation to diagonal form is only possible if there is
no continuous spectrum---this obviously is not the case even in the simple case of a hydrogen atom. The use of such constructs as ``continuous matrices'', in 
which one operates simultaneously with discrete matrices and integral equation kernels, is very difficult to accomplish with full mathematical rigor.
\item Only recently, with the work of Born and Jordan, has the connection to probabilities, visible at the outset in the matrix approach [the square of matric 
elements of the coordinate operator gave atomic transition probabilities], emerged. This connection was expanded to a complete and closed system by Jordan,
but only at the expense of introducing mathematically inadmissible objects. In particular, one obtains eigenfunctions which are ``improper'' functions, such as
the one introduced by Dirac, ``which is supposed to have the following (absurd) properties:''
\begin{equation}
   \delta(x) = 0,\;\;\mathrm{for}\;x\neq 0,\quad\int_{-\infty}^{+\infty}\delta(x)dx = 1.
\end{equation}
\item A common defect of all these methods is the introduction of elements in the calculation which are in principle unobservable and physically
meaningless. While the final probabilities are unique and meaningful, the eigenfunctions which appear at intermediate stages of the calculation are
undetermined up to a phase (in cases of degeneracy, up to an entire unfixed unitary rotation).
\end{enumerate}

  Following this terse (and incisive) review of the state of affairs, von Neumann alerts the reader that the rest of the paper would present a detailed
mathematical preparation (sections 2-11), followed by the application of the formalism to quantum theory in the final sections (12-15). The title
of the first part (preceding section 2), ``The Hilbert space'' (\emph{Der Hilbertsche Raum}), requires some comment. What precisely von Neumann means
is first explained in section 3, following a brief reference to Hilbert's 1906 theory of bilinear forms:
\begin{quote}
 In this case there are good grounds to focus on the following manifold [space]: all sequences $x_{1},x_{2},\ldots$ of complex numbers with finite $\sum_{n=1}^{\infty}
|x_{n}|^{2}$. For this space it is customary to use the name (complex) Hilbert space \autocite[p.\ 7]{VonNeumann:1927a}.
\end{quote}
The term ``customary'' (\emph{\"ublich}) is a little puzzling here, as the term ``Hilbert space'' does not appear in
any of  Hilbert's numerous works on integral equations and bilinear forms in the relevant
period (mainly 1906-1912), nor, to the author's knowledge, in the papers of  other workers (pre-1927) in the field of function spaces (later, functional analysis). 
The particular Hilbert space cited in the quote above (now referred to by mathematicians as $l^2$) was in fact first defined explicitly by Schmidt \autocite*{Schmidt:1908}.
All of the defining properties of a Hilbert space (linear  space, existence of bilinear inner product, Cauchy completeness) were shown by Erhard Schmidt
in his 1908 paper to apply to this space,
the geometrical interpretation of which as an infinite-dimensional vector space he attributes in a footnote to Gerhard Kowalewski. So a more historically accurate
term should perhaps be ``Schmidt-Kowalewksi'' spaces. It is however quite possible that in conversations between mathematicians (even, perhaps, between 
von Neumann and Schmidt himself) the term ``Hilbert space'' was already in informal use. This usage became universal once it was sanctified by von Neumann
in his mathematical papers form 1927 on.
A brief summary of the contents of this groundbreaking paper follows:
\begin{description}
\item[Section 2.]  The formulation of the eigenvalue problem (cited in his introduction as \emph{the basic problem} in quantum theory) is reviewed in
both matrix and wave mechanics. In both cases one has to deal with the determination of the possible values of $w$ where $Hf=wf$, $H$ is a linear
operator, and $f$ is either a sequence (with $H$ a matrix) or a function (with $H$ a differential operator).
\item[Section 3.] The deep analogy between the space of square-summable sequences ($l^2$ now, in von Neumann's notation $\mathfrak{H}_{0}$) 
and square-integrable functions $\varphi(x)$ over some space $\Omega$
 (thus: $\int_{\Omega}|\varphi(x)|^{2}dv <\infty$) is discussed ($L^2$ now, in von Neumann's notation $\mathfrak{H}$), 
with clear reference to the matrix and continuous (wave) formulations of quantum mechanics. (The
rigorous isomorphism of the spaces will be explained in Section 4). The notation $Q(f,g)$ for the complex inner product of two vectors $f$ and $g$ is introduced
(in modern Dirac notation, this would be written $\langle g|f\rangle$---note the inverted order).
\item[Section 4.] The isomorphism of $\mathfrak{H}_{0}$ ($l^2$) with $\mathfrak{H}$ ($L^2$) is shown, citing the Riesz-Fischer theorem.
\item[Section 5.] The axioms defining a (separable) Hilbert space are laid out: linear vector space, complex inner product with associated metric, infinite-dimensional with
a countable basis (separable), and Cauchy-complete.
\item[Section 6.] A plethora of basic theorems, propositions and lemmas following directly from the axioms (Bessel's inequality, Parseval identity, etc.).
\item[Section 7,] Operator calculus: the importance of densely defined operators (that can act sensibly on vectors arbitrarily close to any element of the Hilbert space),
definition of the hermitian adjoint  and symmetric operators (those equal to their hermitian adjoints).
\item[Section 8.] Projection operators (\emph{Einzeloperator}) are introduced: $E$ is a projection operator if it is symmetric ($E=E^{\dagger}$) and idempotent, $E=E^2$.
Several important definitions and theorems are presented involving projection operators.
\item[Section 9.] A universally valid formulation of the eigenvalue problem can now be given: the operator $A$ in question is to be written as an integral over (differential) 
spectral projection operators. If $E(l)$ projects onto the subspace spanned by all eigenvectors of $A$ with eigenvalue less than or equal to $l$ (so $E(-\infty)=0$ and $E(+\infty)=1$),
then
\begin{equation}
\label{specres}
     A = \int_{-\infty}^{+\infty} ldE(l),\quad 1 = \int_{-\infty}^{+\infty} dE(l).
\end{equation}
How this works is illustrated for operators in a finite-dimensional Euclidean space.
\item[Section 10.] The Hilbert-Hellinger theory establishes the validity of Eq.\ (\ref{specres}) for all symmetric bounded operators. For the unbounded operators
predominant in quantum mechanics, the existence of such a spectral resolution is at this point not known in general. However, for the fundamental
(unbounded!) operators $q$ and $p$, the spectral resolution is explicitly constructed, making it clear that there is no fundamental obstruction to this
desirable result for unbounded operators.
\item[Section 11.] A norm $||A||$ is introduced for arbitrary operators $A$: one evaluates the matrix of the operator $A_{\mu\nu}$ with respect to a complete
orthonormal basis in the Hilbert space, then computes the sum of absolute squares: $||A|| := \sqrt{\sum_{\mu,\nu}|A_{\mu\nu}|^{2}} \equiv \sqrt{[A]}$. Typical 
properties attached to a norm are shown to hold, e.g.,  the triangle inequality $||A+B|| \leq ||A||+||B||$. 
\item[Section 12.] The precisely defined operator norms of section 12 are deployed to put Jordan's conditional probabilities on a completely rigorous footing.
The necessary ingredients are just the spectral operators associated with any observables whose value (ranges) are prescribed/measured. As measurements
have necessarily limited precision, it suffices to specify, for any given interval $I=(l_1,l_2)$ (with $l_1<l_2$), the associated spectral projection operator,
for (say) operator $A$,  $E(I)=E(l_2)-E(l_1)$. Then the conditional probability that another observable, associated with operator $B$ (spectral operator $F$), is found to have 
a value in the interval $J$, given that $A$ is known to be in the interval $I$, is just $[E(I)F(J)]=||E(I)F(J)||^{2}$. \emph{This is the Born rule in its most  precise and
mathematically rigorous form.}
\item[Section 13.] The probability rule of section 12 is extended to simultaneous measurements of sets of compatible (i.e., commuting) observables. 
The postulates forming the basis of Jordan's formulation of quantum mechanics in \emph{Neue Begr\"undung} are discussed in the context of
the new, mathematically rigorous framework.  The symmetry of amplitudes (Jordan Postulate 2) becomes a symmetry under ``replacement of all
assertions (\emph{Behauptungen}) with all conditions (\emph{Voraussetzungen}).'' This symmetry follows from the obvious identity $[AB] = [BA]$.

Following Jordan, von Neumann asserts that the law of multiplication of probabilities (for independent events) in not generally valid---rather a weaker law, of ``superposition
of probability amplitudes'', as stated by Jordan, holds. That this is not unexpected, claims von Neumann, is because``the dependency relations
of our probabilities can be arbitrarily complicated.'' What does this mean? It should be remembered that at the time of writing this paper, von Neumann had
not yet read Heisenberg's \autocite*{Heisenberg:1927b}
uncertainty paper (written in Copenhagen, and not published till more than a week after von Neumann's paper was submitted on May 20). The ``dependency relations'' somewhat hazily alluded
to by von Neumann were exposed with brilliant clarity in Heisenberg's discussion of the passage of atoms through multiple Stern-Gerlach devices. In this
thought experiment it is shown that the establishment of the ``assertions'' concerning the intermediate state of the atoms  needed to apply the classical law
for multiplication of probabilities would require measurements of this intermediate state, which would destroy the phase relations of the atomic states. This intimate
interconnection of the atomic phases which requires use of products of complex amplitudes (rather than of positive real probabilities) can, with some charity,
be equated with von Neumann's vague ``dependency relations.'' In any event, von Neumann would soon absorb the lesson of Heisenberg's paper, which is prominently
cited in his next paper \autocite{VonNeumann:1927b}, and, in agreement with Heisenberg, maintain that the classical probability laws (both multiplication and addition) would 
always be perfectly valid, \emph{provided the appropriate measurements needed to establish the claimed conditions and assertions were carried out.}

The law of addition for probabilities of exclusive outcomes was far less problematic, and holds without further ado. Taking for simplicity assertions and conditions
associated with a single observable each, with spectral projections $E$, $F$ respectively, then, if $J^{\prime}$ and $J^{\prime\prime}$ are two non-overlapping
intervals, with $J= J^{\prime}\cup J^{\prime\prime}$, then 
\begin{equation}
[E(I)F(J)] = [E(I)F(J^{\prime}] + [E(I)F(J^{\prime\prime})].
\end{equation}
\item[Section 14.] In this section (entitled ``Applications''---\emph{Anwendungen}) von Neumann discusses the spectrum of the Hamiltonian operator, which is
of course of central physical importance. In general, the spectrum contains both point eigenvalues, some of which may be degenerate, and continuum
eigenvalues. The thorny issue of the a-priori probability of a quantum state can now be settled with complete precision. Von Neumann applies the
term ``quantized state'' for the eigenfunctions associated with the point spectrum, and ``unquantized state'' for the continuum spectrum. The a-priori probability 
associated with an interval $I$ is $[E(I)]$, which is just the dimension of the subspace projected onto by $E(I)$, to wit, the number of independent eigenfunctions
$\varphi_{\nu_1},\varphi_{\nu_2},\ldots$ with eigenvalues in the interval $I$. For an unquantized state, the a-priori probability is zero: the probability of finding the system
in a state with any preassigned \emph{exact} value of the energy in the continuum part of the spectrum is zero. Note that in this entire discussion, the probabilities
referred to are typically relative ones: in the hydrogen atom, for example, there are infinitely many discrete eigenstates, so the assertion is that each individual
state has the same a-priori probability as any other, while an exactly prescribed continuum state has zero probability relative to any of the discrete ones.
There is a short discussion of the calculation of conditional probabilities when two observables are ``causally'' connected (by which von Neumann means, 
functionally connected, $A=f(B)$). Finally, he derives the original Born rule, in the context of the time evolution of eigenstates of a time-dependent Hamiltonian $H(t)$:
if the system is placed in eigenstate $\varphi_{\mu}^{(0)}$ of $H(0)$ at time 0, the probability that it will be found to be in the $\nu$'th eigenstate $\varphi_{\nu}^{(t)}$
of $H(t)$ at time $t$ is $|Q(\varphi_{\mu}^{(0)},\varphi_{\nu}^{(t)})|^{2}$; i.e., just the absolute square of the inner product of the two eigenstates.
\item[Section 15.] The conclusion reiterates the basic result for conditional probabilities in quantum mechanics, in the most general case, where an arbitrary number of
compatible observables ($R_1,R_2,\ldots$, spectral functions $E_1,E_2,\ldots$) are presumed to lie in the intervals $I_1,I_2,\ldots$, and one then wishes to find the probability
that another set of mutually compatible observables ($S_1,S_2,\ldots$, spectral functions $F_1,F_2,\ldots$) will be found to have values in the intervals $J_1,J_2,\ldots$.
The desired conditional probability is given by the beautifully compact and intuitive formula
\begin{equation}
    [E_{1}(I_{1})\cdot E_{2}(I_{2})\cdot\ldots\cdot F_{1}(J_{1})\cdot F_{2}(J_{2})\cdot\ldots].
\end{equation}
The paper concludes with the admonition that mutually incompatible observables (corresponding to noncommuting operators) not be allowed among either the 
conditions (the $R_{i}$ above) or the assertions (the $S_{j}$ above), although the $R_{i}$ and $S_{j}$ may of course not be compatible. As von Neumann
puts it, ``quantum mechanics does not in any way allow the positing of such a question!''
\end{description}

The enormous wealth of new insights and beautifully precise (and powerful!) formalism brought to bear by von Neumann on the conceptual makeup of quantum theory
should not blind us to the two really central, and for the future development of the theory, seminal contributions of this paper:
\begin{enumerate}
\item The paper gives a mathematically unassailable demonstration of the complete identity of matrix and wave mechanics, which is nothing more or less than
the isomorphism between the Hilbert spaces of square-summable series ($l^2$ now; $\mathfrak{H}_{0}$ for von Neumann) and of (Lebesgue) square-integrable
functions ($L^2$ now; $\mathfrak{H}$ for von Neumann). The underlying theorem (Riesz-Fischer) was not due to von Neumann, but he understood clearly its
relevance to establishing the underlying unity of the two regnant forms of the theory. \emph{Any problem of wave (resp.\ matrix) mechanics, can be rephrased,
with no loss of information, as a problem of matrix (resp.\ wave) mechanics!} Even the hydrogen atom, with its annoying mix of point-like and continuous spectrum,
can be completely described by an infinite discrete matrix Hamiltonian, \emph{which contains all the information concerning the normalizable discrete bound states
as well as the non-normalizable continuous eigenstates of the system.}\footnote{One could, for example, use the square-integrable Hermite eigenfunctions
of a three-dimensional oscillator as a complete orthonormal basis, and evaluate the matrix elements of the Coulomb Hamiltonian of the hydrogen atom
in this basis.} Such a matrix is of necessity non-diagonal, if a continuous spectrum is present, but it nonetheless
contains \emph{all} the physics of the system. 
\item Equally importantly, the mathematical expression of the theory correctly identifies its meaningful physical content. The states that are physically
accessible to an electron are the ones with a normalizable (i.e., square-integrable) wave function, corresponding to a ray (unique direction) in Hilbert space.
Electrons are never found localized to exactly a single point (with a wave function proportional to a delta function), or infinitely extended through an infinite
volume universe, with exactly well-defined momentum (the plane waves of collision theory). These are idealizations which, it should not be surprising, are
not accommodated in the physical Hilbert space, where every vector has finite norm. They may be convenient idealizations for certain calculations, but the 
states corresponding to real particles always have some dispersion in position and momentum: they are wave packets of ultimately finite extent, when viewed
in coordinate space. Thus, the importance of von Neumann's projection operators, which are bounded and  always produce states of finite norm; they also correspond to the 
fact that measurements are never infinitely precise, but determine the desired quantities up to finite error.\footnote{One may for convenience introduce a broader
formalism (e.g., rigged Hilbert space) to accommodate continuum eigenstates, but the $L^2$ space of von Neumann still remains the only physically meaningful 
setting for all processes in the quantum world.} Moreover, by phrasing the empirical content of the theory in terms of projection operators rather than states,
the (for von Neumann) annoying phase ambiguity in the state vectors disappears: the arbitrary phase $\vartheta$ in a state vector $e^{i\vartheta}|\varphi\rangle$ (helping
ourself to Dirac notation) disappears once we construct the projection operator onto that state: $P_{\varphi}=e^{i\vartheta}|\varphi\rangle\cdot e^{-i\vartheta}\langle\varphi|
=|\varphi\rangle\langle\varphi|$.
\end{enumerate}

\section{Paper 2: Probability-theoretic Construction of Quantum Mechanics}

 In the second paper of the trilogy presented here, von Neumann \autocite*{VonNeumann:1927b} sets out
 to provide an inductive construction of the kinematic structure of quantum mechanics,
starting from some foundational assumptions of the simplest possible character.
``Kinematic'', because the paper is not concerned with the dynamics of a specific physical system (specifying the degrees of freedom, Hamiltonian, time development 
of states, etc.); instead, the focus is on the statistical properties of the theory, and how these properties constrain the mathematical framing of the theory. What was
lacking in the previous deductive approaches (calculation of the expansion coefficients of the wave function from the Schr\"odinger 
equation, followed by a ``dogmatic'' identification of absolute squares thereof with probabilities), ``a systematic derivation of quantum mechanics starting from empirical facts, or
 probability-theoretical axioms'', would be provided by this paper. Von Neumann is at pains to emphasize his   confidence in the absolute validity of
conventional probability theory---a confidence inspired by his absorption of the arguments of Heisenberg's uncertainty paper \autocite{Heisenberg:1927b}---specifically,
 the validity of \emph{both}
the addition and multiplication laws for probabilities of mutually exclusive or independent events. Jordan's assertion that the laws applied for probability amplitudes,
and not their squares (the probabilities themselves), would be shown to arise from an incomplete analysis of the connection between the processes of measurement
and the underlying quantum mechanical structures. We provide here a brief summary of the paper.
\begin{description}
\item[Section 2.]  What are the foundational assumptions from which von Neumann would manage to reconstruct, almost magically, the full kinematic structure of an arbitrary quantum
system? They turn out to be of two kinds. First, the notion of expectation value (\emph{of an observable}, \emph{in an ensemble of exemplars of a given physical
system}) is introduced, and certain intuitively plausible formal properties of the expectation value are imposed (axioms \textbf{A} and \textbf{B} below). Second, each 
observable is associated with a linear, symmetric (i.e., self-adjoint) operator, with the spectrum of this operator covering the values of the associated observable
which can be obtained by the measurement of the observable on any given exemplar of the system. The mapping from observables to operators is then subjected
to two requirements, again of an intuitively plausible character (axioms \textbf{C} and \textbf{D} below). The entire statistical/transformation theory of Jordan-Dirac is then
shown to be an inevitable consequence of the stated axioms. In particular, the Born rule emerges in an extremely natural manner.

   The probability theory lying in the background of the entire paper is the frequentist theory championed by Richard von Mises \autocite*{VonMises:1928}, with which
von Neumann was certainly familiar (von Mises was professor of applied mathematics at the University of Berlin during von Neumann's time in Berlin). 
Intriguingly, von Neumann also uses language more reminiscent of the information theory approach that would later emerge
with the work of Shannon in the 1940s. He speaks, for example, of degrees of ``knowledge'' (\emph{Kenntnis}) of an ensemble, corresponding to the measurement
of the expectation values of sets of observables.

  The basic objects studied by von Neumann in this paper are ensembles ${\mathfrak{S}_{1},\mathfrak{S}_{2},\ldots.}$ consisting of the concatenation of (in general,
infinitely many) exemplars of a definite physical system (e.g.,  a hydrogen atom) $\mathfrak{S}$. Each $\mathfrak{S}_{i}$ thus represents the physical system 
$\mathfrak{S}$ in a particular quantum state. There is no obstruction to the same state appearing several times in the ensemble. However, the ensembles 
envisaged are \emph{very large}, in the sense that any randomly selected subensemble will retain the same statistical properties as the original ensemble.
One ensemble plays a particularly central role in the argument, as it can serve as the source of all others, via well-defined ``pruning'' processes. This is the
``elementary disordered'' ensemble, in which ``every imaginable state of the system $\mathfrak{S}$ occurs with equal frequency.'' Such an ensemble corresponds
to the situation in which ``one knows nothing about the system.'' All other ensembles $\mathfrak{S}^{\prime}$ 
can be obtained from the elementary one by a process of pruning. The technical
machinery to accomplish this exactly corresponds to the algebra of projection operators developed in paper I. For example, the process of measuring some kinematic
quantity $\mathfrak{a}$, performed on each of the (infinitely many!) members of the elementary ensemble results in a value in the spectrum of $\mathfrak{a}$: by
discarding all systems except those where this value lies in some interval $I$, we obtain a new ensemble which can be associated in an obvious way with the 
projection operator $E(I)$ (where $E$ is the spectral operator associated with $\mathfrak{a}$). The resulting ensemble is ``sharp'' with respect to the observable
$E(I)$---a measurement of $E(I)$ gives the value 1 on every system in the ensemble.

  Von Neumann defines the knowledge we have about an ensemble
 $\mathfrak{S}^{\prime}={\mathfrak{S}_{1}^{\prime},\mathfrak{S}_{2}^{\prime},\ldots}$
 as the collection
 of all possible expectation values $\mathbf{E}(\mathfrak{a})$
 (i.e., the average value obtained when the quantity $\mathfrak{a}$ is measured on every system $\mathfrak{S}_{i}^{\prime}$ with $i=1,2,\ldots$  in the ensemble). 
This includes the possibility of measuring all powers
of a quantity, and therefore, via knowledge of all the moments, reconstruction of the full distribution of the quantity in the ensemble. 

  The inductive reconstruction of quantum kinematics in this paper is based on the following four axioms, the first two of which constrain the expectation values
$\mathbf{E}(\mathfrak{a})$, while the second two assert  basic properties of the operators associated with each observable. 
\begin{description}
\item[A.] If $\mathfrak{a},\mathfrak{b},\mathfrak{c},\ldots$ form a finite or infinite set of quantities, and $\alpha, \beta, \gamma,\ldots$ are real numbers (with convergent sum), then
\begin{equation}
 \mathbf{E}(\alpha\mathfrak{a}+\beta\mathfrak{b}+\gamma\mathfrak{c}+\ldots) = \alpha \mathbf{E} (\mathfrak{a})+\beta \mathbf{E} (\mathfrak{b})+\gamma \mathbf{E} (\mathfrak{c})+\ldots
\end{equation}
\item[B.] If $\mathfrak{a}$ is an intrinsically non-negative quantity, then 
\begin{equation}
 \mathbf{E}(\mathfrak{a}) \geq 0.
   \end{equation}
\item[C.] Let $S,T,\ldots$ (finite or infinite in number) be normal operators,\footnote{A normal operator is one that commutes with its adjoint. This class, which is the most
expansive for which the spectral theory of von Neumann applies, includes self-adjoint and unitary operators.} and $\alpha S+\beta T+\ldots$ likewise normal. If $S,T,\ldots$
are associated to the quantities $\mathfrak{a},\mathfrak{b},\ldots$, then $\alpha S+\beta T+\ldots$ is associated to $\alpha\mathfrak{a}+\beta\mathfrak{b}+\ldots$
\item[D.] Let $S$ be a normal operator, $f(x)$ a real-valued function defined for all real $x$, and let $S$ correspond to the quantity $\mathfrak{a}$. Then $f(S)$
corresponds to the quantity $f(\mathfrak{a})$.
\end{description}

   These innocent assumptions, almost self-evident in appearance, will quite remarkably allow von Neumann to reconstruct the entire measure-theoretic structure of
quantum mechanics. In particular, the Born rule will emerge as just one very special case of the general theory.
\item[Section 3.] Working in the discrete realization of Hilbert space (von Neumann's $\mathfrak{H}_{0}$, modern $l^2$, the space of square-summable sequences),
von Neumann shows that the complete statistical information contained in an ensemble of states of an arbitrary quantum system can be encapsulated in a 
single matrix $U$. The operator represented by this matrix is symmetric (i.e., self-adjoint) and positive (i.e., no negative eigenvalues).\footnote{This is the famous ``density operator'', or ``density matrix'' at the center of modern quantum statistical mechanics.} For \emph{any} observable
represented by an operator $S$, the mean value for the observable when measured on all elements of the ensemble---in other words, the expectation value of the
observable in the ensemble---is given by\footnote{For a detailed reconstruction of von Neumann's derivation of this result, see Duncan and Janssen \autocite*[pp.\ 248--251]{DuncanJanssen:2013} or Duncan and Janssen \autocite*[Vol.\ 2, Sec.\ 17.2]{DuncanJanssen:2019-2023}.}
\begin{equation}
   \mathbf{E}(S) = \mathrm{Tr}(SU).
\end{equation}
\item[Section 4.] Moreover, if the ensemble is ``pure'', or ``uniform'' (von Neumann uses both denotations), with the ensemble consisting of 
identical systems,  there exists a unit vector $\varphi$ in the Hilbert space
such that $U=P_{\varphi}$, the projection operator onto that state vector. For such a pure ensemble, the expectation value of any observable $S$ is just
$Q(\varphi,S\varphi)$ (in modern notation: $\langle\varphi|S|\varphi\rangle$). This is essentially the content of the ``Born rule.''
\item[Section 5.] In this section von Neumann explores the formal implications of simultaneous measurability of several quantities $\mathfrak{a}_{1},\ldots,\mathfrak{a}_{m}$
(associated with operators $S_1,\ldots,S_{m}$). The normal interpretation of this property implies the commutativity of the operators $S_{i}$, but von Neumann
insists on more: the mutual commutativity of the spectral projectors $E_{\mu}(\lambda)$ associated with $S_{\mu}$ (where $\mu=1,2,\ldots,m$). The proof relies on the 
existence of a single observable, with operator $S$, from which the values of all the  $S_{\mu}$ follow. This can be shown explicitly if the $\mathfrak{a}_{\mu}$
have purely discrete spectra \autocite[see][p. 251]{DuncanJanssen:2013}, but is not at all clear when we are dealing with operators---such as the various
components of coordinates or momenta in a multi-dimensional problem---with a partly or completely continous spectrum. Of course, it is perfectly straightforward
to show by explicit construction that the spectral projectors for the various components of the momentum vector (say) commute: this is self evident from the
discussion of the spectral theorem for these operators by von Neumann \autocite*{VonNeumann:1927a}. The commutativity of the spectral projectors (which is named
``complete commutativity''),
\begin{equation}
     [E_{i}(\lambda), E_{j}(\lambda^{\prime}] = 0,\quad(i,j=1,2,\ldots,m),
\end{equation}
is therefore the appropriate expression, according to von Neumann, of the simultaneous measurability of the corresponding quantities. In fact, this is both
mathematically and physically sensible. For unbounded operators, the products $S_{i}S_{j}$ may encounter domain difficulties ($S_{j}$ may produce a vector
not in the domain of $S_{i}$), while the spectral projectors $E_{i}$ and $E_{j}$ are bounded and everywhere defied, so the calculation of products and commutators 
is troublefree. And physically, measurements in the real worlds only establish the value of measured quantities within finite bounds determined by the
limitations of the experimental setup: formally this corresponds to measurement of the projection operator $E_{i}(I)$ (with $I$ an interval).
\item[Section 6.] The exploration of the properties of the statistical operator $U$ associated with an ensemble ${\mathfrak{S}_{1},\mathfrak{S}_{2},\ldots}$
continues, in particular, the properties of ensembles specified by the simultaneous measurement of $m$ compatible (commuting) quantities
$\mathfrak{a}_{1},\mathfrak{a}_{2},\ldots,\mathfrak{a}_{m}$. If, for example, it is known that in every system in the ensemble, a measurement
of $\mathfrak{a}_{1}$ yields a value in the interval $I_1$, of $\mathfrak{a}_{2}$ yields a value in the interval $I_2$, etc., then, von Neumann shows,
the statistical operator $U$ of the ensemble is constrained to commute with the projection operator
\begin{equation}
\label{Eproj}
  E = E_{1}(I_1)E_{2}(I_2)\cdots E_{m}(I_m).
\end{equation}
In certain cases the simultaneous projections may result in $E$ projecting onto a one dimensional subspace, spanned by a vector $\varphi$. In
that case we have simply $E=U=P_{\varphi}$, and the ensemble is a pure one in which every system is in exactly the state $\varphi$.
\item[Section 7.] The arguments of section 6 are extended to provide a complete specification of the density operator $U$ of an ensemble of which it is known
that each of $m$ compatible observables lie in specified intervals $I_1,I_2,\ldots,I_m$ for each system in the ensemble. The previous arguments had
only established $UE=EU=U$, with $E$ given in Eq.\ (\ref{Eproj}). Now, von Neumann is able to show that, up to an unimportant overall real positive
normalization, $U=E$.
\item[Section 8.]  The rigorous and fully general extension of Jordan's specification of conditional probabilities in quantum mechanics is derived: the result
turns out to be a remarkably simple formula involving the trace of products of spectral projection operators. Specifically, the conditional probability that
an ensemble corresponding to the projector $E$ of Eq.\ (\ref{Eproj}) (i.e., $\mathfrak{a}_1$ known to be in interval $I_1$ etc.) will, if another
set of observables $\mathfrak{b}_1,\ldots$ (spectral projectors $F_1,F_2,\ldots$) is measured, be found to  have these observables in intervals $J_1,J_2,\ldots$ is given by
$\mathrm{Tr}(EF)$, where $F=F_{1}(J_1)F_{2}(J_{2})\ldots$
   In this section von Neumann points out an extremely important feature of the statistical operator $U$ of a ensemble: it does not, in general, uniquely
specify the actual individual states appearing in the ensemble. Different mixtures of different states can combine to give an ensemble with exactly the
same statistical operator $U$---an ensemble, in other words, statistically indistinguishable with respect to arbitrary measurements, from the first. This
simple observation would turn out to have striking implications for quantum measurement theory. This ambiguity is only avoided in the case that the
ensemble is pure: every system in the ensemble in the same state $\varphi$.
\item[Section 9] Conclusions: (a) Measurements inevitably disturb the measured object; in the case of compatible measurements, simultaneous measurement
is equivalent to a single measurement, (b) the immediate repetition of a measurement will given the same result, and (c) physical quantities are to
be represented by functional operators (i.e., linear operators in a Hilbert space). Further, the a-causality inherent in quantum theory is not to be attributed
to quantum dynamics, which is fully causal, but rather to limitations in the possibility of simultaneous measurement of incompatible observables:
if an experiment is performed to ascertain the value of an observable in an ``incompatible'' state (i.e., one which is not an eigenvector of the
associated operator) the preexistent state is ``destroyed'' (\emph{zertr\"ummert}), and a new one produced in an unpredictable way, governed only
by statistical laws. This is the first clear expression of the notion of ``collapse of the wave function'' which would emerge as a central component of
the Copenhagen school of quantum measurement theory.
\end{description}

\section{Paper 3: The thermodynamics of quantum mechanical ensembles}

The main objective of the third paper \autocite{VonNeumann:1927c}  is the derivation of  a general formula for the entropy $S$ of an ensemble of $N$ quantum mechanical systems, the statistical properties
 of which are specified by a statistical operator $U$, as defined and studied in the second paper of the 1927 trilogy. The resulting formula for the ``von Neumann
 entropy'' plays a fundamental role in quantum statistical mechanics:
 \begin{equation}
 \label{vonNeumannS}
     S = -Nk\;\mathrm{Tr}(U\ln{(U)}).
     \end{equation}
     The remarkable thing about von Neumann's derivation of this formula is that it is shown to be a direct consequence of the volume dependence of the entropy
     of a \emph{classical ideal gas}! Most of the paper is devoted to establishing some ancillary results extending the classical properties of semipermeable membranes
     to the quantum case. The argument leading to Eq.\ (\ref{vonNeumannS}), when it appears, is very short and appears almost miraculous.
     
     Here is a brief summary of the paper:
         \begin{description}
\item[Section 1.] In the introduction, the essential results  of von Neumann \autocite*{VonNeumann:1927b} are summarized: the notion of an ensemble of quantum systems $\mathfrak{S}$,
     with an associated statistical operator $U$ such that the expectation value of an arbitrary observable associated to operator $R$ is given by $E(R)=\mathrm{Tr}(UR)$.
     Two special ensembles (in a sense, extreme opposites of each other) are identified: (a) pure ensembles, in which all systems are in the same quantum state $\varphi$,
     with statistical operator $U=P_{\varphi}$, and (b) the ``elementary random'' ensemble in which all possible states with equal a-priori weight appear with the same
     statistical weight, corresponding to statistical operator $U=1$. For purely mathematical questions, the reader is referred to Paper 1 \autocite{VonNeumann:1927a}.
\item[Section 2.] The central thermodynamics concepts---heat ($Q$), temperature ($T$), and entropy ($S$)---are introduced, now in the context of ensembles of quantum systems.
     Two ensembles $\mathfrak{S}_{1}$ and $\mathfrak{S}_{2}$ are defined to have a difference of entropy $\Delta S= -Q/T$ if a reversible transformation of $\mathfrak{S}_{1}$
     into $\mathfrak{S}_{2}$ can be accomplished via a transfer of heat $Q$ to an ambient (infinite) heat reservoir at temperature $T$. The additive ambiguity in the entropy can be removed by defining
     the entropy of all pure ensembles to be zero, whence the principal result of the paper for the entropy of an ensemble with statistical operator $U$ (stated here without proof), Eq.\ (\ref{vonNeumannS}).
     The stratagem to be employed derives from a paper by Einstein \autocite*{Einstein:1914b}, in which individual quantum-mechanical systems are enclosed in ``containers'', which are then given the dynamics of
     a \emph{classical} ideal gas.
\item[Section 3.] Assuming the basic result for quantum entropy, Eq.\ (\ref{vonNeumannS}), von Neumann shows that the ensembles with maximum entropy for a given number of systems $N$ and
     prescribed energy $E$ (thus, with $\mathrm{Tr}(U)=1$ and $\mathrm{Tr}(UH)=E$) are those satisfying the Boltzmann distribution, i.e.
     \begin{equation}
     \label{Uform}
      U=\alpha \, e^{\beta H}, 
      \end{equation}
      with $\beta=-1/kT$, where $k$ is Boltzmann's constant.
     The usual (canonical-ensemble) thermodynamics formulas for $E$ and $S$ (as partition sums) are then derived.
\item[Section 4.] In this short transitional section, the normalization constant $\alpha$ in Eq.\ (\ref{Uform}) is set to unity, and the final result obtained: $U=e^{-H/kT}$. This is seen
     to tend to $U=1$ in the limit of infinite temperature, as expected for the elementary random ensemble where all states (of equal a-priori weight) are equally represented in the 
     ensemble.
\item[Section 5.] The quantum mechanical analog of the semipermeable wall of classical thermodynamics is introduced. One imagines a partition with ``doors'' at which one stations
     measurement 
     devices for some quantity $\mathfrak{a}$, associated with some measurable property of the quantum systems contained in each of the ``containers/molecules'' constituting
     the ideal gas of Section 2. If the orthogonal eigenstates of $\mathfrak{a}$ are segregated into two non-overlapping sets $\varphi_{\mu}$ and $\psi_{\nu}$, then a molecule arriving at
     one of the doors and subjected to a measurement of $\mathfrak{a}$ will be allowed to pass through (resp.\ reflected back from) the door according to the measured state being 
     found to be a $\varphi_{\mu}$ (resp.\ $\psi_{\nu}$).
\item[Section 6.] Here, von Neumann derives another important preliminary result: the existence of a reversible quasi-static transformation of one pure ensemble $\mathfrak{S}_{1}$ (with
     statistical operator $P_{\varphi}$) into another pure ensemble $\mathfrak{S}_{2}$ (statistical operator $P_{\psi}$). This transformation is shown to involve (in general) performance of work,
     but no exchange of heat, and is hence isentropic.
\item[Section 7.] Here, the converse of the result of section 5 is demonstrated: if a semipermeable wall perfectly transmits (i.e., allows to pass through) systems in a state $\varphi$, but perfectly
     reflects systems in a state $\psi$, the $\varphi$ and $\psi$ must be orthogonal. Here the basic result (not yet proven) Eq.\ (\ref{vonNeumannS}) is assumed.
\item[Section 8.] The central result of the paper, the von Neumann entropy formula Eq.\ (\ref{vonNeumannS}) is derived, using an argument which can only be described as a remarkable feat of
     mathematical legerdemain. The net effect is to transform the entropy formula for an ideal classical gas (more exactly, the volume-dependent part of this formula) into the formula for
     the entropy of an arbitrary ensemble of $N$ states of a given quantum system.  What von Neumann actually calculates is the change in entropy $S$ entailed by the reversible transformation of 
     a pure ensemble $\mathfrak{S}_{1}=\{\varphi,\varphi,\varphi,\ldots.\}$ into a general mixed ensemble $\mathfrak{S}_{2}$ in which orthogonal states $\psi_{1},\psi_{2},\psi_{3},\ldots, \psi_{N}$
     occur with weights $w_{1},w_{2},w_{3},\ldots, w_{N}$, where $w_{1}+w_{2}+w_{3}+ \ldots + w_{N}=1$. Thus, the statistical operator for the ensemble is
     \begin{equation}
          U = \sum_{\mu} w_{\mu}P_{\psi_{\mu}},
          \end{equation}
          and
          \begin{equation}
          \label{Utrace}
          \mathrm{Tr}(U\ln{U}) = \sum_{\mu}w_{\mu}\ln{w_{\mu}}.
          \end{equation}
     
         The argument is simplicity itself. Recall that we are employing the Einsteinian artifice of enclosing each quantum state in the ensemble in a separate massive box. The collection of these boxes/molecules are
         then supposed to form an ideal classical gas at some temperature $T$, bouncing around in a containing volume  $V$. The entropy of the entire system is thus the sum of the quantum entropy of
         the systems enclosed by the boxes, plus the classical formula for the entropy of an ideal gas. One starts (in the pure ensemble) by inserting walls randomly closing off fractions $w_{1}V$, $w_{2}V,\ldots$
         of the entire volume, thereby enclosing $w_{1}N$, $w_{2}N$,\ldots molecules in each subvolume. The systems (all $\varphi$) in the first box can be isentropically transformed into all $\psi_{1}$, those
         in the second to all $\psi_{2}$, and so on. The first enclosure is then expanded isothermally (this affects only the ``molecules'', not the quantum systems contained in them) from volume $w_{1}V$ back
         to $V$, entailing an increase in entropy $(w_{1}N)k\ln{\{V/(w_{1}V)\}}=-Nkw_{1}\ln{w_{1}}$, and so on for all the other enclosures. Finally, all the enclosures (once again of volume $V$) are isentropically
         merged by clever use of semipermeable walls of the type explained in previous sections, yielding the ensemble $\mathfrak{S}_{2}$. The only change in entropy occurs in the isothermal expansion stage.
         The net result, with the help of Eq.\ (\ref{Utrace}), is the desired entropy formula Eq.\ (\ref{vonNeumannS}).
\item[Section 9.] As already demonstrated in section 6, all pure ensembles have the same entropy, which can conveniently be normalized to zero. As Eq.\ (\ref{Utrace}) displays the change in entropy required in going
         from a pure ensemble $\mathfrak{S}_{1}$ to a general mixed ensemble $\mathfrak{S}_{2}$, it must give directly the entropy of the latter. This is the famous von Neumann entropy formula. With this, the result previously
         announced in section 2 is therefore established.
\item[Section 10.] The Boltzmann distribution for general quantum ensembles is now derived. The problem amounts to finding the particular statistical operator $U$ (normalized to $\mathrm{Tr}(U)=1$) which maximizes the
         entropy Eq.\ (\ref{vonNeumannS}), subject to the constraint that the expectation value of the energy $E$, given by $\mathrm{Tr}(UH)$, is held fixed. The result, after some simple algebra, is, as expected $U=\alpha\exp{(\beta H)}$,
         with $\alpha=1/\mathrm{Tr}(\exp{(\beta H)})$. The interpretation of the Lagrange multiplier $\beta$ requires thermodynamic reasoning.
\item[Section 11.] To fix the interpretation of $\beta$ one calculates the derivative of energy with respect to entropy, finding $\beta = -1/kT$. The thermodynamics of the ensemble is thus fully determined. To conclude the
         paper, von Neumann points out that the results obtained depend on the underlying quantum system having a completely discrete spectrum, with eigenvalues tending to infinity (if they are bounded there must be
         an accumulation point, and the partition sums would diverge). This is guaranteed if the system is enclosed in a finite volume, which is a perfectly natural restriction for thermodynamic systems.
         \end{description}
          The concept of von Neumann entropy would be further deepened by Claude Shannon in his seminal work on information and communication theory in 1948. The density operator was also introduced independently and
     at about the same time as von Neumann by Lev Landau \autocite*{landau:1927}, but in a much more limited context (radiative damping in wave mechanics).
     
      \section{A note on the translation and annotations}
    A literal word-for-word translation of academic scientific German from the 1920s will inevitably appear extremely stilted, and sometimes confusing, when rendered in modern English. I have
    therefore allowed myself a certain degree of freedom in rephrasing von Neumann's original words, while, of course, trying to maintain as far as possible the original sense.  Von Neumann's original footnotes (which  begin from 1 on each new page) have been reproduced as the numerically ordered footnotes in the translation, which (as conventional nowadays) do not restart on each page. Short comments of
    the annotator are enclosed inline in square brackets. More extended annotations and commentary are relegated to alphabetically labelled footnotes (a, b, c, etc.).   

\setcounter{footnoteA}{0}
\setcounter{footnoteB}{0} 
\newpage

\chapter{Paper 1: Mathematical Foundation of Quantum Mechanics}

\section*{J. v. Neumann}

\section*{Submitted by M. Born in the session of May 20, 1927}

\section{Introduction}

 The formulations of quantum mechanics given by Heisenberg, Dirac, Born, Schr\"odinger and Jordan
 \footnoteA{See various works of the given authors in 1926/27 in \emph{Annalen der Physik,
 Zeitschrift f\"ur Physik, Proceedings of the Royal Society}.}
 \footnoteB{Most importantly, \autocite{Heisenberg:1925c},\autocite{BornJordan:1925b},\autocite{BornHeisenbergJordan:1926},
\autocite{Dirac:1926a},\autocite{Dirac:1926b},\autocite{Dirac:1926c},\autocite{Schroedinger:1926c},\autocite{Schroedinger:1926d},
\autocite{Schroedinger:1926e},\autocite{Schroedinger:1926f},\autocite{Schroedinger:1926h},\autocite{Born:1926e},\autocite{Born:1927b},\autocite{Jordan:1927b},\autocite{Jordan:1927c}.}
  have
introduced many quite novel concepts and questions, among which we would like to emphasize the following:\\
\\
 $\alpha.$ It has become apparent that the behavior of an atomic system is somehow connected with a
certain eigenvalue problem (the formulation of which will be treated later, in Sec.\ 12), in particular,
the values of the characteristic quantities describing the system are the eigenvalues themselves.\\
 $\beta.$ In this fashion one has achieved in a satisfying way the long awaited amalgamation of continuous
(classical-mechanical) and discontinuous (quantized) phenomena in the world of atoms: for an eigenvalue spectrum
can indeed just as easily possess continuous as well as discontinuous parts.\footnoteB{The eigenvalue problem as such, in the context of infinite bilinear forms (rather than matrices), and for
linear integral equations (rather than differential equations), was treated exhaustively by Hilbert in a series of papers from 1904-1910; the resulting theory (presented in full in \autocite{Hilbert:1912}), which was restricted to 
bounded forms---and a further subcategory of these, the completely continuous (now, \emph{compact}) forms---also showed the way to deal rigorously with situations in which both
a point and continuous spectrum arose. The  use by Hilbert of the term ``spectrum'' to delineate the set of eigenvalues seems in hindsight remarkably prescient.}\\
 $\gamma.$ Moreover, indications have emerged in the new quantum mechanics that the laws of Nature (or at least
the quantum laws known to us) do not determine atomic processes in an unambiguously causal way, but rather that
the elementary laws merely provide probability distributions, which only degenerate in exceptional cases to
causally sharp ones.\\
 $\delta.$ The eigenvalue problem appears in various manifestations: as the problem of determining eigenvalues of
an infinite matrix (equivalently, the transformation of the matrix to diagonal form), or as the eigenvalue problem
of a differential equation. Nonetheless, the two formulations are equivalent to one another, as the matrix (viewed
as a linear transformation) arises from the differential operator (which gives the left side of
the differential equation\footnoteA{The typical (Schr\"odinger) differential equation of quantum mechanics has the form $H\psi = \lambda\psi$,
where $H$ is a differential operator, $\lambda$ is the eigenvalue parameter, and $\psi$ is the
``wave function'', of which certain regularity and boundary conditions are required, leading to
the eigenvalue problem.} when applied to the ``wave function''), once one goes over from the ``wave function'' to
its expansion coefficients in a complete orthogonal set.\footnoteA{Schr\"odinger, \emph{Annalen der Physik}, Vol.\ 79/8, p.\ 734 (1926).[Schr\"odinger's ``equivalence'' paper.]} (The matrix then provides the corresponding [to the
action of the differential operator] transformation
of the expansion coefficients).\footnoteB{A rigorous proof of the equivalence of wave and matrix mechanics will, as von Neumann shows in Section 4, rely on theorems of Riesz and Fischer going back to the first
decade of the century (note 15).}\\
 $\varepsilon.$ Both methods have their problems. In the matrix approach one has almost always to deal with an unsolvable
 problem: the transformation of the energy matrix to diagonal form. This is indeed only possible when there is no continuous
 spectrum\footnoteA{Hellinger, inaugural dissertation, Section 4 (G\"ottingen, 1907). It is immediately apparent that this
 condition is necessary: the spectrum is invariant under a transformation, and a diagonal matrix has only a discontinuous spectrum (with the
 diagonal elements as the eigenvalues).} in which case the treatment of the problem is uniform (even though in the opposite sense [i.e., discrete rather than continuous] of classical mechanics):
 only discontinuous (quantized) quantities appear therein. (The hydrogen atom---which also contains a continuous spectrum\footnoteA{Schr\"odinger, \emph{Annalen der Physik},
 Vol.\ 79/4, p.\ 367 (1926).[Schr\"odinger's first paper on wave mechanics.]}---cannot
 therefore be correctly treated [by matrix methods]). One can of course help oneself to the use of ``continuous matrices'',\footnoteA{cf.\ primarily Dirac, \emph{Proceedings of the Royal Society}, Vol.\ 113, p.\ 621 (1927).}
 but this procedure (which is actually a simultaneous operation with matrices and integral kernels) can only be carried through
 with great difficulty in a mathematically rigorous fashion, as one would then have to introduce concepts such as infinitely
 large matrix elements or infinitely close diagonal elements.\footnoteB{See note 24.}\\
 $\zeta.$ Initially the probability postulates of the matrix method were not available in the differential equation approach.
 (These will be discussed later in detail). This lacuna was repaired first by Born and then later by Pauli and Jordan; however,
 the complete procedure, as it was built up by Jordan to a closed system,\footnoteA{\emph{Zeitschrift f\"ur Physik}, Vol.\ 40, 11/12, p.\ 809 (1927).
 cf.\ also a paper shortly to appear in \emph{Mathematische Annalen} of Hilbert, Nordheim, and the present author.[\autocite{HilbertvonNeumannNordheim:1928}]} is subject to difficult mathematical problems.In
 particular one cannot avoid the admission of so called ``improper eigenfunctions'', for example, the [delta] function $\delta(x)$ employed
 first by Dirac, which is supposed to have the following (absurd) properties:
 \begin{eqnarray}
 \delta(x) &=& 0\;\; \mathrm{for\;\;} x \neq 0, \nonumber \\
 \int_{-\infty}^{+\infty}\delta(x)dx &=& 1.
 \end{eqnarray}
 A particular difficulty in Jordan's approach is the need to calculate not only the transformation operators (whose integral
 kernels are ``probability amplitudes''), but also the variable region [range] onto which the operators map, i.e., the
 eigenvalue spectrum.\\
 $\vartheta.$ A common deficiency of all of these methods, however, is that they introduce elements into the calculation that
 are in principle unobservable and without physical content: eigenfunctions must be calculated, which as a consequence of
 their normalization remain undetermined up to a constant of absolute value 1 (the ``phase'' $e^{i\varphi}$).\footnoteB{The unavoidable presence of
 complex phases, some, but not all, of which were physically irrelevant, was a subject of confusion in the early days of matrix mechanics: see, for
 example Heisenberg \autocite*[p.\ 883]{Heisenberg:1925c} and Born and Jordan \autocite*[p.\ 875]{BornJordan:1925b}.} Indeed, in the
 case of a $k$-fold degeneracy (i.e., an eigenvalue appearing $k$ times) they remain undetermined up to $k$-dimensional
 orthogonal transformation.\footnoteA{Here we mean the complex orthogonality of the transformation  matrix $\{\alpha_{\mu\nu}\}$:
 \begin{eqnarray}
 \sum_{\rho=1}^{k}\alpha_{\mu\rho}\alpha^{*}_{\nu\rho} &=& 1, \;\mathrm{for}\;\mu=\nu, \nonumber \\
 &=& 0, \;\mathrm{for}\; \mu\neq\nu,
 \end{eqnarray}
  which leaves the ``hermitian identity form'' $\sum_{\rho=1}^{k} x_{\rho}y^{*}_{\rho}$ invariant 
  [von Neumann uses a bar (e.g., $\overline{y_\rho}$) instead of a star (e.g., $y^*_\rho$) to indicate complex conjugation]. 
  In the mathematical literature the usual term for this is ``unitary.''
 }
  The probabilities appearing at the end of the calculation are indeed invariant,
 but it is unsatisfying and unclear why this detour through  unobservable and non-invariant quantities is necessary.
 
  In the present work we shall present an approach which remedies these defects, and which, we believe, summarizes the statistical
 interpretation of quantum mechanics as it now stands in a unified and consistent way.

  A detailed investigation of purely mathematical questions has been avoided throughout, wherever calculational aspects are
  not essential; nonetheless, our exposition should be regarded as mathematically rigorous in all essential respects.
  It could not be avoided devoting the greater part of the work to developing and explaining formal concepts prior to
  showing their application. Accordingly, sections 2-11 have a preparatory character, while the actual theory is presented
  in sections 12-14.

  \section{Hilbert space [motivation]}
   As we saw  in section 1.$\delta$, eigenvalue problems in quantum mechanics appear in two main forms: as eigenvalue problems of
   infinite matrices (or, what amounts to the same thing, bilinear forms), and as eigenvalue problems of differential equations.

   We would like to first examine each of these problems on its own, and then emphasize the common features. Throughout we will
   have to pay attention to long established mathematical facts, which are also not physically novel, inasmuch as they have
   already in essence appeared in the work of Schr\"odinger cited above (note 3) [the ``equivalence'' paper] and in numerous works of Dirac.\footnoteB{Dirac \autocite*{Dirac:1926a,Dirac:1926b,Dirac:1926c,Dirac:1926d,Dirac:1927a}.} Nonetheless
   it is perhaps appropriate to develop everything in a connected way, and it will also be useful to first present results
   which will motivate the abstract conceptual developments of the following sections.

   Let us first consider the matrix formulation. Here we encounter an infinite matrix representing the energy (how one arrives
   at it will be explained later), and our task is to transform it to diagonal form, as the diagonal elements are just the
   energy levels.\footnoteA{cf.\ for example Born, \emph{Probleme der Atomdynamik}, p.\ 86 (Berlin 1926).} Let us assume that this proceeds without complications---i.e., that only a point spectrum is present (cf.\ 1.$\varepsilon$
   and note 4).

   The energy matrix will be written
   \begin{equation}
       H = \{h_{\mu\nu}\} \quad (\mu,\nu=1,2,\ldots);
       \end{equation}
       it is assumed to be hermitian, i.e.
       \begin{equation}
       h_{\mu\nu} = h^{*}_{\nu\mu}.
       \end{equation}
       One seeks a transformation matrix
       \begin{equation}
       S = \{s_{\mu\nu}\} \quad (\mu,\nu=1,2,\ldots)
       \end{equation}
       with the following properties: it is orthogonal [unitary], i.e.
       \begin{equation}
       \sum_{\rho=1}^{\infty} s_{\mu\rho}s^{*}_{\nu\rho} = 1 \;(\mathrm{for}\;\mu=\nu),\;\;0\; (\mathrm{for}\;\mu\neq\nu),
       \end{equation}
       and $S^{-1}HS$ is in diagonal form.

       We will denote by $W$ the matrix $S^{-1}HS$, and the diagonal elements of this (diagonal) matrix will be denoted $w_{1},w_{2},\ldots$
       Thus we have
       \begin{eqnarray}
        HS &=& SW,\\
        \sum_{\rho=1}^{\infty} h_{\mu\rho}s_{\rho\nu} &=& s_{\mu\nu}w_{\nu}.
        \end{eqnarray}
        In other words, the $\nu$'th column of $S$, ($s_{1\nu},s_{2\nu},\ldots$), is multiplied by $w_{\nu}$ through the application of $H$.
        Each column of $S$ is consequently a solution of the eigenvalue problem, which consists of finding just those sequences $x_1,x_2,\ldots$
        which are transformed by $H$ into a multiple of themselves (by some constant $w$). ($x_1,x_2,\ldots$ is then an eigenvector, the
        proportionality factor $w$ an eigenvalue, excluding naturally the trivial solution $0,0,\ldots$. The eigenvalue associated with
        $s_{1\nu},s_{2\nu},\ldots$ is thus $w_{\nu}$.)\footnoteB{Here, von Neumann is extending the standard procedure for diagonalizing a finite-dimensional
        hermitian matrix to the infinite-dimensional case, under the assumption that the associated linear operator has only a discrete point spectrum. It is
        \emph{not} assumed that the operator is bounded---for example, the harmonic (or anharmonic) oscillator can be handled in this way, despite having an unbounded energy
        spectrum.}

        One can now show that these [sequences] are the only solutions: more exactly, there are no eigenvalues different from the $w_1, w_2, \ldots$,
        and if $w$ is an eigenvalue, the eigenvectors associated with it are linear combinations of all the columns $s_{1\nu},s_{2\nu},\ldots$ for
        which $w_{\nu}=w$ (see Appendix 1).

        Consequently, the determination of the matrix $S$ is essentially accomplished as soon as the eigenvalue problem, as formulated above,
        is completely solved.

        In the formulation in terms of differential equations the situation is even clearer: it is from the outset given in terms of an eigenvalue
        problem. One is given a differential operator (for example, for a rotating rigid body, an oscillator, hydrogen atom, one has, respectively
        \begin{eqnarray}
         &&\frac{1}{\sin{\vartheta}}\frac{\partial}{\partial\vartheta}\left(\sin{\vartheta}\frac{\partial}{\partial\vartheta} \ldots\right) +\frac{1}{\sin^{2}{\vartheta}}\frac{\partial}{\partial\varphi^{2}}\cdots\\[.1cm]
         &&\frac{d}{dq^{2}}-\frac{16\pi^{4}}{h^2}\nu_{0}^{2}q^{2}\cdots\\[.1cm]
         &&\frac{\partial^{2}}{\partial x^2}+\frac{\partial^{2}}{\partial y^2}+\frac{\partial^{2}}{\partial z^2}+\frac{8\pi^{2}me^{2}}{h^{2}}\frac{1}{\sqrt{x^2+y^2+z^2}} \cdots
         \end{eqnarray}
         and one seeks a function $\psi$ (in our examples $\psi$ is a function of $\vartheta,\varphi;q;x,y,z$ respectively) for which
         \begin{equation}
           H\psi = w\psi,
           \end{equation}
            i.e., which is transformed into a multiple of itself by the action of $H$, and for which the eigenvalues $w$ are again, up to a factor of
            $8\pi^{2}m/h^2$, the energy levels. Naturally, $\psi$ must satisfy certain regularity conditions, and not vanish identically.\footnoteA{Schr\"odinger, \emph{Annalen der Physik}, Vol.\ 79/4, p.\ 361, and 79/6, p.\ 489 (1926).}

            What is the common feature of all these cases? Apparently this: in each case one is given a space of certain quantities (namely, the
            manifold of all sequences $x_1,x_2,\ldots$; or that of all functions of two angles $\vartheta,\varphi$, or of a single coordinate $q$, or of
            three coordinates $x,y,z$), and a linear operator $H$ in this space. In each case one seeks all solutions of the eigenvalue problem
            associated with the operator $H$, namely all (real) numbers $w$ for which there exists a non-vanishing element $f$ of this space
            satisfying
            \begin{equation}
            Hf = wf.
            \end{equation}
         These eigenvalues $w$ then represent the energy levels [of the system].

         It is now our task to proceed from a unified formulation to a unique problem. This will be achieved once we show that all spaces
         introduced in this way---in particular, absolutely all the spaces to which one is led in the usual formulations of quantum mechanics
         at the present time---are in essence identical to one another; i.e., that they can all be obtained from a single space (which will be
         described in the subsequent sections) by a mere redefinition.

          To this end we will need however  to specify more exactly which sequences (resp.\ functions) can be accepted in the spaces mentioned above---in
          other words, we will need to indicate the regularity- and boundary-conditions, which turn out to be of decisive importance in the
          eigenvalue problem.

\section{Hilbert space [introduction of $l^2$ and $L^2$]}
 We begin with the space \footnote{Here the term \emph{Mannigfaltigkeit} (rather than \emph{Raum}) is used, as was more commonly the case with sequences.}
 of numerical [complex] sequences $x_{1},x_{2},\ldots.$ It seems reasonable to require the convergence of the squared
 sum $\sum_{n=1}^{\infty}|x_{n}|^{2}$. In fact, this holds for the solutions of the eigenvalue problem, as long as there is only a
 point spectrum, which are just the columns $s_{1\nu}, s_{2\nu},\ldots$ of the matrix $S$ (cf.\ Section 2); indeed, as we required above, one has
 \begin{equation}
 \sum_{\rho=1}^{\infty}s_{\mu\rho}s^{*}_{\mu\rho} = \sum_{\rho=1}^{\infty} |s_{\mu\rho}|^{2} = 1
 \end{equation}
 (If there is a continuous spectrum, it is no longer possible to solve the eigenvalue problem only employing eigen-sequences with finite
 $\sum_{n=1}^{\infty}|x_{n}|^{2}$. We shall see below that our procedure nevertheless will allow a comprehensive treatment of the continuous
 spectrum).\footnoteB{Von Neumann is here emphasizing in fact, not often fully appreciated by physicists, that a complete and rigorous treatment of bounded
 linear operators is available even in the case where a \emph{continuous} spectrum is present, in the framework of the \emph{discrete} space of square-summable
 sequences.}

  Even the simplest linear transformations $H$ (i.e., corresponding to matrices $h_{\mu\nu}, \mu,\nu=1,2,\ldots$) fail us when acting on
  sequences with arbitrary $\sum_{n=1}^{\infty}|x_{n}|^{2}$, inasmuch as the series $\sum_{\nu=1}^{\infty}h_{\mu\nu}x_{\nu}$ are not guaranteed
  to converge. (It will become clear that the matrices of quantum mechanics have the property that the squared row sums $\sum_{\nu=1}^{\infty}|h_{\mu\nu}|^{2}$
  are all finite, in which case the convergence of $\sum_{\nu=1}^{\infty}h_{\mu\nu}x_{\nu}$ follows from the finiteness of $\sum_{n=1}^{\infty}|x_{n}|^{2}$.\footnoteA{The series $\sum_{\nu=1}^{\infty}h_{\mu\nu}x_{\nu}$
  in fact converges absolutely as a consequence of the inequality $|h_{\mu\nu}x_{\nu}|\leq\frac{1}{2}|h_{\mu\nu}|^{2}+\frac{1}{2}|x_{\nu}|^{2}$.})\footnoteB{Quantum mechanics is rife with operators which are \emph{not}
  bounded, e.g., momentum, position, energy, etc. What von Neumann is saying here is that the theory can be formulated in such a way that its statistical phenomenological content can be  completely expressed
  relying only on \emph{bounded} operators---specifically, the bounded projection operators associated with the spectral resolution of all the operators---usually, unbounded---representing physical quantities of interest.
  How this can be done is shown in Sections 12--14 of the paper.}

  Finally, it is precisely this limitation of the range of the $x_1, x_2,\ldots$ [namely, $\sum_{n=1}^{\infty}|x_{n}|^{2}<\infty$] which has proven most suitable
  in the theory of infinite matrices\footnoteA{The theory of infinite matrices (resp.\ bilinear forms) is in essence
  due to Hilbert, based on the complete clarification of the relations (and in particular, on the solution of the eigenvalue problem) between the 
  important classes of the so-called completely continuous and the (originating from these) bounded bilinear forms (cf.\ \emph{G\"ottingen Nachrichten, Mathematische-Physikalische Klasse}, 1906, pp.\ 159-227 
 [\autocite{Hilbert:1906a}]).
A large part of the mathematical confusions and difficulties of quantum mechanics derive from the fact that even the simplest operators (or matrix bilinear forms) appearing therein
  do not belong to the class of bounded operators treated by Hilbert. 
  The present author has recently shown how the eigenvalue problem of arbitrary, including therefore unbounded, operators can be unambiguously solved (to 
  appear shortly in \autocite{VonNeumann:1929a}). In this work we will naturally need to investigate the precise formulation of an eigenvalue problem for
  unbounded operators.}
    (and on the appropriate generalization of these which must serve as the foundation for the mathematical construction
  of quantum mechanics).

  There is therefore a good motivation for distinguishing the following space: 
  all sequences of complex numbers $x_1, x_2,\ldots$ with finite
  $\sum_{n=1}^{\infty}|x_{n}|^{2}$. The name (complex) Hilbert space is usual for this space.\footnoteB{The space of square-summable sequences, now referred to as ``$l^2$'', was first studied in detail
  in a seminal paper by Erhard Schmidt \autocite*{Schmidt:1908}. In this paper he introduces the standard geometrical concepts associated with ``Hilbert spaces'' ( a term never used
  by Hilbert)---e.g., inner product, norm, orthogonal projections, etc. Many of the results obtained earlier by Hilbert for bounded bilinear forms and integral kernels were re-derived in
  a simpler form in this language.}

  At this point, let us move on to the so-called \emph{function spaces} considered in Section 2. Here the situation with regard to supplementary
  conditions (i.e., regularity and boundary conditions) is more complicated. It is commonplace to require double differentiability, as well as
  uniqueness, vanishing at infinity or at the boundary of the domain of definition, and the like. How can we arrive at a unified standpoint in
  such a situation?

  Let the domain of definition of the functions in question (in our examples: the $\vartheta,\varphi$-space, i.e., the sphere; the $q$-space, i.e., the
  real line; the $x,y,z$-space, i.e., the usual [3-]space) be called $\Omega$. Of course only uniquely defined functions in $\Omega$ are to be
  considered, to which the operator $H$ can be applied. Consequently, if $H$ is a differential operator of second order (which is generally the case
  in quantum mechanical problems), double differentiability of $\psi$ is essential.  
  
   However, with the reservation that this need only hold at those points
  of $\Omega$ allowed by the application of $H$---for example, at points where factors in $H$ display singularities ($x=y=z=0$, say, in the
  hydrogen atom [where the Coulomb potential diverges]), this [double-differentiability] is not required.\footnoteA{For example, the ground state solution for the hydrogen atom (in Schr\"odinger's notation
  $n=1,l=0$) $\psi(x,y,z) = \exp{(-\sqrt{x^2+y^2+z^2})}$, has a conical singularity at $x=y=z=0$.}
   Indeed, it is well known that it is
  precisely this imposition of regular behavior at such points that results in the appearance of an eigenvalue problem: still, one can (and must)
  achieve this by the introduction of considerably weaker conditions.

  We will therefore require, that $\int_{\Omega}|\psi|^{2}dv$ (denoting with $dv$ respectively the line, surface, and
  volume element of $\Omega$ [in the examples cited previously]) remain finite: this precludes a too rapid divergence of the singularities
  in $H$, and also guarantees a part of the boundary conditions, namely, the vanishing [of $\psi$] at infinity. Success will show that we have
  made the right choice.

  Certain boundary conditions (e.g., vanishing at the boundary of a finite region [cf.\ particle in a box]) are still not encompassed in the previous
  requirements. In order to clarify the role of these, consider the following: in order that an eigenvalue problem be at all possible, we must
  require that the matrices $H$ possess 
  hermitian symmetry
  \begin{equation}
      H = \{h_{\mu\nu}\},\;\;h_{\mu\nu} = h^{*}_{\nu\mu};
      \end{equation}
      and in the case of differential equations, 
      self-adjointness is essential, i.e., that\footnoteB{The importance of enforcing appropriate boundary conditions in order to 
      obtain the desired self-adjoint property of differential operators was already well-known in the mathematical literature, see, e.g., Hilbert \autocite*[pp.\ 39-41]{Hilbert:1912}.}
      \begin{equation}
      \int_{\Omega}\{\psi_{1}\cdot(H\psi_{2})^{*}-(H\psi_{1})\cdot\psi_{2}^{*}\}dv
      \end{equation}
      vanish for all functions $\psi_{1}, \psi_{2}$ that vanish sufficiently fast at the boundary of their domain of definition. (Equivalently:
      the quantity $\{\psi_{1}\cdot(H\psi_{2})^{*}-(H\psi_{1})\cdot\psi_{2}^{*}\}dv$ should be an [exact] differential of some expression built
      from $\psi_{1}, \psi^{*}_{2}$ and their derivatives.) For this to hold the boundary conditions are in certain circumstances absolutely
      important. For example, let $\Omega$ be the line segment [0,1] and $H$ the operator $i\frac{d}{dx}$. We then have [$dv = dx$]
      \begin{eqnarray}
      \int_{\Omega}\{\psi_{1}\cdot(H\psi_{2})^{*}-(H\psi_{1})\cdot\psi_{2}^{*}\}dv &=& \int_{0}^{1}\{\psi_{1}(x)\cdot[-i(\psi_{2}^{\prime}(x))^{*}]
      -i\psi_{1}^{\prime}\cdot\psi_{2}^{*}(x)\}dv \nonumber \\
      &=& -i\int_{0}^{1}\{\psi_{1}(x)(\psi_{2}^{\prime}(x))^{*}+\psi_{1}^{\prime}(x)\psi_{2}^{*}(x)\}dv \nonumber \\
      &=& -i[\psi_{1}(x)\psi_{2}^{*}(x)]_{0}^{1}.
      \end{eqnarray}
      This certainly vanishes, provided $\psi_{1}$ and $\psi_{2}$ vanish at both end-points of $\Omega$, but, lacking imposition of all boundary conditions,
      could very well be nonzero.

      We therefore further require (and this will prove to contain the last essential boundary conditions) that $\psi$ behaves in such a way as to
      preserve the self-adjointness of $H$.

      Altogether, we therefore require: $\psi$ is uniquely defined in $\Omega$, has the property that $H$ can be applied to it and has the property of
      self-adjointness, and $\int_{\Omega}|\psi|^{2}dv$ is finite. The only nontrivial condition---the one responsible for the ``regularity requirement''
      giving rise to the eigenvalue problem---is, as one sees, the last. (One could raise the objection that this requirement excludes access to
      the continuous spectrum, where the eigenfunctions are never square-integrable. We also encountered an analogous situation in the matrix case. It
      will however become apparent, that precisely this constraint allows a particularly fruitful---and mathematically fully rigorous---conception
      of the continuous spectrum, completely faithful to its physical meaning, without having to admit improper forms as eigenfunctions. Cf.\ Section 10.)

 To summarize: we have a space $\mathfrak{H}$ (in the matrix formulation the space of all sequences $x_1,x_2,\ldots$ with finite $\sum_{n=1}^{\infty}|x_{n}|^{2}$;
 in the differential equation formulation the space of all functions $\psi$ defined in $\Omega$ with finite $\int_{\Omega}|\psi|^{2}dv$, where $\Omega$ can
 be a surface of arbitrary dimension, with or without boundary, and bounded or unbounded) the elements of which we denote $f,g,\ldots$ to maintain a unified
 notation. In addition, a linear operator $H$ is given (providing for each $f$ in $\mathfrak{H}$ another element $Hf$, and for which $H(af)=aHf$ ($a$ complex),
 and $H(f+g)=Hf+Hg$), which is also symmetric (resp.\ self-adjoint). This last requirement can also be formulated in a unified way. For by defining [the
 inner product] $Q(f,g) = \sum_{n=1}^{\infty}x_{n}y^{*}_{n}$ (resp.\ $\int_{\Omega}\varphi\psi^{*}dv$), where $f=x_1,x_2,\ldots, g=y_1,y_2,\ldots$ in the matrix case,
 and $f=\varphi, g=\psi$ in the differential equation case, the symmetry/self-adjointness condition takes the form, for allowed $f,g$,
 \begin{equation}
     Q(f, Hg) = Q(Hf, g)
\end{equation}

 We now seek the solutions of the eigenvalue problem ($Hf=wf, f\neq 0, w$ a real constant) for the operator $H$. In this way we obviously in the first instance only
 obtain the point spectrum, but the general formulation achieved in this fashion will subsequently enable us to get control also of the continuous spectrum.

 It is apparent that these formulations differ only in the underlying space $\mathfrak{H}$. Is it not possible also to interpret these spaces somehow in a unified fashion?

\section{[Critique of Dirac-Jordan transformation theory]}

A suggestive approach would be to exploit the analogous construction of the expressions $\sum_{n=1}^{\infty}|x_{n}|^{2}$ and $\int_{\Omega}|\psi|^{2}dv$ (similarly
$\sum_{n=1}^{\infty}x_{n}y^{*}_{n}$ and $\int_{\Omega}\varphi\psi^{*}dv$) to say: given a space $R$, which can be discrete or continuous (the ``space'' 1,2,\ldots,
or a $\Omega$), we consider all functions on $R$ (all sequences $x_1,x_2,\ldots$, or, all functions on $\Omega$).  There exists an ``integration over $R$''
($\sum_{n=1}^{\infty}x_{n}$, or $\int_{\Omega}\psi dv$), and we admit only such ``proper'' functions into consideration the absolute square of which possesses
a finite ``integral over $R$'' ($\sum_{n=1}^{\infty}|x_{n}|^{2}$ finite [the space $l^2$, in modern notation] or $\int_{\Omega}|\psi|^{2}dv$ finite [$L^{2}[\Omega]$,
in modern notation]. Naturally, in certain cases $R$ might be of mixed type, containing for example both discrete points and line intervals.

In essence this is the approach followed by Dirac with great persistence and unquestioned success in a number of ground breaking works in quantum mechanics.\footnoteA{In several
articles of 1926/27 in the \emph{Proceedings of the Royal Society}.}
If we nevertheless attempt to find a different solution, it is just for the reason that the analogy sketched above necessarily remains a superficial one,
as long as one adheres to normal standards of mathematical rigor.

For example, in the case where $R$ is the ``space'' 1,2,\ldots, so that $\mathfrak{H}$ contains the sequences $x_1,x_2,\ldots$, all linear operators $H$ are
representable by matrices $\{h_{\mu\nu}\}$:
\begin{eqnarray}
  H(x_1,x_2,\ldots) &=& (y_1,y_2,\ldots), \\
  y_{n} &=& \sum_{\nu=1}^{\infty}h_{\mu\nu}x_{\nu}.
  \end{eqnarray}
   If one tries to perform an analogous representation in the case where $\Omega$ is the interval [0,1] (so $\mathfrak{H}$ contains all functions
   defined on this interval), then one must demand that every linear operator $H$ be represented by an integral kernel $\varphi(x,y)$:
   \begin{equation}
     H\varphi(x) = \int_{0}^{1}\varphi(x,y)\varphi(y)dy.
     \end{equation}
     This fails already for the simplest operators (e.g., the ``identity'' operator, which takes every function to itself). If one still wishes to ``feign'' the
     correctness of this false assertion, one is forced, with Dirac, to consider ``improper'' integral kernels [e.g., the Dirac delta ``function''].\footnoteB{Von Neumann
     is at pains throughout his work on quantum mechanics to appeal to functional analytic methods and results which were on a firm mathematical footing in the
     mid-1920s. A rigorous theory of ``distributions'', including  Dirac's $\delta(x)$, would emerge, starting in the 1930s with the work of Sobolev, and reaching 
     maturity in the period 1945-1950 with the work of Laurent Schwartz \autocite*{Schwartz:1945,Schwartz:1950}. Schwartz was initially motivated to extend the solution space of differential equations by allowing arbitrary derivatives of
     continuous functions. A fully developed theory based on continuous linear functionals over suitable topological function spaces was presented in his 1950 treatise.
      As von Neumann shows in the present paper, Hilbert space provides a completely adequate arena to construct a rigorous operator theory
     containing the full statistical content of quantum mechanics---the use of ``improper functions'' (later, distributions), is not necessary.}

     We shall embark on a different route, which at heart lies at the foundation of Schr\"odinger's ``equivalence proof'' of the matrix [e.g., matrix mechanics] and differential
     equation [e.g., wave mechanics] formulations (cf.\ note 3), and can be derived from long known mathematical material. In order to describe this approach, some preparatory
     remarks are required.

     Let $\Omega$ once again be a surface (1,2,\ldots dimensional; with or without boundary; finite or infinite), $dv$ the differential element of integration
     on $\Omega$. We consider all complex-valued functions $f,g,\ldots$ on $\Omega$ with finite $\int_{\Omega}|f|^{2}dv, \ldots$ as constituting the ``space'' $\mathfrak{H}$.
     The integral $\int_{\Omega}fg^{*}dv$ (which provides the only context in which integration over $\Omega$ will appear in the following) will be written
     for brevity $Q(f,g)$ [modern notation: $(g,f)$ or (Dirac) $\langle g|f\rangle$, with the anti-linear element on the left]. For $Q(f,f)$ (the finiteness of which specifies $\mathfrak{H}$!) we write simply $Q(f)$ [modern notation $(f,f)$,
     $\langle f|f\rangle$, or $||f||^{2}$].

     Two functions $f,g$ of $\mathfrak{H}$ are called orthogonal if
     \begin{equation}
        Q(f,g) = 0,
        \end{equation}
     and a system $f_1,f_2,\ldots$ is a normalized orthogonal [orthonormal] system if
     \begin{eqnarray}
         Q(f_{\mu},f_{\nu}) &=& 1,\;\;\mathrm{for\;}\mu=\nu,\nonumber \\
         &=& 0,\;\;\mathrm{for\;}\mu\neq\nu.
         \end{eqnarray}
     Finally, a complete orthonormal system is one to which no further $f$ can be added while preserving the orthonormality conditions.(This is clearly
     equivalent to the statement that there is no function $f$, not identically zero, orthogonal to all $f_{n}$.)

      One can show that there exist complete orthonormal systems in $\mathfrak{H}$ (see Section 5 for a more exact
      discussion of this and all further unproven assertions [in the remainder of this Section]). Let $\varphi_{1},\varphi_{2},\ldots$ be one such.
      If $f$ is a function in $\mathfrak{H}$, we shall call
      \begin{equation}
        c_{\mu} = Q(f,\varphi_{\mu}), \quad (\mu=1,2,\ldots)
        \end{equation}
        the expansion coefficients of $f$ (with regards to $\varphi_{1},\varphi_{2},\ldots$). The sum $\sum_{\mu=1}^{\infty}|c_{\mu}|^{2}$ is always
        finite (indeed, it equals $Q(f)$, which is the so-called Parseval formula), and the series $\sum_{\mu=1}^{\infty}c_{\mu}\varphi_{\mu}$
        converges in a certain sense to $f$. The series in fact need not converge in any particular point of $\Omega$; but when one follows
        the deviation of the $N$'th partial sum of $f$ (namely, $\sum_{n=1}^{N}c_{\mu}\varphi_{\mu}-f$) over all of $\Omega$, and characterizes
        the net deviation through the integral of its absolute square (thus, $Q(\sum_{n=1}^{N}c_{\mu}\varphi_{\mu}-f)$, which, the smaller it is,
        the closer $\sum c_{\mu}\varphi_{\mu}$ approximates $f$ over all of $\Omega$, and not just at special points of $\Omega$), then this integral
        converges for $N\rightarrow\infty$ to 0 (i.e.,
        \begin{equation}
\lim_{N\rightarrow\infty}Q\left(\sum_{n=1}^{N}c_{\mu}\varphi_{\mu}-f\right)=0.
\end{equation}
(This form of
        convergence is denoted ``convergence in the mean''. It is less stringent than point-wise convergence, but is the more appropriate
        concept when dealing with orthogonal expansions.)

        The converse statement is also correct (Fischer-Riesz theorem):\footnoteA{G\"ottingen Nachrichten, Mathematische-Physikalische Klasse, 1907, pp.\ 116--122.}\\ 
        If $c_1,c_2,\ldots$ is an arbitrary sequence of complex numbers with
        finite $\sum_{\mu=1}^{\infty}|c_{\mu}|^{2}$, then (function)-sum $\sum_{\mu=1}^{\infty}c_{\mu}\varphi_{\mu}$ converges in the mean
        (i.e., there exists a $f$ in $\mathfrak{H}$ such that $Q(f-\sum_{\mu=1}^{N}c_{\mu}\varphi_{\mu})\rightarrow 0$ as $N\rightarrow\infty$),
        and the sum $f$ has the expansion coefficients $c_1,c_2,\ldots.$

        We have therefore a one-to-one mapping between functions $f$ in $\mathfrak{H}$ and sequences $c_1,c_2,\ldots$ with finite $\sum_{\mu=1}^{\infty}|c_{\mu}|^2$.
        This mapping is manifestly linear, i.e., if $f$ corresponds to $c_1,c_2, \ldots$, then $af$ ($a$ a complex constant) corresponds to $ac_{1},ac_{2},\ldots$;
        and if $f$ (resp.\ $g$) corresponds to $c_1,c_2, \ldots$ (resp.\ $d_1,d_2,\ldots$), then $f+g$ corresponds to $c_{1}+d_{1},c_{2}+d_{2},\ldots$\footnoteA{From the relation 
        $|u+v|^{2}\leq |u+v|^{2}+|u-v|^{2}=2|u|^{2}+2|v|^{2}$ one sees directly that the finiteness of $\sum_{\mu=1}^{\infty}|c_{\mu}|^{2}, \sum_{\mu=1}^{\infty}|d_{\mu}|^{2}$ (resp.
        $\int_{\Omega}|f|^{2}dv, \int_{\Omega}|g|^{2}dv$) implies the finiteness of $\sum_{\mu=1}^{\infty}|c_{\mu}+d_{\mu}|^{2}$ (resp.\ $\int_{\Omega}|f+g|^{2}dv$).} Moreover
        $Q(f,g)$ is given by $\sum_{\mu=1}^{\infty}c_{\mu}d^{*}_{\mu}$, which is the generalized Parseval formula
        \begin{eqnarray}
          c_{\mu} &=& Q(f,\varphi_{\mu}),\;\;d_{\mu} = Q(g,\varphi_{\mu}),\;\;(\mu=1,2,\ldots) \\
          Q(f,g) &=& \sum_{\mu=1}^{\infty}c_{\mu}d^{*}_{\mu},
          \end{eqnarray}
          to be demonstrated in the next Section.

          Precisely since we denoted $\sum_{\mu=1}^{\infty}x_{\mu}y^{*}_{\mu}$ (in the space of sequences $x_1,x_2,\ldots$) by $Q(x,y)$ ($x$ for $x_1,x_2,\ldots$, $y$
          for $y_1,y_2,\ldots$), we can assert: $\mathfrak{H}$ can be mapped in a one-to-one fashion to the space of all sequences $x_1,x_2,\ldots$ with finite
          $\sum_{\mu=1}^{\infty}|x_{\mu}|^{2}$ in such a way that the operations $af$ ($a$ a complex constant), $f+g$, $Q(f,g)$---i.e., all operations
          previously used in the description of quantum mechanics---are transformed into themselves (i.e., to their analogs in the space of sequences).
          Consequently, all function spaces $\mathfrak{H}$  (whatever the associated $\Omega$) are distinguished from the space of sequences $x_1,x_2,\ldots$ with finite
        $\sum_{\mu=1}^{\infty}|x_{\mu}|^{2}$ (and likewise, with one another) solely by the denotation of elements and operations, while agreeing completely
        with each other in all their properties.

        In other words: even without the introduction of ``continuous matrices'' and ``improper forms'', the various sequence- and function-spaces which underly
        quantum mechanics are in essence, and with absolute mathematical rigor, identical.\footnoteB{This is, in essence, the desired proof of equivalence of
        matrix and wave mechanics, satisfying ``usual levels of mathematical rigor'' which von Neumann found wanting in the Dirac-Jordan approach.}

        \section{[Hilbert space: formal axiomatic definition]}

        The space of all complex valued number sequences $x_1,x_2,\ldots$ with finite $\sum_{\mu=1}^{\infty}|x_{\mu}|^{2}$ will be
        called complex infinite-dimensional Euclidean space or 
        complex Hilbert space; we will denote this space $\mathfrak{H}_{0}$
        [in modern notation $\mathfrak{H}_{0}$ is written $l^2$].

        As we saw in previous sections, all function spaces $\mathfrak{H}$ have formal properties which coincide with each other and with
        the space $\mathfrak{H}_{0}$: the differences lie solely in the notation used for, or the interpretation of, the elements of the space.
        This suggests that we can characterize all these spaces by their common properties, and define any space possessing all the specified
        characteristic properties as an abstract Hilbert space. One can therefore imagine all the particular sequence- and function- spaces
        as arising from the particular notation attached to its elements (by interpreting them as sequences $x_1,x_2, \ldots$ or functions $\psi$
        on a space $\Omega$), rather as one can arrive at a space-time interpretation in relativity theory of a metrically homogeneous 4-dimensional
        ``world'' by choice of a special coordinate system (namely, by specifying world points as number quadruplets).

        We thus have the task of describing abstract Hilbert space on the basis of its ``inner'' properties: i.e., those that can be formulated
        independently of the interpretation of the elements of the space as sequences or functions. This will be accomplished by specifying
        5 properties, from which all others follow, as we shall show. These 5 properties refer to an ``abstract Hilbert space $\bar{\mathfrak{H}}$'',
        with elements $f,g,\ldots,$ in which the operations $af$ ($a$ a complex constant), $f+g$, $Q(f,g)$ are defined ($af, f+g$ are
        again elements of $\bar{\mathfrak{H}}$, while $Q(f,g)$ is a complex number). These operations (in particular $Q(f,g)$) should now however
        be viewed independently of their definition in Section 3: we require of them just the 5 properties to be specified. Of course, these
        properties will be satisfied (cf.\ Appendix 2) if we replace $\bar{\mathfrak{H}}$ with either its ``discrete realization'' $\mathfrak{H}_{0}$ [$l^2$]
        (i.e., space of all sequences $x_1,x_2,\ldots$ with finite $\sum_{n=1}^{\infty}|x_{n}|^{2}$, $Q(x,y) = \sum_{n=1}^{\infty}x_{n}y^{*}_{n}$),
        or with its ``continuous realization'' $\mathfrak{H}$ (space of all functions $\varphi$ defined on $\Omega$ with finite $\int_{\Omega}|\varphi|^{2}dv,
        Q(\varphi,\psi) = \int_{\Omega}\varphi\psi^{*}dv$ [modern $L^2[\Omega]$]). To visualize the concrete meaning of our axioms, one has just to replace the
        abstract $\bar{\mathfrak{H}}$ with one of the special realizations cited above.

        We shall now give the 5 characteristic properties A.--E.\ of abstract (complex) Hilbert space $\bar{\mathfrak{H}}$ mentioned above. In each of them we shall
        immediately derive the simplest consequences; at the end shall follow the proof that these properties fully determine $\bar{\mathfrak{H}}$,
        in the sense that $\bar{\mathfrak{H}}$ can be uniquely constructed---given invariance of the operations $af, f+g, Q(f,g)$---from
        $\mathfrak{H}_{0}$ (the idea of the proof was already sketched in Section 4 for the function-spaces $\mathfrak{H}$). That the spaces
        $\mathfrak{H}_{0},\mathfrak{H}$ can in fact be regarded as $\bar{\mathfrak{H}}$'s (i.e., possess the properties A.--E.)
        will be demonstrated in Appendix 2.

        Our five axioms 
        A.--E.\ for the abstract complex Hilbert space $\bar{\mathfrak{H}}$ read as follows:\footnoteB{Schmidt's \autocite*{Schmidt:1908} treatment of the fundamental Hilbert space $l^2$ (not referred to as such!)
        includes discussion of the properties (not identified as ``axioms'') A.--C. and E. (Cauchy-completeness). The property D. (separability) is not discussed explicitly. Schmidt refers throughout to a ``function space'', i.e., the space of complex
        functions  defined on the integers, thus complex sequences $c_1,c_2,\ldots$ The inner product of two such functions is written $(f;g^{*})$ (instead of von Neumann's $Q(f,g)$), 
        and the norm as $||f||$ (instead of von Neumann's $\sqrt{Q(f)}$)---in both cases, somewhat closer to modern usage!. However, both authors place the anti-linear (complex conjugated) element in the inner product on
        the right, rather than the left, as done at present. All the standard inequalities, including the triangle inequality  (\ref{distineq}),
        Bessel's inequality (Section 6, Theorem 2), etc. are derived by Schmidt. He credits (note 8, op.cit.) Gerhard Kowalewski (1876-1950) for the geometric interpretation of the concepts and theorems discussed: specifically, for the treatment of his ``functions''
        as ``vectors in a space of infinitely many dimensions.''}\\
        \\
        \textbf{A}. $\bar{\mathfrak{H}}$ is a linear space. 
        
        Namely: $\bar{\mathfrak{H}}$ possesses an addition  $f+g$ and a multiplication $af$
        ($f,g$ in $\bar{\mathfrak{H}}$, $a$ a complex number, $f+g, af$ in $\bar{\mathfrak{H}}$). These operations satisfy the known rules
        followed by the analogous operations with vectors (in particular: existence of a [null vector] 0, commutativity and associativity of
        addition, distributivity and associativity of [scalar] multiplication).

        A fundamental concept which can be grounded solely on the basis of \textbf{A}. is that of linear independence: any [set of] elements $f_1,f_2,\ldots,f_k$
        of $\bar{\mathfrak{H}}$ are said to be linearly independent if
        \begin{equation}
          a_{1}f_{1}+a_{2}f_{2}+\cdots+a_{k}f_{k} = 0
          \end{equation}
          implies $a_1 = a_2 = \cdots = a_{k} = 0$. A further concept so grounded is that of a subspace of $\bar{\mathfrak{H}}$ generated by a
          subset $\mathfrak{M}$: this is the set of all $a_{1}f_{1}+a_{2}f_{2}+\cdots+a_{k}f_{k}$, where $a_1,a_2,\ldots a_k$ are arbitrary complex
          numbers, and $f_1,f_2,\ldots,f_k$ are arbitrary members of $\bar{\mathfrak{H}}$.\\
        \\
        \textbf{B}. $\bar{\mathfrak{H}}$ is a metric space, with a metric derived from a bilinear form $Q(f,g)$. Namely: there exists a function
        $Q(f,g)$ (defined for all $f,g$ in $\bar{\mathfrak{H}}$, and complex numbers as values) with the following properties:
        \begin{enumerate}
        \item $Q(af,g) = aQ(f,g)$, ($a$ a complex constant).
        \item $Q(f_{1}+f_{2},g) = Q(f_1,g)+Q(f_2,g)$.
        \item $Q(f,g) = (Q(g,f))^{*}$.\\
        From 1, 2 (given 3) we can conclude:\\
        1$^{\prime}$. $Q(f,ag) = a^{*}Q(f,g)$.\\
        2$^{\prime}$. $Q(f,g_{1}+g_{2}) = Q(f,g_1)+Q(f,g_2)$.\\
        1, 2, 1$^{\prime}$, 2$^{\prime}$ express the bilinear hermiticity of $Q$, 3 the symmetry. From 3 it follows that $Q(f,f)$ is
        always real, but we shall further require (writing again $Q(f)$ for $Q(f,f)$):
        \item $Q(f)\geq 0$, and only -0, if $f=0$. From 1-4 one concludes without difficulty that
        \begin{equation}
         |Q(f,g)| \leq \sqrt{Q(f)Q(g)}
         \end{equation}
         always holds, and that also\footnoteA{One easily calculates (with $a,b$ real constants; Re($z$), Im($z$) the real and imaginary parts of the complex number $z$
         [von Neumann uses the notation $\mathfrak{R}$ and $\mathfrak{I}$ for real and imaginary parts, respectively]) 
         \begin{equation}
         Q(af+bg) =a^{2}Q(f) + ab (Q(f,g)+Q(g,f))+b^{2}Q(g) = a^{2}Q(f)+2ab\mathrm{Re}(Q(f,g))+b^{2}Q(g).
         \end{equation}
         The left side is always $\geq 0$, the right is a quadratic form in $a,b$, so must have its discriminant $\leq 0$, whence
         \begin{equation}
          \mathrm{Re}(Q(f,g))^{2} - Q(f)Q(g) \leq 0,\;\;|\mathrm{Re}(Q(f,g))| \leq \sqrt{Q(f)Q(g)}.
          \end{equation}
          If we now replace $f$ with $e^{i\varphi}f$ ($\varphi$ real), the right side is unchanged, while the left side becomes
          \begin{equation}
          \label{Qphi}
          |\mathrm{Re}(e^{i\varphi}Q(f,g))| =|\cos{(\varphi)}\mathrm{Re}(Q(f,g))-\sin{(\varphi)}\mathrm{Im}(Q(f,g))|.
          \end{equation}
          As the maximum of this quantity is $\sqrt{\mathrm{Re}(Q(f,g))^{2}+\mathrm{Im}(Q(f,g))^{2}}=|Q(f,g)|$ [the right hand side of (\ref{Qphi}) can be regarded as the
          inner product of the two-vectors $(\cos{\varphi},-\sin{\varphi})$ and $(\mathrm{Re}(Q(f,g)),\mathrm{Im}(Q(f,g))$, which reaches its maximum value when these two vectors are aligned---as the
          first is a unit vector, the inner product then just becomes the magnitude of the second vector] we indeed have
          \begin{equation}
           |Q(f,g)| \leq \sqrt{Q(f)Q(g)}.
           \end{equation}
           Further, one has [$a$ complex] $Q(af) = aa^{*}Q(f) = |a|^{2}Q(f),\;\sqrt{Q(af)} = |a|Q(f)$, and
           \begin{eqnarray}
           Q(f+g) &=& Q(f)+Q(f,g)+Q(g,f)+Q(g) = Q(f)+2\mathrm{Re}(Q(f,g))+Q(g)  \\
           &\leq& Q(f)+2\sqrt{Q(f)Q(g)}+Q(g) = (\sqrt{Q(f)}+\sqrt{Q(g)})^{2}, \\
           \label{distineq}
           \sqrt{Q(f+g)} &\leq& \sqrt{Q(f)}+\sqrt{Q(g)}.
           \end{eqnarray}
          }
         \begin{eqnarray}
         \sqrt{Q(af)} &=& |a|\sqrt{Q(f)} \\
         \sqrt{Q(f+g)} &\leq& \sqrt{Q(f)}+\sqrt{Q(g)}.
         \end{eqnarray}
         \end{enumerate}
         The  last two relations motivate the consideration of $\sqrt{Q(f)}$ as the absolute value of $f$, and $\sqrt{Q(f-g)}$
         as the distance between $f$ and $g$.\footnoteA{From relation (\ref{distineq}) follows the fundamental postulate [triangle inequality] for
         every distance relation:
         \begin{equation}
          \mathrm{distance} (f,h) \leq \mathrm{distance}(f,g) + \mathrm{distance} (g,h).
          \end{equation}
          Our distance function $\sqrt{Q(f-g)}$ is therefore given in $\mathfrak{H}_{0}$ (resp.\ $\mathfrak{H}$, where we indeed know the [inner product] $Q$)
          by $\sqrt{\sum_{n=1}^{\infty}|x_{n}-y_{n}|^{2}}$ (resp.\ $\sqrt{\int_{\Omega}|f-g|^{2}dv}$); namely, the reasonable generalization of the distance concept
          in the usual euclidean spaces.}

         In this way $Q$ indeed provides $\bar{\mathfrak{H}}$ with a metric, a concept of separation. Consequently expressions
         such as ``continuous'', ``bounded'', ``arbitrarily close'' etc. acquire a meaning in $\bar{\mathfrak{H}}$.\\
         \\
         \textbf{C}. $\bar{\mathfrak{H}}$ has infinitely many dimensions: i.e., there are an arbitrarily large number of linearly independent
         elements of $\bar{\mathfrak{H}}$.\\
         \\
         \textbf{D}. There exists an everywhere dense sequence in $\bar{\mathfrak{H}}$. Namely: there exists a sequence $f_1,f_2,\ldots$ such that
         elements of the sequence appear in an arbitrary neighborhood of any element $f$ of $\bar{\mathfrak{H}}$.\\
         \\
         \textbf{E}. The Cauchy convergence requirement holds in $\bar{\mathfrak{H}}$. Namely: Every sequence $f_1,f_2,\ldots$ in $\bar{\mathfrak{H}}$
         satisfying the Cauchy convergence condition (i.e., for every $\varepsilon >0$ there is a  $N=N(\varepsilon)$ such that $N\leq m\leq n$ implies
         $\sqrt{Q(f_{m}-f_{n})}\leq\varepsilon$) is convergent (i.e., there exists a $f$ in $\bar{\mathfrak{H}}$ so that
          for every $\varepsilon >0$ there is a  $N=N(\varepsilon)$ such that $N\leq m$ implies $\sqrt{Q(f_{m}-f)}\leq \varepsilon$).

          Temporarily postponing to Appendix 2 the question of whether $\mathfrak{H}_{0}, \mathfrak{H}$ [i.e., $l^2$ and $L^{2}[\Omega]$] satisfy these conditions,
           we will instead proceed to derive the most immediate consequences of 
           A.--E., specifically the often repeated claim of
           constructibility of $\bar{\mathfrak{H}}$ from $\mathfrak{H}_{0}$, in the course of which we shall also acquire an insight into
           the structure of orthogonal systems in $\bar{\mathfrak{H}}$ which will be critical for our later considerations.

           In the next Section we shall give (for the sake of completeness) a rigorous discussion of these ideas using known mathematical
           procedures---one can skip the details, paying attention only to the main results.

           \section{[Basic properties of orthogonal systems in Hilbert space]}

           We must first give some simple definitions of a general nature. 
           
           $f$ is an accumulation point of a subset $\mathfrak{M}$ of
           $\bar{\mathfrak{H}}$, if points of  $\bar{\mathfrak{H}}$ [-->$\mathfrak{M}$] can be found arbitrarily close (cf.\ the remark following axiom \textbf{B})
           to $f$. A subset $\mathfrak{M}$ of  $\bar{\mathfrak{H}}$ is closed if it contains all its accumulation points; it is everywhere dense,
           if every $f$ in $\bar{\mathfrak{H}}$ is an accumulation point [of $\mathfrak{M}$]; it is dense in [another subset] $\mathfrak{N}$, if
           every point of $\mathfrak{N}$ is an accumulation point [of $\mathfrak{M}$].

           $\mathfrak{M}$ is a linear subspace if it is mapped to itself by linear operations: thus, if $f_1,f_2,\ldots,f_k$ is in $\mathfrak{M}$,
           so is $a_{1}f_{1}+a_{2}+f_{2}+\ldots a_{k}f_{k}$.

           Two elements $f,g$ are orthogonal, if $Q(f,g)=0$. $\mathfrak{M}$ is an orthonormal system, if for all $f,g$ in $\mathfrak{M}$
           \begin{eqnarray}
             Q(f,g) &=& 1,\;\;\mathrm{for}\;\;f=g, \\
             &=& 0,\;\;\mathrm{for}\;\;f\neq g.
             \end{eqnarray}
            $\mathfrak{M}$ is a complete orthonormal system if no further $f$ can be added to it while preserving the orthogonal character of the set.
            Equivalently: if no $f$ (other than $f=0$) is orthogonal to all $g$ in $\mathfrak{M}$.

            We can now proceed to the proof of the previously announced theorems:
            
            \textbf{Theorem 1}. Every orthonormal system $\mathfrak{M}$ is either finite or a sequence; every complete orthonormal system is a sequence.
            
            \textbf{Proof}: Let $\mathfrak{M}$ be an orthonormal system, $f_1,f_2,\ldots$ the everywhere dense sequence [cf.\ axiom \textbf{D}]. For every
            pair $f,g$ of $\mathfrak{M}$ we have
            \begin{equation}
              Q(f-g) = Q(f) - 2\mathrm{Re}(Q(f,g)) + Q(g) = 1-0+1 = 2,
              \end{equation}
             so their  separation is $\sqrt{2}$. We can associate to each element $f$ of $\mathfrak{M}$ some element of the [everywhere dense] sequence
             $f_1,f_2,\ldots$ which is nearer to it than $\sqrt{2}/2$, from which it follows that two separate elements of the dense sequence are
             associated to two different elements of $\mathfrak{M}$. Consequently, $\mathfrak{M}$ has at most as many elements as the given
             dense sequence, which was to be proven.

             If on the other hand $\mathfrak{M}$ is complete, we wish to show that $\mathfrak{M}$ cannot be finite, i.e., that to any finite
             set $\varphi_{1},\varphi_{2},\ldots,\varphi_{k}$ there is an orthogonal $f\neq 0$. But we cannot find $k+1$ linearly independent elements
             in the linear subspace spanned by $\varphi_{1},\varphi_{2},\ldots,\varphi_{k}$, so $\bar{\mathfrak{H}}$ must contain [axiom $\mathbf{C}$]
             an element $f$ outside of this space. Accordingly
             \begin{equation}
                f-c_{1}\varphi_{1}-c_{2}\varphi_{2}-\cdots c_{k}\varphi_{k}
                \end{equation}
                is never $=0$: indeed, by choosing $c_{n}=Q(f,\varphi_{n})$ we can make it orthogonal to all $\varphi_{1},\ldots\varphi_{k}$, [thus,
                any $\mathfrak{M}$ with a finite basis cannot be complete], which was to be proven.Q.E.D.
               
               \textbf{Theorem 2}. Let $\varphi_{1},\varphi_{2},\ldots$ be an orthonormal system [not necessarily complete!]. Then every series
                \begin{equation}
                   \sum_{n=1}^{\infty} Q(f,\varphi_{n})Q(g,\varphi_{n})^{*}
                   \end{equation}
                   is absolutely convergent, and in particular [for $f=g$], one has $\sum_{n=1}^{\infty}|Q(f,\varphi_{n})|^{2}\leq Q(f)$.
                   
            \textbf{Proof}: Choosing $c_{n}=Q(f,\varphi_{n}), n=1,2,\ldots$, we have
            \begin{eqnarray}
            Q(\sum_{n=1}^{N}c_{n}\varphi_{n}-f) &=& Q(f) -\sum_{n=1}^{N}2\;\mathrm{Re}(Q(f,c_{n}\varphi_{n})) +\sum_{m,n=1}^{N}Q(c_{m}\varphi_{m},c_{n}\varphi_{n}) \nonumber \\
            &=& Q(f) -\sum_{n=1}^{N}2\;\mathrm{Re}(c_{n}^{*}Q(f,\varphi_{n}))+\sum_{m,n=1}^{N}c_{m}c^{*}_{n}Q(\varphi_{m},\varphi_{n}) \nonumber \\
            &=& Q(f)-2\sum_{n=1}^{N}|c_{n}|^{2}+\sum_{n=1}^{N}|c_{n}|^{2} = Q(f)-\sum_{n=1}^{N}|c_{n}|^{2}.
            \end{eqnarray}
            As the left side is always $\geq 0$, it follows that [for any $N$]
            \begin{equation}
              \sum_{n=1}^{N}|c_{n}|^{2} \leq Q(f),
              \end{equation}
              thereby establishing the convergence of the series
              \begin{equation}
              \sum_{n=1}^{\infty}|c_{n}|^{2} = \sum_{n=1}^{\infty}|Q(f,\varphi_{n})|^{2},
              \end{equation}
              and also that the sum of the series is $\leq Q(f)$ (the second assertion).

              The first assertion of the theorem follows from the obvious identities\footnoteB{The first identity follows from
              \begin{equation}
               |Q(\frac{f\pm g}{2},\varphi_{n})|^{2} = \frac{1}{4}(|Q(f,\varphi_{n})|^{2}\pm 2\mathrm{Re}(Q(f,\varphi_{n})Q(g,\varphi_{n})^{*})+|Q(g,\varphi_{n})|^{2}),
               \end{equation}
               by subtracting the right hand sides with opposite choices of sign. Similarly with the second identity.\label{polidentity}}
              \begin{eqnarray}
              \mathrm{Re}(Q(f,\varphi_{n})Q(g,\varphi_{n})^{*}) &=& |Q(\frac{f+g}{2},\varphi_{n})|^{2}-|Q(\frac{f-g}{2},\varphi_{n})|^{2}, \\
              \mathrm{Im}(Q(f,\varphi_{n})Q(g,\varphi_{n})^{*}) &=& |Q(\frac{f+ig}{2},\varphi_{n})|^{2}-|Q(\frac{f-ig}{2},\varphi_{n})|^{2},
              \end{eqnarray}
              as the sum over each term on the right hand sides converges absolutely [by the previous argument].Q.E.D.
              
              \textbf{Theorem 3}. Let $\varphi_{1},\varphi_{2},\ldots$ be an orthonormal system. The series $\sum_{n=1}^{\infty}c_{n}\varphi_{n}$
              converges\footnoteA{Note, that this is convergence in $\bar{\mathfrak{H}}$! Therefore, if one
              considers a continuous realization $\mathfrak{H}$ of $\bar{\mathfrak{H}}$ (the space of all functions $f$ defined in $\Omega$ with
              finite $\int_{\Omega}|f|^{2}dv$), then this refers not to point-wise convergence, but convergence in the mean.}
               if and only if $\sum_{n=1}^{\infty}|c_{n}|^{2}$ is finite.
               
              \textbf{Proof}: By axiom \textbf{E} [Cauchy-completeness of $\bar{\mathfrak{H}}$] the convergence of $\sum_{n=1}^{\infty}c_{n}\varphi_{n}$ is
              equivalent to the requirement that for every $\varepsilon>0$ there is a $N=N(\varepsilon)$, so that for $N\leq m\leq n$
              \begin{equation}
              \sqrt{Q(\sum_{p=1}^{n}c_{p}\varphi_{p}-\sum_{p=1}^{m}c_{p}\varphi_{p})} \leq \varepsilon.
              \end{equation}
        But it now follows that [for $m<n$]
        \begin{eqnarray}
         Q(\sum_{p=1}^{n}c_{p}\varphi_{p}-\sum_{p=1}^{m}c_{p}\varphi_{p})&=& Q(\sum_{p=m+1}^{n}c_{p}\varphi_{p}) \nonumber \\
         &=& \sum_{p,q=m+1}^{n}c_{p}c^{*}_{q}Q(\varphi_{p},\varphi_{q}) \nonumber \\
         &=& \sum_{p=m+1}^{n}|c_{p}|^{2} \nonumber \\
         &=& \sum_{p=1}^{n}|c_{p}|^{2}-\sum_{p=1}^{m}|c_{p}|^{2},
         \end{eqnarray}
         whereby we have exactly the convergence condition [this is the Cauchy completeness property for the real numbers] for the series $\sum_{p=1}^{\infty}|c_{p}|^{2}$.Q.E.D.
                  
         \textbf{Corollary.} For this $f$ [$ = \sum_{n=1}^{\infty}c_{n}\varphi_{n}$], one has $Q(f,\varphi_{p}) = c_{p}$.
         
         \textbf{Proof:} Certainly we have, for $p\leq N$,
         \begin{equation}
          Q(\sum_{n=1}^{N}c_{n}\varphi_{n}, \varphi_{p}) = \sum_{n=1}^{N}c_{n}Q(\varphi_{n},\varphi_{p}) = c_{p}.
          \end{equation}
          Now $Q(f^{\prime},g^{\prime})$, on account of its bilinearity, and the inequality $|Q(f^{\prime},g^{\prime})|\leq\sqrt{Q(f^{\prime})Q(g^{\prime})}$,
          is a continuous function of $f^{\prime}$ and $g^{\prime}$.\footnoteB{In other words, if $f_{n}\rightarrow f, n\rightarrow\infty$ (which means $Q(f-f_{n})\rightarrow 0$), then $Q(f_{n},g)\rightarrow Q(f,g)$.
          And similarly for a sequence $g_{n}\rightarrow g$.} So we can let $N\rightarrow\infty$, and obtain $Q(f,\varphi_{p}) = c_{p}$ [i.e., defining
          $f^{N}=\sum_{n=1}^{N}c_{n}\varphi_{n}$, by continuity $c_{p}=\lim_{N\rightarrow\infty}Q(f^{N},\varphi_{p}) = Q(\lim_{N\rightarrow\infty}f^{N},\varphi_{p})=Q(f,\varphi_{p})$].Q.E.D.
          
          \textbf{Theorem 4.} Let $\varphi_{1},\varphi_{2},\ldots$ be an orthonormal system. The series
          \begin{equation}
            f^{\prime} = \sum_{n=1}^{\infty}c_{n}\varphi_{n},\;\; c_{n} = Q(f,\varphi_{n}),
            \end{equation}
            is convergent for all $f$, and $f-f^{\prime}$ is orthogonal to all $\varphi_{1},\varphi_{2},\ldots$
            
            \textbf{Proof:} Follows directly from theorems 2 and 3.
            
           \textbf{Theorem 5.} Any one of the following three conditions is both necessary and sufficient for an orthonormal system $\varphi_{1},\varphi_{2},\ldots$
         to be complete:\\
         $\alpha$. The linear space spanned by $\varphi_{1},\varphi_{2},\ldots$ is everywhere dense.\\
         $\beta$. For all $f$, $f=\sum_{n=1}^{\infty}c_{n}\varphi_{n}, c_{n}=Q(f,\varphi_{n})$.\\
         $\gamma$. For all $f,g$, $Q(f,g) = \sum_{n=1}^{\infty}Q(f,\varphi_{n})Q(g,\varphi_{n})^{*}$.
                  
                  \textbf{Proof:} [In the following, ``completeness'' is short for ``completeness of the orthonormal $\varphi_{1},\varphi_{2}, \ldots$''] First, completeness implies
         $\beta$, as the $f-f^{\prime}$ of Theorem 4 must vanish [by definition, there is no nonzero element orthogonal to all elements of a complete set].

         Secondly, $\beta$ implies $\alpha$, as the sequence $f^{N}, N=1,2, \ldots$, $f^{N} = \sum_{p=1}^{N}c_{p}\varphi_{p}$ has $f$ as its accumulation point, and
         all the $f^{N}$ belong to the linear space spanned by the $\varphi_{1},\varphi_{2},\ldots$

         Thirdly, $\beta$ implies $\gamma$, as (cf.\ Eq.(35))
         \begin{equation}
         Q(\sum_{n=1}^{N}c_{n}\varphi_{n}-f) = Q(f) - \sum_{n=1}^{N}|c_{n}|^{2},
         \end{equation}
          and for $N\rightarrow\infty$ (as $Q(f)$, as remarked above, is continuous), the left hand side tends to zero, we have
          \begin{equation}
          \sum_{n=1}^{\infty}|c_{n}|^{2} = Q(f),\;\;\sum_{n=1}^{\infty}|Q(f,\varphi_{n})|^{2} = Q(f).
          \end{equation}
          If we now sequentially replace $f$ by $\frac{f+g}{2}$, $\frac{f-g}{2}$, and $\frac{f+ig}{2}$, $\frac{f-ig}{2}$,
          and subtract twice, one obtains, as in the proof of Theorem 2, the representation $\gamma$.[See note \ref{polidentity}.]

          Fourth, $\alpha$ implies completeness. For suppose a $f$ exists that is orthogonal to all $\varphi_{1},\varphi_{2},\ldots$: then it is also
          orthogonal to the entire space spanned by them, and as the latter is everywhere dense, orthogonal to itself, i.e., $Q(f)=0, f=0$.

          Fifth, $\gamma$ implies completeness: for if $f$ is orthogonal to all $\varphi_{1},\varphi_{2},\ldots$, then $\gamma$ implies (putting $g=f$) that
          $Q(f)=0, f=0$.

          We therefore have the following logical scheme: 
          $$
  \mathrm{completeness} \; \rightarrow \; \beta \; 
\begin{array}{ccc}
 \; & \!\! \alpha \!\!  &  \; \\
\nearrow  &  & \searrow  \\
\searrow  &  &  \nearrow \\
\;  & \!\! \gamma \!\! & \;
\end{array}
\; \mathrm{completeness}
$$
          All four assertions [i.e., completeness, $\alpha$, $\beta$, $\gamma$] are therefore logically equivalent.Q.E.D.
                    
                    \textbf{Theorem 6.} To every sequence $f_1,f_2,\ldots$ corresponds an orthonormal system $\varphi_{1},\varphi_{2},\ldots$ (both sequences may terminate after
          a finite number of elements) that spans the same linear space.
          
           \textbf{Proof:} Denote the first nonzero element of $f_1,f_2, \ldots$ as $g_1$; the first element of $f_1,f_2, \ldots$ $\neq a_{1}g_{1}$ as $g_2$; the
          first element of $f_1,f_2, \ldots$ $\neq a_{1}g_{1}+a_{2}g_{2}$ as $g_3$; and so on. The $g_1,g_2,\ldots$ are obviously linearly independent, and span
          the same linear space as $f_1,f_2, \ldots$ By the well-known procedure of [Schmidt]-orthogonalization\footnoteA{This procedure [introduced in \autocite[Ch.\ 1, Sec.\ 5]{Schmidt:1908}] runs as follows:
          \begin{eqnarray}
           \gamma_{1} &=& g_{1},\;\;\varphi_{1} = \frac{1}{\sqrt{Q(\gamma_{1})}}\gamma_{1}, \\
           \gamma_{2} &=& g_{2} - Q(g_{2},\varphi_{1})\varphi_{1},\;\;\varphi_{2} = \frac{1}{\sqrt{Q(\gamma_{2})}}\gamma_{2}, \\
           \gamma_{3} &=& g_{3} - Q(g_{3},\varphi_{1})\varphi_{1}-Q(g_{3},\varphi_{2})\varphi_{2},\;\;\varphi_{3} = \frac{1}{\sqrt{Q(\gamma_{3})}}\gamma_{3},\ldots
           \end{eqnarray}
           The $\varphi_{1},\varphi_{2},\ldots$ clearly span the same linear space as the $g_{1},g_{2},\ldots$, and are normalized, and orthogonal to each other.}
            one can transform the $g_1,g_2,\ldots$ into an
          orthonormal system.Q.E.D.
           
                    \textbf{Corollary.} There exist complete orthonormal systems.
                    
          \textbf{Proof:} By axiom \textbf{D}, choose an everywhere dense sequence $f_1,f_2,\ldots$ Replace it by an orthonormal system as in Theorem 6. This is
          complete, by Theorem 5$\alpha$.Q.E.D.\\

          If we now take an arbitrary complete orthonormal system $\varphi_{1},\varphi_{2},\ldots$, and to every $f$ in $\bar{\mathfrak{H}}$ assign the
          sequence $c_1,c_2,\ldots$ ($c_{n}=Q(f,\varphi_{n}), n=1,2,\ldots$), then the sequence $c_1,c_2,\ldots$ belongs to $\mathfrak{H}_{0}$ by Theorem 2,
          and determines in its turn $f$ by Theorem 5$\beta$. This one-to-one representation of $\bar{\mathfrak{H}}$ covers all of $\mathfrak{H}_{0}$
          by Theorem 3. The invariance of the [vector] addition and [scalar] multiplication in this mapping is trivial, the invariance of $Q$ follows
          from Theorem 5$\gamma$.

          In this way we have indeed demonstrated that every space $\bar{\mathfrak{H}}$ obeying \textbf{A.--E.} must agree in all its properties with
          the usual Hilbert space $\mathfrak{H}_{0}$. (This mapping from $\bar{\mathfrak{H}}$ to $\mathfrak{H}_{0}$ is naturally analogous to that
          sketched in Section 3 from [the function space] $\mathfrak{H}$ to $\mathfrak{H}_{0}$.)

          \section{Operator Calculus}

          Following the considerations of previous Section, we can use abstract (complex) Hilbert space as the foundation for further
          investigations.  Appeal to its various representations (the discrete, as well as the various continuous ones,
          cf.\ the beginning of Section 5) will be seen in the forthcoming Sections (7-11, 13) to be absolutely unnecessary: all of
          our general developments are independent of these. We shall only need to concern ourselves with these aspects later,
          with the physical applications.

          The first thing  we must develop in $\bar{\mathfrak{H}}$ is the operator calculus---we already know (cf.\ the end of Section 2)
          the fundamental importance it has for quantum mechanics.

          A function in $\bar{\mathfrak{H}}$ that is defined in certain (perhaps all) points of $\bar{\mathfrak{H}}$, and has points of $\bar{\mathfrak{H}}$
          as values, is called an operator. An operator $T$ is linear, if, first, it is defined in a linear manifold (which by no means needs to be
          closed), and if, second, it always satisfies
          \begin{equation}
           T(a_{1}f_{1}+a_{2}f_{2}+\cdots +a_{k}f_{k}) = a_{1}Tf_{1}+a_{2}Tf_{2}+\cdots +a_{k}Tf_{k},
           \end{equation}
           where $a_1,a_2,\ldots a_k$ are complex constants.

          As we already observed in Section 4, the expressions ``continuous'' and ``bounded in the sphere of radius 1 centered on 0'' are meaningful
          for operators, and for linear operators obviously imply the same thing.\footnoteA{The second condition means: $Q(f)\leq 1$ implies $Q(Tf)\leq C$ ($C$ a
          constant). Given that $T(af)=aTf$, this implies
          \begin{eqnarray}
           Q(Tf) &\leq& CQ(f),\\
           Q(Tf-Tg) &\leq& CQ(f-g),
           \end{eqnarray}
           in other words, continuity. Conversely: for continuous $T$ there exists an $\varepsilon>0$ such that $Q(f)<\varepsilon$ implies
           $Q(Tf)\leq 1$, from which it immediately follows that for $Q(f)\leq 1$ we have $Q(Tf) \leq \frac{1}{\varepsilon}$[$\equiv C$].} 
            In this case the Hilbert expression ``bounded'' is the common one.
          If we consider the discrete representation $\mathfrak{H}_{0}$, then every bounded linear operator corresponds to an infinite matrix, and
          the associated bilinear form (of infinitely many variables) belong to that class of bounded bilinear forms, for which Hilbert has completely
          solved the eigenvalue problem (cf.\ note 12)\footnoteB{The 1912 treatise of Hilbert summarizes this work, see, for example Hilbert \autocite*[p.\ 15]{Hilbert:1912}. 
          The spectral theory of bounded forms was first presented in Hilbert \autocite*{Hilbert:1906a}.}.
           The operators of  interest in quantum mechanics do not (cf.\ same note) however belong to this class.

            In the continuous realizations $\mathfrak{H}$ of Hilbert space one indeed, for certain bounded operators $T$, has at hand the integral
            kernel representation analogous to matrices [in the discrete case]:
            \begin{equation}
              Tf(P) = \int_{\Omega}\varphi(P,Q)f(Q)dv_{Q},
              \end{equation}
              where the integration [in $\Omega$, with volume element $dv_{Q}$] is over $Q$; however, such a representation fails already with the
              simplest of all operators, the ``identity operator'' (which transforms $f$ to $f$). Our considerations will be accomplished without employment
              of either a matrix or integral kernel representation.

              It is essential to recall some simple operators, and rules for calculation with operators. One has
              \begin{equation}
                 0f = 0,\;\;\; 1f = f.
                 \end{equation}
              (0 thus appears in three forms: as a number, as a point in $\bar{\mathfrak{H}}$, and as the null operator; but there is hardly any danger
              of confusing them.) Moreover we have:
              \begin{eqnarray}
               (aT)f &=& a\;Tf \;\;(a\;\mathrm{a\;complex\;constant}), \\
               (R+T)f &=& Rf + Tf,   \\
               (RT)f &=& R(Tf).
               \end{eqnarray}
                It is generally known that addition of operators is commutative and associative, both types of multiplication [scalar and operator] are
                distributive and associative, but in general not commutative, and that 0 and 1 play the role of null and unity.

         Finally we observe the following: if $T$ is a linear operator, and for a particular element $f$ of $\bar{\mathfrak{H}}$ there exists
         a $\bar{f}$, so that (given existence of $Tg$) we always have
         \begin{equation}
             Q(\bar{f}, g) = Q(f, Tg), 
             \end{equation}
             then we denote this $\bar{f}$ by $T^{\dagger}f$ 
             [von Neumann uses an asterisk (e.g., $T^*$) rather than a dagger to denote the hermitian conjugate of an operator]. It is uniquely determined provided that the $g$ for which $Tg$ exists are everywhere dense.
             Indeed, if this were the case for two elements $\bar{f}_{1},\bar{f}_{2}$ we would have
             \begin{equation}
             Q(\bar{f}_{1},g) = Q(\bar{f}_{2},g),\;\; Q(\bar{f}_{1}-\bar{f}_{2},g) = 0,
             \end{equation}
             for an everywhere dense set of $g$, hence [by continuity of $Q$] for all $g$, and therefore [setting $g=\bar{f}_{1}-\bar{f}_{2}$]
             \begin{equation}
             Q(\bar{f}_{1}-\bar{f}_{2}) = 0,\;  \bar{f}_{1} = \bar{f}_{2}.
             \end{equation}
             This requirement (of meaningful action on an everywhere dense set of $g$) will now be imposed on all operators to be considered---it
             will indeed be satisfied by all the operators of real interest to us.\footnoteA{The meaning of this requirement can be made clear with the following two examples:\\
             Let $\Omega$ be the interval $(0,1)$, $T$ the derivative operation. $Tf$ does not always exist (not all $f(x)$ are differentiable), but differentiable functions on this interval
             are certainly dense, as any function can be arbitrarily well approximated---in the mean---by polynomials (certainly differentiable!).\\
              Let $\Omega$ be the interval $(-\infty,\infty)$, $T$ multiplication by $x$. $Tf$ does not always exist ($\int_{-\infty}^{\infty}|f(x)|^{2}dx$ can be finite, while
              $\int_{-\infty}^{\infty}x^{2}|f(x)|^{2}dx$ is not). However, these $f$ [for which $Tf$ exists] are everywhere dense: all $f$, that vanish identically outside an (arbitrarily large,
              but finite) interval, belong to this set.}

             $T^{\dagger}$ is also, as one immediately sees, also a linear operator; also for $T^{\dagger}$ we shall presume that it acts meaningfully
             on an everywhere dense set of elements $f$. It is defined by the equation
             \begin{equation}
             Q(f,Tg) = Q(T^{\dagger}f,g).
             \end{equation}
             On the basis of the properties of $Q$ (Section 4B) one can easily show (with the necessary assumptions of meaningfulness):
             \begin{eqnarray}
              (aT))^{\dagger} &=& aT^{\dagger}  \\
              (R+T)^{\dagger} &=& R^{\dagger} + T^{\dagger} \\
              (RT)^{\dagger} &=& T^{\dagger}R^{\dagger} \\
              T^{\dagger\dagger} &=& T.
              \end{eqnarray}

              A linear operator $T$ is called \emph{symmetric} (more accurately, complex hermitian) if $T=T^{\dagger}$. Using the above equations
              one can verify immediately the following assertions:\\
              If $T$ is symmetric, so is $aT$ if and only if $a$ is real.\\
               If $R, T$ are symmetric then so is $R+T$;\\
               on the other hand, $RT$ is
              symmetric if and only if $R$ and $T$ commute ($RT=TR$).\\
               0 and 1 [null and identity] are both symmetric.

\section{Projection Operators}

After these general considerations we now wish to consider an important special class of operators: the
projection operators,\footnoteA{The term [German \emph{Einzeloperator} (literally, ``single-operator'')] is derived from Hilbert's designation for the analogous bilinear
forms: \emph{Einzelform}.} 
defined as follows: a symmetric [modern: hermitian] linear operator $E$ is a projection
operator (abbreviated P.\ Op. [German, \emph{Einzeloperator, kurz E. Op.}]) if $E=E^2$. One sees at once that 0 and 1
are P. Ops. Moreover, if $E$ is a P. Op., so is $1-E$, as $(1-E)^{2}=1-2E+E^2 = 1-2E+E = 1-E$.

We now wish to derive a few theorems concerning projection operators. While we shall not need all of these results
in the following discussion, they are nonetheless useful in throwing the right light on the essence of
this fundamental concept. In this respect the following two facts are of primary interest:
\begin{enumerate}
\item A P.\ Op.\ maps certain $f$ into $f$, and others into 0. The first set constitutes the interior, the second set
the exterior of the given P.\ Op.\ \footnoteB{In modern terminology, ``interior'' and exterior'' would be denoted the ``range'' and ``kernel'' respectively of the given projection operator.}
Every $f$ in $\bar{\mathfrak{H}}$ can be decomposed in precisely one way as $g+h$,
with $g$ in the interior and $h$ in the exterior of $E$. Thus, $Ef$ arises from $f$ by discarding its ``outer component'' $h$.
\item There is a certain [partial] ordering of size among projection operators: namely, according to the extension of its
interior (or, inversely, according to the extension of its exterior).
\end{enumerate}
Keeping these results in view one may skip the following discussion (in which the definition of the interior and exterior of $E$
are used).

One has obviously
\begin{eqnarray}
 Q(Ef, Eg) &=& Q(f, E^{\dagger}Eg) = Q(f, E^2 g) = Q(f, Eg), \\
 Q(Ef, Eg) &=& Q(E^{\dagger}Ef, g) = Q(E^{2}f, g) = Q(Ef, g),
 \end{eqnarray}
 and in particular [setting $g=f$]
 \begin{equation}
  Q(f, Ef) = Q(EF).
  \end{equation}
  From this follows  
  
  \textbf{Theorem 1.} One has always $0 \leq Q(EF) \leq Q(f)$.
  
  \textbf{Proof:} The first inequality is trivial, the second holds because $1-E$ is also a P. Op.:
  \begin{eqnarray}
     Q(f, Ef) &=& Q(f) - Q(f, (1-E)f), \\
     Q(Ef) &=& Q(f) - Q((1-E)f) \leq Q(f).Q.E.D.
     \end{eqnarray}

Therefore every P.\ Op.\ is continuous, indeed all P. Ops. are uniformly continuous.

\textbf{Theorem 2.} If $E, F$ are P. Ops. then $EF$ is a P.\ Op.\ if and only if they commute ($EF=FE$). $E+F$ is then a P.\ Op.\ if and only if
$EF=0$ (or equivalently $FE=0$). $F-E$ is then a P.\ Op.\ if and only if $EF=E$ (or equivalently $FE=E$).

\textbf{Proof:} In the case of the product $EF$, $EF=FE$ is already both necessary and sufficient for symmetry [hermiticity] to hold [as
$(EF)^{\dagger}=F^{\dagger}E^{\dagger}=FE$]; but as from this also follows
\begin{equation}
 (EF)^{2} = EFEF = EEFF = E^{2}F^{2} = EF,
 \end{equation}
 the proof is completed in this case.

  For $E+F, F-E$ the symmetric property is assured, one need only check the relation $T^2 =T$ [$T=E+F$ or $F-E$].
Take first $E+F$. If $E+F$ is a P. Op.,
\begin{eqnarray}
   Q(f, Ef) + Q(f, Ff) &=& Q(f, (E+F)f), \\
   Q(Ef)+Q(Ff) &=& Q((E+F)f) \leq Q(f);
   \end{eqnarray}
   and from $Ef=f$ it then follows that
   \begin{equation}
     Q(Ff) \leq 0,  \;\; Ff = 0.
     \end{equation}
Now for any $g$ it is always the case that $EEg = Eg$, so $FEg = 0$, whence $FE = 0$.

Conversely, if $FE=0$, so also $EF=0$ (as $FE$ is a P. Op., therefore $F,E$ commute), and
\begin{equation}
(E+F)^{2} = E^2 + EF + FE + F^2 = E + 0 + 0 + F = E+F,
\end{equation}
and consequently $E+F$ is a P.\ Op.\ In other words, $FE=0$ is necessary and sufficient [for $E+F$ to be a P. Op.]; and
as the roles of $E$ and $F$ are fully symmetric, the same holds for $EF=0$.

Next, consider $F-E$. It is  a P.\ Op.\ if  and only if $1-(F-E)=E+(1-F)$ is a P.\ Op.\ For this to hold it is however
both necessary and sufficient (as $E$ and $1-F$ are P. Ops.) that
\begin{equation}
E(1-F) = 0,\;\;E=EF,
\end{equation}
and also
\begin{equation}
 (1-F)E = 0,\;\;E=FE.Q.E.D.
 \end{equation}

 We now introduce the following definitions: if $E+F$ is a P. Op., we say that $E$ and $F$ are \emph{orthogonal} [German: \emph{fremd}]; if $F-E$ is a P. Op.
 we write $E\leq F$. Theorem 2 thus provides simple necessary and sufficient conditions for both situations. One sees right away that
 $\leq$ is an ordering relation for the P. Ops.: i.e., that it satisfies $E\leq E$; $E\leq F$ and $F\leq E$ imply $E=F$; and $E\leq F$ and $F\leq G$
 imply $E \leq G$. In this ordering 0 precedes, 1 follows all $E$. $E\leq F$ is equivalent to the statement that $E$ and $1-F$ are
 orthogonal, or to $1-F\leq 1-E$.

 We should now investigate more closely the interior and exterior sets of $E$. That both are closed linear manifolds follows from the linearity
 and continuity of $E$. Further one sees immediately that they are simply interchanged under the mapping $E$ to $1-E$. The interior of 1
 and the exterior of 0 contains all $f$ of $\bar{\mathfrak{H}}$; the exterior of 1 and the interior of 0 consists on the other hand only of the null
 element 0.
 
\textbf{Theorem 3.} The interior of $E$ consists of elements of the form $Eg$ ($g$ arbitrary), the exterior of $E$ those of form $(1-E)g$.

\textbf{Proof:} The second assertion follows from the first, simply by replacing $E$ with $1-E$. The first is proven as follows: if $f$ isin the
interior of $E$, then $f=Ef$, which is of the desired form; conversely, if $f=Eg$, then $Ef = EEg = Eg = f$, i.e., $f$ is in the interior of $E$.Q.E.D.

\textbf{Theorem 4.} $f$ lies in the interior (resp.\ exterior) of $E$ if and only if $Q(Ef) = Q(f)$ (resp.\ $=0$).

\textbf{Proof:} The second part is trivial, as $Q(Ef)=0$ if and only if $Ef=0$. The first part follows from this by interchange of $E$ and $1-E$ [namely:
$Q(f-Ef)=0$ iff $f=Ef$; however, a short calculation shows that $Q(f-Ef)=2(Q(f)-Q(Ef))=0$ iff $Q(Ef)=Q(f)$].Q.E.D.

\textbf{Theorem 5.} Each $f$ can be decomposed in a unique fashion as $g+h$ where $g$ (resp.\ $h$) is in the interior (resp.\ exterior) of $E$. This
unique decomposition is in fact $Ef+(1-E)f$.

\textbf{Proof:} That $Ef+(1-E)f$ accomplishes the desired decomposition is clear; the uniqueness follows from the fact that the interior and exterior of $E$
are linear spaces which share only the null element 0.Q.E.D.

\textbf{Theorem 6.} The interior (resp.\ exterior) of $E$ consists of those, and only those, elements of $\bar{\mathfrak{H}}$ which are orthogonal to every
element of the exterior (resp.\ interior) of $E$.

\textbf{Proof:} If $f$ (resp.\ $g$) lie in the interior (resp.\ exterior) of $E$, they must be orthogonal:
\begin{equation}
   Q(f,g)= Q(Ef, g) = Q(f, Eg) = Q(f,0) = 0.
   \end{equation}
   Our assertion follows obviously from this, together with Theorem 5.Q.E.D.
   
\textbf{Theorem 7.} $E,F$ are orthogonal P. Ops. if and only if each $f$ in the interior of $E$ lies in the exterior of $F$.

\textbf{Proof:} The condition is necessary and sufficient, as it amounts to saying that for all $f$ in the interior of $E$ one has $Ff=0$, i.e.
we have always $FEg = 0$ [for arbitrary $g$], i.e., $FE=0$, whence $E$ and $F$ are orthogonal.Q.E.D.

\textbf{Theorem 8.} $E\leq F$ is equivalent to the statement that the interior of $E$ is a subset of the interior of $F$, or also, that the exterior of $F$ is
a subset of the exterior of $E$.

\textbf{Proof:} $E\leq F$ means: $E$ and $1-F$ are orthogonal [as $E=EF$]. One can then apply Theorem 7 first to $E$ and $1-F$, then to $1-F$ and $E$.Q.E.D.

\textbf{Theorem 9.} $E\leq F$ is equivalent to the statement that for all $f$, $Q(Ef)\leq Q(Ff)$.

\textbf{Proof:} The condition is sufficient, for from it we deduce that if $f$ lies in the exterior of $F$,
\begin{equation}
  Q(Ef) \leq Q(Ff) = 0,\;\;Ef=0,
  \end{equation}
  i.e., $f$ is in the exterior of $E$, showing that $E\leq F$.  It is necessary, as from $E\leq F$ [which means $E=EF$] we deduce [cf.\ Theorem 1]:
  \begin{equation}
   Q(Ef)  = Q(EFf) \leq Q(Ff).Q.E.D.
   \end{equation}

   We are now in a position to formulate the eigenvalue problem of a general symmetric operator with the help of  projection operators: indeed,
   in contrast to Section 3, in such a way that the continuous spectrum is correctly treated.

   \section{The eigenvalue problem}

As we mentioned previously, we find the most obvious formulation of the eigenvalue problem unsatisfying. It goes as follows:\\
Let $T$ be a linear symmetric [hermitian] operator. One seeks all real numbers $l$ and all $f\neq 0$ in $\bar{\mathfrak{H}}$, for which
\begin{equation}
   Tf = lf.
   \end{equation}
   The $l$ are the eigenvalues, the $f$ the eigenfunctions.

   This formulation has two essential defects:
   
   First,  the eigenfunctions $f$ are not uniquely determined objects, even when one normalizes them via $Q(f)=1$: a factor
   of absolute value 1 remains undetermined. In fact, in the case of a degenerate eigenvalue (i.e., a $l$ associated with several
   linearly independent eigenfunctions) there is even the freedom of an arbitrary orthogonal [unitary] transformation of the
   eigenfunctions [associated with the given $l$].
  
  Secondly, this formulation fails for the continuous spectrum. If we go back to the $\mathfrak{H}_{0}$ (resp.\ $\mathfrak{H}$) realizations,
   it becomes apparent that there are possibly eigensequences $x_1,x_2,\ldots$ (resp.\ eigenfunctions $f(P)$) which however do not belong to
   $\mathfrak{H}_{0}$ (resp.\ $\mathfrak{H}$), i.e.
   \begin{equation}
      \sum_{n=1}^{\infty}|x_{n}|^{2}\;\;(\mathrm{resp.}\;\;\int_{\Omega}|f(P)|^{2}dv)
      \end{equation}
      is infinite.\footnoteA{This suggests the objection, that we have arbitrarily restricted excessively admission to $\mathfrak{H}_{0}$ (resp.\ $\mathfrak{H}$---therefore of $\bar{\mathfrak{H}}$);
      perhaps one should also admit functions to $\mathfrak{H}$ with infinite $\int_{\Omega}|f(P)|^{2}dv$. But this objection is not sound: we shall show that for quantum mechanics what is important
      for the continuous spectrum is not the eigenfunctions, but quite different features, for the description of which our $\bar{\mathfrak{H}}$ provides the most
      suitable space. For the preceding cases the admission of all functions $f(P)$ would not even have sufficed, which can be clarified with two examples we now give. In one case, the indicated
      expansion of the function space helps, but in the other it does not.
            
            First: let $\Omega$ be the real line $(-\infty,\infty)$, $T$ the symmetric [self-adjoint] operator, well known from quantum mechanics,
      \begin{equation}
         p = \frac{h}{2\pi i}\frac{d}{dx}\cdots.
         \end{equation}
         As one sees immediately, the spectrum consists of the entire interval $(-\infty,\infty)$, and the eigenvalue $l$ is associated with the eigenfunction $\exp{(2\pi ilx/h)}$. As $\int_{-\infty}^{\infty}|\exp{(2\pi ilx/h)}|^{2}dx = \int_{-\infty}^{\infty}dx = \infty$,
         this eigenfunction does not belong to $\mathfrak{H}$.
         
         Second: Let $\Omega$ be as before, $T$ the similarly very important symmetric [self-adjoint] operator $q = x\cdots$. It now appears (and will be rigorously established in Section 10) that again
         every $l$ between $-\infty$ and $\infty$ is an eigenvalue, but now the eigenfunction must vanish for all $x\neq l$ [as $xf(x)=lf(x)$ implies $(x-l)f(x)=0$]. Such a function is equivalent in any integral to zero,
         so must be regarded as [identically] zero: in other words, there are no eigenfunctions. (Naturally one can consider the Dirac $\delta(x-l)$ (cf.\ Section 1$\zeta.$) as an eigenfunction, but it is ``improper.''
         
         One sees therefore that there is no way to delineate a mathematically unobjectionable function space, in which the continuous spectrum can be handled directly.}
  
      We are therefore forced to seek another formulation. We shall try to approximate $\bar{\mathfrak{H}}$ by using spaces of known character!
      For this purpose we shall use the discrete realization $\mathfrak{H}_{0}$: the space of all sequences $x_1,x_2,\ldots$ with finite
      $\sum_{n=1}^{\infty}|x_{n}|^{2}$. We can regard this space as the limiting case of 1-, 2-, 3-,\ldots dimensional complex Euclidean spaces,
      as the dimension grows beyond bound.

      We therefore denote $k$-dimensional (complex) Euclidean space by  $\mathfrak{R}^{(k)}$: the space of all [finite] sequences $x_{1},x_{2},\ldots,x_{k}$.
      It is natural to define $Q(x,y)$ in $\mathfrak{R}^{(k)}$ as $\sum_{n=1}^{k}x_{n}y^{*}_{n}$ (i.e., as the inner product of the vectors $x,y$).
      Then every linear operator $\mathbf{A}$ in $\mathfrak{R}^{(k)}$ can be specified through a matrix $\{a_{\mu\nu}\}$, inasmuch as $\mathbf{A}$ transforms
      the vector $x$ into the vector $y$ as follows
      \begin{equation}
         y_{\nu} = \sum_{\mu=1}^{k} a_{\mu\nu}x_{\mu},\;\;\;(\nu=1,2,\ldots,k),
         \end{equation}
        and  symmetry [hermiticity (of $\mathbf{A}$), namely $Q(x,\mathbf{A}y)=Q(\mathbf{A}x,y)$] is equivalent to the property
        \begin{equation}
        a_{\mu\nu} = a^{*}_{\nu\mu}.
        \end{equation}

      As is well known, the hermitian form associated to $\mathbf{A}$,
      \begin{equation}
         \mathbf{A}(x|y) = Q(\mathbf{A}x|y) = Q(x|\mathbf{A}y) = \sum_{\mu,\nu=1}^{k}a_{\mu\nu}x_{\mu}y^{*}_{\nu},
         \end{equation}
         can always be brought into ``principal axis form'':
         \begin{equation}
         \mathbf{A}(x|y) = \sum_{p=1}^{k}l_{p}(\alpha_{p1}x_{1}+\cdots\alpha_{pk}x_{k})(\alpha_{p1}y_{1}+\cdots\alpha_{pk}y_{k})^{*},
         \end{equation}
         where the matrix $\{\alpha_{\mu\nu}\}$ is orthogonal (unitary), i.e., the ``identity form''
         \begin{equation}
          Q(x|y) = \sum_{n=1}^{k}x_{n}y_{n}^{*}
          \end{equation}
          is transformed into itself [by the matrix $\{\alpha_{\mu\nu}\}$]:
          \begin{equation}
          \sum_{n=1}^{k}x_{n}y_{n}^{*} = \sum_{p=1}^{k}(\alpha_{p1}x_{1}+\cdots+\alpha_{pk}x_{k})(\alpha_{p1}y_{1}+\cdots+\alpha_{pk}y_{k})^{*}.
          \end{equation}

          The ``eigenvalues'' $l_{1},l_{2},\ldots,l_{k}$ are known to be uniquely specified, up to reordering; not so the ``eigenvectors''
          $\alpha_{1\nu},\alpha_{2\nu},\ldots,\alpha_{k\nu},\;(\nu=1,2,\ldots,k)$. For each of them there is still the freedom of a factor
          of absolute value one (the ``phase''), and, if several eigenvalues coincide (``degeneracy''), even the freedom of an
          orthogonal (unitary) transformation of the associated eigenvectors [to a given eigenvalue] among themselves.

       Nevertheless, it is easy to arrive at a unique standard form for $\mathbf{A}(x|y)$:
       
       We shall label the distinct values among the
   $l_{1},l_{2},\ldots,l_{k}$ (ordered by magnitude) $L_{1},L_{2},\ldots,L_{q}$, $(q\leq k)$. Further, define
   \begin{equation}
    \mathbf{E}(l;x|y) = \sum_{L_{p}\leq l}(\alpha_{p1}x_{1}+\cdots+\alpha_{pk}x_{k})(\alpha_{p1}y_{1}+\cdots+\alpha_{pk}y_{k})^{*}.
    \end{equation}
    One can easily establish that the $L_{p}$ and $\mathbf{E}(l;x|y)$ are uniquely determined, despite the above mentioned indeterminacy of the $l_{p}$ and
    $\alpha_{\mu\nu}$. The matrix associated with [the bilinear form] $\mathbf{E}(l;x|y)$ we write as
    \begin{equation}
       \mathbf{E}(l) = \{e_{\mu\nu}(l)\}.
       \end{equation}
       
       Evidently
       \begin{equation}
       \mathbf{A}(x|y) = \sum_{p=1}^{q}L_{p}\{\mathbf{E}(L_{p};x|y)-\mathbf{E}(L_{p-1};x|y)\},
       \end{equation}
       where $L_{0}$ is chosen as an arbitrary number $<L_{1}$.As $\mathbf{E}(l;x|y)$, viewed as a function of $l$, is constant on
       the intervals $l<L_{1}$, $L_{p-1}\leq l <L_{p},(p=2,\ldots,q)$, $L_{q}\leq l$, this can also be written
       \begin{equation}
         \mathbf{A}(x|y) = \int_{-\infty}^{\infty}ld\mathbf{E}(l;x|y),\;\;Q(x|\mathbf{A}y) = \int_{-\infty}^{\infty}ldQ(x|\mathbf{E}(l)y).
         \end{equation}
         (The integral $\int_{-\infty}^{\infty}$ is the so called Stieltjes integral:  see Appendix 3 for further details.)

         The $\mathbf{E}(l;x|y)$ are the so called projection forms [German:\emph{Einzelformen}, or ``identity forms''], namely, they are
         symmetric [hermitian] $e_{\mu\nu}(l) = e_{\nu\mu}(l)^{*}$, and their matrices satisfy\footnoteA{That $\mathbf{E}(l;x|y)$ is an identity form can be readily verified by direct
         calculation, as it corresponds to the bilinear form
         \begin{equation}
           \sum L(x)L^{*}(y),
           \end{equation}
           where the $L(x)$ are a set of mutually orthogonal and normalized linear combinations of $x_{1},x_{2},\ldots,x_{k}$.}
         \begin{equation}
           \mathbf{E}(l)^{2} = \mathbf{E}(l).
       \end{equation}
           We can therefore regard the corresponding operators
           \begin{eqnarray}
           \mathbf{E}(l)x &=& y,\\
           y_{\nu} &=& \sum_{\mu=1}^{k}e_{\mu\nu}(l)x_{\mu}
           \end{eqnarray}
           as P. Ops. in the space $\mathfrak{R}^{k}$.

           Further, one can immediately see that for $l<L_{1}$ (resp.\ $\geq L_{q}$),
           \begin{equation}
            \mathbf{E}(l;x|y) = 0,\;\;\mathrm{resp.}\; =\sum_{p=1}^{k}x_{p}y_{p}^{*},
            \end{equation}
            i.e., $\mathbf{E}(l)=$ the null (resp.\ the identity) operator; and that it follows from $l\leq l^{\prime}$ that
            \begin{equation}
             \mathbf{E}(l;x|x) \leq \mathbf{E}(l^{\prime};x|x).
             \end{equation}

             The matrix $\mathbf{E}(l)$ is therefore everywhere constant, save at the eigenvalues, where it makes discontinuous jumps:
             the same holds for $\mathbf{E}(l;x|x)$, which is moreover monotonically increasing [for fixed $x$, as a function of $l$].

         This formulation can be taken over without difficulty to the space $\mathfrak{H}_{0}$, and indeed to [the abstract Hilbert space] $\bar{\mathfrak{H}}$.
         Here again we seek, for each symmetric [hermitian] linear operator $T$ a family of P. Ops. $E(l)$, so that
         \begin{equation}
           Q(f, Tg) = \int_{-\infty}^{\infty} ldQ(f, E(l)g)
           \end{equation}
           always holds. This state of affairs can more concisely be expressed as follows:
           \begin{equation}
            Tg = \int_{-\infty}^{\infty}ld(E(l)g),\;\; T = \int_{-\infty}^{\infty} ldE(l).
            \end{equation}

    Here we require always that, for $l\leq l^{\prime}$, $Q(E(l)f)\leq Q(E(l^{\prime})f)$, i.e., $E(l)\leq E(l^{\prime})$
    (analogously to the monotonicity of $\mathbf{E}(l;x|x)$!). $\mathbf{E}(l)$ had at most [a finite number] $k$ of points of discontinuity,
    but as we now must let $k$ become infinity, so that $\mathfrak{R}^{k}$ approximates $\mathfrak{H}_{0}$, we cannot demand the same of $E(l)$.
    On the other hand, the fact that $\mathbf{E}(l)$ begins as zero (for small $l$) and ends as unity (for large $l$) can be interpreted [in
    the infinite dimensional case] by analogy as follows: for $l\rightarrow +\infty$ (resp.\ $-\infty$), $E(l)f\rightarrow f$ (resp.\ 0).

 Finally, the jumps of $\mathbf{E}(l)$ are always on the left side of the points of discontinuity (i.e., it is upper semi-continuous, namely
 a sum of the form $\sum_{L_{p}\leq l}$, not $\sum_{L_{p}<l}$). We will require the corresponding property for $E(l)$: for $l^{\prime}\rightarrow l,l^{\prime}>l$
 [$l^{\prime} \rightarrow l$],
 one also has $E(l^{\prime})f\rightarrow E(l)f$.
 
 Let us summarize our requirements:
 If $T$ is a symmetric linear operator on $\bar{\mathfrak{H}}$, we say that $T$ is represented in eigenvalue form when we have found a set
 of P.\ Op.($l$) (for every real $l$), for which the following conditions hold:
 
 \textbf{1.}\hspace*{1.0in} $\quad Q(f, Tg) = \int_{-\infty}^{\infty}ldQ(f, E(l)g)$,\\
  or, concisely,
 \begin{equation}
  Tg = \int_{-\infty}^{\infty} ld\{E(l)g\},\;\; T = \int_{-\infty}^{\infty}ldE(l).
  \end{equation}
  
   \textbf{2a.} $l\leq l^{\prime}$ implies $E(l)\leq E(l^{\prime})$.
   
   \textbf{2b.} For $l\rightarrow +\infty$ (resp.\ $-\infty$), $E(l)f\rightarrow f$ (resp.\ 0).
   
   \textbf{2c.} For $l^{\prime} \downarrow l$, $E(l^{\prime})f\rightarrow f$.
   
   Under these circumstances, we call $E(l)$ the ``partition of unity associated with $T$.''\footnoteA{In essence this is the form in which the eigenvalue problem for bounded bilinear forms
   has been solved by Hilbert. To be sure, we  consistently suppress the separation of the point and continuous spectrum usual in the mathematical literature. Moreover, the operator
   $T=\int_{-\infty}^{\infty}ldE(l)$ does not always act sensibly; one can show that $Tf$ exists (in $\bar{\mathfrak{H}}$), if and only if the integral $\int_{-\infty}^{\infty}l^{2}dQ(E(l)f)$ is finite.}
   
\section{[The eigenvalue problem: further discussion]}

  The definition just given of the eigenvalue representation [now: spectral resolution] of a symmetric linear operator requires naturally some
  additional critical considerations. 
  
  First, it is not immediately clear whether every symmetric linear operator possesses an eigenvalue representation, and whether it is uniquely
  determined. For bounded operators the theorems of Hilbert establish that one and only one such representation exists (see note 12); for unbounded
  operators only the uniqueness has been established.\footnoteA{For the case of real symmetric operators the author could show that one, and only one, solution exists;
  for complex symmetric (hermitian) operators one conjectures the same, but certain difficulties stand in the way of the proof. Cf.\ the article mentioned in note 12,
  to appear in \emph{Mathematische Annalen}.}\footnoteB{Two years later, von Neumann \autocite*{VonNeumann:1929a} 
  published a comprehensive treatment of
  the eigenvalue problem for a large class of densely defined operators (hypermaximal hermitian---now, self-adjoint---operators $R$), not necessarily bounded, in which a spectral decomposition
  $R=\int \lambda dE(\lambda)$ can be rigorously established. The basic tool of von Neumann's proof is the use of a Cayley transform to map such operators to
  unitary (hence, bounded) ones to which the Hilbert spectral theory apples.}
  
  Further, it is certainly desirable to arrive at a direct interpretation of the P. Ops. $E(l)$. To this end, let us consider the simple case in
  which $T$ has a point spectrum [only]: eigenvalues $l_{1},l_{2},\ldots$ with associated eigenfunctions $\varphi_{1},\varphi_{2}\ldots$ These are well known
  to constitute (e.g., in a continuous realization, and therefore, also in $\bar{\mathfrak{H}}$) a complete orthonormal system, whence
  \begin{equation}
      f = \sum_{n=1}^{\infty} Q(f,\varphi_{n})\varphi_{n},
      \end{equation}
      and, as they are eigenfunctions [of $T$],
      \begin{equation}
      Tf = \sum_{n=1}^{\infty}Q(f,\varphi_{n})l_{n}\varphi_{n}.
      \end{equation}
      (As we are concerned with a merely preliminary orientation, we shall not consider further here any questions of convergence.)
      From this it further follows that
      \begin{equation}
       Tf = \int_{-\infty}^{+\infty} ld[\sum_{l_{n}\leq l}Q(f,\varphi_{n})\varphi_{n}].
       \end{equation}
       This is however just the desired eigenvalue representation [i.e., spectral resolution], with
       \begin{equation}
          E(l)f = \sum_{l_{n}\leq l}Q(f,\varphi_{n})\varphi_{n};
          \end{equation}
          $E(l)$ clearly has the properties 1.-2. of Section 9.

 The [projection] operator $E(l)$ therefore serves the following function: one expands $f$ in the eigenfunctions of $T$ ($\varphi_{1},\varphi_{2}, \ldots$) and omits all terms with eigenvalues greater than $l$.
 The interior of $E(l)$ therefore consists of all functions $f$ in whose expansion only eigenfunctions with eigenvalues $\leq l$ appear, i.e., it is the set of linear combinations of all eigenfunctions
 with eigenvalues $\leq l$.
 
 Now this is also the meaning of the $E(l)$ if a continuous spectrum is present: the interior of $E(l)$ (which in fact determines $E(l)$, as follows from the results of Section 6) consists of all linear
 combinations (which can also include integrals int eh case of a continuous spectrum) of eigenfunctions of $T$ with eigenvalues $\leq l$.\footnoteA{The first example of note 24 shows how linear combinations of eigenfunctions
 which do not belong to $\mathfrak{H}$ (i.e., those with infinite $\int_{\Omega}|\varphi|^{2}dv$) can be constructed which do lie in $\mathfrak{H}$:\\
 The eigenfunctions $\exp{(2\pi ilx/h)}$ are not square-integrable, but linear combinations of them are indeed so:
 \begin{equation}
     \int_{l_1}^{l_2}\exp{(2\pi ilx/h)}dl = \frac{h}{2\pi i}\left(e^{2\pi il_{2}x/h}-e^{2\pi il_{1}x/h}\right)\cdot\frac{1}{x},
     \end{equation}
     where the last factor $\frac{1}{x}$ guarantees the square-integrability.}
 
  This is clearly not yet rigorous, but in many cases it provides a clue to the determination of the $E(l)$, as one may know or have an indication of the eigenfunctions of the continuous spectrum:
  the exact definition at the conclusion of Section 9 then allows a verification, whether the $E(l)$ so obtained are the correct ones. To clarify this we would like to give two examples. We consider
  a continuous realization, with $\Omega$ the interval $(-\infty,+\infty)$. Let $T$ be the operator
  \begin{equation} p = \frac{h}{2\pi i}\frac{d}{dx}\cdots \mathrm{(resp.}\;\; q = x\cdots),
  \end{equation}
  as in note 24.
  
  In the first case we have the eigenfunctions $e^{2\pi ilx/h}$, the eigenfunction-expansion of a function $f$ is therefore the Fourier integral
  \begin{equation}
     f(x) = \int_{-\infty}^{+\infty}\left[\int_{-\infty}^{+\infty}e^{-2\pi ilz/h}f(z)dz\right] e^{2\pi ilx/h}dl.
     \end{equation}
     $E(l)$ effects the ``excision of the remainder'':
     \begin{equation}
     E(l)f(x) = \int_{-\infty}^{l}\left[\int_{-\infty}^{+\infty}e^{-2\pi il^{\prime}z/h}f(z)dz\right] e^{2\pi il^{\prime}x/h}dl^{\prime}.
     \end{equation}
     Indeed, we have:
     \begin{eqnarray}
    \int_{-\infty}^{\infty} ld \left\{E(l)f(x) \right\} &=&  \int_{-\infty}^{\infty} ld   \left\{ \int_{-\infty}^{l} \left[\int_{-\infty}^{\infty}e^{-2\pi il^{\prime}z/h}f(z)dz \right] e^{2\pi il^{\prime}x/h} \right\} \\
      &=& \int_{-\infty}^{\infty}l\; \left[ \int_{-\infty}^{\infty}e^{-2\pi ilz/h} f(z)dz \right] e^{2\pi ilx/h}dl. \\
      &=& \frac{h}{2\pi i}f^{\prime}(x).
      \end{eqnarray}
      
In the second case [coordinate operator $q$] we would look for eigenfunctions as those functions that are non-vanishing only for $x=l$ [as $qf(x0 := xf(x) = lf(x)$ implies $(x-l)f(x)=0$, so either $x=l$ or $f(x)=0$],
which is impossible. However, we should still expect that linear combinations of all eigenfunctions belonging to eigenvalues $\leq l$ should simply be those functions which can only be nonzero for $x\leq l$.
$E(l)$ can therefore be defined as follows:
\begin{eqnarray}
    E(l)f(x) &=& f(x),\;\;\mathrm{if}\;\;x\leq l,\\
    &=& 0,\;\;\mathrm{if}\;\;x>l.
    \end{eqnarray}
In this case one easily verifies that
\begin{equation}
  \int_{-\infty}^{\infty} l d\left[ E(l)f(x) \right] = xf(x).
  \end{equation}
  
  Finally, let us consider yet another aspect of the eigenvalue representation [spectral resolution]. It allows a very simple representation of powers of $T$. Indeed, one has
  \begin{equation}
   T^n = \int_{-\infty}^{\infty} l^{n}dE(l).
   \end{equation}
   For $n=0$ this is trivial ($T^{0}=1$), for $n=1$ true by definition. We shall demonstrate it for $n=2$, the proof is analogous for all higher $n$ (i.e., one deduces the identity for $n+1$ given $n$ exactly
   as we do below for $n=2$ given $n=1$). One has:
   \begin{eqnarray}
    Q(f,T^{2}g) &=& Q(Tf, Tg) \\
    &=& \int_{-\infty}^{\infty} ldQ(Tf, E(l)g) \\
    &=& \int_{-\infty}^{\infty} ld \left[ \int_{-\infty}^{\infty} l^{\prime}dQ(E(l^{\prime})f,E(l)g) \right] \\
    &=& \int_{-\infty}^{\infty} ld \left[ \int_{-\infty}^{\infty} l^{\prime}dQ(f, E(l^{\prime})E(l)g) \right] \\
    &=& \int_{-\infty}^{\infty} ld \left[ \int_{-\infty}^{\infty} l^{\prime}dQ(f, E(\mathrm{min}(l,l^{\prime}))g) \right].
    \end{eqnarray}
For $l^{\prime}>l$ we have $\mathrm{min}(l,l^{\prime})$ is constant (=$l$), so it suffices to run the inner integral over the interval $(-\infty,l)$:\footnoteA{For Stieltjes integrals one has the relation
\begin{equation}
\int_{A}^{B}u(l)d\left[\int_{A}^{l} v(l^{\prime})dw(l^{\prime})\right] = \int_{A}^{B} u(l)v(l)dw(l),
\end{equation}
corresponding to the relation
\begin{equation}
\int_{A}^{B} u(l)\frac{d}{dl}\left[\int_{A}^{l}v(l^{\prime})dl^{\prime}\right]dl = \int_{A}^{B} u(l)v(l)dl
\end{equation}
for conventional [Riemann] integrals.}
\begin{eqnarray}
    Q(f, T^2 g) &=& \int_{-\infty}^{\infty} ld \left[ \int_{-\infty}^{l} l^{\prime}dQ(f,E(l^{\prime})g) \right] \\
    &=& \int_{-\infty}^{\infty} l\cdot ldQ(f, E(l)g) \\
    &=& \int_{-\infty}^{\infty} l^{2}dQ(f, E(l)g),
    \end{eqnarray}
    or, in our abbreviated notation,
    \begin{equation}
    T^{2} = \int_{-\infty}^{\infty} l^{2}dE(l),
    \end{equation}
    as was to be proved.
    
   We will also make use of the following notation: since $l\leq l^{\prime}$ implies $E(l)\leq E(l^{\prime})$, and therefore  $E(l^{\prime})-E(l)$ is a P. Op., which we shall denote $E(l, l^{\prime}) \equiv E(I)$, where
    $I$ is the interval $(l,l^{\prime})$.
    
    \section{The absolute value of an operator}
    
    Before proceeding to physical applications, we still need to develop one last set of concepts which we shall be using. 
    
    Let $A$ be a linear operator (which, as well as [its adjoint] $A^{\dagger}$, acts meaningfully on a dense subset
    of $\bar{\mathfrak{H}}$, cf.\ Section 7). We take two complete orthonormal sets $\varphi_{1},\varphi_{2},\ldots$ and $\psi_{1},\psi_{2},\ldots$, where $A\psi_{1}, A\psi_{2}, \ldots$ exist,\footnoteA{By selecting an everywhere dense sequence $f_1,f_2,\ldots$
    from the everywhere dense set of $f$ where $Af$ is defined, and then applying Section 6, Theorem 6, we obtain the desired orthonormal system  $\psi_{1},\psi_{2},\ldots$}
    and set
    \begin{equation}
         [A;\varphi_{\mu},\psi_{\nu}] = \sum_{\nu,\nu=1}^{\infty} |Q(\varphi_{\mu},A\psi_{\nu})|^{2}.
         \end{equation}
         The sum may be finite or infinite, but is always meaningful, as only non-negative terms appear in it.\footnoteB{This quantity is now referred to as the \emph{Hilbert-Schmidt norm} (squared) $||A||^{2}_{HS}$ of the operator $A$. It was
         exploited in Schmidt's \autocite*{Schmidt:1907} reformulation of Hilbert's integral equation theory, in the integral form \ref{sqintkern}. The finiteness of the Hilbert-Schmidt norm identifies the special class of ``compact'' operators, which
         have a purely point spectrum and square-summable eigenvalues.}
         
         One then has
         \begin{equation}
         \sum_{\mu=1}^{\infty} |Q(\varphi_{\mu},A\psi_{\nu})|^{2} = Q(A\psi_{\nu}), 
         \end{equation}
         \begin{equation}
         [A;\varphi_{\mu},\psi_{\nu}]= \sum_{\nu=1}^{\infty} Q(A\psi_{\nu}),
         \end{equation}
         i.e., the dependence of the $[A;\varphi_{\mu},\psi_{\nu}]$ on the $\varphi_{\mu}$ is only an apparent one. As a consequence of
       \begin{equation}
         \sum_{\mu,\nu}^{\infty}|Q(\varphi_{\mu},A\psi_{\nu})|^{2} = \sum_{\mu,\nu}^{\infty}|Q(A^{\dagger}\varphi_{\mu},\psi_{\nu})|^{2} = \sum_{\mu,\nu}^{\infty}|Q(\psi_{\nu},A^{\dagger}\varphi_{\mu})|^{2},
         \end{equation}
    we have (provided all $A^{\dagger}\varphi_{\mu}$ exist, cf.\ note 30)
    \begin{equation}
    \label{AbsAAdagger}
     [A;\varphi,\psi] = [A^{\dagger};\psi,\varphi].
     \end{equation}
     The same  holds for the $\psi_{\nu}$, so that $[A;\varphi,\psi]$ depends solely on $A$ [and not on the choice of orthonormal systems $\varphi_{\mu},\psi_{\nu}$]:
     \begin{equation}
        [A;\varphi,\psi] = [A].
        \end{equation}
        The equality (\ref{AbsAAdagger}) consequently implies
        \begin{equation}
         [A] = [A^{\dagger}].
         \end{equation}
         The quantity $\sqrt{[A]}$ will be called the absolute value of the operator $A$.\footnoteB{Now called the Hilbert-Schmidt norm.}  We will now study this quantity in greater detail.
Our first task will be to establish the meaning of $[A]$ in the various representations of [the abstract space] $\bar{\mathfrak{H}}$.

  In the discontinuous realization $\mathfrak{H}_{0}$,  $A$ can be represented by a matrix
  \begin{equation}
      \{a_{\mu\nu}\},\;\;a_{\mu\nu} = a_{\nu\mu}^{*}\;\;\;\;(\mu,\nu=1,2,\ldots).
      \end{equation}
      The points [i.e., vectors, represented as row sequences] (1,0,0,\ldots), (0,1,0,\ldots), (0,0,1,\ldots),\ldots form, as one easily sees, a complete orthonormal system. [The operator] $A$ correspondingly transforms these to
          \begin{equation}
      (a_{11}, a_{21}, a_{31},\ldots),(a_{12},a_{22},a_{32},\ldots),(a_{13},a_{23},a_{33},\ldots),\ldots
      \end{equation}
      Denoting these vectors by $\psi_{1},\psi_{2}, \ldots$, we find
      \begin{equation}
      [A] = \sum_{\nu=1}^{\infty} Q(A\psi_{\nu}) = \sum_{\nu=1}^{\infty}\left[\sum_{\mu=1}^{\infty}|a_{\mu\nu}|^{2}\right] = \sum_{\mu,\nu=1}^{\infty} |a_{\mu\nu}|^{2}.
      \end{equation}
      Therefore: $[A]$ is the sum of the absolute squares of all elements of the matrix [of $A$ relative to this orthonormal basis]. 
      
      In the continuous realization $\mathfrak{H}$ [i.e., $L^2$] we shall evaluate $[A]$ only in the special case that the operator $A$ can be represented by an 
      integral  kernel $\varphi$:
      \begin{equation}
         Af(P) = \int_{\Omega} \varphi(P,Q)f(Q)dv_{Q}.
         \end{equation}
         Then one finds:
         \begin{eqnarray}
          [A] &=& \sum_{\nu=1}^{\infty} Q(A\psi_{\nu}) \\
          &=& \sum_{\nu=1}^{\infty}\int_{\Omega} |A\psi_{\nu}(P)|^{2}dv_{P} \\
          &=& \sum_{\nu=1}^{\infty} \int_{\Omega}\left|\int_{\Omega}\varphi(P,Q)\psi_{\nu}(Q)dv_{Q}\right|^{2}dv_{P}  \\
          &=&\int_{\Omega} \left[\sum_{\nu=1}^{\infty}\left|\int_{\Omega}\varphi(P,Q)\psi_{\nu}(Q)dv_{Q}\right|^{2}\right]dv_{P}  \\
          \label{sqintkern}
          &=&\int_{\Omega}\int_{\Omega}|\varphi(P,Q)|^{2}dv_{P}dv_{Q}.
          \end{eqnarray}
          So in this case $[A]$ is the integral of the absolute square of the integral kernel [see note r]. 
          
          Next, we shall derive the most important properties of $[A]$.\\
         \indent
          \textbf{Theorem 1.} $[A]\geq 0$ always, and $[A]=0$ only if $A=0$. (The same holds consequently for $\sqrt{[A]}$.)
          
          \textbf{Proof:} $[A]\geq 0$ is obvious. Let $f$ be an arbitrary element of $\bar{\mathfrak{H}}$ for which $Af$ exists [i.e., $f$ in domain of $A$: $A$ is assumed to be densely defined in this proof]. Either $f$ is zero,
          in which case $Af=0$, or we may define $\varphi= f/\sqrt{Q(f)}$, with $Q(\varphi)=1$. Setting $\psi_{1}=\varphi$, we may complete $\varphi$ to a complete orthonormal set $\psi_{1},\psi_{2},\ldots$ (where all $A\psi_{\nu}$ exist!).
          As $[A]=0$,
          \begin{equation}
             \sum_{\nu=1}^{\infty} Q(A\psi_{\nu}) = 0, \;\; Q(A\psi_{1}) \leq 0 \Rightarrow A\psi_{1}=0
             \end{equation} 
             whence $A\varphi=0$, $Af = 0$. Accordingly $A$ must = 0. Q.E.D.
             
             \textbf{Theorem 2.} The following relations hold:
             \begin{eqnarray}
                 \sqrt{[aA]} &=& |a|\sqrt{[A]},   \\
                 \sqrt{[A+B]} &\leq& \sqrt{[A]} +\sqrt{[B]},   \\
                 \sqrt{[AB]} &\leq& \sqrt{[A]}\;\sqrt{[B]}.
                 \end{eqnarray}
                 
           \textbf{Proof:} The first formula is trivial. The second follows from the identity
           \begin{equation}
              Q((A+B)\psi_{\nu}) - Q(A\psi_{\nu}) - Q(B\psi_{\nu}) = 2\mathrm{Re}\;Q(A\psi_{\nu},B\psi_{\nu}).
              \end{equation}
              Taking absolute values, we have, using the inequality
              \begin{equation}
               |\mathrm{2Re}\;Q(A\psi_{\nu},B\psi_{\nu})|\leq 2\sqrt{Q(A\psi_{\nu})Q(B\psi_{\nu})},
               \end{equation}
                and after summation $\sum_{\nu=1}^{\infty}$,\footnoteA{For all non-negative $a_{n},b_{n}$ one has the inequality $\sqrt{a_{1}b_{1}}+\cdots\sqrt{a_{n}b_{n}}\leq\sqrt{(a_{1}+\cdots+a_{n})(b_{1}+\cdots+b_{n})}$.
                [Squaring both sides, the assertion follows from the inequality $2\sqrt{a_{i}b_{i}a_{j}b_{j}}\leq a_{i}b_{j}+a_{j}b_{i}$, which itself follows, after squaring, from $(a_{i}b_{j}-a_{j}b_{i})^{2}\geq 0$.]}
              \begin{eqnarray}
                [A+B] - [A] - [B] &\leq& \left|\sum_{\nu=1}^{\infty} 2\mathrm{Re}\;Q(A\psi_{\nu},B\psi_{\nu})\right|  \\
                &\leq&  2\sum_{\nu=1}^{\infty}\sqrt{Q(A\psi_{\nu})Q(B\psi_{\nu})}\\
                &\leq& 2\sqrt{\sum_{\nu=1}^{\infty}Q(A\psi_{\nu})\sum_{\nu=1}^{\infty}Q(B\psi_{\nu})} \\
                &=& 2\sqrt{[A][B]}.
       \end{eqnarray}
          Thus
          \begin{eqnarray}
              [A+B]&\leq& [A]+[B]+2\sqrt{[A][B]} = (\sqrt{[A]}+\sqrt{[B]})^{2} \\
              \sqrt{[A+B]} &\leq& \sqrt{[A]}+\sqrt{[B]}.
              \end{eqnarray}
      The third relation can be proved as follows:
      \begin{eqnarray}
        [AB] &=& \sum_{\mu,\nu=1}^{\infty}\left|Q(\varphi_{\mu},AB\psi_{\nu})\right|^{2} \\
        &=& \sum_{\mu,\nu=1}^{\infty}\left|Q(A^{\dagger}\varphi_{\mu}, B\psi_{\nu})\right|^{2} \\
        &\leq& \sum_{\mu,\nu=1}^{\infty}Q(A^{\dagger}\varphi_{\mu})Q(B\psi_{\nu}) \\
        &=& \sum_{\mu=1}^{\infty} Q(A^{\dagger}\varphi_{\mu})\cdot\sum_{\nu=1}^{\infty}Q(B\psi_{\nu}) \\
        &=& [A^{\dagger}][B] = [A][B],\\
        \sqrt{[AB]} &\leq& \sqrt{[A]}\sqrt{[B]}.Q.E.D.
        \end{eqnarray}
        
        \textbf{Theorem 3.} The equation
        \begin{equation}
          [A+B] = [A]+[B]
          \end{equation}
          holds, as long as one of the four equations below is satisfied:
          \begin{equation}
          AB^{\dagger} = 0,\;\;A^{\dagger}B = 0,\;\;BA^{\dagger} = 0,\;\;B^{\dagger}A = 0.
          \end{equation}
          
          \textbf{Proof:} For $A^{\dagger}B=0$ we have
          \begin{equation}
          Q((A+B)\psi_{\nu})-Q(A\psi_{\nu})-Q(B\psi_{\nu}) = 2\mathrm{Re}\;Q(A\psi_{\nu},B\psi_{\nu}) = 2\mathrm{Re}\;Q(\psi_{\nu},A^{\dagger}B\psi_{\nu}) = 0,
          \end{equation}
          which, after summation $\sum_{\nu=1}^{\infty}$ yields
          \begin{equation}
          [A+B]-[A]-[B] = 0 \Rightarrow [A+B] = [A] + [B].
          \end{equation}
          We can replace $A,B$ with $A^{\dagger},B^{\dagger}$, as then from $A^{\dagger}B=0$ we obtain the sufficient condition $A^{\dagger\dagger}B^{\dagger}=AB^{\dagger}=0$; further, we can
          interchange $A$ and $B$, obtaining [the sufficient conditions] $B^{\dagger}A=0$ and $BA^{\dagger}=0$.Q.E.D.
          
          It is clear that
          \begin{equation}
          \sum_{\mu,\nu=1}^{\infty}\left|Q(A\varphi_{\mu}),B\psi_{\nu})\right|^{2} = \sum_{\mu,\nu=1}^{\infty}\left|Q(\varphi_{\mu}),A^{\dagger}B\psi_{\nu})\right|^{2} = [A^{\dagger}B].
          \end{equation}
          The expression on the left (manifestly dependent on $A, B$, but independent of the choice of the $\varphi$ and $\psi$) depends only on $A^{\dagger}B$. It will later play a significant role, and
          we will
          therefore introduce an independent notation for it, to wit:\footnoteB{Von Neumann uses the notation $[A, B]$, which invites an almost unavoidable identification with the commutator for modern readers. We have
          therefore elected to use angle brackets instead. This makes it reminiscent of an inner product, and indeed, we shall shortly see that it satisfies many of the expected properties for such.}
          \begin{equation}
            \langle A, B \rangle \equiv [A^{\dagger}B].
            \end{equation}
            In particular, from $(A^{\dagger}B)^{\dagger} = B^{\dagger}A^{\dagger\dagger} = B^{\dagger}A$ it follows that
            \begin{equation}
              \langle A, B\rangle = \langle B, A\rangle.
              \end{equation}
          From Theorem 2 it follows that:
          \begin{equation}
           \langle aA, bB\rangle = |ab|^{2}\langle A, B\rangle.
           \end{equation}
           Likewise, from Theorem 3 we have
           \begin{equation}
           \label{linbrack}
            \langle A, B+C\rangle = \langle A, B\rangle + \langle A, C\rangle,
            \end{equation}
            when any of the following four conditions is satisfied:
            \begin{eqnarray}
              A^{\dagger}B(A^{\dagger}C)^{\dagger} &=& A^{\dagger}BC^{\dagger}A = 0, \\
              (A^{\dagger}B)^{\dagger}A^{\dagger}C &=& B^{\dagger}AA^{\dagger}C = 0, \\
              A^{\dagger}C(A^{\dagger}B)^{\dagger} &= & A^{\dagger}CB^{\dagger}A = 0, \\
              (A^{\dagger}C)^{\dagger}A^{\dagger}B &=& C^{\dagger}AA^{\dagger}B = 0.
              \end{eqnarray}
              In particular, $BC^{\dagger}=0$ or $CB^{\dagger}=0$ are sufficient [for (\ref{linbrack}) to hold].
         
        We will primarily be concerned with expressions of the type $\langle E, F\rangle$, where $E, F$ are P. Ops. If $E, F$ commute, then $EF$ is also a P. Op., in which case $\langle E, F\rangle$ has a simple geometrical
        interpretation. For in that case
        \begin{equation}
          \langle E, F\rangle = [E^{\dagger}F] = [EF],
          \end{equation}
           and we are therefore just concerned with the meaning of $[E]$, where $E$ is a P.\ Op.\ It is easy to construct a complete orthonormal system of $\psi_{\nu}$ that lies exclusively in the interior or exterior of $E$,\footnoteA{Let
           $f_{1},f_{2},\ldots$ (resp.\ $g_{1},g_{2},\ldots$) be an everywhere dense sequence in the interior (resp.\ exterior) of $E$. By application of Section 6, Theorem 6 we can construct from each orthonormal systems $\rho_{1},\rho_{2},\ldots$
           (resp.\ $\sigma_{1},\sigma_{2},\ldots$). From Section 8, Theorem 6, $\rho_{1},\sigma_{1},\rho_{2},\sigma_{2},\ldots$ is also an orthonormal system, and from Section 8, Theorem 5 it spans an everywhere [in $\bar{\mathfrak{H}}$] dense space,
           i.e., it is complete.} 
           in which case
           \begin{eqnarray}
            [E] &=& \sum_{\nu=1}^{\infty} Q(E\psi_{\nu}) \\
            &=& \sum_{\psi_{\nu}\;\mathrm{in\;interior\;of\;}E} Q(\psi_{\nu}) +  \sum_{\psi_{\nu}\;\mathrm{in\;exterior\;of\;}E} Q(0) \\
            &=& \sum_{\psi_{\nu}\;\mathrm{in\;interior\;of\;}E} 1 \\
            &=& \mathrm{number\; of}\; \psi_{\nu} \;\mathrm{in\;interior\;of\;}E.
            \end{eqnarray}
            But this is just the dimension of the interior of $E$. Consequently, we have, for example, $[1]=\infty$ (as the interior of the identity ! encompasses the whole space $\bar{\mathfrak{H}}$), and $[0]=0$.
            
             Theorems 1 and 2 show that $\sqrt{[A]}$ can indeed be treated as the absolute value of operators. Nevertheless, it is clearly only useful in the immediate neighborhood of 0: $\sqrt{[1]}$ is already infinite.
             
             \section{The statistical postulate of quantum mechanics.}
             
             We are now in a position to undertake the program of particular interest here: the mathematically unobjectionable unification of statistical quantum mechanics. To this end we shall start by considering the
             simplest possible case, one with unambiguously clear results, and then translate these (known) results into our [mathematical] language: this will then provide the clue to the method which should be
             employed in the general case.

            We consider therefore the following extremely simple case arising in the continuous realization [i.e., in $\mathfrak{H}$, or $L^2$]: 
            
            Let $\Omega$ be $k-$dimensional Euclidean space, and consider a
            quantum-mechanical system for which the Schr\"odinger equation reads
            \begin{equation}
             H\psi - l\psi = 0.
             \end{equation}
              As is well known,\footnoteA{Cf.\ for example note 3.} the symmetric linear operator $H$ arises in the following manner: one takes the classical-mechanical expression for the energy as a function of the coordinates $q_{1},q_{2},\ldots,q_{k}$ and the
              momenta $p_{1},p_{2},\ldots,p_{k}$, and replaces each $q_{\mu}$ with the operator $q_{\mu}\cdot\ldots$ [i.e., multiplication by $q_{\mu}$], and each $p_{\mu}$ by the operator $\frac{h}{2\pi i}\frac{\partial}{\partial q_{\mu}}\cdot\ldots$.
              (Here there arises some ambiguity, as the $q_{\mu},p_{\nu}$ commute multiplicatively with one another as normal numbers, in contrast to the operators $q_{\mu}\cdot\ldots$ and $\frac{h}{2\pi i}\frac{\partial}{\partial q_{\mu}}\cdot\ldots$
              Clearly one has the constraint on $H$ that it be symmetric; however, this does not suffice to determine it uniquely. This constitutes one of the essential deficiencies in quantum theory.)
              
              Now let us assume that only a non-degenerate (i.e., consisting of only simple eigenvalues) point spectrum is present, and call the eigenvalues $l_{1},l_{2},\ldots,$ and the corresponding (normalized) eigenfunctions
              $\varphi_{1},\varphi_{2},\ldots$ 
              
              We have already determined in Section 10 the partition of unity corresponding to this $H$. It is:
              \begin{equation}
                E(l)f = \sum_{l_{n}\leq l} Q(f,\varphi_{n})\cdot\varphi_{n}.
                \end{equation}
                
                Further, we need the partition of unity $F_{\mu}(l)$ corresponding to the operator $q_{\mu}$. This is a trivial generalization of the final example considered in Section 10 (where $k$ was simply 1), namely
                \begin{eqnarray}
                  F_{\mu}(l)f(q_{1},\ldots,q_{k}) &=& f(q_{1},\ldots,q_{k})\;\;\mathrm{for}\;q_{\mu}\leq l, \\
                  &=& 0\;\;\mathrm{for}\;q_{\mu}>l.
                  \end{eqnarray}
               (This can be shown by exactly the same qualitative consideration as previously; once again, one easily establishes the result
               \begin{equation}
                  q_{\mu}f(q_{1},\ldots,q_{k}) = \int_{-\infty}^{\infty} l d\{ F_{\mu}(l)f(q_{1},\ldots,q_{k})\},
                  \end{equation}
                  which holds by definition [of the $F_{\mu}(l)$].   
                  
                  The probability postulate of Pauli and Dirac\footnoteA{Cf.\ for example the work of Jordan cited in note 7.}\footnoteB{The  identification of (\ref{Bornrule}) with a probability for observing a particle in a given region of space
                  goes back to Born's examination of collision phenomena in Born \autocite*{Born:1926a,Born:1926b}. It is now universally referred to as the Born rule. Further elaborations of the probabilistic interpretation of
                  wave functions by Pauli (primarily in private correspondence with Heisenberg) were acknowledged by Jordan \autocite*{Jordan:1927b}; 
                  Dirac's \autocite*{Dirac:1927a} development of  statistical transformation theory occurred more or less simultaneously with, and independently
                  of, Jordan's.}
                  applicable in this case now takes the form: if the system is in the $n$-th quantum state $(l_{n},\varphi_{n})$, then the probability that the coordinates
                  $q=(q_{1},\ldots,q_{k})$ lie in the $k$-dimensional region $K$ is (writing $dq$ for $dq_{1}\cdots dq_{k}$)
                  \begin{equation}
                  \label{Bornrule}
                   \int_{K} |\varphi_{n}(q)|^{2}dq.
                   \end{equation}
                   We can further generalize this Ansatz. If we only know that the energy lies in the interval $I$, then the probability (apart from normalization factors) is given by
                   \begin{equation}
                     \sum_{l_{n}\in I}\int_{K}|\varphi_{n}(q)|^{2}dq.
                     \end{equation}
                      Indeed: if only a single eigenvalue lies in $I$, this follows from our earlier assertion; while if several are present, it follows from this case if we assume (as is usual) the single non-degenerate quantum states are to
                      be regarded as a priori equally probable. 
                      
                  This expression, in the event that $K$ (resp.\ $I$) is defined by the inequalities $q_{1}^{\prime}<q_{1}\leq q_{1}^{\prime\prime}, q_{2}^{\prime}< q_{2}\leq q_{2}^{\prime\prime},\ldots,q_{k}^{\prime}< q_{k}\leq q_{k}^{\prime\prime}$
                  (resp.\ $l^{\prime}< l\leq l^{\prime\prime}$), can however be written also in the following fashion [to improve readability, we replace von Neumann's notation $F_1 \cdot \ldots$ etc.\ by $F_1 * \ldots$ etc.]
                  \begin{eqnarray}
                     \sum_{l_{n}\in I}\int_{K}|\varphi_{n}(q)|^{2}dq &=& \sum_{l_{n}\in I}\int_{q_{1}^{\prime}}^{q_{1}^{\prime\prime}}\cdots\int_{q_{k}^{\prime}}^{q_{k}^{\prime\prime}}|\varphi_{n}(q_{1},\ldots,q_{k})|^{2}dq_{1}\cdots dq_{k}  \nonumber \\
                     &=&  \sum_{l_{n}\in I}\int_{-\infty}^{\infty}\cdots\int_{-\infty}^{\infty}|F_{1}(q_{1}^{\prime},q_{1}^{\prime\prime})*\ldots*F_{k}(q_{k}^{\prime},q_{k}^{\prime\prime})\varphi_{n}(q_{1},\ldots,q_{k})|^{2}dq_{1}\cdots dq_{k}  \nonumber \\
                     &=& \sum_{l_{n}\in I}\int_{\Omega}|F_{1}(q_{1}^{\prime},q_{1}^{\prime\prime})*\ldots*F_{k}(q_{k}^{\prime},q_{k}^{\prime\prime})\varphi_{n}(q)|^{2}dq  \nonumber \\
                     &=& \sum_{n=1}^{\infty} \int_{\Omega}|F_{1}(q_{1}^{\prime},q_{1}^{\prime\prime}*\ldots*F_{k}(q_{k}^{\prime}q_{k}^{\prime\prime})\cdot E(l^{\prime},l^{\prime\prime})\varphi_{n}(q)|^{2}dq \nonumber \\
                     &=& \sum_{n=1}^{\infty} Q(F_{1}(q_{1}^{\prime},q_{1}^{\prime\prime})*\ldots*F_{k}(q_{k}^{\prime},q_{k}^{\prime\prime})\cdot E(l^{\prime},l^{\prime\prime})\varphi_{n}(q))  \nonumber \\
                     &=& [F_{1}(q_{1}^{\prime},q_{1}^{\prime\prime})*\ldots*F_{k}(q_{k}^{\prime},q_{k}^{\prime\prime})\cdot E(l^{\prime},l^{\prime\prime})].
                     \end{eqnarray}
                     Now one may easily check that the $F_{\mu}(q_{\mu}^{\prime},q_{\mu}^{\prime\prime})$ commute with one another, from which one further concludes that
                     \begin{eqnarray}
                     \sum_{l_{n}\in I}\int_{K}|\varphi_{n}(q)|^{2}dq &=& \langle (F_{1}(q_{1}^{\prime},q_{1}^{\prime\prime})*\ldots*F_{k}(q_{k}^{\prime},q_{k}^{\prime\prime})^{\dagger}, E(l^{\prime},l^{\prime\prime})\rangle \nonumber \\
                     &=& \langle F_{k}(q_{k}^{\prime},q_{k}^{\prime\prime})*\ldots*F_{1}(q_{1}^{\prime},q_{1}^{\prime\prime}), E(l^{\prime},l^{\prime\prime})\rangle \nonumber \\
                     &=& \langle F_{1}(q_{1}^{\prime},q_{1}^{\prime\prime})*\ldots*F_{k}(q_{k}^{\prime},q_{k}^{\prime\prime}),E(l^{\prime},l^{\prime\prime})\rangle.
                    \end{eqnarray}       
                    In other words: if the energy lies in $I$, the (relative) probability that $q_{1}$ lies in $J_{1}$, $q_{2}$ lies in $J_{2}$, etc. is given by
                    \begin{equation}
                      \langle E(I), F_{1}(J_{1})*\ldots*F_{k}(J_{k})\rangle.
                      \end{equation} 
                      
                      But inversely, by a postulate of Jordan, the quantity
                      \begin{equation}
                     \sum_{l_{n}\in I}\int_{K}|\varphi_{n}(q)|^{2}dq.
                     \end{equation}
                      may also be regarded as the probability that the $l_{n}$, i.e., the energy, lies in $I$, if $q$  lies in $K$ (i.e., $q_{1}$ in $J_{1}$,\ldots,$q_{k}$ in $J_{k}$---cf.\ note 7). Or more exactly: the Pauli-Jordan assertion is relevant only in
                      the limit of infinitely small $K$ (where, after dividing by the proportionality factor of the ``volume of $K$'' one is left with the probability [density] $\sum_{l_{n}\in I}|\varphi_{n}(q)|^{2}$). Nevertheless, a correct result also emerges
                      in the opposite limit $K=\Omega$ (i.e., the entire space):
                      \begin{equation}
                      \sum_{l_{n}\in I}\int_{\Omega}|\varphi_{n}(q)|^{2}dq = \sum_{l_{n}\in I} 1 = \mathrm{number\;of}\;l_{n}\;\mathrm{in}\;I,
                      \end{equation}
                      i.e., all quantized (non-degenerate) states are a-priori equiprobable, and un-quantized states are impossible---both of which assertions belong to the basic assumptions of quantum theory.
                      
                      We therefore conclude: if $q_{1}$ lies in $J_1$,\ldots, $q_{k}$ in $J_{k}$, then the (relative) probability that the $l_{n}$ (the energy) lies in $I$ is given by the same expression as previously, which we will in this case write as
                      follows:
                      \begin{equation}
                        \langle F_{1}(J_{1})*\ldots*F_{k}(J_{k}), E(I)\rangle.
                        \end{equation}
                        
                        \section{The statistical postulate of quantum mechanics (contd.)}
                      
                      The results just obtained suggest the following postulate: let $R_{1},R_{2},\ldots,R_{i}$ and $S_{1},S_{2},\ldots,S_{j}$ be symmetric linear operators, representing particular sensible physical quantities. (We cannot more closely 
                      examine here the exact meaning of this last concept of ``representation of a quantity by an operator'', which at the moment is absolutely fundamental  in quantum mechanics; cf.\ the discussions at the beginning of 
                      preceding sections on the relation of the classical-mechanical Hamiltonian function and the ``energy operator'' $H$.)  The partitions of unity associated with the $R_{\mu}, \;\mu=1,2,\ldots,i$ (resp.\ $S_{\nu},\; \nu=1,2,\ldots,j$)
                      will be denoted  $E_{\mu}(l)$ (resp.\ $F_{\nu}(l)$).\footnoteB{The mathematical apparatus developed at great length in this paper finally is brought to bear on the central interpretational problem of quantum mechanics:
                      how are the formal elements of the theory linked to phenomenologically observable quantities? The key is the use of projection operators. On the formal side, they possess the following desirable properties, identified
                      by von Neumann from the outset as absent in the Dirac-Jordan theory: (a) they are unambiguous (no phase ambiguities), (b) they are \emph{bounded operators}, which are defined on all elements of the Hilbert space and for which 
                      multiplication operations can be rigorously defined---in particular, there is no appeal to ``improper functions'', such as the delta function. On the phenomenological side, they correspond naturally to the inescapable limitations
                      of actual physical measurements, which establish the value of observed quantities only in finite ranges, and not with infinitesimal exactitude. Von Neumann's elegant rephrasing of the Born rule in the language of
                      projection operators is the direct precursor to von Neumann \autocite*{VonNeumann:1927b}, where the entire statistical formalism outlined here is built up inductively from a few, apparently innocuous, basic assumptions on measurements carried
                      out on ensembles of quantum systems.}
                      
                      We assume that all the $E_{\mu}(l)$ commute with each other, as do all the $F_{\nu}(l)$.\footnoteB{Here, the notion of ``simultaneous measurement of compatible observables'', which had first been exhibited clearly
                      in Heisenberg's \autocite*{Heisenberg:1927b} ``uncertainty paper'', is introduced: von Neumann wants to derive the conditional probabilities which apply for the most general measurement which are (at least
                      in principle) possible.}
                       This property will be called the ``complete commutability'' of the $R_{1},R_{2},\ldots,R_{i}$ (likewise, of the $S_{1},S_{2},\ldots,S_{j}$).
                      (For complete commutability of two operators $T^{\prime}, T^{\prime\prime}$, the conventional commutability property, is clearly necessary; it is also sufficient, provided at least one of the operators is bounded. If both
                      are unbounded, certain difficulties of a formal nature arise, which we shall not discuss here. For all operators appearing in quantum mechanics this is the case.)
                      
                      We now make the following physical assumption:
                      \begin{quote}
                       If the quantities represented by $R_{1},\ldots,R_{i}$ have assumed values which lie in the intervals $I_{1},\ldots,I_{i}$, then the (relative) probability that quantities represented by $S_{1},\ldots,S_{j}$ lie in the intervals
                       $J_{1},\ldots,J_{j}$ is given by
                       \begin{equation}
                       \langle E_{I_{1}}*\ldots*E_{I_{i}}, F_{J_{1}}*\ldots*F_{J_{j}}\rangle = \langle E_{I_{1}}*\ldots*E_{I_{i}}*F_{J_{1}}*\ldots*F_{J_{j}}\rangle.
                       \end{equation}
                       The equality holds because the $E_{I_{\nu}}$ all commute with one another.
                       \end{quote}
                       We shall now draw some conclusions from this assumption, in order to indicate its usefulness:
                       
                       $\alpha.$ Both the premises ($(R_{1},\ldots,R_{i}$) as well as the conclusions  $(S_{1},\ldots,S_{j}$) may be freely interchanged, without altering the probability distribution. This follows from the commutability with each other
                       of the $E_{\mu}(l)$ (resp.\ the $F_{\nu}(l)$), whence also of differences of the $E_{\mu}(l)$ (resp.\ the $F_{\nu}(l)$).
                       
                       $\beta.$ Interchange of all premises with all conclusions changes nothing (i.e., the probability distribution behaves as if it arises from a priori probabilities). This follows from the generally valid formula $\langle A,B\rangle = \langle B,A\rangle$.
                       
                       $\gamma.$ Tautologous premises or conclusions may be                       
                   freely introduced or removed (i.e., those for which the interval is $(-\infty,\infty)$). For they occasion just the appearance or removal of a factor of
                       \begin{equation}
                          E_{\mu}(-\infty,\infty)\;\mathrm{or}\;F_{\nu}(-\infty,\infty) = 1-0 = 1
                          \end{equation}
                          
                      $\delta.$ The multiplication law for probabilities does not in general hold (rather, a weaker law holds, corresponding to the ``superposition of probability amplitudes'' of Jordan---see note 7---which we will not discuss
                      further here).\footnoteB{In von Neumann \autocite*{VonNeumann:1927b}, this assertion (of the invalidity of the multiplication law in quantum mechanics) is withdrawn: the rules of probability are mathematical certainties, and their
                      application to quantum mechanics, if carried out with sufficient care, is unobjectionable.} This is hardly surprising, given that the dependency relations of our probabilities can be arbitrarily complicated; furthermore, one is dealing with relative probabilities.
                      
                      $\varepsilon.$ The addition law for probabilities is valid. (Indeed, this is valid in the conventional probability calculus without regard to dependencies.) We must show that from $J_{j}^{\prime}+J_{j}^{\prime\prime}=J_{j}$
                      follows
                      \begin{eqnarray}
                      &&[E_{1}(I_{1})*\ldots*E_{i}(I_{i})*F_{1}(J_{1})*\ldots*F_{j}(J_{j}^{\prime})]+[E_{1}(I_{1})*\ldots*E_{i}(I_{i})*F_{1}(J_{1})*\ldots*F_{j}(J_{j}^{\prime\prime})] \nonumber \\
                      &=& [E_{1}(I_{1})*\ldots*E_{i}(I_{i})*F_{1}(J_{1})*\ldots*F_{j}(J_{j})].
                      \end{eqnarray}
                      (From $\alpha.$ and $\beta.$ we can certainly confine ourselves to the case where the last interval, $J_{j}$, can be split [into two subintervals].) This can also be written
                      \begin{equation}
                         [AF_{j}(J_{j}^{\prime})] +  [AF_{j}(J_{j}^{\prime\prime})] =  [AF_{j}(J_{j})].
                      \end{equation}
                      Using Theorem 3, Section 11, this is certainly the case provided
                      \begin{eqnarray}
                          AF_{j}(J_{j}^{\prime}) +  AF_{j}(J_{j}^{\prime\prime}) &=&  AF_{j}(J_{j}), \\
                          AF_{j}(J_{j}^{\prime})(AF_{j}(J_{j}^{\prime\prime}))^{\dagger} &=& AF_{j}(J_{j}^{\prime})F_{j}(J_{j}^{\prime\prime})A^{\dagger} = 0.
                          \end{eqnarray}
                          The first equation follows from $F_{j}(J_{j}^{\prime})+F_{j}(J_{j}^{\prime\prime})=F_{j}(J_{j})$, but the second follows as well, as then we must have that $F_{j}(J_{j}^{\prime})$ and $F_{j}(J_{j}^{\prime\prime})$ are
                          orthogonal [projection operators] (cf.\ Theorem 2, Section 8). However, if $J_{j}^{\prime},J_{j}^{\prime\prime},J_{j}$ are the intervals $(l^{\prime},l^{\prime\prime}), (l^{\prime\prime},l^{\prime\prime\prime}),
                          (l^{\prime},l^{\prime\prime\prime})$ [with $l^{\prime}\leq l^{\prime\prime}\leq l^{\prime\prime\prime}$], then it is clear that
                          \begin{eqnarray}
                            F_{j}(J_{j}^{\prime}) &=& F_{j}(l^{\prime\prime})-F_{j}(l^{\prime}),  \\
                            F_{j}(J_{j}^{\prime\prime}) &=& F_{j}(l^{\prime\prime\prime})-F_{j}(l^{\prime\prime}),  \\ 
                            F_{j}(J_{j}) &=& F_{j}(l^{\prime\prime\prime})-F_{j}(l^{\prime}).
                            \end{eqnarray}
                            
                            $\vartheta.$ Our expression for probabilities is invariant with respect to canonical transformations. By a canonical transformation we mean the following (cf.\ references in note 7):
                            let $U$ be a linear operator with the property $UU^{\dagger} = U^{\dagger}U = 1$: in this case we call $U$ ``orthogonal'' [modern term: unitary]. A canonical transformation consists
                            of replacing every linear operator $R$ by $URU^{\dagger}$. With regard to such a transformation, the operations $aR, R+S, RS, R^{\dagger}$ are invariant, and therefore also the properties of being
                            symmetric [self-adjoint], of being a projection operator, and, between projection operators, the relations $\leq$ and of orthogonality. Further, $Q$ [the inner product] is invariant if $\bar{\mathfrak{H}}$ is
                            mapped onto itself by $U$:
                            \begin{equation}
                               Q(Uf, Ug) = Q(U^{\dagger}Uf, g) = Q(f,g),
                               \end{equation}
                               so that $U\varphi_{1},U\varphi_{2},\ldots$ is a complete orthonormal system if $\varphi_{1},\varphi_{2},\ldots$ is one, from which the invariance of [the operator absolute value] $[A]$ follows. Additionally, the invariance
                               of the eigenvalue [spectral] representation is manifest (cf.\ Section 9). 
                               
                               As none of the concepts used by us is changed [by a canonical transformation], the same holds for the (relative) probabilities derived on the basis of these.
                            \section{Applications}
                            
                            We now consider some physical applications. First of all, [take] the case of Schr\"odinger's equations. These were already discussed in the case of non-degenerate systems in Section 12,
                            and if we also admit degeneracies---that is, multiple eigenvalues---nothing in the results there obtained is changed. The a priori probabilities of the single [i.e., non-degenerate] eigenvalues, for
                            example, are (cf.\ the characterization given in Section 12)
                            \begin{eqnarray}
                            \langle F_{1}(-\infty,\infty)*\ldots*F_{k}(-\infty,\infty), E(I)\rangle &=& \langle 1, E(I) \rangle = [E(I)]  \nonumber \\
                            &=& \mathrm{dimension\;of\;the}\;f(q)\;\mathrm{lying\;in\;the\;interior\;of}\;E(I).\nonumber \\
                            &&
                            \end{eqnarray}
                            In the interior of $E(I)$ however, as we know, are found all linear combinations
                            \begin{equation}
                             a_{1}\varphi_{\nu_{1}}(q)+a_{2}\varphi_{\nu_{2}}(q)+\cdots,
                             \end{equation}
                             where $l_{\nu_{1}},l_{\nu_{2}},\ldots$ are the eigenvalues of $H$ lying in the interval $I$ (multiple eigenvalues counted separately), and $\varphi_{\nu_{1}},\varphi_{\nu_{2}},\ldots$ are the associated
                             eigenfunctions. The dimension indicated is therefore just the number of eigenvalues in $I$.
                             
                             This result also implies that the a priori probability of a quantized state is the multiplicity of the eigenvalues for that state, and unquantized states are impossible. 
                             
                             Secondly, let us consider the case of sharp and causal dependencies.[Von Neumann's use of the word ``causal'' in the following is somewhat jarring. Here there is no implication of
                             of temporal sequence: von Neumann is simply referring to a situation in which the determination of the value of one quantity immediately determines, through a unique functional
                             relationship, the value of the other. Thus, a measurement of the distance $r$ of an electron from the nucleus of a hydrogen atom would immediately determine the electrostatic
                             potential $-e^{2}/r$, and vice-versa.] We set $i=j=1$ and assume that a certain quantity is given, while we seek another, which is a function of the first---i.e., is causally
                             determined by it. The corresponding operators we can denote $R$ (resp.\ $S$), with $S=f(R)$, where $f(x)$ is a real function. We assume, that $f(x)$ is monotone increasing. (The restrictions $i=j=1$,
                             and the monotonicity of $f(x)$ are not essential, but are assumed, only for the sake of orientation, to simplify the calculation). 
                             
                             Let the partition of unity associated with $R$ be $E(l)$:
                             \begin{equation}
                                 R = \int_{-\infty}^{\infty} ldE(l).
                                 \end{equation}
                                 As we showed in Section 10, one then has
                                 \begin{equation}
                                 R^{n} = \int_{-\infty}^{\infty}l^{n}dE(l),
                                 \end{equation}
                                 whence
                                 \begin{equation}
                                 S = f(R) = \int_{-\infty}^{\infty}f(l)dE(l) = \int_{-\infty}^{\infty}l^{\prime}dE(g(l^{\prime})),
                                 \end{equation}
                                 where $g$ is the inverse of $f$ [i.e., $f(g(l)) = l$]. Accordingly, $E(g(l))$ is the partition of unity associated with $S$.
                                 
                                 If $J$ is the interval $(l^{\prime},l^{\prime\prime})$, then we will denote the interval $(g(l^{\prime}),g(l^{\prime\prime})$ by $g(J)$. One thus has [for the partition of unity associated with $S$]:
                                 \begin{equation}
                                 F(J) = E(g(J)),
                                 \end{equation}
                                 and consequently,
                                 \begin{equation}
                                 \langle E(I), F(J)\rangle = [E(I)F(J)] = [E(I)E(g(J)] = E(I\cap g(J)).
                                 \end{equation}
                                 (Here $I\cap g(J)$ is the intersection of the intervals $I$ and $g(J)$; one easily verifies the identity which ensures always $E(I)E(J^{\prime}) = E(I\cap J^{\prime})$.)
                                 
                                 If the intervals $I$, $g(J)$ have no intersection, this is equal to zero; but $I\cap g(J)$ is empty just when there exists no $x$ in $I$ for which $f(x)$ is in $J$. Otherwise said: values
                                 in contradiction to a causal connection do not appear.
                                 
                                 If $I\cap g(J)$ is not empty, then it corresponds to the interval $I^{\prime}$ of all $x$ in $I$, for which $f(x)$ lies in $J$, i.e., those values allowed on the basis of the causal connection.
                                  We therefore have the probability relation:
                                  \begin{eqnarray}
                                  [E(I^{\prime})] &=& \mathrm{dimension\;of\;the}\;f\;\mathrm{lying\;in}\;E(I^{\prime}),\;\mathrm{or,\;as\;we\;know,} \nonumber \\
                                  &=& \mathrm{number\;of\;eigenvalues\;of\;R\;lying\;in}\; I^{\prime}.
                                  \end{eqnarray}
                                  In summary: in the context of causal connections one obtains, if $R$ is quantized, the usual quantum-theoretical result. (Besides, it is clear that this probability becomes infinite if 
                                  $R$ has a continuous spectrum which intrudes in $I^{\prime}$.)
                                  
                                  With these two examples we believe that we have demonstrated that statistical quantum mechanics, despite its probabilistic character, is very well capable of making sharp and  binding
                                  assertions, as long as the occasion for such arises, e.g., in the case of absolute quantum prohibitions [e.g., selection rules] and causal conditions.
                                  
                                  Finally, we consider the quantum-mechanical treatment of time-dependent systems as given by Born.\footnoteA{This example I owe to an observation of L. Nordheim.} The Hamiltonian function of the system depends explicitly on time, as does the
                                  associated operator $H(t)$ (time is handled as a number, not as an operator ``quantity''). Let us exclude degeneracies, and denote the eigenfunctions at time $t_0$ by
                                  \begin{equation}
                                    \varphi_{1}^{(0)},\varphi_{2}^{(0)},\ldots
                                    \end{equation}
                                    and at time $t$ by
                                    \begin{equation}
                                      \varphi_{1}^{(t)},\varphi_{2}^{(t)},\ldots
                                      \end{equation}
                                      
                                  Then, according to Born,\footnoteA{\emph{Zeitschrift f\"ur Physik}, Vol.\ 38, p.\ 803; Vol.\ 40, p.\ 167 (1926).} the probability that the system is in the $\nu$-th state at time $t$, 
                                  given that it was in the $\mu$-th state at time $t_0$, is $|c_{\mu,\nu}(t)|^{2}$, where
                                  $c_{\mu,\nu}(t)$ is the $\mu$-th expansion coefficient of $\varphi_{\nu}^{(t)}$ with respect to the $ \varphi_{1}^{(0)},\varphi_{2}^{(0)},\ldots$:
                                  \begin{equation}
                                     \varphi_{\mu}^{(t)} = \sum_{\nu=1}^{\infty}c_{\mu,\nu}(t)\varphi_{\nu}^{(0)}.
                                     \end{equation}
                                     
                                  Indeed, suppose that $E_{0}(l)$ is the partition of unity for $H(t_0)$, and $E_{t}(l)$ that  for $H(t)$. As we saw in Section 10, the interior of $E_0(l)$ (resp.\ $E_{t}(l)$) consists
                                  of the linear combination of all $\varphi_{\mu}^{(0)}$ (resp.\ $\varphi_{\mu}^{(t)}$) with eigenvalues $\leq l$. Consequently the interior of $E_{0}(I)$ (resp.\ $E_{t}(I)$) consists of the
                                  linear combination of all $\varphi_{\mu}^{(0)}$ (resp.\ $\varphi_{\mu}^{(t)}$) with eigenvalues lying in the interval $I$. 
                                  
                                  Now suppose that [the interval] $I_{\mu}$ contains a single eigenvalue of $H(0)$, the $\mu$-th, and $J_{\nu}$ a single eigenvalue of $H(t)$, the $\nu$-th.   Then the
                                  interior of $E_{0}(I_{\mu})$ (resp.\ $E_{t}(J_{\nu})$) consists of [the one dimensional space consisting of] multiples of $\varphi_{\mu}^{(0)}$ (resp.\ $\varphi_{\nu}^{(t)}$), so that
                                  \begin{eqnarray}
                                     E_{0}(I_{\mu})f &=& Q(f,\varphi_{\mu}^{(0)})\cdot\varphi_{\mu}^{(0)}, \\
                                     E_{t}(J_{\nu})f &=& Q(f, \varphi_{\nu}^{(t)})\cdot\varphi_{\nu}^{(t)}.
                                     \end{eqnarray}
                                  We now calculate $\langle E_{0}(I_{\mu}), E_{t}(J_{\nu})\rangle$, using as a complete orthonormal system $\varphi_{1}^{(0)},\varphi_{2}^{(0)},\ldots$:
                                  \begin{eqnarray}
                                       \langle E_{0}(I_{\mu}), E_{t}(J_{\nu})\rangle &=& [E_{t}(J_{\nu})E_{0}(I_{\mu})]  \\
                                       &=& \sum_{\rho=1}^{\infty} Q(E_{t}(J_{\nu})E_{0}(I_{\mu})\varphi_{\rho}^{(0)})  \\
                                       &=& Q(E_{t}(J_{\nu})\varphi_{\mu}^{(0)})  \\
                                       &=& Q(Q(\varphi_{\mu}^{(0)},\varphi_{\nu}^{(t)})\cdot\varphi_{\nu}^{(t)})  \\
                                       &=& |Q(\varphi_{\mu}^{(0)},\varphi_{\nu}^{(t)})|^{2},
                                       \end{eqnarray}                    
                                  which is precisely Born's expression (given that $c_{\mu,\nu}(t) = Q(\varphi_{\nu}^{(t)},\varphi_{\mu}^{(0)}) = Q^{*}(\varphi_{\mu}^{(0)},\varphi_{\nu}^{(t)})$).
                                  
                                  \section{Summary}
                                  
                                 It is apparent that our postulate of the double nature of dynamics (continuous--discontinuous) is justified by the association of a partition of unity into projection operators,
                                 $E(l^{\prime},l^{\prime\prime})= E(l^{\prime\prime})-E(l^{\prime}), l^{\prime}\leq l^{\prime\prime}$, to each physical quantity (and to the system from which it derives).
                                 
                                 The function $E(l)$ is monotone nondecreasing (in the sense appropriate for projection operators), and this monotonicity displays all the characteristics familiar from the study
                                 of atoms: its growth from 0 (for $l=-\infty$) to 1 (for $l=\infty$) can proceed in discrete jumps (quantized states), or continuously (un-quantized states), and in between there can also
                                 be intervals where it is constant (forbidden states). The assertion ``the quantity associated with $R$ has a value lying in the interval $l^{\prime}<x\leq l^{\prime\prime}$'' is represented in
                                 out calculus by the projection operator $E(l^{\prime},l^{\prime\prime})$.
                                 
                                 In the event that several assertions are made, e.g., if the values of the quantities $R_1,R_2,\ldots,R_{i}$ lie in the intervals $I_{1},I_{2},\ldots,I_{i}$, while the values of the quantities $S_{1},S_{2},\ldots,S_{j}$
                                 lie in the intervals $J_{1},J_{2},\ldots,J_{j}$, one must form the product
                                 \begin{equation}
                                  E_{1}(I_{1}*\ldots*E_{i}(I_{i})*F_{1}(J_{1})*\ldots*F_{j}(J_{j}),
                                  \end{equation}
                                  (which is the form in which the multiplication law for probabilities is to be taken over [in quantum mechanics]); the absolute square [of this product]
                                  \begin{equation}
                                  [ E_{1}(I_{1}*\ldots*E_{i}(I_{i})*F_{1}(J_{1})*\ldots*F_{j}(J_{j})] = \langle E_{1}(I_{1})*\ldots*E_{i}(I_{i}), F_{1}(J_{1})*\ldots*F_{j}(J_{j})\rangle
                                  \end{equation}
                                  is then the (relative) probability of the simultaneous validity [of these assertions]. Note that in this calculation it is not at all necessary to introduce a cause/effect distinction. Such a distinction
                                  will subsequently appear automatically as a consequence of the commutation relations: if all the $E_{\mu}(I_{\mu})$ on the one hand, and the $F_{\nu}(J_{\nu})$ on the other, all commute with each
                                  other, then the product decomposes automatically into these two groups. Within each of these groups the order of factors is insignificant ($\alpha.$ in Section 13), as is the order of the two groups taken
                                  as a whole ($\beta.$ in Section 13), i.e., what one regards as the cause and what one regards as effect.
                                  
                                  Of course the commutation relations do not on their own necessarily determine the cause--effect distinction uniquely: if, for example, a $E_{\rho}$ commutes with all the $F_{\nu}$, it can be freely assigned
                                  to either one of the groups.
                                  
                                  On the other hand, certain assignments are excluded from the outset: it is not possible in all cases to enforce the absolutely sharp connection which as  a rule applies for mutually commuting quantities (cf.\ the
                                  second example in Section 14), in particular to observe (as one does in classical mechanics) all coordinates and their conjugate momenta, and to make them individually and collectively ``causes'', i.e., $R_{\mu}$s.
                                  (Dirac was the first to point out the impossibility of such a procedure.) For it is well known that the operators of a coordinate and its conjugated momentum do not commute
                                   ($q_{\mu}\cdot \ldots$ and $\frac{h}{2\pi i}\frac{\partial}{\partial q_{\mu}}\cdot \ldots$), so that they must unavoidably be separated: one must be the cause, the other the effect (resp.\ the observed and the prescribed).
                                  Even observing everything does not help: quantum mechanics (which encompasses practically everything that we now know with exactitude about atoms) simply affords no way to frame the question!
                                  
                                  It should be mentioned in conclusion that the material presented here does not by any means exhaust the applicability of our method. We will return to this on another occasion, as well as to certain formal
                                  mathematical questions which remain unresolved here. 
                                  
                                  \section{Appendix 1}
                                  Let $w$ be an eigenvalue [of $H$], and $(x_1,x_2,\ldots)$ a sequence which is transformed by $H$ into itself, multiplied by $w$. By application of $S^{-1}$ let this sequence be transformed to $(y_1,y_2,\ldots)$. Under
                                 these circumstances one has that $(y_{1},y_{2},\ldots)$ is transformed by $S^{-1}HS$ into itself, multiplied by $w$. Now $S^{-1}HS = W$ [=diagonal matrix] transforms $(y_1,y_2,\ldots)$ into $(w_{1}y_{1},w_{2}y_{2},\ldots)$,
                                  so all $y_{\mu}$ must =0 unless $w_{\mu}=w$. Consequently, if $w$ differs from all $w_1,w_2,\ldots$, then all $y_{\mu}$, and therefore all $x_{\mu}$, must be zero---hence, $w$ cannot be an eigenvalue. 
                                  
                                  Conversely, if $w$ is equal to some of the $w_{\mu}$, let's say $w_{\mu^{\prime}},w_{\mu^{\prime\prime}},\ldots,$ then $y_{\mu^{\prime}},y_{\mu^{\prime\prime}},\ldots,$ can take arbitrary values (all other
                                 $y_{\mu}=0$). The $(x_{1},x_{2},\ldots)$ arise from  $(y_{1},y_{2},\ldots)$ by application of $S$, so
                                  \begin{equation}
                                    x_{\mu} = \sum_{\mu=1}^{\infty} s_{\mu\nu}y_{\nu} = s_{\mu\mu^{\prime}}y_{\mu^{\prime}}+s_{\mu\mu^{\prime\prime}}y_{\mu^{\prime\prime}}+\cdots.
                                    \end{equation}
                                   Therefore $(x_{1},x_{2},\ldots.)$ can in fact be an arbitrary linear combination of the columns $(s_{1\mu^{\prime}},s_{2\mu^{\prime}},\ldots)$, $(s_{1\mu^{\prime\prime}},s_{2\mu^{\prime\prime}},\ldots)$,\ldots, and of no others.
                                    
                                    \section{Appendix 2}
                                    We wish to show that the space $\mathfrak{H}$ of all functions $f(P)$ defined on $\Omega$ ($\Omega$ an arbitrary $k$-dimensional surface in $l$-dimensional space, $P$ an arbitrary point
                                    in $\Omega$, $dv$ the differential volume element in $\Omega$) with finite $\int_{\Omega}|f(P)|^{2}dv$ satisfies the conditions A.--E. of Section 4. For the standard Hilbert space (all
                                    sequences $x_{1},x_{2},\ldots$ with finite $\sum_{n=1}^{\infty}|x_{n}|^{2}$) the proof is superfluous, as, according to Section 6 [last paragraph], every space with the properties A.--E. has the properties of $\mathfrak{H}_{0}$,
                                    whence $\mathfrak{H}_{0}$ itself must satisfy A.--E.\footnoteB{See the final paragraph of Section 6 (not Section 5, as referenced by von Neumann). The complete isomorphism of $\mathfrak{H}$ with
                                    $\mathfrak{H}_{0}$ is a part of the rigorous mathematical framing of the complete equivalence of wave and matrix mechanics.}
                                    
                                    The definition of the operations $af$ and $f+g$ is clear; that $f$ belonging to $\mathfrak{H}$ implies $af$ belonging to $\mathfrak{H}$ is obvious, while $f,g$ belonging to $\mathfrak{H}$ implies that
                                    $f+g$ belongs to $\mathfrak{H}$ as a consequence of the inequality given in note 16.[A. is now established.] We must define $Q(f,g)$ as $\int_{\Omega}f(P)g^{*}(P)dv$, the motivation for which was given in Section 3. This
                                    gives a finite number for all $f,g$ of $\mathfrak{H}$ as
                                    \begin{equation}
                                      |f(P)g^{*}(P)| \leq \frac{1}{2}|f(P)|^{2}+\frac{1}{2}|g(P)|^{2},
                                      \end{equation}
                                      and $\int_{\Omega}|f(P)|^{2}dv$ and $\int_{\Omega}|g(P)|^{2}dv$ are finite.[The second axiom, B., now follows directly.]
                                   
                                   Let us now proceed to the verification of the requirements A.--E.! 
                 
                                   A. and B. are manifestly satisfied.
                                   
                                   C.: [Infinite-dimensionality] We shall choose $k$ non-overlapping regions $\mathfrak{M}_{1},\mathfrak{M}_{2},\ldots,\mathfrak{M}_{k}$ in $\Omega$. Let $f_{p}(P)$ be [the characteristic function of $\mathfrak{M}_{p}$]
                                   1 if $P$ is in $\mathfrak{M}_{p}$, 0 otherwise (for $p=1,2,\ldots,k$. The $f_{p}(P)$ ($p=1,2,\ldots,k$) are linearly independent, as if
                                   \begin{equation}
                                      a_{1}f_{1}+\cdots+a_{k}f_{k} = 0,
                                      \end{equation}
                                      then, in particular, for points $P$ in $\mathfrak{M}_{p}$,
                                      \begin{equation}
                                       a_{1}f_{1}(P)+\cdots+a_{k}f_{k}(P) = 0
                                       \end{equation}
                                       implies $a_{p}=0$, for any $p=1,2,\ldots,k$.  [As $k$ can be chosen arbitrarily large, the space $\mathfrak{H}$ must be infinite dimensional.]
                                       
                                   D.: [Separability]  A general verification of this property for all spaces $\Omega$ is not really possible without going further into the precise definition of general [i.e., measurable] spaces and the
                                   so-called Lebesgue measure.\footnoteA{Cf.\ for example, Carath\'eodory, \emph{Vorlesungen \"uber reele Funktionen}, Berlin-Leipzig 1918, Chaps.\ 5-9.} We shall not do so here, but will give examples in two
                                   characteristic cases of everywhere dense sequences $f_{1},f_{2},\ldots$ in $\mathfrak{H}$.
                                     First, let $\Omega$ be the ``$n$-dimensional cube'' of all points $x_{1},x_{2},\ldots,x_{n}$ satisfying
                                     \begin{equation}
                                      0\leq x_{\nu}\leq 1,\;\;(\nu=1,2,\ldots,n).
                                      \end{equation}
                                      $\mathfrak{H}$ is then the space of all (complex) functions $f$ defined on $\Omega$ with finitely integrable absolute square. We can expand all such functions $f(P)=f(x_1,x_2,\ldots,x_{n})$, following Fourier,
                                      as follows
                                      \begin{equation}
                                      f(x_{1},x_{2}, \ldots,x_{n}) =\sum_{r_{1},r_{2}, \ldots,r_{n}=-\infty}^{\infty}c_{r_{1},r_{2},\ldots,r_{n}}e^{2\pi i(r_{1}x_{1}+r_{2}x_{2}+\cdots+r_{n}x_{n})}.
                                      \end{equation}
                                 Now, the partial sum
                                 \begin{equation}
                                 \label{Fouriersum}
                                     \sum_{r_{1},r_{2}, \ldots,r_{n}=-N}^{N}c_{r_{1},r_{2},\ldots,r_{n}}e^{2\pi i(r_{1}x_{1}+r_{2}x_{2}+\cdots+r_{n}x_{n})}                                      
                                \end{equation}
                                converges in the mean to $f$ for $N\rightarrow\infty$.\footnoteA{A general and direct construction of an everywhere dense sequence in [a general Hilbert space] $\mathfrak{H}$ will be given in the work
                                cited in note 12.}   
                                   There therefore exists, arbitrarily close to an arbitrary element $f$ of $\mathfrak{H}$, a function of the form (\ref{Fouriersum}), with rational $c_{r_{1},r_{2},\ldots,r_{n}}$. But it is well known that these [coefficients]
                                   can be written in the form of a sequence.\footnoteA{One can, as is well known, write any assembly of finitely many rational numbers (which is the case here) in the form of a sequence.}
                                   
                                 Second: let $\Omega$ be the $n$-dimensional (real) space of all points $x_{1},x_{2},\ldots,x_{n}$ [$-\infty <x_{i}<\infty$], $\mathfrak{H}$ the corresponding [function-space]. We consider an arbitrary (differentiable)
                                  function $\varphi(x)$ which maps the interval $(0,1)$ to the real line $(-\infty,\infty)$, and let $\psi(y)$ be its inverse (for example, $\varphi(x)=\ln{\frac{x}{1-x}},\; \psi(y)=\frac{e^{y}}{e^{y}+1}$).
                                    Then one has generally that
                                    \begin{equation}
                              \int_{-\infty}^{\infty}\cdots\int_{-\infty}^{\infty}f(x_{1},\ldots,x_{n})dx_{1}\ldots dx_{n} = \int_{0}^{1}\cdots\int_{0}^{1}f(\varphi(u_{1}),\ldots,\varphi(u_{n}))\varphi^{\prime}(u_{1})*\ldots*\varphi^{\prime}(u_{n})du_{1}*\ldots*du_{n}.
                              \end{equation}
                              Accordingly, we only need to choose $f_{1},f_{2},\ldots$ so that they are everywhere dense in the $\mathfrak{H}$ of the first example, whence the $g_{1},g_{2},\ldots$, defined as 
                              \begin{equation}
                              \label{vNmap}
                              g_{\mu}(x_{1},\ldots,x_{n}) = \frac{f_{\mu}(\psi(x_{1}),\ldots,\psi(x_{n}))}{\varphi^{\prime}(\psi(x_{1}))*\ldots*\varphi^{\prime}(\psi(x_{n}))}
                              \end{equation}
                              are then everywhere dense in our [second] example.\footnoteB{There is a small subtlety in this argument:  denseness requires closeness with respect to the metric in each space, which involves the integral
                              of the absolute square, possibly multiplied by a positive measure factor.  One needs to define a
                              mapping between square integrable functions $g(x_{i})\in L^{2}[(-\infty,\infty)^{n}]$ and $f(u_{i})\in L^{2}[(0,1)^{n}]$, which can be done by a small modification of von Neumann's formula (\ref{vNmap}). The
                              correct association is given by setting
                              \begin{equation}
                                g_{\mu}(x_{1},\ldots,x_{n}) = \frac{f_{\mu}(\psi(x_{1}),\ldots,\psi(x_{n}))}{\sqrt{\varphi^{\prime}(\psi(x_{1}))*\ldots*\varphi^{\prime}(\psi(x_{n}))}}
                                \end{equation}
                                Note that the function $\varphi(u)$ (e.g., $\ln{u/(1-u)}$) is monotonic so $\varphi^{\prime}$ is strictly positive, allowing the square-root. Alternatively, if we stick with von Neumann's mapping, we arrive at the
                              Hilbert space for square-integrable functions on $(-\infty,\infty)^{n}$ with a non-trivial measure $dv=\prod_{i}\varphi^{\prime}(\psi(x_{i}))dx_{i}$.}
                              
                              E.: Let $f_{1},f_{2},\ldots$ be a sequence of functions in $\mathfrak{H}$, and for every $\varepsilon>0$ let there be a $N=N(\varepsilon)$, so that $N\leq m\leq n$ implies
                              \begin{equation}
                              \int_{\Omega}|f_{m}(P)-f_{n}(P)|^{2}dv \leq \varepsilon.
                              \end{equation}
                              Choose a monotone increasing sequence of integers $N_1,N_2,\ldots$ such that
                              \begin{equation}
                                 N_{\nu} \geq N(\frac{1}{8^{\nu}}).
                                 \end{equation}
                                 We then have
                                 \begin{equation}
                                 \int_{\Omega}|f_{N_{\nu+1}}(P)-f_{N_{\nu}}(P)|^{2}dv \leq \frac{1}{8^{\nu}}.
                                 \end{equation}
                                  The set of points $P$ for which
                                  \begin{equation}
                                   |f_{N_{\nu +1}}(P)-f_{N_{\nu}}(P)| \geq \frac{1}{2^{\nu}}
                                   \end{equation}
                                   therefore forms a region of measure (surface-extension\footnoteA{This is to be understood as the Lebesgue measure, see note 37.}) $\leq \frac{1}{2^{\nu}}$. 
                                   
                                   [Proceeding in this way,]
                                   the set of points for which we do \emph{not}
                                   have
                                   \begin{eqnarray}
                                    |f_{N_{\nu +1}}(P)-f_{N_{\nu}}(P)| &\leq& \frac{1}{2^{\nu}} \\
                                    |f_{N_{\nu +2}}(P)-f_{N_{\nu+1}}(P)| &\leq& \frac{1}{2^{\nu+1}} \\
                                     |f_{N_{\nu +3}}(P)-f_{N_{\nu+2}}(P)| &\leq& \frac{1}{2^{\nu+2}} \\
                                     \ldots\ldots\ldots\ldots
                                     \end{eqnarray}
                                     must have a total measure 
                                     \begin{equation}
                                     \leq \frac{1}{2^{\nu}}+\frac{1}{2^{\nu+1}}+\frac{1}{2^{\nu+2}}+\cdots = \frac{1}{2^{\nu-1}}.
                                     \end{equation}
                                     But everywhere, where the above inequalities hold, the sequence $f_{N_{1}}(P),f_{N_{2}}(P),\ldots$ must converge [being a Cauchy sequence of complex numbers]: the points of non-convergence therefore
                                     form a region of measure $\leq 1/2^{\nu-1}$. This holds for all $\nu$, hence the region of non-convergence must have measure zero. The limit of $f_{N_{\nu}}(P)$ therefore exists everywhere (with the exception
                                     of a set of measure zero): let us call it $f(P)$.
                                        
                                        If $m\geq N=N(\varepsilon)$, then for $N_{\nu}\geq m$, (i.e., for all but at most a finite number of the $N_{\nu}$),
                                        \begin{equation}
                                        \int_{\Omega}|f_{m}(P)-f_{N_{\nu}}(p)|^{2}dv \leq \varepsilon;
                                        \end{equation}
                                        so we can let $\nu\rightarrow\infty$: thus,
                                        \begin{equation}
                                        \int_{\Omega} |f_{m}(P)-f(P)|^{2}dv \leq \varepsilon.
                                        \end{equation}
                                        Thus: $f$ belongs to $\mathfrak{H}$, and the sequence  $f_1,f_2,\ldots$ converges to $f$.
                             
          \section{Appendix 3}
          The Stieltjes integral
          \begin{equation}
          \label{Stieltint}
            \int_{a}^{b}f(x)d\varphi(x)
            \end{equation}
            is defined,  whether $a,b$ is a finite or infinite interval, if $f(x)$ is a continuous function in this interval (and also at the endpoints $a,b$!), and $\varphi(x)$ is a monotone non-decreasing function in the same interval (finite at
            $a,b$)---or the difference of two such functions (i.e., a function of bounded variation). It is defined,\footnoteA{Stieltjes, \emph{Recherches sur les fractions continues}, \emph{Annales de
            la Facult\'e des Sciences de Toulouse}, 1894/95, Chapter 6. A short treatment can be found, for example, in
            Carlemann, \emph{\'Equations Int\'egrales \`a noyeau r\'eel singulier}, Uppsala 1923, pp.\ 7-9.} in analogy to the Riemann integral, as the limit under arbitrarily fine divisions
            $x_{0},x_{1},\ldots,x_{n}$ of the interval $(a,b)$ ($a=x_{0}\leq x_{1}^{\prime}\leq x_{1} \leq x_{2}^{\prime}\leq\cdots\leq x_{n-1}\leq x_{n-1}^{\prime}\leq x_{n}=b$) of
            \begin{equation}
              \sum_{n=1}^{N} f(x_{n}^{\prime})(\varphi(x_{n}-\varphi(x_{n-1})).
              \end{equation}
              
              Without going into this more closely here, let us indicate as follows a geometrical realization of this quantity (for monotone $\varphi$): draw in the $x-y$ plane the curve
              \begin{equation}
              \label{Stieltcurve}
                 x = \varphi(u),\;\;y=f(u),\;\;(a\leq u\leq b).
                 \end{equation}
                 If $x$ falls in a gap, because $\varphi(u)$ discontinuously jumps over it, the neighboring $[\varphi(u),]f(u)$ points are connected horizontally. The integral
                 \begin{equation}
                        \int_{a}^{b}f(u)d\varphi(u)
            \end{equation}     
                          is then the area of the plane between this curve (\ref{Stieltcurve}), the $x-$axis, and the abscissas $x=\varphi(a)$ and $x=\varphi(b)$ (see Figures 1--3).  
                          
 \begin{figure}
\includegraphics[width=0.9\hsize]{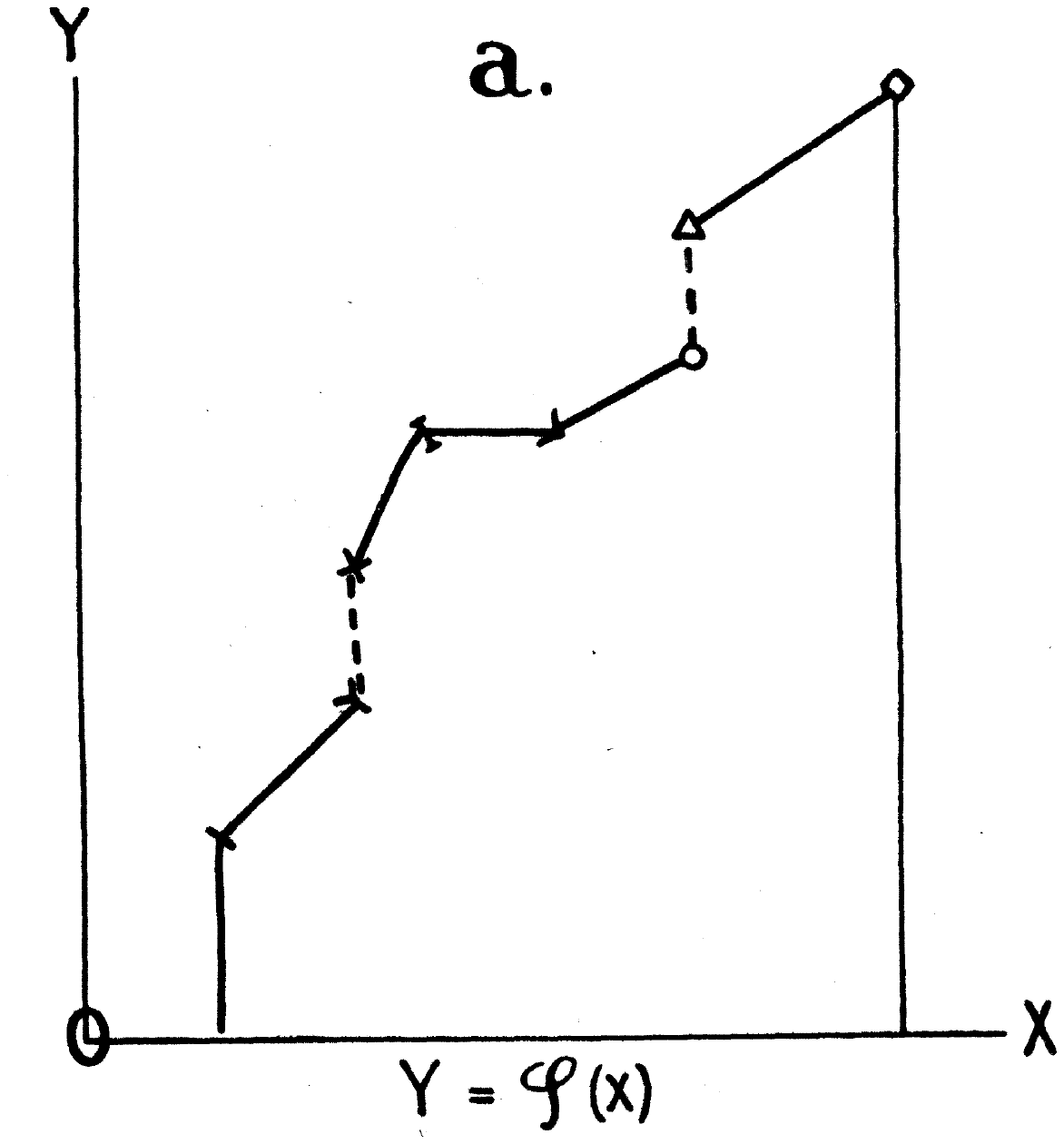}
\caption{von Neumann's Figure a., showing the Stieltjes measure function $\varphi(x)$ as a function of $x$. The function is monotone non-decreasing, and continuous save for a finite number (two) of jump discontinuities.}
\label{phi}
\end{figure}
 \begin{figure}
\includegraphics[width=0.9\hsize]{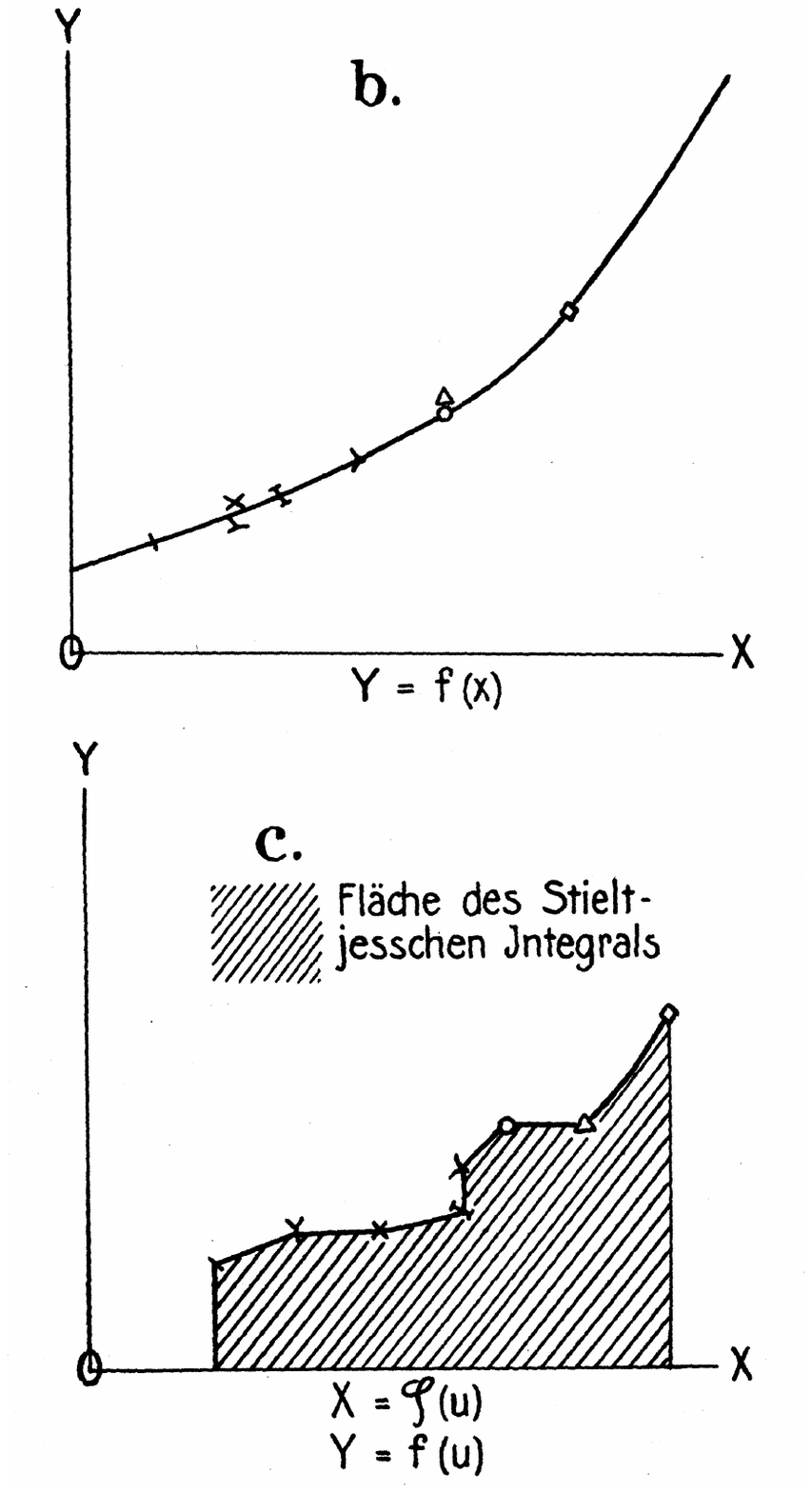}
\caption{von Neumann's Figures b,c., showing (b) the  function $f(x)$ in (\ref{Stieltint}) as a function of $x$, and (c) the function of (\ref{Stieltcurve}) defining the area (= German \emph{Fl\"ache}) represented by the Stieltjes integral.}
\label{Stielt}
\end{figure}   

\setcounter{footnoteA}{0}
\setcounter{footnoteB}{0} 
\newpage

\chapter{Paper 2: Probability-theoretic Construction of Quantum Mechanics}

\section*{J.\ v.\ Neumann}

\section*{Submitted by M. Born in the session of November 11, 1927}

\section{Introduction}
 The recent development of quantum mechanics has led, as is well known, to two versions which differ in principle in the conceptual formulation of the results, designated the ``wave theory'', and
 the ``transformation-'' or ``statistical-theory.'' It is the second of these which will be of primary concern to us here.
 
 The statistical theory was initiated by Born, Pauli and London, and has been completed by Dirac and Jordan.\footnoteA{In particular, see several works of P. A. M. Dirac in the  1926/27 issues of
 the \emph{Proceedings of the Royal Society} (particularly Vol.\ 113, 1927), as well as P. Jordan, \emph{Zeitschrift f\"ur Physik}, Vol.\ 40, 11/12 (1927), and Vol.\ 44, 1/2 (1927). See also J. v. Neumann,
 \emph{G\"ottingen Nachrichten}, session of May 20, 1927.}\footnoteB{Dirac \autocite*{Dirac:1927a}, Jordan \autocite*{Jordan:1927b,Jordan:1927c}, von Neumann \autocite*{VonNeumann:1927a}.} Primarily this theory allows one to answer questions of the following form:
 
  Let us specify a particular physical quantity in a definite physical system. What values can it assume? What are the a priori probabilities of these values? How do these probabilities change, in the
  event that the values of certain other quantities (previously measured) are specified?
  
   As such, such questions are in no way unusual in classical mechanics, but there it is always possible to make the statistical question a ``sharp'' one, that is to say, to reach the point where a particular
   physical quantity assumes a particular value with probability 1, and all other values with probability 0. To achieve this, one has only to measure sufficiently many quantities: namely, if the system has
   $f$ degrees of freedom, $2f$ independent quantities (e.g., $f$ coordinates and their conjugate momenta).
             
             That this is \emph{not} the case in quantum mechanics constitutes one of its most characteristic features, for it is impossible to measure certain quantities simultaneously\footnoteA{That is to say, the 
             measurement of one influences the other, and vitiates the validity of a previous measurement of the latter. cf.\ for example Dirac, \emph{Proceedings of the Royal Society}, Vol.\ 112 (1926),
             and Heisenberg, \emph{Zeitschrift f\"ur Physik}, Vol.\ 43, 3/4 (1927).}\footnoteB{Dirac \autocite*{Dirac:1926c}, Heisenberg's \autocite*{Heisenberg:1927b} ``uncertainty paper''.}: for example,  measurements 
             of a quantity and of its conjugated momentum are \emph{always} incompatible.
                               
     The usual approach employed up to now in statistical quantum mechanics was in essence deductive: the absolute square of certain expansion coefficients of the wave function,\footnoteA{The designation
     ``wave function'' stems from the wave theory of Schr\"odinger; Dirac views it as a row of a certain transformation matrix, while Jordan calls it the ``probability amplitude.''} or of the wave function
     itself, was more or less dogmatically equated to a probability---and agreement with experience ascertained  afterwards. A systematic derivation of quantum mechanics from the phenomenological facts
     or basic probability-theoretic  assumptions, in other words, an inductive derivation, has not been given. Additionally, its relation to the usual probability calculus was insufficiently clarified: the validity
     of the basic laws [of probability calculus] (the addition and multiplication laws for calculating probabilities) was not adequately discussed.\footnoteA{Thus, for example, according to Jordan (\emph{Zeitschrift
     f\"ur Physik}, Vol.\ 40, 11/12) both the addition and the multiplication rule apply to the ``probability amplitudes'', and not to their absolute squares, which are the probabilities themselves. In opposition
     to this, cf.\ Heisenberg, \emph{Zeitschrift f\"ur Physik}, Vol.\ 43, 3/4.}
     
     In the following work an inductive construction of this type will be sought. We assume throughout the absolute validity of the usual probability calculus. It will become clear that this [calculus] is not
     only compatible with quantum mechanics, but even (in combination with some not very deep factual and formal assumptions---cf.\ the summary in Section 9, 1--3) suffices for its unique derivation.
     Indeed, we shall be in a position to erect the entire ``time-independent'' quantum mechanics on this foundation. 
     
       It will turn out to be useful to posit certain mathematical concepts as given from the outset---e.g., those collected in Sections 1-11 of the author's work entitled ``Mathematical Foundation of Quantum Mechanics.'' \footnoteA{\emph{G\"ottingen
       Nachrichten}, session of May 20, 1927.}\footnoteB{Von Neumann \autocite*{VonNeumann:1927a}.}
       We shall therefore make use of the contents of this work, all the more because  it contains (Section 13) a formulation  of the statistical predictions [of quantum mechanics] which deviates from (and
       is more general than) the usual one, which we will need in any case. This work will be referred to as ``M.B.Q.''
       
       \section{Basic assumptions}
        In the following, we denote by $\mathfrak{S}$ a permanently fixed physical system. (In the following, many so designated systems, 
        such as $\mathfrak{S}^{\prime},\mathfrak{S}_{1},\mathfrak{S}_{2},\ldots,\mathfrak{S}^{\prime}_{1},\mathfrak{S}^{\prime}_{2},\ldots$, will appear.\footnoteB{The formulation of probability theory
        followed by von Neumann here adheres closely to the methodology developed over many
        years by Richard von Mises and propounded the following year in a  book in which he summarized his frequentist approach  \autocite{VonMises:1928}. The terminology of von Neumann differs slightly from von Mises: the latter uses the term ``\emph{Kollektiv}'' for an ensemble (von Neumann's \emph{Gesamtheit}), ``\emph{Merkmal}'' in
        von Mises becomes a ``\emph{physikalische Gr\"osse}'' (physical quantity) in von Neumann.} All of
        these will have the physical structure of $\mathfrak{S}$, only their states will vary---cf.\ note 6.) As we are concerned with the application of statistical
        methods, we imagine this system to be absolutely isolated from the environment (an isolation only occasionally interrupted by measurements, i.e., 
        external interventions), and presented in very (perhaps infinitely) many exemplars $\mathfrak{S}_{1},\mathfrak{S}_{2},\ldots$ The physical quantities
        defined in $\mathfrak{S}$ we shall denote $\{\mathfrak{a},\mathfrak{b},\ldots\}$; in order to obtain the statistics of the quantity $\mathfrak{a}$ in the
        ensemble $\mathfrak{S}_{1},\mathfrak{S}_{2},\ldots$, one subjects each of the systems $\mathfrak{S}_{1},\mathfrak{S}_{2},\ldots$ to the experiment
        ``measurement of $\mathfrak{a}$'', and makes a note of the resulting distribution of values. (Note that, in accordance with the fundamental principles
        of probability theory,\footnoteB{Cf.\ Von Mises \autocite*[pp.\ 24-25]{VonMises:1928}: ``the limiting value [for a probability] must  remain unchanged, if one 
        arbitrarily selects any sub-ensemble of the full ensemble and considers just these.''} one obtains the same statistics from examining, instead of $\mathfrak{S}_{1},\mathfrak{S}_{2},\ldots$, an arbitrarily chosen subset
        of the ensemble, as long as it has very many members, even though it may be arbitrarily small in comparison to $\{\mathfrak{S}_{1},\mathfrak{S}_{2},\ldots\}$;
        the ensemble $\{\mathfrak{S}_{1},\mathfrak{S}_{2},\ldots\}$ is altered as little as we please by such a ``statistical levy.'')  In particular, the statistics will be
        termed ``sharp'' if the same value is measured in all the systems $\mathfrak{S}_{1},\mathfrak{S}_{2},\ldots$---i.e., the entire distribution consists of a single number.
        
        In any event, the foundation of any statistical investigation consists in having at one's disposal an ``elementary random'' ensemble $\{\mathfrak{S}_{1},\mathfrak{S}_{2},\ldots\}$,
        in which ``all conceivable states of the system $\mathfrak{S}$ appear equally frequently''; the distribution of values of this ensemble must then be that ascribed to a
        system $\mathfrak{S}$, of whose state one knows absolutely nothing.\footnoteA{Naturally, one knows the physical structure of $\mathfrak{S}$, which should be specified from
        the outset---$\mathfrak{S}$ is perhaps a harmonic oscillator, or a hydrogen atom, etc.; what one does not know is the present state of same,} Even the statistical properties of such systems $\mathfrak{S}^{\prime}$,
        of which one knows something, are essentially dependent on [the elementary random ensemble] $\{\mathfrak{S}_{1},\mathfrak{S}_{2},\ldots\}$: if, for example, the only thing that one knows about
        $\mathfrak{S}^{\prime}$ is that the value of the quantity $\mathfrak{a}$ lies in the interval $I$, then one must take the ensemble $\{\mathfrak{S}_{1},\mathfrak{S}_{2},\ldots\}$, 
        carry out an experiment on each member $\mathfrak{S}_{1},\mathfrak{S}_{2},\ldots$, which decides if $\mathfrak{a}$ lies in the interval $I$, or not; those $\mathfrak{S}_{1},\mathfrak{S}_{2},\ldots$ for which
        the first situation holds form an ensemble $\{\mathfrak{S}_{1}^{\prime},\mathfrak{S}_{2}^{\prime},\ldots\}$, which is to be taken as determining the statistics of our system $\mathfrak{S}^{\prime}$. A system
        $\mathfrak{S}^{\prime}$, of which we have some information, therefore always represents a particular ensemble $\{\mathfrak{S}_{1}^{\prime},\mathfrak{S}_{2}^{\prime},\ldots\}$ which arises from the
        [fundamental] ensemble $\{\mathfrak{S}_{1},\mathfrak{S}_{2},\ldots\}$ in a well-defined manner.
    
       All ``knowledge'' about $\mathfrak{S}^{\prime}$, i.e., of an ensemble $\mathfrak{S}_{1}^{\prime},\mathfrak{S}_{2}^{\prime},\ldots$, corresponds to an association of an expectation value to
       each quantity $\mathfrak{a}$---namely, the average value of the distribution of values [for $\mathfrak{a}$] that occur in  $\{\mathfrak{S}_{1},\mathfrak{S}_{2},\ldots\}$. Conversely, ``knowledge'' of
       $\mathfrak{S}^{\prime}$ (i.e., the statistical composition of $\{\mathfrak{S}_{1}^{\prime},\mathfrak{S}_{2}^{\prime},\ldots\}$) is completely described once the association
       \begin{equation}
           \mathfrak{a} \longleftrightarrow \mathrm{Expectation\;value\;of}\;\mathfrak{a} = \mathbf{E}(\mathfrak{a})
           \end{equation}
           is given for \emph{all} quantities $\mathfrak{a}$.\footnoteA{Naturally, the distribution of values of $\mathfrak{a}$ is not determined solely by the specification of the average value.
           Inasmuch however as we also know the average value of all powers of $\mathfrak{a}$, i.e., all ``moments'' of the distribution, the latter is fully determined.}
           Consequently, it is conceptually equivalent, and formally more convenient, to examine this association. Before we investigate it further, however, 
           we must study somewhat more closely the quantities $\mathfrak{a},\mathfrak{b},\ldots$ in a system $\mathfrak{S}$.
           
           In quantum mechanics, when one considers simultaneously two quantities $\mathfrak{a},\mathfrak{b}$,  there is a fundamental distinction between the cases where they can be simultaneously observed, or not.
           In the case where they can, there must be an experiment measuring $\mathfrak{a}$ and an experiment measuring $\mathfrak{b}$, of such a kind that the experiments do not disturb one another (cf.\ the
           remarks at the beginning of Section 1). Alternatively, if we combine both experiments into a single one, there should exist a single experiment which measures $\mathfrak{a}$ as well as $\mathfrak{b }$.
           Given such an experiment we may very well presume that  a third quantity $\mathfrak{d}$ is being measured, the knowledge of which involves that of both $\mathfrak{a}$ and $\mathfrak{b}$, i.e., one of
           which both $\mathfrak{a}$ and $\mathfrak{b}$ are functions.\footnoteA{At this point we deviate for the first time in an essential way from classical mechanics. There, the reduction of two, perhaps independent,
           quantities (which is very easily compatible with simultaneous measurability) to a third, of which they should both be functions, is completely unnatural. This reduction is also completely incompatible with
           the concept of ``degrees of freedom'', as $\mathfrak{a},\mathfrak{b}$ (if independent) correspond to two degrees of freedom, $\mathfrak{d}$ to just one.
           
             It is perhaps also appropriate here to point to the fact, that quantum mechanics is also lacking a concept of ``degrees of freedom.''  For example, a hydrogen atom perturbed by an inhomogeneous electric field
             (with the nucleus held fixed at the origin!) can be described either by three independent quantities, the coordinates of the electron, but also by just one: the total energy (as in this case there are no degeneracies!).}
             
             The concept ``function of a quantity $\mathfrak{a}$'', $f(\mathfrak{a})$ (where $f(x)$ is a function defined for all real numbers, and only for these) is always meaningful: $f(\mathfrak{a})$ is a quantity which
             is measured whenever $\mathfrak{a}$ is, with the value $f(w)$ ascribed to it whenever the value $w$ is observed for $\mathfrak{a}$. The situation is different for two (or more) quantities, as these may possibly
             not be simultaneously measurable. Nevertheless one may still in this case define the sum [of quantities not simultaneously measurable], and for the following reason. In order to learn the expectation value
             of a sum, one need only add the expectation values of the two summands, without regard to the detailed properties of the individual distributions (in particular, to questions of independence). Now, as only the expectation 
             values of quantities played a role in the preceding description of quantum-mechanical systems, we can therefore regard $\mathfrak{a}+\mathfrak{b}$ as a quantity, even if $\mathfrak{a},\mathfrak{b}$  are not
             simultaneously measurable.\footnoteB{The simplest example being the total energy $H=T+V$ for a mechanical system possessing both kinetic and potential energy 
             (cf.\ note 9 below).}
              In an analogous fashion we can indeed define the sum of arbitrarily many (even infinitely) quantities $\mathfrak{a},\mathfrak{b},\mathfrak{c}, \ldots$: for the expectation value of
             the sum (in agreement with basic principles of the probability calculus) is defined as the sum of the expectation values. A further (trivial) generalization is that the components of the sum can be multiplied by
             arbitrary real constants.
             
              Functions other than the sum\footnoteB{Here, by ``sum'', one should more broadly admit ``linear combinations'',  as becomes clear in postulate \textbf{A.} below.
              Von Neumann is pointing to the difficulties which arise when one considers
              kinematical quantities involving \emph{products} of non-commuting quantities.} of two (possibly not simultaneously observable!) quantities have however absolutely no sense, 
              inasmuch as the expectation value of another function of $\mathfrak{a},\mathfrak{b}$
              cannot be expressed in terms of the expectation values of $\mathfrak{a}$ and $\mathfrak{b}$. 
              
              We are now in a position to formulate completely what we shall demand of the expectation value functions $\mathbf{E}(\mathfrak{a})$, which determine a ``knowledge'' of $\mathfrak{S}^{\prime}$, i.e., of the ensemble
              $\{\mathfrak{S}_{1}^{\prime},\mathfrak{S}_{2}^{\prime},\ldots\}$. Namely:
              \begin{quote}
\begin{description}
\item[\;\;A.]  If $\mathfrak{a},\mathfrak{b},\mathfrak{c},\ldots$ are a finite or infinite set of quantities, and $\alpha,\beta,\gamma,\ldots$ are real numbers, then\footnoteA{Observe the deep implications of this property of
              quantum statistics: suppose we  have an electron with coordinates $q_{1},q_{2},q_{3}$, momenta $p_{1},p_{2},p_{3}$, and mass $m$. The three quantities
              \begin{equation}
              \alpha(q_{1}^{2}+q_{2}^{2}+q_{3}^{2}),\;\;\frac{1}{2m}(p_{1}^{2}+p_{2}^{2}+p_{3}^{2}),\;\;\alpha(q_{1}^{2}+q_{2}^{2}+q_{3}^{2})+\frac{1}{2m}(p_{1}^{2}+p_{2}^{2}+p_{3}^{2}),
              \end{equation}
              (i.e., the squared distance from the origin in appropriate units of the free electron, the kinetic energy of the free electron, and the energy of the electron trapped in a harmonic oscillator [potential])
              have completely different spectra: the first two each have a continuous spectrum, the third a discrete spectrum; moreover, no two of them are simultaneously measurable. Nonetheless, the sum
              of the expectation values of the first two is equal to the expectation value of the third. (Additionally, it is convenient, in order to find finite expectation values, to consider fixed ``states'' of the system, cf.\ Section 4.)} 
              \begin{equation}
              \mathbf{E}(\alpha\mathfrak{a}+\beta\mathfrak{b}+\gamma\mathfrak{c}+\cdots) = \alpha\mathbf{E}(\mathfrak{a})+\beta\mathbf{E}(\mathfrak{b})+\gamma\mathbf{E}(\mathfrak{c})+\cdots
              \end{equation}
\item[\;\;B.] If $\mathfrak{a}$ is a quantity which is intrinsically non-negative, then
              \begin{equation}
               \mathbf{E}(\mathfrak{a}) \geq 0.
               \end{equation}         
\end{description}
\end{quote}

           The motivation for \textbf{A.} has already been discussed; \textbf{B.} is also an obvious property of all expectation values. A topic so far omitted needs further commentary, namely, the normalization requirement: if
           the quantity $\mathfrak{a}$ always has the value 1, then we must also have $\mathbf{E}(\mathfrak{a})=1$. Absent this we are confined to considering only relative expectation values, given up to a proportionality
           factor. The reasons for this generalization can best be illustrated with the following example.
           
           Let $\mathfrak{a}$ be a quantity, $I$ an interval on the real line, and $f(x)$ the function defined to equal 1 in $I$ and 0 otherwise. The expectation value for $f(\mathfrak{a})$ is then manifestly the probability that
           the value of $\mathfrak{a}$ lies in the interval $I$. If now, for example, $\mathfrak{a}$ can assume all real values, and all with the same probability, then if we were to normalize the total probability (i.e., that of
           the interval $-\infty,+\infty$) to 1, the probability that the value of $\mathfrak{a}$ lies in $I$, for any finite interval $I$, is equal to zero. On the other hand, we have no such problem with the relative probabilities
           of two finite intervals. Conclusions of this type are naturally much more important than the requirement that the total probability be equal to 1: it is therefore useful to give up on the latter (and allow the total
           probability to be perhaps $+\infty$), and satisfy ourselves with relative expectation values and relative probabilities---which can then in interesting (and \emph{not} impossible) situations vanish!\footnoteA{Indeed, in
           systems in which both the available space and the potential energy are finite, this complication usually will not appear. Still, it is appropriate to deal with it given that it becomes apparent in the most typical
           systems in quantum mechanics: a free electron, hydrogen atom, etc.}
           
           We must add to the preceding considerations of a conceptual and foundational nature the following remark of a formal character: a theory is impossible, as long as we have not found a formal (i.e., accessible
           to calculation)  equivalent for the quantities $\mathfrak{a},\mathfrak{b},\mathfrak{c},\ldots$ of the system $\mathfrak{S}$. But we know very well from quantum mechanics which mathematical objects are to be
           associated with the physical quantities: they are in fact the so-called linear symmetric [i.e., hermitian] functional operators, which act on complex-valued functions (the wave functions) defined in the state space of the system $\mathfrak{S}$,
           transforming them into other such functions. For the time being we shall not discuss why this is so; later, in the course of our considerations, we shall find a direct interpretation of the operator assigned to a
           [physical] quantity (cf.\  Section 4, $\gamma$).
           
           Now, the complex valued functions defined in the state space of $\mathfrak{S}$ form a realization of the abstract Hilbert space $\mathfrak{H}$;\footnoteA{From here on, we will make essential use of the concepts and
           terminology of M.B.Q.  In Section 4 of M.B.Q.\ it is stated, and in Section 6 and Appendix 2 proved (Theorem of Fischer and F. Riesz), that the stated function space is a realization of $\mathfrak{H}$ (i.e., the operations of addition, multiplication by a
           constant and the so-called inner product---in the notation of M.B.Q.:$f+g, a.g, Q(f,g)$---are mapped uniquely and reversibly to $\mathfrak{H}$). Abstract Hilbert space is introduced in Sections 5, 6.} we can therefore replace
           the linear symmetric functional operators acting on these with linear symmetric operators of $\mathfrak{H}$. But an operator is only useful for quantum mechanics if its ``eigenvalue problem'' is solved, or, at least, solvable.\footnoteA{For
           the possible values of the associated quantity are indeed the eigenvalues, and to calculate the probabilities one must also know the eigenfunctions! In addition, the representation in M.B.Q.\ (Section 14) presumes
           knowledge of the partition of unity (defined there in Section 9) associated to the operator, i.e., the solution of the eigenvalue problem.}
           Consequently, we must assign to every physical quantity a linear symmetric operator which admits an ``eigenvalue representation'' [i.e., spectral resolution].\footnoteA{For further information on operators in 
           $\mathfrak{H}$, see Sections 7,8 in M.B.Q.\ The eigenvalue problem is discussed in Sections 9, 10.}  We call such operators, for the sake of brevity, normal: it is highly likely, that every linear symmetric operator is normal.\footnoteA{The
           present state of affairs is as follows: A linear symmetric operator, which allows at all an eigenvalue representation, allows  only one such (for such an operator there is an exact partition of unity, cf.\ M.B.Q., Section 9);
           all bounded (i.e., continuous) operators, further, all differential operators of second order with sufficiently regular coefficients, indeed absolutely all operators which have ever appeared in quantum mechanics, admit an
           eigenvalue representation. The existence of an eigenvalue representation for real symmetric linear operators was shown in general by the author (to appear shortly in \emph{Mathematische Annalen}). Certain mathematical
           difficulties appear however in the general proof for the complex linear symmetric operators which concern us here. Cf.\ also M.B.Q., notes 12 and 27.}\footnoteB{The general spectral theory for self-adjoint---more generally,
           normal---operators was presented in von Neumann \autocite*{VonNeumann:1929a,VonNeumann:1929b}. In these works, von Neumann identified the appropriate subclass of linear operators with ``nice'' spectral properties as those (designated ``normal'')
           which commute with their adjoint, $AA^{\dagger}=A^{\dagger}A$ \autocite[p.\ 115]{VonNeumann:1929a}. This includes, in addition to the self-adjoint operators, which are trivially normal (as $A=A^{\dagger}$),
           the important class of unitary (and thus, automatically bounded) operators $U$ for which $UU^{\dagger}=U^{\dagger}U=1$.} We do not however need to concern ourselves with this question here.
           With regard to all this we shall therefore assume:\\
           Every physical quantity $\mathfrak{a}$ of the system $\mathfrak{S}$ corresponds to a linear symmetric operator, and conversely.\footnoteA{We will make rather little use of the converse case, which is surely uncontroversial.}
            How this association is accomplished in detail is immaterial for us,\footnoteA{Also, until quite recently there was no prescription that determined it \emph{in general}!} we will only employ
           two properties, in particular:
           \begin{quote}
\begin{description}
\item[\;\;C.] Let $S,T,\ldots$ (in finite or infinite number) as well as $\alpha S+\beta T+\cdots$ be normal operators. If $S,T,\ldots$ are associated with the quantities $\mathfrak{a},\mathfrak{b},\ldots$, then $\alpha S+\beta T+\cdots$
           is associated with the quantity $\alpha\mathfrak{a}+\beta\mathfrak{b}+\cdots$.\footnoteA{In the highly unlikely, but at present un-contradicted, circumstance (see note 14) that $S,T,\ldots$ are normal, but
           $S+T+\cdots$ not, this will prevent us from forming the quantity $\mathfrak{a}+\mathfrak{b}+\cdots$. This restriction will however be unimportant for us.}
\item[\;\;D.] Let $S$ be a normal operator, $f(x)$ a real-valued function defined for all real $x$, and let $S$ be associated with the quantity $\mathfrak{a}$. Then $f(S)$\footnoteA{If $f(x)=x^n$ then the meaning
           of $f(S)$ is clear: it is just the $n$-th iteration of $S$, $S^n$; and from this one obtains also the meaning of $f(S)$ for polynomial $f(x)$. For arbitrary $f(x)$ one gets [$f(S)$] most
           conveniently following M.B.Q., Section 14, second example: if $S$ has the partition of unity $E(\lambda)$ (i.e., $S=\int_{-\infty}^{+\infty}\lambda dE(\lambda)$ [see note h]) then
           \begin{equation}
           \label{fspecres}
             f(S) = \int_{-\infty}^{+\infty} f(\lambda)dE(\lambda).
             \end{equation}
             The normal property of $S$ is therefore presumed; the normal property of $f(S)$ can be easily demonstrated.}\footnoteB{For the reader familiar with Dirac notation, the spectral resolution $S=\int \lambda dE(\lambda)$
             (more generally, $f(S) = \int f(\lambda)dE(\lambda)$) is
             more familiar in the form $S =\int \lambda|\lambda\rangle\langle\lambda|d\lambda$ (resp.\ $f(S) = \int f( \lambda)|\lambda\rangle\langle\lambda|d\lambda$), 
             where the states are continuum normalized $\langle\lambda|\lambda^{\prime}\rangle =\delta(\lambda-\lambda^{\prime})$. The ``partition of unity'' referred to by von Neumann is the completeness property 
             associated with the spectral resolution of any normal operator, $\int dE(\lambda)=\int|\lambda\rangle\langle\lambda| = 1$. If there is a point spectrum
             then a discrete sum is present in addition to the integral, $\sum_{n}\lambda_{n}|\lambda_{n}\rangle\langle\lambda_{n}|$ (with $\langle\lambda_{n}|\lambda_{m}\rangle = \delta_{nm}$). One easily establishes in this version Eq.\ (\ref{fspecres})
             for any power or polynomial function of $S$. Below, we encounter projection operators $E(I)$ which project onto the subspace spanned by eigenvectors $|\lambda\rangle$ of an observable with eigenvalues
             $\lambda$  in some interval $I$: in Dirac notation
             this would be $E(I) = \int_{I} |\lambda\rangle\langle\lambda|d\lambda$.}
              is associated with the quantity $f(\mathfrak{a})$.
\end{description}
\end{quote}
           Both of these requirements are very plausible, and correspond to common practice, making a more detailed discussion superfluous.
           
           \section{General form of the expectation values}
           
           On the basis of \textbf{A.--D.} in Section 2 we can construct all possible expectation-value functions $\mathfrak{a}\longleftrightarrow \mathbf {E}(\mathfrak{a})$, i.e., all possible ``knowledge'' about $\mathfrak{S}^{\prime}$ or statistical
           ensembles $\{\mathfrak{S}^{\prime}_{1},\mathfrak{S}_{2}^{\prime},\ldots\}$. For the purpose of calculation one associates a normal operator to the expectation value of the physical quantity corresponding to the operator: if $S$ is
           the operator, then we call this (relative) expectation value $\mathbf{E}(S)$. $\mathbf{E}(S)$ must satisfy the conditions \textbf{A.} and \textbf{B.}, whereby also \textbf{C.} and \textbf{D.} must also be taken into consideration.
           
           In order to facilitate calculation,  we will replace [the function space] $\mathfrak{H}$ by its realization $\mathfrak{H}_{0}$ [$l^2$], the conventional Hilbert space,\footnoteA{i.e., the space of all sequences $x_1,x_2,\ldots$ with
           finite $\sum_{n=1}^{\infty}|x_{n}|^{2}$, cf.\ M.B.Q., Section 5.} in other words we map [$\mathfrak{H}_{0}$] one-to-one to [$\mathfrak{H}$] (which is obviously possible in many ways). A linear operator $S$ is then described
           by a matrix $\{s_{\mu\nu}\} (\mu,\nu=1,2,\ldots)$, whereby generally
           \begin{eqnarray}
              S(x_1, x_2, \ldots) &=& (y_1, y_2, \ldots), \\
              y_{\mu} &=& \sum_{\nu=1}^{\infty} s_{\mu\nu}x_{\nu}.
              \end{eqnarray}
            The symmetry [hermitian] property means, as one sees right away, $s_{\mu\nu}=s_{\nu\mu}^{*}$. We now introduce the following operators: $A_{\mu}$, with the matrix $s_{\mu\mu} =1$, otherwise $s_{\rho\sigma}=0$;
            $B_{\mu\nu}, (\mu<\nu)$, with the matrix $s_{\mu\nu}=s_{\nu\mu}=1$, otherwise $s_{\rho\sigma}=0$; $C_{\mu\nu}, (\mu<\nu)$, with the matrix $s_{\mu\nu}=i$, $s_{\nu\mu}=-i$, otherwise $s_{\rho\sigma}=0$. They are
            all clearly linear symmetric [hermitian], and also normal (the $A$ have the simple eigenvalue 1, the $B$ and $C$ have the simple eigenvalues 1 and -1; otherwise only the infinitely [degenerate eigenvalue] 0). In general
            one has (we use the notation $\mathrm{Re}(z)$ (resp.\ $\mathrm{Im}(z)$) for $u$ (resp.\ $v$) when $z=u+iv$):
            \begin{equation}
                S = \sum_{\mu} s_{\mu\mu}\cdot A_{\mu} + \sum_{\mu<\nu}\mathrm{Re}(s_{\mu\nu})\cdot B_{\mu\nu} + \sum_{\mu<\nu}\mathrm{Im}(s_{\mu\nu})\cdot C_{\mu\nu},
                \end{equation}
                and consequently one has
                \begin{equation}
                \mathbf{E}(S) =  \sum_{\mu} s_{\mu\mu}\cdot \mathbf{E}(A_{\mu}) + \sum_{\mu<\nu}\mathrm{Re}(s_{\mu\nu})\cdot \mathbf{E}(B_{\mu\nu}) + \sum_{\mu<\nu}\mathrm{Im}(s_{\mu\nu})\cdot \mathbf{E}(C_{\mu\nu}).
                \end{equation}
                If we now set
                \begin{eqnarray}
                u_{\mu\mu} &=& \mathbf{E}(A_{\mu}), \\
                u_{\mu\nu} &=& \frac{1}{2}\mathbf{E}(B_{\mu\nu})+\frac{1}{2}i\mathbf{E}(C_{\mu\nu}),\;\;\mu<\nu,  \\
                u_{\nu\mu} &=& \frac{1}{2}\mathbf{E}(B_{\mu\nu})-\frac{1}{2}i\mathbf{E}(C_{\mu\nu}),\;\;\mu<\nu,
                \end{eqnarray}
                then $u_{\mu\nu}=u_{\nu\mu}^{*}$, and one has also
                \begin{equation}
                \label{expectEU}
                  \mathbf{E}(S) = \sum_{\mu\nu} s_{\mu\nu}u_{\mu\nu}^{*}.
                  \end{equation}
                  The matrix $\{u_{\mu\nu}\}$ clearly corresponds to a linear symmetric [hermitian] operator, which we shall denote $U$.\footnoteB{Here, the ``density operator'', nowadays usually denoted $\rho$, is introduced into quantum theory.
                  It would become an indispensable tool in quantum statistical mechanics, in particular, in the decoherence approach to quantum measurement theory. The result Eq.\ (\ref{expectEU}) is 
                  now written $ \mathbf{E}(S) = {\mathrm Tr}(\rho S)$ (by hermiticity $u_{\mu\nu}^{*}=u_{\nu\mu}=\rho_{\nu\mu}$, so the right hand side of  (\ref{expectEU}) equals $\sum_{\mu}(S\rho)_{\mu\mu} =\mathrm{Tr}(S\rho)=\mathrm{Tr}(\rho S)$).}
                
                We should further consider the implications of \textbf{B.} If $\varphi$ is an arbitrary point in the abstract Hilbert space $\mathfrak{H}$, with $Q(\varphi)$=1 (i.e., on the unit sphere), then we define
                the operator $P_{\varphi}$ (``projection in the direction of $\varphi$'') by
                \begin{equation}
                \label{PphiQ}
                  P_{\varphi}f = Q(f,\varphi)\cdot\varphi;
                  \end{equation}
                  it is clearly linear, symmetric and normal (a simple eigenvalue 1, otherwise infinitely degenerate 0). $P_{\varphi}$ is a projection operator, i.e., $P_{\varphi}^{2}=P_{\varphi}$. In the $\mathfrak{H}_{0}$ [$l^2$] realization,
                 should $\varphi$ correspond to $(x_{1},x_{2},\ldots)$ ($\sum_{n=1}^{\infty}|x_{n}|^{2}=1$), then $P_{\varphi}$ clearly has the matrix $\{x_{\mu}x_{\nu}^{*}\}$.\footnoteB{Thus, if $f=(y_{\mu}), \varphi=(x_{\mu})$, 
                 then $(P_{\varphi}f)_{\mu} = \sum_{\nu}x_{\mu}x_{\nu}^{*}y_{\nu}= Q(f,\varphi)\varphi_{\mu}$, which is Eq.\ (\ref{PphiQ}). Recall that in von Neumann's notation for the inner product, $Q(y,x)=\sum_{\nu}y_{\nu}x_{\nu}^{*}$ is
                 anti-linear in the \emph{second} argument, in contrast to modern usage. Thus $Q(f,g)$ (von Neumann) $\rightarrow (g,f)$ or $\langle g|f\rangle$ (modern).}
                 
                 If $P_{\varphi}$ corresponds to the quantity $\mathfrak{a}$, then $P_{\varphi}=P_{\varphi}^{2}$ corresponds to quantity $\mathfrak{a}^{2}$, i.e., to an intrinsically non-negative one. Therefore $\mathbf{E}(P_{\varphi})\geq 0$, i.e.
                \begin{equation}
                \sum_{\mu\nu}x_{\mu}^{*}x_{\nu}u_{\mu\nu}^{*} = \sum_{\mu\nu}u_{\nu\mu}x_{\nu}x_{\mu}^{*} \geq 0,
                \end{equation}
                where the left hand side is, as one easily verifies, equal to $Q(\varphi, U\varphi)$. For all $\varphi$ with $Q(\varphi)=1$ we must therefore have $Q(\varphi, U\varphi)\geq 0$, i.e.
                \begin{equation}
                 Q(f, Uf) \geq 0
                 \end{equation}
                 for any $f$ in $\mathfrak{H}$ whatsoever. An operator $U$ with this property is said to be non-negative definite.\footnoteA{If $f$ corresponds to the point $x_{1},x_{2},\ldots$ in $\mathfrak{H}_{0}$, this means
                 \begin{equation}
                   \sum_{\mu\nu}u_{\mu\nu}x_{\mu}x_{\nu}^{*} \geq 0,
                   \end{equation}
                   i.e., the hermitian form associated with the matrix $\{u_{\mu\nu}\}$ must be non-negative definite; this characterization we carry over to the operator $U$.}
                   
                   We now show the converse: every [non-negative] definite linear symmetric operator $U$ determines (via the formula $\mathbf{E}(S) = \sum_{\mu\nu}s_{\mu\nu}u^{*}_{\mu\nu}$) a statistics complying with the requirements \textbf{A.}, \textbf{B.}
                   That \textbf{A.} is satisfied is evident (after consideration of \textbf{C.}); it remains to establish \textbf{B.} 
                   
                   If the quantity $\mathfrak{a}$ is intrinsically non-negative, we can form the quantity $\mathfrak{b}=\sqrt{\mathfrak{a}}$, to which belongs the operator $T$. As $\mathfrak{a}=\mathfrak{b}^{2}$, we have the operator $S=T^2$,
                   therefore one which is [non-negative] definite.\footnoteA{For $Q(f, Sf) = Q(f,T^{2}f)=Q(Tf,Tf) \geq 0$.} We must show the following: if the [non-negative] definite operators $S$, $U$ have the matrices $\{s_{\mu\nu}\}, \{u_{\mu\nu}\}$,
                   then
                   \begin{equation}
                       \sum_{\mu\nu} s_{\mu\nu}u_{\mu\nu}^{*} \geq 0.
                       \end{equation}
                       It suffices to show that for all $N=1,2,\ldots$
                       \begin{equation}
                       \sum_{\mu,\nu=1}^{N} s_{\mu\nu}u_{\mu\nu}^{*} \geq 0,
                       \end{equation}
                       and then to let $N\rightarrow\infty$, considering that the (finite dimensional!) hermitian forms
                       \begin{equation}
                         \sum_{\mu,\nu=1}^{N}s_{\mu\nu}x_{\mu}x_{\nu}^{*},\;\; \sum_{\mu,\nu=1}^{N}s_{\mu\nu}x_{\mu}x_{\nu}^{*},
                         \end{equation}
                         are non-negative definite, given that the same holds for the corresponding infinite-dimensional ones. One easily convinces oneself that $\sum_{\mu\nu=1}^{N} s_{\mu\nu}u_{\mu\nu}^{*}$ is an orthogonal invariant (in $N$-dimensional
                         space)---consequently we can bring $\sum_{\mu,\nu=1}^{N}s_{\mu\nu}x_{\mu}x_{\nu}^{*}$ into diagonal form. All $s_{\mu\nu}, \mu\neq\nu$ are then 0, so $\sum_{\mu,\nu=1}^{N}s_{\mu\nu}u_{\mu\nu}^{*}$ is a sum of products
                         of diagonal elements. But diagonal elements of non-negative definite hermitian forms are never negative, so our sum must always be $\geq 0$. This concludes the proof.
                         
                         We have therefore achieved the following result: all possible ``knowledge'' of a system $\mathfrak{S}^{\prime}$, i.e., all possible statistical ensembles $\{\mathfrak{S}_{1}^{\prime},\mathfrak{S}_{2}^{\prime},\ldots\}$, correspond
                         in one-to-one fashion to the [non-negative] definite linear symmetric operators $U$. The correspondence is described by the expectation value formula
                         \begin{equation}
                          \mathbf{E}(S) = \sum_{\mu\nu}s_{\mu\nu}u_{\mu\nu}^{*},
                          \end{equation}
                          where $S,U$ have the matrices $\{s_{\mu\nu}\}, \{u_{\mu\nu}\}$.
                          
                          One sees from our formula that the modification of $U$ by a [multiplicative] factor only changes $\mathbf{E}(s)$ by the same, constant factor [necessarily real and positive, to preserve the non-negative definite property],
                          and therefore, as only relative expectation values concern us, is of no importance. 
                          Any other [type of] alteration of $U$ is however significant. Moreover, we cannot have $\mathbf{E}(S)=0$ identically, i.e., $U=0$, as then all (relative) expectation values and probabilities vanish, and there would be no
                          [definable] statistics.\footnoteA{The connection between $\mathbf{E}(S)$ and $U$ is obviously dependent on the nature of the realization of $\mathfrak{H}$ in $\mathfrak{H}_{0}$, i.e., the mapping of $\mathfrak{H}$
                          to $\mathfrak{H}_{0}$, and this is possible in (infinitely) many ways. One can easily show [see note $l$] that this dependence is only an apparent one, i.e., that $U$ is invariant---but this will not concern us here.}\footnoteB{The freedom 
                          of choice of orthonormal basis
                          in $\mathfrak{H}_{0}$ [$l^2$] is reflected by a unitary operator $V$ satisfying $V^{\dagger}V=1$: thus, if $\{\varphi_{n}\}$ is an orthonormal basis, so is $\{V\varphi_{n}\}$. If $s,u$ are the matrices of $S,U$ with respect to the first basis, the matrices 
                          relative to the second are $\tilde{s}=V^{\dagger}sV, \tilde{u}=V^{\dagger}uV$. However, Tr$(\tilde{s}\tilde{u})$ = Tr($V^{\dagger}sVV^{\dagger}uV)$ = Tr($su)$.}
                          
               \section{States} 
               
               In Section 3 we determined all possible statistical ensembles  $\{\mathfrak{S}_{1}^{\prime},\mathfrak{S}_{2}^{\prime},\ldots\}$  and associated them with the [non-negative] definite linear symmetric operators $U$. In this Section
               we shall determine the ``pure'', or ``uniform'' ensembles, i.e., those in which all the systems $\mathfrak{S}_{1}^{\prime},\mathfrak{S}_{2}^{\prime},\ldots$   are in the same state. Thereby we shall also have found all the
               states, in which the system $\mathfrak{S}$ can be found. In this connection the following should be emphasized: in classical mechanics (which, in the first place, one can develop on a statistical basis just as well as
               quantum mechanics) every quantity $\mathfrak{a}$ has a sharp distribution [i.e., a well-defined value] in a ``uniform'' ensemble $\{\mathfrak{S}_{1}^{\prime},\mathfrak{S}_{2}^{\prime},\ldots\}$ in a system $\mathfrak{S}^{\prime}$ in a completely
               known state. That is to say, every $\mathfrak{a}$ has a value which it assumes with absolute certainty. The situation in quantum mechanics is, as is well known  (and as we shall show)  quite different: in every state
               $\mathfrak{S}^{\prime}$ there are quantities $\mathfrak{a}$ whose distribution is not sharp, i.e., the value of which is still subject to chance.
               
               How can one characterize a ``uniform'' ensemble $\{\mathfrak{S}_{1}^{\prime},\mathfrak{S}_{2}^{\prime},\ldots\}$? In the following way: such an ensemble cannot arise by the merging of two other ensembles, unless both of these
               are the same [statistically] as it. (A non-uniform ensemble can always be constructed out of two others that differ from it---which hardly needs any further discussion.) We need to formalize this property. 
               
                If $\{\mathfrak{S}_{1}^{\prime},\mathfrak{S}_{2}^{\prime},\ldots\}$ arises from $\{\mathfrak{S}_{1}^{*},\mathfrak{S}_{2}^{*},\ldots\}$ and $\{\mathfrak{S}_{1}^{**},\mathfrak{S}_{2}^{**},\ldots\}$, in the sense that it arises from the combination
                of these in some fixed proportion, then one has for the (relative) expectation values:
                \begin{equation}
                   \mathbf{E}(S) = \eta\mathbf{E}^{*}(S)+\vartheta\mathbf{E}^{**}(S),\;\;(\eta>0,\vartheta>0),
                   \end{equation}
                   and for the (non-negative definite, linear symmetric) operators which (by Section 3) are associated with these [ensembles]:
                   \begin{equation}
                   \label{Ucomb}
                     U = \eta U^{*} + \vartheta U^{**}.
                     \end{equation}
                $U$ characterizes therefore an uniform ensemble, i.e., a completely determined state of $\mathfrak{S}$, if and only if Eq.\ (\ref{Ucomb}) implies that $U^{*},U^{**}$ differ only by constant factors from $U$. Or, if we absorb $\eta$ (resp.\ $\vartheta$)
                in $U^{*}$ (resp.\ $U^{**}$), then from
                \begin{equation}
                   U = U^{*}+U^{**}
                   \end{equation}
                   (with $U^{*},U^{**}$ non-negative definite linear symmetric), the proportionality of $U^{*}, U^{**}$ and $U$ should follow.
                
                We now assert: the only non-negative definite, linear symmetric operators $U\neq 0$ with the preceding property are (up to unimportant constant positive factors) are the $P_{\varphi}$ ($\varphi$ in
                $\mathfrak{H}$, $Q(\varphi)=1$, cf.\ definition in Section 3). Indeed, let $U$ possess the property cited above. We then choose a $f$ from $\mathfrak{H}$ with $Uf\neq 0$. This implies that $Q(f,Uf)>0$,\footnoteA{For
                non-negative $U$ we have in general
                \begin{equation}
                \label{Uinequal}
                |Q(f,Ug)| \leq \sqrt{Q(f,Uf)Q(g,Ug)},
                \end{equation}
                which is demonstrated in exactly  the same way as the analogous relation for $Q(f,g)$ in M.B.Q., note 17. Therefore, if $Q(f,Uf)$ is not greater than zero, it must equal zero, so for any $g$ (where
                $Ug$ exists!), $Q(f,Ug)= Q(Uf,g) =0$. Therefore we must have $Uf=0$.}\footnoteB{For non-negative $U (=\int_{0}^{\infty}\lambda dE(\lambda))$ 
                we can define the square-root operator $T(=\int_{0}^{\infty}\sqrt{\lambda}dE(\lambda))$, with $U=T\cdot T$.
                Then Eq.\ (\ref{Uinequal}) in note 23 follows, by replacing $f\rightarrow Tf, g\rightarrow Tg$, in the inequality $|Q(f,g)|\leq\sqrt{Q(f)Q(g)}$ established in M.B.Q., note 17. Note that as $T$ is hermitian, $Q(Tf,Tg)=Q(f,T\cdot Tg)=Q(f,Ug)$.}
               and we can define:
                \begin{eqnarray}
                \label{defUstar}
                U^{*}g &:=& \frac{Q(g, Uf)}{Q(f, Uf)}\cdot Uf, \\
                U^{**}g &:=& Ug - \frac{Q(g, Uf)}{Q(f, Uf)}\cdot Uf.
                \end{eqnarray}
                That $U=U^{*}+U^{**}$ is clear; further, we have
                \begin{eqnarray}
                \label{posUstar}
                Q(g,U^{*}g) &=& Q(g, Uf)\frac{Q(Uf, g)}{Q(Uf,f)} = \frac{|Q(g,Uf)|^{2}}{Q(f,Uf)} \\
                \label{posUstarstar}
                Q(g, U^{**}g) &=& Q(g, Ug) - Q(g, Uf)\frac{Q(Uf, g)}{Q(Uf,f)} \nonumber \\
               &=& \frac{Q(g, Ug)Q(f, Uf)-|Q(g, Uf)|^{2}}{Q(f,Uf)}.
                \end{eqnarray}
                The [non-negative] definiteness of $U^{*}$ follows directly from the first equation [(\ref{posUstar})], that of $U^{**}$ from the second [(\ref{posUstarstar})], taking into account the inequality in note 23; moreover, it is clear
                that $U^{*}$ and $U^{**}$ are linear symmetric. Consequently, $U^{*}$ must be proportional to $U$, but since
                \begin{equation}
                U^{*}f = Uf \neq 0,
                \end{equation}
                we must have $U^{*}=U$.\footnoteB{In other words, by the definition of a pure ensemble, we must have $U^{*}=C\cdot U,\; C>0$; but there exists a $f\neq 0$ for which (by Eq.\ (\ref{defUstar})) $U^{*}f = Uf$, so $C=1$.} If we now define
                \begin{equation}
                  \varphi = \frac{1}{\sqrt{Q(Uf)}}\cdot Uf,\;\;\alpha = \frac{Q(Uf)}{Q(f,Uf)},
                  \end{equation}
                  then\footnoteB{From (\ref{defUstar}) we have, for any $g$ in $\mathfrak{H}$, using $Uf = \sqrt{Q(Uf)}\varphi$,
                  \begin{eqnarray}
                  U^{*}g &=& Ug \\
                  &=& \frac{Q(g,\varphi)\sqrt{Q(Uf)}}{Q(f,Uf)}\cdot \sqrt{Q(Uf)}\varphi \\
                  &=& \frac{Q(Uf)}{Q(f,Uf)}Q(g,\varphi)\varphi \\
                  &=& \alpha Q(g,\varphi)\varphi.
                  \end{eqnarray}
                  Thus, $U=U^{*}=\alpha P_{\varphi}$.}
                  \begin{equation}
                  U = U^{*} = \alpha P_{\varphi},\;\; Q(\varphi) = 1.
                  \end{equation}
                 In other words, $U$ has, up to a constant positive factor $\alpha$, the desired form.
                 
                 Conversely , let $U=P_{\varphi}$, $Q(\varphi)=1$. Further, let
                 \begin{equation}
                 U = U^{*}+U^{**},
                 \end{equation}
                 with $U^{*}, U^{**}$ non-negative definite. Whenever $Uf=0$, then, because
                 \begin{equation}
  0\leq Q(f,U^{*}f) \leq Q(f,U^{*}f) + Q(f,U^{**}f) = Q(f, Uf) = 0,
\end{equation}
\begin{equation}
Q(f,U^{*}f) = 0,
\end{equation}
                 $U^{*}f=0$, by note 23. Now, since $Q(f,\varphi)=0$, [equivalently] $Uf=0$, $U^{*}f=0$ follows; thus, for any $g$,
                 \begin{equation}
                   Q(U^{*}f,g) = Q(f, U^{*}g) = 0.
                   \end{equation}
                   Namely: $U^{*}g$ is orthogonal to every $f$ that is orthogonal to $\varphi$, and must therefore be proportional to $\varphi$. In particular, we must have $U^{*}\varphi=\alpha\varphi$.
                   Further, for every $f$, $f=Q(f,\varphi)\cdot\varphi+f^{\prime}$, where $f^{\prime}$ is orthogonal to $\varphi$, from which it follows that\footnoteB{That $U^{*}f^{\prime}=0$ follows from the fact
                   that $Uf^{\prime}=P_{\varphi}f^{\prime}=0$ for $f^{\prime}$ orthogonal to $\varphi$. Hence, by the argument just given, $U^{*}f^{\prime}$ must also vanish.}
                   \begin{equation}
                   U^{*}f = Q(f,\varphi)\cdot U^{*}\varphi + U^{*}f^{\prime}  = Q(f,\varphi)\cdot\alpha\varphi = \alpha P_{\varphi}f = \alpha Uf,
                   \end{equation}
                   i.e., $U^{*}=\alpha U$. Therefore, $U^{*}$ is proportional to $U$, therefore also $U^{**}=U-U^{*}$, completing the proof.
                   
                   The fully determined states, or uniform ensembles, therefore correspond to the elements $\varphi$ of $\mathfrak{H}$ lying on the unit sphere ($Q(\varphi)=1$). The correspondence is fixed by the fact that the operator $U$ that
                   determines expectation values is equal to $P_{\varphi}$: from this, we can immediately calculate $\mathbf{E}(S)$. It suffices to use the $\mathfrak{H}_{0}$ [i.e., discrete, $l^2$] realization, in which $S$ has the matrix $\{s_{\mu\nu}\}$,
                   and $\varphi$ is the point $(x_1,x_2,\ldots)$. Then $P_{\varphi}$ has the matrix $\{x_{\mu}^{*}x_{\nu}\}$, and one finds:
                   \begin{equation}
                   \label{expectS}
                     \mathbf{E}(S) = \sum_{\mu\nu}s_{\mu\nu}x_{\mu}x_{\nu}^{*} = Q(\varphi,S\varphi).
                     \end{equation}
                     In other words, the quantity $\mathfrak{a}$, to which the operator $S$ corresponds, has, in the state associated with the point $\varphi$ of $\mathfrak{H}$, the expectation value $Q(\varphi,S\varphi)$.\footnoteB{This concludes von Neumann's
                     inductive derivation of the Born rule.}
                
                Next, we wish to establish some properties of this expression [i.e., Eq.\ (\ref{expectS})]:
                               
                $\alpha.$ Two [points in Hilbert space] $\varphi,\psi$ clearly determine the same expectation values (up to a positive constant factor), i.e., the same state, if and only if they differ merely by a constant
                value (which, given that $Q(\varphi)=Q(\psi)=1$, must have absolute value 1 [thus, is a complex phase only]).
                
                $\beta.$ If a quantity is always equal to 1, then its corresponding operator is $1$ (the unit operator, according to \textbf{D.}), and, according to the formula above, has an expectation value calculated as
                \begin{equation}
                Q(\varphi, 1\varphi) = Q(\varphi,\varphi) = 1.
                \end{equation}
                This means that we have the correct normalization, and our eigenvalues and probabilities are absolute, not relative.
                
                $\gamma.$ The formula $\mathbf{E}(S) = Q(\varphi,S\varphi)$ implies a direct interpretation of the operator $S$ associated with a quantity $\mathfrak{a}$: it (or, the hermitian form corresponding to its matrix,) gives, in 
                all imaginable states of the system, the (absolute) expectation value of $\mathfrak{a}$. The connection [between quantity and operator] is therefore the most simple one imaginable.
                
                $\delta.$ The necessary and sufficient condition that the statistical distribution of the quantity $\mathfrak{a}$ (with operator $S$) be sharp is that $\varphi$ be an eigenfunction of $S$; the [sharp] value
                of $\mathfrak{a}$ is then the eigenvalue of $\varphi$. This can be shown in the following fashion:
                
                In order that only the single value $w$ be possible for the quantity $\mathfrak{a}$, it must be the case that the probability that the value of $\mathfrak{a}$ lies in the interval $I$ is 1, or 0,
                depending on whether $w$ lies in $I$, or not. If $f(x)$ be the function defined to be equal to 1 in $I$ and zero otherwise [i.e., the characteristic function\footnoteB{See, for example, Rudin \autocite*[p.\ 11]{Rudin:1987}.}, usually written $\chi_{I}(x)$, of the interval $I$],
                 then this probability is
                the expectation value of $f(\mathfrak{a})$. As $f(S)$ is equal to $E(I)$ (where $E(\lambda)$ is the partition of unity associated with $S$, using the terminology of M.B.Q., section 9), this expectation value
                is equal (taking $I$ to be the interval $(w^{\prime},w^{\prime\prime})$) to\footnoteB{Taking $I=(w^{\prime},w^{\prime\prime})$,
                \begin{equation}
                  f(S) = \int_{-\infty}^{\infty} f(\lambda) dE(\lambda) = \int_{-\infty}^{\infty} \chi_{I}(\lambda)dE(\lambda) = \int_{w^{\prime}}^{w^{\prime\prime}} dE(\lambda) = E(w^{\prime\prime}) - E(w^{\prime}) := E(I).
                  \end{equation}}
                \begin{eqnarray}
                 Q(\varphi, E(I)\varphi) &=& Q(\varphi, E(w^{\prime\prime})\varphi) - Q(\varphi, E(w^{\prime})\varphi) \nonumber \\
                 &=& Q(E(w^{\prime\prime})\varphi) - Q(E(w^{\prime})\varphi).
                 \end{eqnarray}
               This should equal 1 or 0, according to whether $w$ lies in $I$ or not. Thus,
               \begin{equation}
                 Q(E(w^{\prime})\varphi) = 0,\quad E(w^{\prime})\varphi = 0,\quad [w^{\prime}<w],
                 \end{equation}
                 or,
                 \begin{equation}
                  Q(E(w^{\prime})\varphi) = 1 = Q(\varphi),\quad E(w^{\prime})\varphi = \varphi,\quad [w^{\prime}\geq w].
                  \end{equation}
                  But this implies that $S\varphi = w\varphi$, which was to be proven.\footnoteA{To see this, note that for every $g$,
                  \begin{equation}
                  Q(S\varphi,g) = \int_{-\infty}^{\infty}\lambda dQ(E(\lambda)\varphi,g) = wQ(\varphi,g),\;\;Q(S\varphi-w\varphi,g) = 0,
                  \end{equation}
                  from which $S\varphi - w\varphi = 0$ follows. On the other hand, from $S\varphi = w\varphi$ we have
                  \begin{equation}
                  Q((S-w\cdot 1)\varphi) = \int_{-\infty}^{\infty} (\lambda-w)^{2}dQ(E(\lambda)\varphi) = 0;
                  \end{equation}
                  as the integrand is never negative and the expression following the [differential] $d$ is non-decreasing, it [$Q(E(\lambda)\varphi)$] must be constant for $\lambda>w$ and $\lambda<w$, where the integrand is positive [and non-zero].
                  But $Q(E(\lambda)\varphi)$ tends to 1 (resp.\ 0) for $\lambda\rightarrow +\infty$ (resp.\ $\lambda\rightarrow -\infty$). Therefore it is equal to 1 (resp.\ 0) for $\lambda>w$ (resp.\ $\lambda<w$), whence follows $E(\lambda)\varphi = \varphi$ for
                  $\lambda>w$, $E(\lambda)\varphi=0$ for $\lambda<w$.
                  
                  Dirac appears to have been
                   the first to call attention to the important fact stated in $\delta.$ See \emph{Proceedings of the Royal Society}, Vol.\ 112 (1926).}\footnoteB{The reference in note 24 is probably to Dirac \autocite*[p.\ 639]{Dirac:1927a}, which is in
                  \emph{Proc. Roy. Soc.}, Vol.\ 113.}
              
              As there are certainly operators $S$, of which $\varphi$ is not an eigenfunction, so there must, for any state, exist quantities whose statistical distribution is not sharp.
              
              \section{Measurements and States}
              
              The end
              result of Section 4 allows us to determine all expectation values and probabilities in a system that finds itself in a completely known state. Nonetheless, this is a digression from our main objective. Indeed, our
              knowledge of a system $\mathfrak{S}^{\prime}$---the structure of a statistical ensemble $\{\mathfrak{S}^{\prime}_{1},\mathfrak{S}^{\prime}_{2},\ldots\}$---is never described by specification of the state, or even of the
              $\varphi$ belonging to it, but rather, in general, through the results of experiments carried out on the system. We must therefore try to arrive at a treatment of such descriptions using the formal methods at our disposal.
              
              First, we  establish the conditions under which  $m$ quantities $\mathfrak{a}_{1},\mathfrak{a}_{2},\ldots,\mathfrak{a}_{m}$ (with operators $S_{1},S_{2},\ldots,S_{m}$) can be simultaneously observed. In principle we already
              answered this question in Section 2: [this is the case] if their measurement, taken together, can be viewed as a single measurement, that then measures a quantity $\mathfrak{a}$ (with the operator $S$). In this case
              however $\mathfrak{a}_{1},\mathfrak{a}_{2},\ldots,\mathfrak{a}_{m}$ must be functions of $\mathfrak{a}$, i.e., $S_{1},S_{2},\ldots,S_{m}$ are functions of $S$ (cf.\ note 8). When therefore, given $m$ normal
              operators $S_{1},S_{2},\ldots,S_{m}$, does there exist a normal operator $S$, and $m$ functions $f_{1}(x),f_{2}(x),\ldots,f_{m}(x)$, such that $S_{\mu}=f_{\mu}(S), \mu=1,2,\ldots,m$?
              
              Two functions of $S$ commute,\footnoteA{If (for real $x$!) $h(x)=f(x)g(x)$, then one has, as one easily demonstrates (and should expect for any definition of functions of operators)
              \begin{equation}
                 f(S)\cdot g(S) = h(S).
                 \end{equation}
                 } so all $S_{1},S_{2},\ldots,S_{m}$ must commute [with one another]: but even more must be the case. If $E_{1}(\lambda),E_{2}(\lambda),\ldots,E_{m}(\lambda)$ are the partitions of unity
              corresponding to $S_{1},S_{2},\ldots,S_{m}$, then $E_{\mu}(\lambda)$ is a function of $S_{\mu}$,\footnoteA{For [$f(x) = \vartheta(\lambda-x)$] $f(x)$ = 1 (resp.\ 0) if $x<\lambda$ (resp.\ $x\geq\lambda$), $f(S_{\mu})=E_{\mu}(\lambda)$.} hence of $S$, 
              so all $E_{\mu}(\lambda)$ ($\mu=1,2,\ldots,m$, $\lambda$ arbitrary) must commute with each other. If these
              relation hold among $S_{1},S_{2},\ldots,S_{m}$, then we say, following Section 13 of M.B.Q., that they \emph{completely commute} [our emphasis].\footnoteA{We allow ourselves to repeat here what was stated in M.B.Q.\ on complete
              commutativity.  It implies the usual commutativity ($S_{\mu}S_{\nu}=S_{\nu}S_{\mu}$), and, for bounded $S_{\mu}$, follows from this. For unbounded $S_{\mu}$ the equivalence is unproven. Indeed, this problem arises from the
              fact that it is difficult to define sensibly the usual commutativity (as $S_{\mu}f$ sometimes fails to exist, cf.\ M.B.Q., note 27). Still, complete commutativity is the correct generalization of commutativity for arbitrary (normal) operators.}\footnoteB{The
              slippery issue of commutativity for unbounded operators is examined in Appendix 3 of von Neumann \autocite*{VonNeumann:1929b}.}
              
              The converse of this assertion is also correct: if $S_{1},S_{2},\ldots,S_{m}$ are completely commutative, i.e., the [spectral resolutions] $E_{\mu}(\lambda)$ ($\mu=1,2,\ldots,m$, $\lambda$ arbitrary) commute, then there exists a normal
              operator $S$, of which the $S_{1},S_{2},\ldots,S_{m}$ are functions. We will not go here into the proof, which involves only formal difficulties.\footnoteA{The author intends to return to this and other related questions on
              functions of operators in a later mathematical work [see von Neumann \autocite*{VonNeumann:1929a,VonNeumann:1929b}].}
              
              We have therefore shown: $m$ quantities $\mathfrak{a}_{1},\mathfrak{a}_{2},\ldots,\mathfrak{a}_{m}$ are simultaneously measurable, if, and only if their operators are completely commutative. Complete commutativity is more or less
              the same---cf.\ note 27---as usual commutativity, which was first identified by Dirac as the criterion for simultaneous measurability (cf.\ citation in note 24). And in this case, there always exists a quantity $\mathfrak{a}$, the
              measurement of which involves their measurement [i.e., the $\mathfrak{a}_{1},\mathfrak{a}_{2},\ldots,\mathfrak{a}_{m}$].
              
              We have so far established when several quantities can be simultaneously measured with arbitrary precision. We have yet to explore how to proceed if we are only interested in some quantities to a limited extent, for example, 
              if we only wish to know if they lie in a specified interval, or not. It might, for example,  be of interest to decide if $\mathfrak{a}_{1},\mathfrak{a}_{2},\ldots,\mathfrak{a}_{m}$ lie (respectively) in $I_{1},I_{2},\ldots,I_{m}$, and further, to
              measure $\mathfrak{b}_{1},\mathfrak{b}_{2},\ldots,\mathfrak{b}_{n}$ exactly. Let $f_{\mu}(x)$ be the function which equals 1 (resp.\ 0) if $x$ lies (resp.\ does not lie) in $I_{\mu}$. Then the quantity $f_{\mu}(\mathfrak{a}_{\mu})$ is
              1 (resp.\ 0) if $\mathfrak{a}$ lies (resp.\ does not lie) in $I_{\mu}$. Otherwise stated: we must determine the $f_{\mu}(\mathfrak{a}_{\mu})$ and the $\mathfrak{b}_{\nu}$ with arbitrary precision.
              
              Let the operators $S_{\mu},T_{\nu}$ correspond to the quantities $\mathfrak{a}_{\mu},\mathfrak{b}_{\nu}$, with the corresponding partitions of unity $E_{\mu}(\lambda), F_{\nu}(\lambda)$; the quantity $f_{\mu}(\mathfrak{a}_{\mu})$
              then has the operator $f_{\mu}(S_{\mu})=E_{\mu}(I_{\mu})$. The partition of unity for $E_{\mu}(I_{\mu})$  (which is a projection operator!) is clearly as follows: for $\lambda <0$, it equals 0; for $0\leq \lambda <1$, it equals $1-E_{\mu}(I_{\mu})$;
              for $\lambda \geq 1$, it equals 1. The complete commutativity of the $f_{\mu}(S_{\mu})$ and $T_{\nu}$ (which we must require) amounts to saying that the $E_{\mu}(I_{\mu})$ commute with each other and with all the $F_{\nu}(\lambda)$,
              and the latter with each other. In this way the required properties are achieved. 
              
              We will further show that if  $\mathfrak{a}_{1},\mathfrak{a}_{2}$ are two simultaneously observable quantities, with operators $S_{1},S_{2}$, then their product (which is clearly meaningful on account of their simultaneous measurability)
              has the operator $S_{1}S_{2}$. For $\mathfrak{a}_{1},\mathfrak{a}_{2}$ must be functions of a quantity $\mathfrak{a}$ (with operator $S$), namely, $\mathfrak{a}_{1}=f(\mathfrak{a})$, $\mathfrak{a}_{2}=g(\mathfrak{a})$,
              whence  $\mathfrak{a}_{1}\mathfrak{a}_{2}= h(\mathfrak{a})$ ($h(x)=f(x)g(x)$), so that $S_{1}=f(S), S_{2}=g(S)$, and the operator corresponding to $\mathfrak{a}_{1},\mathfrak{a}_{2}$ is $h(S)=f(S)g(S)=S_{1}S_{2}$ (cf.\ note 25).
              Naturally, by complete commutativity, we have $S_{1}S_{2}=S_{2}S_{1}$, i.e., commutative multiplication. This result carries over immediately to an arbitrary (finite) number of simultaneously measurable factors.\footnoteA{Instead
              of a product we could just as well have considered any other function of simultaneously measurable quantities, i.e., completely commutative operators. However, do not wish to go into such matters any further than necessary: for
              our purposes the sum and the product suffice.}
              
              \section{Measurements and States (contd.)}
              
              Let us now assume that we have simultaneously determined that the $m$ quantities $\mathfrak{a}_{1},\mathfrak{a}_{2},\ldots,\mathfrak{a}_{m}$ have values lying in the intervals $I_{1},I_{2},\ldots,I_{m}$---whether \emph{only} [our emphasis]
               these determinations are simultaneously possible, or indeed all the $\mathfrak{a}_{1},\mathfrak{a}_{2},\ldots,\mathfrak{a}_{m}$ are simultaneously measurable, is not important.
               
               If all we know of a system $\mathfrak{S}^{\prime}$ is this---i.e., if an ensemble $\{\mathfrak{S}_{1}^{\prime},\mathfrak{S}_{2}^{\prime},\ldots\}$ is selected from the fundamental ensemble defined in Section 2, in such a way that the
               measurement specified above is executed on each element [of the fundamental ensemble], with those giving a positive result collected into the ensemble $\{\mathfrak{S}_{1}^{\prime},\mathfrak{S}_{2}^{\prime},\ldots\}$---what then is
               the statistics of $\mathfrak{S}^{\prime}$ as regards $\{\mathfrak{S}_{1}^{\prime},\mathfrak{S}_{2}^{\prime},\ldots\}$? In other words: which [nonnegative]-definite linear symmetric operator characterizes (following Section 3) this statistics?
               
               We will only be able to answer this question fully in later Sections; here we carry out the preparatory work and answer a partial question, namely, when does the knowledge cited above suffices for a complete determination
               of the state of $\mathfrak{S}^{\prime}$, and, if so, what is the state?
               
               Every element of $\{\mathfrak{S}_{1}^{\prime},\mathfrak{S}_{2}^{\prime},\ldots\}$ has been subjected to a measurement, that gave for the quantity $f_{\mu}(\mathfrak{a}_{\mu})$ the value 1, and correspondingly for $g_{\mu}(\mathfrak{a}_{\mu})$
               (where $g(x)=0$ in $I_{\mu}$, otherwise = 1, $f(x)+g(x)=1$, and therefore $g_{\mu}(S_{\mu})=1-f_{\mu}(S_{\mu})=1-E_{\mu}(I_{\mu})$) the value 0. Indeed $f_{\mu}(\mathfrak{a}_{\mu})$ and $g_{\mu}(\mathfrak{a}_{\mu})$ are
               from the outset only capable of assuming the values 0 and 1. If the measurement [of $g_{\mu}(\mathfrak{a}_{\mu})$] is repeated the value 0 must be obtained in every case,\footnoteA{A measurement is in fact in principle
               an intervention (from which the ``a-causal'' character of quantum mechanics arises, cf.\ Heisenberg, \emph{Zeitschrift f\"ur Physik}, Vol.\ 43, 3/4, 1927); however, we must assume that
               the alteration proceeds in the interest of the experiment, i.e., that as soon as the experiment is carried out, the system is in a state, in which the \emph{same} measurement can be executed without any
               further alteration of the system. Otherwise stated: if the same measurement is carried out twice (with nothing happening in the interim!), the same result is obtained.} i.e., the statistical distribution of $g_{\mu}(\mathfrak{a}_{\mu})$
               is sharp, with the value 0. Accordingly, the expectation value (also, the relative one!) of $g_{\mu}(\mathfrak{a}_{\mu})$ must vanish, and since $g_{\mu}(\mathfrak{a}_{\mu})$ has the operator $1-E_{\mu}(I_{\mu})$, we must have
               \begin{equation}
                 \mathbf{E}(1-E_{\mu}(I_{\mu})) = 0.
                 \end{equation}
               Certain conclusions about $U$ follow in fact from this equation.
               
               Let $\varphi$ lie in the exterior of $E_{\mu}(I_{\mu})$, i.e., in the interior of $1-E_{\mu}(I_{\mu})$, and let $Q(\varphi)=1$. 
               Then the projection operator $P_{\varphi}$ satisfies $P_{\varphi}\leq 1-E_{\mu}(I_{\mu})$.\footnoteA{From M.B.Q., Section 8, Theorem 8. Indeed,
               the interior of $P_{\varphi}$ consists clearly of the [vectors] $f$ proportional to $\varphi$, and these lie (with $\varphi$) in the interior of $1-E_{\mu}(I_{\mu})$.} Accordingly $P_{\varphi}$ and $1-E_{\mu}(I_{\mu})-P_{\varphi}$ are projection operators,
               hence normal and [non-negative]-definite.\footnoteA{Every projection operator is non-negative definite as $E=E^2$, cf.\ note 21} Therefore,
               \begin{eqnarray}
               0\leq \mathbf{E}(P_{\varphi})&\leq& \mathbf{E}(P_{\varphi})+\mathbf{E}(1-E_{\mu}(I_{\mu})-P_{\varphi}) = \mathbf{E}(1-E_{\mu}(I_{\mu})) = 0,\\
             \mathbf{E}(P_{\varphi}) &=& 0.
             \end{eqnarray}
             In Section 3 however we calculated $\mathbf{E}(P_{\varphi})$: it is equal to $Q(\varphi,U\varphi)$. From $Q(\varphi,U\varphi)=0$ follows (cf.\ note 23) $U\varphi=0$.
              In other words, in the exterior of $E_{\mu}(I_{\mu})$ the requirement $Q(\varphi)=1$  implies $U\varphi =0$,
             whence for any vector $\varphi$ we have $U\varphi=0$. Consequently, for all $g$ in $\mathfrak{H}$,
             \begin{equation}
             \label{UEeqU}
               U(1-E_{\mu}(I_{\mu}))g = 0,\;\;UE_{\mu}(I_{\mu})g = Ug,\;\;UE_{\mu}(I_{\mu}) = U.
               \end{equation}
               As the $E_{1}(I_{1}),E_{2}(I_{2}),\ldots,E_{m}(I_{m})$ must commute, 
               \begin{equation}
                E = E_{1}(I_{1})* E_{2}(I_{2})*\ldots* E_{m}(I_{m})
                \end{equation}
                must be a projection operator. By applying the above equation [Eq.\ (\ref{UEeqU})] for $\mu=1,2,\ldots,m$, we get $UE=U$.
                
                Hence, $U, E, UE$ are linear symmetric operators, which implies that $U$ and $E$ commute [as $UE = (UE)^{\dagger} = E^{\dagger}U^{\dagger} = EU$]; thus,
                \begin{equation}
                 EU = UE = U.
                 \end{equation}
                 At this level of generality we cannot at this point go any further.
                 
                 We consider two special cases. First, suppose $E=0$. Then $U = UE = 0$, i.e., all expectation values vanish, which makes no sense [\emph{was unsinnig ist}].
                  Such results of a measurement can consequently never appear in the real world---indeed, we shall see
                 later that they have probability 0.
                 
                 Second, let the interior of $E$ be one-dimensional, i.e., it consists of a single $\varphi\neq 0$, which we normalize by $Q(\varphi)=1$ (whereby clearly there still remains the freedom of a constant factor of absolute value 1). Then, as one
                 easily can see, $E=P_{\varphi}$. And from this it follows for $U$:
                 \begin{eqnarray}
                   U&=& EUE = P_{\varphi}UP_{\varphi}, \\
                   Uf &=& P_{\varphi}UP_{\varphi}f = Q(f,\varphi)\cdot Q(U\varphi,\varphi)\cdot\varphi = Q(U\varphi,\varphi)\cdot P_{\varphi}f,\\
                   U&=& Q(U\varphi,\varphi)\cdot P_{\varphi}.
                   \end{eqnarray}
              $Q(U\varphi,\varphi)$ is a constant factor that, as $U, P_{\varphi}$ are [positive]-definite, must be positive;
                 accordingly, we can drop it, and obtain $U=P_{\varphi}$. In other words, we now have the statistical [operator] corresponding to
                the state $\varphi$! We can therefore say: $\mathfrak{S}^{\prime}$ is in the state $\varphi$, the ensemble $\{\mathfrak{S}_{1}^{\prime},\mathfrak{S}_{2}^{\prime},\ldots\}$ is uniform.
                
                In this way we can indicate measurements which unambiguously determine the state $\mathfrak{S}^{\prime}$.\footnoteB{At this point, the concept of a complete set of compatible quantities, the simultaneous
                determination of which uniquely specifies a state---i.e., a unit vector in the Hilbert space---enters quantum mechanics. A classic example is familiar to students of any introductory course: the specification
                of a unique bound state of the hydrogen atom by simultaneous measurement of $H,L^{2},L_{z},S_{z}$ (resp.\ energy, squared total orbital angular momentum, component of orbital angular moment in $z$ direction, component
                of spin angular momentum in $z$ direction), typically characterized by the quantum numbers $n,l,m,m_{s}$. The concept of a complete set of compatible quantities is the natural successor in quantum mechanics to the
                specification of a system in classical mechanics in terms of pairs of conjugate variables (e.g., coordinate/momentum pairs).}
                 If $S$ is an operator with a point spectrum consisting purely of simple eigenvalues, and no continuous spectrum---we
                denote such an operator ``absolutely discrete and non-degenerate''---and if further the real line is covered with a sequence of intervals $I_{1},I_{2},\ldots$ such that each interval contains exactly
                one eigenvalue of $S$, then the following measurement suffices clearly to determine unambiguously the state of the system: let $\mathfrak{a}$ be the quantity associated with $S$, then one has to determine in which of the
                intervals $I_{1},I_{2},\ldots$ the value of $\mathfrak{a}$ lies. Indeed, if the eigenvalues $w_1,w_2, \ldots$, with eigenfunctions $\varphi_{1},\varphi_{2},\ldots$, lie in the corresponding intervals $I_{1},I_{2},\ldots$, and the partition of unity
                for $S$ is $E(\lambda)$, then $E(I_{n})$ is equal to $P_{\varphi_{n}}$ (cf.\ M.B.Q., section 10).
                
                Note that this measurement does not even have to be absolutely exact: if the $w_{1},w_{2},\ldots$ lie sufficiently far apart, then the intervals $I_{1},I_{2},\ldots$ can also be chosen arbitrarily large.\footnoteA{Again, one sees how profound
                the effect of the intervention on a system $\mathfrak{S}$ required by a measurement is:  it is forced by the measurement to transition to one of the states $\varphi_{1},\varphi_{2},\ldots$ The entire freedom left to the system consists
                of the choice of one of the elements of the complete orthonormal system $\varphi_{1},\varphi_{2},\ldots$}
                
                \section{Statistical connection of diverse measurements}
                
                Let $\mathfrak{a}_{1},\mathfrak{a}_{2},\ldots,\mathfrak{a}_{m}$, $I_{1},I_{2},\ldots,I_{m}$, as well as $S_{1},S_{2},\ldots,S_{m}$, $E_{1}(\lambda),E_{2}(\lambda),\ldots,E_{m}(\lambda)$ be as in Section 6, but now we no longer require that the
                interior of
                                \begin{equation}
                E = E_{1}(I_{1}) * E_{2}(I_{2}) * \ldots * E_{m}(I_{m})
                \end{equation}
                be one-dimensional. We now wish to carry through, in complete generality, the considerations of Section 6, specifically, the determination of the $U$ associated with the statistics implied by the statement ``the values of $\mathfrak{a}_{1},
                \mathfrak{a}_{2},\ldots,\mathfrak{a}_{m}$ lie in the intervals $I_{1},I_{2},\ldots,I_{m}$''.  
                
                First, a remark: let $\mathfrak{b}_{1},\mathfrak{b}_{2},\ldots,\mathfrak{b}_{n}$ be $n$ other quantities, $J_{1},J_{2},\ldots,J_{n}$ intervals, such that the  determination that $\mathfrak{b}_{1}$ lies in $J_{1}$,
                $\mathfrak{b}_{2}$ in $J_{2}$,\ldots,$\mathfrak{b}_{n}$ in $J_{n}$ is simultaneously possible. We seek a quantity whose (relative) expectation value in any ensemble is the same as the (relative) probability that
                 $\mathfrak{b}_{1},\mathfrak{b}_{2},\ldots,\mathfrak{b}_{n}$ lie, respectively, in $J_{1},J_{2},\ldots,J_{n}$.
                 
                 Let $f_{\nu}(x)$ the function which is equal to 1 in $J_{\nu}$, 0 otherwise. Then the quantities $f_{\nu}(\mathfrak{b}_{\nu})$ are simultaneously measurable, and take the values 0 or 1. Our condition is fulfilled when 
                 they are all equal to 1. We can therefore construct the quantity $f_{1}(\mathfrak{b}_{1})*f_{2}(\mathfrak{b}_{2})\cdots f_{n}(\mathfrak{b}_{n})$: this has the values 0 or 1, and our condition is characterized by
                 saying that it is equal to 1. This quantity has therefore the same (relative) probability as the (relative) expectation value that $\mathfrak{b}_{1},\mathfrak{b}_{2},\ldots,\mathfrak{b}_{n}$ lie, respectively, in $J_{1},J_{2},\ldots,J_{n}$.
                 
                 If $\mathfrak{b}_{1},\mathfrak{b}_{2},\ldots,\mathfrak{b}_{n}$ have operators $T_{1},T_{2},\ldots,T_{n}$, with partitions of unity $F_{1}(\lambda),F_{2}(\lambda),\ldots,F_{n}(\lambda)$, then $f_{\nu}(\mathfrak{b}_{\nu})$ has the operator 
                 $f_{\nu}(T_{\nu})=F_{\nu}(J_{\nu})$. Consequently the quantity introduced just above, $f_{1}(\mathfrak{b}_{1})\cdot f_{2}(\mathfrak{b}_{2})\cdots f_{n}(\mathfrak{b}_{n})$, has the operator (cf.\ Section 5)
                 \begin{equation}
                    F = F_{1}(J_{1}) * F_{2}(J_{2}) * \cdots * F_{n}(J_{n}).
                    \end{equation}
                  Thus we confirm an assertion made in the preceding Section:  if $F=0$, then its expectation value in any ensemble, i.e., the probability of the measurements specified above, equals 0.
                 
                 We now return to our main task. The statistics of the ensemble $\mathfrak{S}_{1}^{\prime},\mathfrak{S}_{2}^{\prime},\ldots$ must be independent of how the measurements used to generate it were carried out: in other words,
                 of the manner in which it was established that the values of $\mathfrak{a}_{1},\mathfrak{a}_{2},\ldots,\mathfrak{a}_{m}$  lie in $I_{1},I_{2},\ldots,I_{m}$. If $\mathfrak{b}$ is a quantity, the measurement of which can
                 be carried out simultaneously with these determinations (namely, with the measurements of the $f_{\mu}(\mathfrak{a}_{\mu})$, where $f_{\mu}(x)$ is 1 in $I_{\mu}$ and 0 otherwise), then it is possible that we could 
                 have also thereby  have measured $\mathfrak{b}$. In particular, if the operator of $\mathfrak{b}$ is absolutely discrete and non-degenerate (cf.\ Section 6), then its measurement implies the exact determination of the
                 state---if the eigenfunctions of this operator are $\varphi_{1},\varphi_{2},\ldots$, then the state must find itself [after measurement of $\mathfrak{b}$] in one of the states $\varphi_{1},\varphi_{2},\ldots$ As the $f_{\mu}(\mathfrak{a}_{\mu})$
                 commute with $\mathfrak{b}$, it follows, as one easily verifies,\footnoteB{Let, as below, the operator of $\mathbf{b}$ be $T$: thus $T\varphi_{n}=w_{n}\varphi_{n}$, where the eigenvalues $w_{n}$ are non-degenerate. Then
                 \begin{equation}
                   f_{\mu}(\mathbf{a}_{\mu})T\varphi_{n} = w_{n} f_{\mu}(\mathbf{a}_{\mu})\varphi_{n} = T f_{\mu}(\mathbf{a}_{\mu})\varphi_{n}.
                   \end{equation}
                   Thus $ f_{\mu}(\mathbf{a}_{\mu})\varphi_{n}\propto\varphi_{n}$ (as every eigenvalue of $T$ is non-degenerate), and $\varphi_{n}$ is an eigenvector of $ f_{\mu}(\mathbf{a}_{\mu})$.}
                  that the $\varphi_{1},\varphi_{2},\ldots$ are eigenfunctions of the operators for $f_{\mu}(\mathfrak{a}_{\mu})$. The quantities $f_{\mu}(\mathfrak{a}_{\mu})$
                 therefore have sharp distributions in these states. And the ensemble  $\mathfrak{S}_{1}^{\prime},\mathfrak{S}_{2}^{\prime},\ldots$  is obtained by selection [from the fundamental random ensemble] of just those systems in
                 which these (sharp) values of the $f_{\mu}(\mathfrak{a}_{\mu})$ are all equal to 1.
                 
                 The ensemble $\mathfrak{S}_{1}^{\prime},\mathfrak{S}_{2}^{\prime},\ldots$ is therefore simply a combination of some of the states $\varphi_{1},\varphi_{2},\ldots$ (namely, those in which all of the quantities $f_{\mu}(\mathfrak{a}_{\mu})$
                 have the sharp value 1), i.e., those belonging to the corresponding uniform ensembles (cf.\ Section 4). The proportions in which particular $\varphi_{1},\varphi_{2},\ldots$ appear in the set $\mathfrak{S}_{1}^{\prime},\mathfrak{S}_{2}^{\prime},\ldots$ 
                 will be given by a subsequent measurement of $\mathfrak{b}$ on this ensemble (as $\mathfrak{b}$ was already measured in the course of the generation of $\mathfrak{S}_{1}^{\prime},\mathfrak{S}_{2}^{\prime},\ldots$, and in the interim nothing
                 has happened, this measurement must yield the same result, cf.\ note 30). We can however already predict the result of this measurement: if $\varphi_{1},\varphi_{2},\ldots$ have the eigenvalues [for the
                 operator associated with the quantity $\mathfrak{b}$] $w_{1},w_{2},\ldots$, and the intervals
                 $J_{1},J_{2},\ldots$ are so chosen that $J_{p}$ (for $p=1,2,\ldots$) contains the eigenvalue $w_{p}$ and no other, then one obtains the state $\varphi_{p}$ if $\mathfrak{b}$ lies in $J_{p}$. Let $\mathfrak{b}$ have the operator $T$
                 with partition of unity $F(\lambda)$. Then the (relative) probability of this result is equal to the (relative) expectation value of a quantity with the operator $F(J_{p})=P_{\varphi_{p}}$. And this expectation value is, as we know,
                 equal to $Q(\varphi_{p},U\varphi_{p})$.
                 
                  We can now use these ideas to determine $U$, by choosing $\mathfrak{b}$ suitably. Let $\psi_{1},\psi_{2},\ldots$ be an orthonormal system in the interior of $E$, spanning an everywhere dense linear manifold therein.\footnoteA{The terminology
                  is from M.B.Q., Sections 5,6.} This system can be extended to a complete orthonormal system [of the full Hilbert space] $\psi_{1}\psi_{2},\ldots,\chi_{1}\chi_{2},\ldots$, with the property that every one of its members
                   lies either in the interior or the exterior of every $E_{\mu}(I_{\mu})$
                  ($\mu=1,2,\ldots,m$).\footnoteA{Form the $2^m$ projection operators $G_{1}\cdot G_{2}\cdots G_{m}$, in which each $G_{\mu}$ is either $E_{\mu}(I_{\mu})$ or $1-E_{\mu}(I_{\mu})$. From the interior of each of
                  these we choose an orthonormal system, that spans an everywhere dense linear manifold (following M.B.Q., Section 6, Theorem 6, in analogy to the corollary there). For $E_{1}(I_{1})* E_{2}(I_{2}) *\ldots *E_{m}(I_{m})$
                  we retain $\psi_{1},\psi_{2},\ldots$ The union of all these systems is, according to the Theorems of M.B.Q., Section 6, a complete orthonormal system (as our $2^m$ projection operators are pairwise orthogonal, and sum to 1),
                  of which $\psi_{1},\psi_{2}, \ldots$ is a part. And every element of this system lies in the interior of some $G_{1}\cdot G_{2}\cdots G_{m}$, i.e., for all $\mu=1,2,\ldots,m$ in the interior of $G_{\mu}=E_{\mu}(I_{\mu})$ or $1-E_{\mu}(I_{\mu})$,
                  or, equivalently, in the interior or the exterior of $E_{\mu}(I_{\mu}$). We thus achieve the desired properties.} We call this system $\varphi_{1},\varphi_{2},\ldots,$ and choose arbitrary real numbers $w_{1},w_{2},\ldots$ (distinct from
                  one another and with no finite accumulation point\footnoteB{This implies, by the Bolzano-Weierstrass theorem, that the $w_{n}$ be unbounded.}),
                   and form the (normal) operator with eigenfunctions  $\varphi_{1},\varphi_{2},\ldots,$ corresponding to eigenvalues $w_{1},w_{2},\ldots$. Call this operator $T$: it is absolutely discrete
                  and non-degenerate, and it corresponds to the quantity $\mathfrak{b}$. As all its eigenfunctions lie in either the interior or the exterior of every $E_{\mu}(I_{\mu})$, i.e., are eigenfunctions of all of these, $T$ completely commutes with
                  all $E_{\mu}(I_{\mu})$. That is to say: the determinations of whether the $\mathfrak{a}_{\mu}$ lie in the respective intervals $I_{\mu}$ can be executed simultaneously with the measurement of $\mathfrak{b}$.
                  
                  Our previous argument therefore applies to this $\mathfrak{b}$. The $\varphi_{1},\varphi_{2},\ldots$ fall into two groups: $\psi_{1},\psi_{2},\ldots$ and $\chi_{1},\chi_{2},\ldots$; in the former group lies every element in the interior of $E$,
                  in the latter, all that lie in the exterior of at least one $E_{\mu}(I_{\mu})$, and hence also of $E$.\footnoteB{The procedure used here works perfectly well if $E$ projects onto a subinterval of the continuous spectrum of some
                  observable. One picks any complete orthonormal set $\varphi_{n}$ (for a one-dimensional problem, these might be the Hermite functions of a harmonic oscillator, regardless of what the actual physical system might be). Then
                  the $\psi_{n}$ (resp.\ $\chi_{n}$) are obtained by subjecting all the  states $E\varphi_{n}$ (resp.\ $(1-E)\varphi_{n}$) to Schmidt orthogonalization, resulting in orthonormal sets spanning the interior (resp.\ exterior) of $E$.}
                   Accordingly, $Q(\varphi,E\varphi)$ takes the value 1 in the first group, the value 0 in the second, i.e., it is the first group,
                  for which the quantities $\mathfrak{a}_{\mu}$ lie in the $I_{\mu}$. Thus, our ensemble    $\mathfrak{S}_{1}^{\prime},\mathfrak{S}_{2}^{\prime},\ldots$ is an assembly of states $\varphi_{1},\varphi_{2},\ldots,$ with the (relative) weights
                  $Q(\varphi_{p},U\varphi_{p})$.\footnoteA{The $\chi_{q}$ have of course weight 0, as, given that $U=UE$, $Q(\chi_{q},U\chi_{q})= Q(\chi_{q},UE\chi_{q}) =0$.}       
                  
                  The statistics of the state $\varphi_{p}$ is described by the operator $P_{\varphi_{p}}$ (cf.\ Section 4), and in fact with the correct normalization. The statistics of the ensemble  $\mathfrak{S}_{1}^{\prime},\mathfrak{S}_{2}^{\prime},\ldots$
                  is described by the operator $U$, in the same (possibly incorrect) normalization, in which the weights of the states $\varphi_{p}$ are equal to $Q(\varphi_{p},U\varphi_{p})$. If we examine the expectation value of the
                  quantity which has the operator $P_{\chi}$ ($Q(\chi)=1$, $\chi$ otherwise arbitrary), we must have
                  \begin{equation}
                  \label{chiQchi}
                  Q(\chi,U\chi) = \sum_{p}Q(\varphi_{p},U\varphi_{p})Q(\chi,P_{\varphi_{p}}\chi) = \sum_{p}Q(\varphi_{p},U\varphi_{p})|Q(\varphi_{p},\chi)|^{2}.
                  \end{equation}
                  Thus---observing first that in this equation, apart from $U$, only the orthonormal system $\varphi_{1},\varphi_{2},\ldots$ appears, which is quite arbitrary, as long as it lies in the interior of $E$ and therein spans an 
                  everywhere dense manifold---we can, taking into account that $EU=UE=U$, conclude after a simple calculation that $U$ agrees with $E$, up to an overall constant factor.\footnoteA{This can be shown, for example, as follows.
                  Let $\varphi_{1},\varphi_{2}$ be two normalized and orthogonal elements of the interior of $E$. We will extend these to an orthonormal system $\varphi_{1},\varphi_{2},\varphi_{3},\ldots.$ in the interior of $E$, that spans (as can easily
                  be arranged) an everywhere dense linear manifold there, and set $\chi=(\varphi_{1}+\varphi_{2})/\sqrt{2}$. Our equation [Eq.\  (\ref{chiQchi})] is then applicable, and indeed gives
                  \begin{eqnarray}
                   Q(\frac{1}{\sqrt{2}}(\varphi_{1}+\varphi_{2}),U\frac{1}{\sqrt{2}}(\varphi_{1}+\varphi_{2}))&\!\!\!=\!\!\!& Q(\varphi_{1},U\varphi_{1})\cdot\frac{1}{2}+Q(\varphi_{2},U\varphi_{2})\cdot\frac{1}{2}, \\
                   Q((\varphi_{1}+\varphi_{2}),U(\varphi_{1}+\varphi_{2})) &\!\!\!=\!\!\!& Q(\varphi_{1},U\varphi_{1})+Q(\varphi_{2},U\varphi_{2}), \\
                   {\mathrm{Re}}\;Q(\varphi_{1},U\varphi_{2}) &\!\!\!=\!\!\!& 0.
                   \end{eqnarray}
                   As $\varphi_{1},i\varphi_{2}$ are also orthogonal and normalized, and lie in the interior of $E$, we similarly obtain ${\mathrm{Im}} \;Q(\varphi_{1},U\varphi_{2}) = 0$, hence $Q(\varphi_{1},U\varphi_{2}) = 0$. In other words,
                   for $f,g$ in the interior of $E$, $Q(f,g)=0$ implies $Q(f,Ug)=0$ if $Q(f)=Q(g)=1$, hence also without this latter constraint. Now $Ug$, for $g$ in the interior of $E$, always lies in the interior of $E$ (as $Ug=EUg$),
                   and is also orthogonal to all $f$ in the interior of $E$ that are orthogonal to $g$: hence [$Ug$] is also proportional to $g$. We have shown that $Ug = \alpha_{g}\cdot g$, for $g$ in the interior of $E$.
                   
                    If $g_{1},g_{2}$ lie in the interior of $E$ ($g_{1}\neq 0, g_{2}\neq 0$ and are not orthogonal, then
                    \begin{equation}
                      Q(g_{1},Ug_{2}) = \alpha_{g_{2}}^{*} Q(g_{1},g_{2}),\quad Q(Ug_{1}, g_{2}) = \alpha_{g_{1}}Q(g_{1},g_{2}),
                      \end{equation}
                      so $\alpha_{g_{1}}=\alpha_{g_{2}}^{*}$. So in the first place $\alpha_{g_{2}}=\alpha_{g_{2}}^{*}$ [choosing $g_{2}=g_{1}$], i.e., $\alpha_{g_{2}}$ is real, and secondly, $\alpha_{g_{1}}=\alpha_{g_{2}}$.
                      If, on the other hand, $g_{1}$ and $g_{2}$ are orthogonal, then neither is orthogonal to $g = g_{1}+g_{2}$, whence $\alpha_{g_{1}}=\alpha_{g_{2}}=\alpha_{g}$. Thus, $\alpha_{g}$ is constant in the
                      interior of $E$ for $g\neq 0$ (for $g=0$ it is arbitrary), so we have $Ug = \alpha g$. For arbitrary $f$ [in $\mathfrak{H}$], $Ef$ lies in the interior of $E$, and $Uf = UEF = \alpha Ef$. Therefore $U=\alpha E$, as asserted.
                   }
                   This factor must, as $U, E$ are positive-definite, be
                  positive, and we can drop it from $U$ without further ado, and set $U=E$.
                  
                  \section{[Statistical consequences of measurement]}
                  
                  The final result of Section 8 is of decisive importance: it enables us to establish the statistical consequences of the result of a measurement.
                  
                  If $\mathfrak{a}_{1},\mathfrak{a}_{2},\ldots,\mathfrak{a}_{m}$ and $\mathfrak{b}_{1},\mathfrak{b}_{2},\ldots,\mathfrak{b}_{n}$  are $m+n$ quantities, and $I_{1},I_{2},\ldots,I_{m}$, $J_{1},J_{2},\ldots,J_{n}$ are $m+n$ intervals;
                  if further all determinations whether $\mathfrak{a}_{\mu}$ lies in $I_{\mu}$ ($\mu=1,2,\ldots,m$) are simultaneously possible, and likewise all determinations whether $\mathfrak{b}_{\nu}$ lies in $J_{\nu}$ ($\nu=1,2,\ldots,n$)
                  (the two groups of course do not have to be simultaneously [measurable]), then we can ask: if we have determined that all $\mathfrak{a}_{\mu}$ lie in $I_{\mu}$ ($\mu= 1,2,\ldots,m$), what is the (relative) probability
                  that all the $\mathfrak{b}_{\nu}$ lie in $J_{\nu}$ ($\nu=1,2,\ldots,n$)?
                  
                  Let the operators associated with $\mathfrak{a}_{\mu}$ (resp.\ $\mathfrak{b}_{\nu}$) be $S_{\mu}$ (resp.\ $T_{\nu}$), their partitions of unity $E_{\mu}(\lambda)$ (resp.\ $F_{\nu}(\lambda)$). Under our assumptions
                  the $E_{\mu}(I_{\mu})$ must commute with each other, as must the $F_{\nu}(J_{\nu})$. Consequently, the operators
                  \begin{equation}
                     E = E_{1}(I_{1})* E_{2}(I_{2}) *\ldots *E_{m}(I_{m}),\quad F = F_{1}(J_{1})* F_{2}(J_{2}) * \ldots * F_{n}(J_{n}),
                     \end{equation}are projection operators.
                     
                     The statistics generated by the assertion that ``all $\mathfrak{a}_{\mu}$ lie in $I_{\mu}$ ($\mu=1,2,\ldots,m$)'' is described by the operator $E$ (cf.\ Section 7), the (relative) probability of a measurement result ``all
                     $\mathfrak{b}_{\nu}$ lie in $J_{\nu}$ ($\nu=1,2,\ldots,n$)'' is equal to the (relative) expectation value of a quantity, whose operator is $F$ (cf.\ beginning of Section 7). Now let us go over as one pleases\footnoteB{The freedom
                     implied here simply refers to the choice of an arbitrary complete orthonormal set of unit vectors $\varphi_{1},\varphi_{2},\ldots$ in $\mathfrak{H}$, which will be mapped to the canonical basis $(1,0,0,\ldots),(0,1,0,\ldots), \ldots$etc. 
                     in $\mathfrak{H}_{0}=l^2$. See note $l$.}to the [discrete] realization
                     $\mathfrak{H}_{0}$ [$l^2$] of $\mathfrak{H}$, where $E,F$ have the matrices $\{e_{\rho\sigma}\},\{f_{\rho\sigma}\}$. Then the desired (relative) probability (or relative expectation value, as stated above) is
                     \begin{equation}
                       \mathbf{E}(F) = \sum_{\rho\sigma} e_{\rho\sigma}f_{\rho\sigma}^{*}.
                       \end{equation}
                       We need to bring this expression into a simple closed form.
                       
                       The points $(1,0,0,\ldots), (0,1,0,\ldots),\ldots$ of $\mathfrak{H}_{0}$ form a complete orthonormal system, which we shall denote in $\mathfrak{H}$ as $\varphi_{1},\varphi_{2},\ldots$.  As $E,F$ are projection operators, their matrices are
                       equal to their squares. From this it follows that\footnoteA{In the notation of M.B.Q., Section 11.}\footnoteB{As in our translation of M.B.Q., we have replaced, in the final equation, von Neumann's confusing use of square brackets, $[E, F]$, 
                       in Eq.\ (\ref{EFprob}) with angle brackets, to avoid 
                       confusion with commutators. In Eq.\ (\ref{usesymm}) the hermitian property of projection operators is employed, e.g., $e_{\tau\sigma}=e_{\sigma\tau}^{*}$.}
                       \begin{eqnarray}
                       \sum_{\rho\sigma} e_{\rho\sigma}f_{\rho\sigma}^{*} &=& \sum_{\rho\sigma\tau\omega} e_{\rho\tau}e_{\tau\sigma}f_{\rho\omega}^{*}f_{\omega\sigma}^{*} \\
                       \label{usesymm}
                       &=&  \sum_{\rho\sigma\tau\omega} e_{\rho\tau}f_{\rho\omega}^{*}\cdot e_{\sigma\tau}^{*}f_{\sigma\omega} \\
                       &=& \sum_{\tau\omega}|\sum_{\rho}e_{\rho\tau}f_{\rho\omega}^{*}|^{2} \\
                       &=& \sum_{\tau\omega}|Q(E\varphi_{\tau}, F\varphi_{\omega})|^{2} \\
                       &:=& \langle E, F \rangle.
                       \end{eqnarray}.
                     The desired relative probability is therefore equal to\footnoteB{This elegant and simple expression contains the entire content of the Jordan-Dirac transformation theory: one simply has to insert the appropriate spectral
                     representations for the projection operators associated with the various observables specified in the preparation (the ``$E(I)$s'') and measurement (the ``$F(J)$s'') of the system. If this is done using Dirac notation (see note h),
                     the results obtained, in particular, by \autocite{Dirac:1927a} are easily recognized.}
                     \begin{equation}
                     \label{EFprob}
                     \langle E, F\rangle = \langle E_{1}(I_{1})* E_{2}(I_{2})*\ldots * E_{m}(I_{m}),\;F_{1}(J_{1})* F_{2}(J_{2})*\ldots * F_{n}(J_{n})\rangle,
                     \end{equation}
                     which is precisely the expression proposed in M.B.Q., Section 13, for it. There (Sections 12,14) it was shown that this  contains as special cases the usual probability formulas of quantum mechanics. Our result is therefore in
                     agreement with experimental data [\emph{Erfahrung}].
                     
                     We are now able to investigate the statistical properties of the fundamental ensemble $\{\mathfrak{S}_{1},\mathfrak{S}_{2},\ldots\}$ (cf.\ the beginning of Section 2) in which all possible states of $\mathfrak{S}$  equiprobably
                     and randomly appear (the paradigm of a system $\mathfrak{S}$ of whose state we are entirely ignorant), i.e., to determine the a priori probabilities of the individual states. We seek the operator $U$ which describes the
                     statistics of this ensemble: from the preceding discussion it must be equal to 1 [identity operator].\footnoteA{One might argue this simply because the product $E=E_{1}(I_{1})\cdot E_{2}(I_{2})\cdots E_{m}$ simply has no
                     factors, as no measurements were performed. Or, if one finds this argument unconvincing, for the following reason: let $\mathfrak{a}$ be an arbitrary quantity, $S$ its operator, $I$ the interval $(-\infty,+\infty)$,
                     $E(\lambda)$ the partition of unity associated with $S$. We can perhaps pretend that $\mathfrak{a}$ was measured and its value was found to be in $I$ (which simply means that nothing happened); then we have $E = E(I) =1$.}
                     The (relative) probability that $\mathfrak{S}$ transitions to the state $\varphi$---for example, that in the measurement of an absolutely discrete and non-degenerate quantity $\mathfrak{a}$ which possesses the eigenfunction
                     $\varphi$, the eigenvalue of $\varphi$ is found [as the result of the measurement]---is (where the operator of $\mathfrak{a}$ is $S$, its partition of unity $E(\lambda)$, $I$ an interval which contains only the eigenvalue of
                     $S$ corresponding to $\varphi$), as $E(I)=P_{\varphi}$,
                     \begin{equation}
                       \mathbf{E}(P_{\varphi}) = Q(\varphi,\mathbf{1}\varphi) = Q(\varphi,\varphi) = 1.
                       \end{equation}
                       In other words, all $\varphi$ have the same (relative a priori) probability 1.\footnoteA{The following circumstance is worthy of note: The possible states of $\mathfrak{S}$ form the unit-sphere (up to constant factors of absolute
                       value 1) of the Hilbert space $\mathfrak{H}$, hence a continuous manifold, and, in fact,
                       an infinite-dimensional one. Nonetheless we need not concern ourself in the specification of a priori probabilities with definitions of volume
                       elements (which are probably not even possible  in a Hilbert space)---as, for example, in classical mechanics---as the only (i.e., in an experiment) $\varphi$ that come into consideration always form a discrete manifold (e.g., all the elements of a complete orthonormal system). So we can assign a positive number as the a priori weight for every $\varphi$, with no further elaboration.}
                       
                       Moreover, one can  also clearly obtain the statistics of $\{\mathfrak{S}_{1},\mathfrak{S}_{2},\ldots\}$ ([namely] $U=1$) by taking any complete orthonormal system $\varphi_{1},\varphi_{2},\ldots$, and mixing these states in the relation
                       1:1:\ldots. (as a consequence of $1=P_{\varphi_{1}}+P_{\varphi_{2}}+\ldots$) to form the uniform [elementary] ensemble.
                       
                       We can proceed similarly for ensembles $\{\mathfrak{S}_{1}^{\prime},\mathfrak{S}_{2}^{\prime},\ldots\}$ that are characterized by the measurement  of quantities $\mathfrak{a}_{1},\mathfrak{a}_{2},\ldots,\mathfrak{a}_{m}$ (whose
                       values lie in the intervals $I_{1},I_{2},\ldots,I_{m}$). Here $U=E$, where $E$ is an operator to be constructed according to [the procedure described in] Section 7. A state $\varphi$ in this case has the
                       (relative) probability $Q(\varphi,E\varphi)$: this is maximal (=1) for $\varphi$ in the interior of $E$ and minimal (=0) for $\varphi$ in the exterior of $E$ (such states are thus impossible [i.e., do not appear in the ensemble]).
                       Further this ensemble can be obtained by mixing the uniform ensembles corresponding to the states $\varphi_{1},\varphi_{2}, \ldots$ in the proportion 1:1:\ldots, where  $\varphi_{1},\varphi_{2}, \ldots$ are an orthonormal system in the
                       interior of $E$, there spanning an everywhere dense linear manifold (because then, as one easily shows, $E=P_{\varphi_{1}}+P_{\varphi_{2}}+\ldots$).
                       
                       These results establish moreover that one can generate the same statistical ensembles by mixing quite different states; or, otherwise put, that mixtures composed in quite different ways can behave identically
                       statistically (and therefore, in every respect).\footnoteB{The fact that the statistical properties of ensembles composed of quite different component states can nonetheless be identical has profound
                       consequences for quantum measurement theory: see, for example, \autocite[sec.\ 5.1]{BassiGhirardi:2003}. A simple example can be given already for two state systems, a ``quantum cat'', for example, with
                       possible states $|A\rangle$ (``cat alive''), $|D\rangle$ (``cat dead''). The density operator corresponding to an ensemble of 50\% live and 50\% dead cats is $U=\frac{1}{2}|A\rangle\langle A|+\frac{1}{2}|D\rangle\langle D|$.
                       The superposition states $|B\rangle=\frac{1}{\sqrt{2}}(|A\rangle+|D\rangle),|C\rangle=\frac{1}{\sqrt{2}}(|A\rangle-|D\rangle)$ provide a perfectly acceptable orthonormal basis for our two state system. One easily verifies
                       that the density operator given previously can also be written as $U=\frac{1}{2}|B\rangle\langle B|+\frac{1}{2}|C\rangle\langle C|$, i.e., as a composition of individual states in which each individual instance involves
                       a cat which is in a superposition state. As the cited authors put it, this means that ``\emph{the statistical operator describing a statistical mixture, describes at the same
                       time infinitely many inequivalent statistical mixtures}''; consequently, theories of what happens in the measurement process have to describe processes that ``affect directly the wave function, not only
                       the statistical operator.'' \autocite{BassiGhirardi:2003}} 
                        Further, we see that a state (a uniform ensemble) is really only determined once the interior of $E$ is one dimensional.
                   
                      In conclusion, we determine the normalization of the statistics characterized by $U=E$, by determining the (relative) expectation value of the operator $\mathbf{1}$, associated with the quantity $1$ (cf.\ Section 5, $\beta$).
                      It is:
                      \begin{equation}
                       \mathbf{E}(\mathbf{1}) = \langle E, \mathbf{1}\rangle = [E] = \mathrm{dimension\;of\;the\;interior\;of}\;E.
                       \end{equation}
                  Thus, the normalization is automatically correct if and only if $E$ has a one-dimensional interior, i.e., the state of $\mathfrak{S}$ is completely determined. If the interior of $E$ is finite dimensional, the normalization can be corrected
                  by division by $[E]$. If it is infinite dimensional, however, the complication mentioned in Section 2 is really present: all expectation values are multiplied by $+\infty$, and hence not normalizable.
                  
                  \section{Summary}
                  
                  The aim of the preceding work was to show that quantum mechanics is not only compatible with the conventional probability calculus, but that, given this calculus---together with some plausible relevant assumptions---is even the
                  only possible solution. The foundational assumptions were the following:
                  \begin{enumerate}
                  \item Every measurement changes the measured object, and two measurements always therefore mutually affect each other---although it may be that both can be replaced by a single one [if the measurements are simultaneously 
                  possible!].
                  \item However, the nature of the change effected by a measurement is such, that this measurement remains valid, i.e., if one repeats the measurement \emph{immediately}, one obtains the same result.
                  \end{enumerate}
                  In addition, there is a formal assumption:
                  \begin{enumerate}
                  \setcounter{enumi}{2}
                  \item  the physical quantities are---given their adherence to some simple formal rules---described by functional operators.
                  \end{enumerate}
                  These principles already inevitably [\emph{unweigerlich}] lead to  quantum mechanics and its associated statistics.
                  
                  One should also take note of the fact that the statistical, ``a-causal'' nature of quantum mechanics is only due to the (intrinsic!) limitations of measurement (cf.\ the paper of Heisenberg cited in notes 2 and 4), for a system
                  left to itself (which is not disturbed by any measurement) develops in time in a fully causal way: if one knows its state $\varphi$ at time $t=t_{0}$, then one can calculate the state at all later times using the time-dependent
                  Schr\"odinger equation.\footnoteA{If $H$ is the energy operator, the time-dependent Schr\"odinger equation is
                  \begin{equation}
                      H\varphi_{t} = \frac{h}{2\pi i}\frac{\partial}{\partial t}\varphi_{t},
                      \end{equation}
                      where $\varphi_{t}$ is the state of the system at time $t$. From this it follow immediately that
                      \begin{equation}
                      \varphi_{t} = e^{\frac{2\pi i}{h}(t-t_{0})H}\varphi_{t_{0}}.
                      \end{equation}
                      The operator $\exp{\frac{2\pi i}{h}(t-t_{0})H}$ is to be built from $H$ and the function $f(x)=\exp{\frac{2\pi i}{h}(t-t_{0})x}$ exactly as described in note 18 for real-valued $f(x)$. As this $f(x)$ is not real-valued (which does
                      not affect the applicability of the definition given there), $f(H)$ is not symmetric [i.e., hermitian], rather it is, since $|f(x)|=1$, ``orthogonal'' [unitary].
                      } Once experiments come into play however, one cannot circumvent the statistical character: for each experiment there is indeed a state, which is adapted to it, and in which the result is unique (the experiment generates
                      directly just such states, if they were not previously present); but for each state there are ``non-adapted'' experiments, the execution of which destroys the [measured] state,\footnoteB{The infamous
                      ``collapse of the wave function'', which would become the keystone of the Copenhagen interpretation of quantum mechanics, makes its first appearance at this point.}  and generates, according to the laws of chance, adapted states.
                      
                      It remains to show that our statistics really satisfies all the rules of the probability calculus.  This is however unnecessary, as we know that the statistics of an ensemble selected by certain measurements is determined solely
                      by the projection operator $E$ associated with these [measurements] (so, with the same operator that allows us to calculate the probabilities of the results of these measurements in any ensemble). The dependence of this
                      operator on the measurements [which define it] was already discussed in all essential details in M.B.Q., Section 14 (and is trivial). The situation becomes quite transparent given that we determined, at the end of Section 8,
                      the composition of  the ensembles by the aggregation of single [i.e., pure] states.
                      
                      \setcounter{footnoteA}{0}
\setcounter{footnoteB}{0} 
\newpage

\chapter{Paper 3: The thermodynamics of quantum-mechanical ensembles.}

\section*{J.\ v.\ Neumann}

\section*{Submitted by M. Born in the session of November 11, 1927}  

\section{Introduction}
 In my paper ``Probability-theoretic construction of quantum mechanics''\footnoteA{\emph{G\"ottinger Nachrichten}, session of November 11, 1927. This work, which will be referred to
 frequently in the following, will be cited as W.A.Q.\ 
 [von Neumann \autocite*{VonNeumann:1927b}]} it was shown that the statistics of quantum mechanics follows inescapably from some simple and \emph{purely qualitative}
 foundational physical  assumptions,\footnoteA{Namely, the following:
 \begin{enumerate}
 \item Two different measurements always disturb one another, i.e., two quantities can be measured simultaneously if and only if their measurement can be replaced by a single
 measurement of a third quantity---that is, when both are functions of a third quantity.
 \item If the same quantity is measured successively twice in the same system, the same result is obtained in both cases.
 \end{enumerate}
 1.\ corresponds to the explanation given by Heisenberg for the a-causal behavior in quantum physics (\emph{Zeitschrift f\"ur Physik} Vol.\ 43, p.\ 178, 1927)  
 [the ``uncertainty'' paper, Heisenberg \autocite*{Heisenberg:1927b}]; 2.\ expresses the fact that\
 this behavior at least to some extent mimics a sort of causality.} as well as the following formal principle: the physical quantities $\mathfrak{a}$ of a given system $\mathfrak{S}$
 are in one to one correspondence with the (hermitian)-symmetric operators of the Hilbert space.\footnoteA{My paper ``Mathematical foundations of quantum mechanics'' (\emph{G\"ottinger
 Nachrichten}, session of May 20, 1927, referred to in the following as M.B.Q.\ 
 [von Neumann \autocite*{VonNeumann:1927b}]) contains further details about the mathematical concepts to be used here: everything to come refers back to
 this work. Cf.\ also the end of section 2 of W.A.Q.\ where the operators in question are presumed to be ``normal'', cf.\ note 14 of W.A.Q.} This last formal assumption is the quantum mechanical
 replacement for the largest component of a mechanical model\footnoteA{More specific indications, e.g., which operator corresponds to the quantity ``energy'' are not required for \emph{general}
 considerations of statistical quantum mechanics. Accordingly, the energy operator does not at all appear in W.A.Q.---in this work, it will, admittedly, play an important role.}: the fact that one arrives
 at the same model for all conceivable systems $\mathfrak{S}$ demonstrates the superior simplicity of quantum mechanics in comparison with classical mechanics. For detailed results we refer to
 this work [W.A.Q.]; here let us review some results and concepts from it which we shall soon employ. These are the following:
 
 $\alpha.$ An ensemble $\mathfrak{G}$ of (very many) systems $\mathfrak{S}$\footnoteA{These systems all have the
 same structure (denoted just by $\mathfrak{S}$), but can be in different states.} is characterized statistically by the association of an expectation value---or average--- $\mathbf {E}(\mathfrak{a})$ to each
 quantity $\mathfrak{a}$. If we place certain simple requirements, which follow from the concept of an ``expectation value'', on this association, one finds the following:

The ensembles $\mathfrak{G}$ are in one to one correspondence with the [non-negative-]definite operators $U$, in the sense that in the ensemble $\mathfrak{G}$ with operator $U$ one always has\footnoteA{if the quantity $\mathfrak{a}$
  has the operator $R$, we will denote $\mathbf{E}(\mathfrak{a})$ also by $\mathbf{E}(R)$.
     The concept of the trace was not introduced in W.A.Q., but will be absolutely essential in the present work. If the operator $A$ has the matrix $\{a_{\mu\nu}\}$, then
     \begin{equation}
        \mathrm{Tr} A = \sum_{\mu}a_{\mu\mu}.
        \end{equation}
        One sees at once that our formula for $\mathbf{E}(R)$ agrees with the one at the end of section 3 of W.A. Q. Moreover,  the concept of trace is invariant, i.e., independent of which complete orthonormal system $\varphi_{1},\varphi_{2}, \ldots$
        is used to associate to the operators $A$ their matrices $\{a_{\mu\nu}\}$. Indeed, if one employs a second complete orthonormal system $\psi_{1},\psi_{2}, \ldots$, one finds
        \begin{equation}
        \label{invtrace}
        \mathrm{Tr} A = \sum_{\mu} a_{\mu\mu} = \sum_{\mu} Q(\varphi_{\mu},A\varphi_{\mu}) = \sum_{\mu,\nu} Q(\varphi_{\mu},\psi_{\nu})Q(\psi_{\nu},A\varphi_{\mu}).
        \end{equation}
        The right hand side is independent of the $\psi_{\nu}$ (as the left [i.e., the term after the second equality in (\ref{invtrace})] is),
         and is transformed to its complex conjugate by interchange of $A,\varphi_{\mu},\psi_{\nu}$ with $A^{\dagger},\psi_{\nu},\varphi_{\mu}$, whence it must 
        also be independent of $\varphi_{\mu}$ [typo in original: $\psi_{\nu}$]. Thus $\mathrm{Tr} A$ depends
        only on $A$, as was asserted.
        }
  \begin{equation}
    \mathbf{E}(R) = \mathrm{Tr}(UR).
    \end{equation}
    (If $\mathrm{Tr}(U)$ is not equal to 1, these expectation values are relative,\footnoteA{For a quantity which is identically equal to 1, has the operator 1, hence the (relative) expectation value $\mathrm{Tr}\; U$.}
     in which case $U$ is determined by $\mathfrak{G}$ only up to a constant positive factor---and only meaningful to the
    extent of this factor.)
    
    $\beta.$ An ensemble is called pure, or uniform, if it cannot be generated by the mixture of two ensembles different from it. (This is the most reliable criterion ensuring that $\mathfrak{G}$ consists
    solely of [systems] $\mathfrak{S}$ in the same state---more so than the ``sharpness'' of the statistics of $\mathfrak{G}$, i.e., the vanishing of the variance for all quantities $\mathfrak{a}$.)
    
    Now the pure ensembles are just those for which $U$ is equal to a $P_{\varphi}$ ($P_{\varphi}$ is the projector in the direction $\varphi$\footnoteA{i.e., $P_{\varphi}f = Q(f,\varphi)\cdot\varphi$.}, $\varphi$ is a
    point on the unit sphere in Hilbert space, $Q(\varphi)=1$).\footnoteB{We remind the reader of von Neumann's inner product notation: using the nowadays more common Dirac notation, $Q(\varphi,\psi)=\langle\psi|\varphi\rangle,
    Q(\varphi)=\langle\varphi|\varphi\rangle$. Lower case Greek letters ($\varphi,\psi, \ldots$) are used to denote Hilbert space vectors, in lieu of $|\varphi\rangle, |\psi\rangle$ etc.} Thus, the states of [the system] $\mathfrak{S}$ correspond to the $\varphi$ with $Q(\varphi)=1$, and should be represented by these $\varphi$. In any
    case, a state (or, a pure ensemble) determines its $\varphi$ only up to a constant of absolute value 1. The formula for $\mathbf{E}(R)$ simplifies to\footnoteA{This is the value of the hermitian form associated with $R$ at
    the point $\varphi$.}
    \begin{equation}
      \mathbf{E}(R) = Q(\varphi, R\varphi).
      \end{equation}
      
      $\gamma.$ An ensemble $\mathfrak{G}$ is said to be  elementary and random if it is unchanged by any measurement: namely, if a measurement of any quantity $\mathfrak{a}$ is performed on every element of
      $\mathfrak{G}$ (whereby all possible eigenfunctions of the operator of $\mathfrak{a}$ must appear), and then all [post-measurement systems] are reassembled into an ensemble, then the ensemble $\mathfrak{G}$
     is regenerated.\footnoteA{One is concerned here with a sort of absolute equilibrium state, which is suited to reflect the a priori weights of the individual states of $\mathfrak{S}$. We will encounter this ensemble 
     on many occasions in the following.}
    There exists one and only one elementary random ensemble, for which $U=1$. 
    
    Employing these theorems certain statistical formulas, which I had previously indicated in M. B. Q., were obtained in W.A.Q.: they contain the generally accepted probability postulates of quantum mechanics. 
    We will not further discuss  these, and certain other results (simultaneous measurability of quantities, etc.).
    
    \section{[Introduction of thermodynamic concepts: statement of principal result]}
    
    The basic random ensemble investigated in W.A.Q.\ is an absolute equilibrium state (cf.\ note 10): it therefore corresponds, if we are allowed to use the language of thermodynamics, to infinite temperature (cf.\ section 4), and
    the ensembles studied there [i.e., W.A.Q.], arising from experiments, for which $U$ is the identity operator, are generated from it. In this work ensembles of finite temperature will be studied, i.e., a proper thermodynamics of
    these ensembles will be erected. To this end, the concepts of entropy and temperature must be introduced for these ensembles.
    
    In doing this we shall encounter a characteristic distinction: in the previous work one had infinite temperature, energy was available in unlimited amounts, and played so inessential a role that its operator was not even mentioned.
    Here, the energy is limited, and therefore is critical in all our considerations (an analog of the Boltzmann formula will emerge).
    
    A thermodynamic view of quantum mechanical ensembles can also be achieved through the following considerations: if $\mathfrak{S}$ is in a pure state $\varphi$, then it evolves in a strictly causal way in accordance with
    the time-dependent Schr\"odinger equation (cf.\ e.g., W.A.Q.\ section 9, and note 41), and, by application of appropriate potential energies, in all respects reversible. As soon however as a measurement is made of a
    quantity $\mathfrak{a}$ which is not sharp in $\varphi$, it [the system] splits into infinitely many pure states (the eigenfunctions of the operator corresponding to $\mathfrak{a}$), and this step is irreversible. As is well known,
    the energy difference of two ensembles $\mathfrak{G}_{1}, \mathfrak{G}_{2}$ is determined as follows: one transforms $\mathfrak{G}_{1}$ reversibly into $\mathfrak{G}_{2}$ (naturally, both must consist of exactly the same number, $N$,
    of systems $\mathfrak{S}$); if this process occurs with the single requirement of an energy transfer to an (infinite) heat reservoir at temperature $\mathbf{T}$ corresponding to the quantity of heat $\mathbf{Q}$, then $\mathfrak{G}_{2}$
    has an entropy differing from that of $\mathfrak{G}_{1}$ by the amount
    \begin{equation}
        \mathbf{S} = -\frac{\mathbf{Q}}{\mathbf{T}}.
        \end{equation}
        In order to find a reversible transformation of this sort, we will take an approach which goes back to Einstein.\footnoteA{\emph{Verhandlungen der Deutschen Physikalischen Gesellschaft}, Vol.\ 12, p.\ 820, 1914. It was  already clearly
        recognized by L. Szilard (\emph{Zeitschrift f\"ur Physik} Vol.\ 32, p.\ 777, 1925) that, in the non-quantum-theoretic case, the procedure initiated by Einstein (loc. cit.) affords a superior method for establishing the foundations
        of statistical thermodynamics; the work there promised has not appeared in print.}        
        
        We take every single element of $\mathfrak{G}$ (i.e., the systems $\mathfrak{S}$), isolate it from the environment, and enclose it in a box. We attach a weight to each box, the mass of which greatly exceeds that of the
        system $\mathfrak{S}$ (so that changes in $\mathfrak{S}$ barely alter the mass of the entire object, and hence are not directly perceptible), and denote the object so obtained a ``molecule.'' We enclose all the molecules
        in a very large box (with the total volume of the boxed systems very small in comparison to that of the large box),\footnoteA{The
        molecules are supposed to lie outside of any gravitational field, and hence can be force-free moving bodies.}\footnoteB{The article referred to here is Einstein \autocite*{Einstein:1914b}, in which Einstein provides a thermodynamic derivation of the
        Planck distribution for an ensemble of one dimensional harmonic oscillators. The basic setup, which von Neumann takes over directly, and generalizes, is described clearly by Einstein:
        \begin{quote}
         Consider a chemically uniform gas, in which every molecule carries a resonator [footnote: By ``resonator'' we here mean in general a carrier of internal energy of an otherwise unspecified character]. The energy of the 
         resonator cannot assume arbitrary values, but only certain discrete values $\varepsilon_{\sigma}$\ldots I will now assume that two molecules are chemically distinct, i.e., as in principle separable using semipermeable walls,
         if their resonator energies $\varepsilon_{\sigma}$ and $\varepsilon_{\tau}$ are unequal \autocite[pp.\ 820--821]{Einstein:1914b}.
         \end{quote}
        Each of these resonators is also called a ``molecule'': they are allowed to move around, confined to a volume $V$ and in thermodynamic equilibrium at temperature $T$.
          Einstein then considers the total entropy of this ``gas of resonator molecules'': it turns out to be the sum of the usual entropy of an ideal gas of $N=\sum_{\sigma}n_{\sigma}$ molecules, plus the entropy due to the 
          assignment of internal quantum states to the molecules, which can be regarded as an ensemble of quantum systems ($n_{\sigma}$ of which are in state of energy $\varepsilon_{\sigma}$) in their own right, ignoring
          their motion in the containing box. It is  crucial that the internal dynamics  (assignment of a definite quantum state to each molecule) is completely decoupled from the gas dynamics (bouncing of the molecules off the walls,
          or off each other). Von Neumann  generalizes this system by imagining an arbitrary quantum system $\mathfrak{S}$ (this could itself be a quantum gas confined to a box!) with a complete denumerable set $\varphi_{\mu}$ of    
          possible quantum states. An ensemble of such systems $\mathfrak{G}$ is then realized physically by enclosing each element of the ensemble (characterized by a definite state $\varphi_{\mu}$) in a box, calling it a
          ``molecule'', and allowing the molecules to have their own dynamics of the usual ideal gas type: i.e., they bounce around in a large box, reaching thermodynamic equilibrium at some temperature $T$. The analogy to Einstein's
          system is clear: unlike Einstein, the individual systems are not required to be simple harmonic oscillators, but can be arbitrary quantum systems---even multi-particle systems, such as a gas contained in a finite volume.
          
          The reference to ``semipermeable walls'' in the quote from Einstein's 1914 article above points to another critical part of the thermodynamic reasoning which von Neumann will employ to arrive at his entropy
          formula. Thought experiments involving partitions which selectively allow chemically distinct molecules to pass through while blocking others were commonly used to establish important thermodynamic properties:
          for example, the additivity of entropy when two equal volumes of distinct gases (at the same pressure and temperature) are reversibly and \emph{isentropically} combined (or separated) by using pistons attached to such
          walls, as for example in Planck's 1922 text on thermodynamics \autocite[Sec.\ 236]{Planck:1922}. The device of a semipermeable wall will be adapted by von Neumann to the quantum case to allow for the selective transmission of a quantum system in 
          a particular selected pure state through a partition, while blocking all other states of the system (orthogonal to the selected one).  Semipermeable walls also play a central role in Szilard's discussion of the Maxwell demon
          problem (Szilard \autocite*{Szilard:1929}, see also below, footnote 18).} and bring this [outer] box into contact with an infinite heat reservoir at temperature $\mathbf{T}$.
         This entire arrangement we shall call the gas at temperature $\mathbf{T}$ corresponding to $\mathfrak{G}$.
        If the weights applied to the molecules and the volume of the large box (depending on $\mathbf{T}$) are sufficiently large, we can regard this gas as ideal.

       We will now devise a reversible process that transforms the gas of $\mathfrak{G}_{1}$ into the gas of $\mathfrak{G}_{2}$ (taking care that both gases are gas-kinetically identical, and differ only in the ``cores'' $\mathfrak{S}$
       of the molecules, which are not directly observable from outside), and determine in this way the entropy difference of $\mathfrak{G}_{1}$ and $\mathfrak{G}_{2}$. In this way it can be seen that all pure ensembles have the
       same entropy, which is less than the entropy of all other [i.e., not pure] ensembles. It thus seems convenient to normalize their entropy to zero, in which case the entropy of the ensemble $\mathfrak{G}$ with the [non-negative] definite
       operator $U$ is found to be
       \begin{equation}
       \label{EntropyTrace}
         \mathbf{S} = -Nk \mathrm{Tr}(U\ln{U}).
         \end{equation}
      [Here]   $N$ is the number of elements of $\mathfrak{S}$, $k$ the Boltzmann constant. Here we assume $\mathrm{Tr}(U)=1$, i.e., that the expectation values of $\mathfrak{G}$ are absolute: in contrast to W.A.Q.\ this assumption
      cannot here be avoided.\footnoteA{In the construction of $U\ln{U}$ we have assumed without mention that $U$ is normal (cf.\ W.A.Q., end of section 2). This is permitted, as a [non-negative] definite $U$ with $\mathrm{Tr}(U)=1$
      must be bounded, even completely continuous. We shall show the first assertion below, the second follows, for example, from Hilbert (\emph{G\"ottingen Nachrichten} 1906, p.\ 203, Theorem 6), and has moreover the consequence,
      that $U$ (with the aid of an appropriate complete orthonormal system) can be represented as a diagonal matrix (cf.\ for example, Hilbert, \emph{G\"ottingen Nachrichten} 1906, p.\ 201, Theorem 5).
      
      First, let $Q(\varphi)=1$. Then $\varphi=\varphi_{1}$ can be extended to a complete orthonormal system, and
      \begin{equation}
       Q(\varphi,U\varphi) \leq \sum_{\mu}Q(\varphi_{\mu},U\varphi_{\mu}) = \mathrm{Tr}(U) = 1.
       \end{equation}
       We therefore have, in general,
       \begin{equation}
         Q(f, Uf) \leq Q(f).
         \end{equation}
         As $U$ is [non-negative] definite, we have
         \begin{equation}
         Q(f, Ug) \leq \sqrt{Q(f,Uf)Q(g,Ug)} \leq \sqrt{Q(f)Q(g)},
         \end{equation}
         where the proof of the second inequality follows from W.A.Q., note 23; now setting $f = Ug$,
         \begin{equation}
           Q(Ug) \leq \sqrt{Q(Ug)Q(g)},\quad Q(Ug)\leq Q(g).
           \end{equation}
           Therefore $U$ is indeed bounded. }
           
           \section{[Derivation of Boltzmann distribution]}
           
           It is advisable to discuss here an obvious objection to the procedure suggested above. 
           
           What we determined above [cf.\ Eq.\ (\ref{EntropyTrace})] was the change in entropy in a reversible transformation for
           the gas corresponding to $\mathfrak{G}_{1}$ to that corresponding to $\mathfrak{G}_{2}$. What interests us however is the transition of the ensemble $\mathfrak{G}_{1}$ (assumed stationary) to another ensemble
           $\mathfrak{G}_{2}$ (also stationary)---the introduction of boxes, weights, heat reservoirs is secondary, indeed contrary to the question posed originally. 
           
           Nonetheless, this objection is not a dangerous one.
           If the heat reservoir had temperature 0, everything would be fine, as the gas  of $\mathfrak{G}$ would be stationary, and the various weights and heat reservoirs inessential accessories. But the temperature
           $\mathbf{T}$ doesn't appear at all in our expression for entropy $\mathbf{S}$: we can therefore set it arbitrarily close to zero, and our result will hold also in the limit $\mathbf{T}=0$. 
           If this argument fails to convince (as the assumed ideal character of the gas seems doubtful), one can argue more exactly as follows. 
           
           The gas of $\mathfrak{G}_{1}$ is gas-kinetically identical to that of $\mathfrak{G}_{2}$,
           so the reversible warming of the former from temperature 0 to temperature $\mathbf{T}$ exactly compensates  the reversible cooling of the latter from $\mathbf{T}$ to 0. We can therefore first warm the ensemble
           $\mathfrak{G}_{1}$ reversibly to temperature $\mathbf{T}$, then transform it to that of $\mathfrak{G}_{2}$, and then reversibly cool to [temperature] 0: the first and third steps compensate each other, and the
           total change of entropy is just that of the second step, namely, the $\mathbf{S}$ given above.\footnoteA{Here it is no longer of consequence that the gases are not ideal at low temperatures. For as the gases
           $\mathfrak{G}_{1}$ and $\mathfrak{G}_{2}$ are gas-kinetically identical, it is irrelevant what equation of state holds in the temperature interval from 0 to $\mathbf{T}$. Only at
           temperature $\mathbf{T}$ is ideality assumed, and at that point it can naturally be achieved.}

        After we have found the entropy of a general ensemble $\mathfrak{G}$, we can undertake the determination of the equilibrium states of $\mathfrak{G}$ (for a given mean energy). In other words, one seeks, among
        all ensembles $\mathfrak{G}$ with given mean energy $\mathbf{E}$\footnoteA{Note that it is in the spirit of the statistical method to prescribe the statistical mean of the energy, not the energy itself. In other words,
        one requires that $\mathbf{E}(H) = \mathbf{E}$ ($H$ the energy operator), and not that $H$ is ``sharp'' in $\mathfrak{G}$ with the value $\mathbf{E}$,   i.e., that every $\mathfrak{S}$ of $\mathfrak{G}$ is in an eigenstate of $H$
        with the eigenvalue $\mathbf{E}$.} the one with the maximal entropy: every other ensemble with the same mean energy can be transformed into this one (indeed, with entropy increase, hence irreversibly), while this
        ensemble cannot [be so transformed] into any other.
        
        Formally this means the following: under the conditions ($H$ is the energy operator)
        \begin{equation}
        \label{auxconds}
           \mathrm{Tr} (U) =1,\quad \mathrm{Tr}(UH) = \mathbf{E},
           \end{equation}
           we wish to maximize (with $U$ [non-negative] definite!)
           \begin{equation}
           -Nk\;\mathrm{Tr}(U\ln{U}).
           \end{equation}
           The result is\footnoteB{See section 10 for the explicit derivation of this result.}
           \begin{equation}
            U = \alpha\exp{\beta H},
            \end{equation}
            where the (real) numbers $\alpha,\beta$ are to be determined by enforcing the conditions [Eq.\ (\ref{auxconds})]. Consequently, $\alpha,\beta$ are functions of $\mathbf{E}$, as is the entropy
            \begin{equation}
            \mathbf{S} = -Nk\;\mathrm{Tr}(U\ln{U}).
            \end{equation}
            
            Further, it is convenient to follow the usual procedure to determine the temperature of the heat reservoir with which the ensemble $\mathfrak{G}$ can remain in equilibrium. It turns out to be\footnoteA{Note that $\mathbf{E}$ is
            not the total energy, but the mean energy of $\mathfrak{G}$. The total energy is (for very large $N$) $N\mathbf{E}$. On the other hand, $\mathbf{S}$ is the total entropy: it is hardly possible to divide it among the individual elements
            of $\mathfrak{G}$, as each one has a state $\varphi$, and so corresponds to a pure state---and these have entropy 0.}
            \begin{equation}
              \mathbf{T} = N\frac{d\mathbf{E}}{d\mathbf{S}} = N\frac{d\mathbf{E}}{d\beta}/\frac{d\mathbf{S}}{d\beta} = -\frac{1}{k\beta}.
              \end{equation}
              $\mathbf{T}$ is suitably regarded as the temperature of $\mathfrak{G}$; with its help, one can find expressions for $\alpha,\beta$, and thence also for $U,\mathbf{E},\mathbf{S}$. One finds:
              \begin{eqnarray}
              \label{equilU}
              U &=& \frac{1}{\mathrm{Tr}(\exp{(-\frac{H}{k\mathbf{T}}}))}\cdot\exp{(-\frac{H}{k\mathbf{T}})},   \\
              \label{equilE}
              \mathbf{E} &=& \frac{\mathrm{Tr}(H\exp{(-\frac{H}{k\mathbf{T}})})}{\mathrm{Tr}(\exp{(-\frac{H}{k\mathbf{T}}}))},  \\
              \label{equilS}
              \mathbf{S} &=& \frac{N}{\mathbf{T}}\frac{\mathrm{Tr}(H\exp{(-\frac{H}{k\mathbf{T}})})}{\mathrm{Tr}(\exp{(-\frac{H}{k\mathbf{T}}}))} + Nk\ln{\mathrm{Tr}(\exp{(-\frac{H}{k\mathbf{T}})}}).
              \end{eqnarray}
              
              If we replace $H$ with a matrix, and write this in diagonal form, with diagonal elements (the eigenvalues of $H$) denoted $w_{1},w_{2},\ldots$,\footnoteB{Von Neumann here, rather confusingly, uses the notation $w_{\mu}$ for
              the quantized energy eigenvalues of the system $\mathfrak{S}$, rather than the more natural (and nowadays ubiquitous) $\varepsilon_{\mu}$: shortly, the $w_{\mu}$ will reappear as the weights in a mixed state, i.e., as the
              eigenvalues of the statistical operator $U$.} then
              \begin{eqnarray}
              \label{equilEdiag}
              \mathbf{E} &=& \frac{\sum_{\mu}w_{\mu}\exp{(-\frac{w_{\mu}}{k\mathbf{T}})}}{\sum_{\mu}\exp{(-\frac{w_{\mu}}{k\mathbf{T}}})},  \\
              \label{equilSdiag}
              \mathbf{S} &=& \frac{N}{\mathbf{T}}\frac{\sum_{\mu}w_{\mu}\exp{(-\frac{w_{\mu}}{k\mathbf{T}})}}{\sum_{\mu}\exp{(-\frac{w_{\mu}}{k\mathbf{T}}})} + Nk\ln{\sum_{\mu}\exp{(-\frac{w_{\mu}}{k\mathbf{T}})}},
              \end{eqnarray}
              in other words, we  have arrived at the usual formulas that follow from Boltzmann's laws.
              
              \section{[Thermodynamic interpretation of the elementary random ensemble; abstract of remaining sections]}
              We can now really prove the assertion of Section 2, that the elementary random ensemble corresponds to temperature $\mathbf{T}=\infty$. Then, relaxing now the requirement that $\mathrm{Tr}(U)=1$, and multiplying
              $U$ by a positive constant, the temperature $\mathbf{T}$ corresponds to an ensemble with the operator
              \begin{equation}
                 U = \exp{(-\frac{H}{k\mathbf{T}})},
                 \end{equation}
                 and for $\mathbf{T}\rightarrow\infty$ this tends to $\exp{(0)}=1$, i.e., to the elementary random ensemble.
                 
                 Note the analogy of this thermodynamics of quantum-mechanical ensembles to the cavity filled with radiation. Here, as in that case, a well-defined entropy can be associated with every macroscopically (i.e., statistically)
                 described state; a unique temperature however can only, in either case, be attributed to certain distinguished states, namely, those which have, for a given energy, the maximal entropy, i.e., are equilibrium states.
                 
                 We now proceed to the derivation of the results announced in Sections 2 and 3. To do this, we must first (Sections 5--7) address some preparatory considerations, as certain conceptual features of the usual [classical]
                 thermodynamics---such as the semipermeable wall---cannot directly be transferred to the quantum-mechanical case. As soon as this is taken care of, the proof of our main conclusions will follow (Sections 8--11).
                 
                 \section{General Preparatory Matters}
                 
                 It is common in thermodynamics to work with the so-called semipermeable walls, i.e., with those which allow molecules in a state $z_1$ to pass through unhindered, while reflecting molecules which are in a
                 state $z_2$.\footnoteA{This is just the thermodynamic definition for the situation when the states $z_1, z_2$ are distinguishable.} Now this, as we shall show, is in general, i.e., for arbitrary states, not possible.
                 We will now describe a case in which it \emph{is} possible, and then (in Section 7) establish that this is essentially the only such case.
                 
                 Let $\varphi_{1},\varphi_{2},\ldots,\psi_{1},\psi_{2},\ldots$ be the elements of a (not necessarily complete) orthonormal system (both sequences may contain only a finite number of states). We will complete them to a complete
                 orthonormal system $\varphi_{1},\varphi_{2},\ldots,\psi_{1},\psi_{2},\ldots,\chi_{1},\chi_{2},\ldots$, and associate to them three sequences of distinct numbers $u_{1},u_{2},\ldots,v_{1},v_{2},\ldots,w_{1},w_{2},\ldots$. These [numbers] can moreover be chosen
                 in such a way that even measurements with finite errors can distinguish with certainty the $u_{\mu}$ from the $v_{\nu}$ and the $w_{\rho}$. We then form the operator
                 \begin{equation}
                   R = \sum_{\mu}u_{\mu}P_{\varphi_{\mu}} + \sum_{\nu}v_{\nu}P_{\psi_{\nu}} + \sum_{\rho}w_{\rho}P_{\chi_{\rho}},
                   \end{equation}
                   with eigenfunctions $\varphi_{1},\varphi_{2},\ldots,\psi_{1},\psi_{2},\ldots,\chi_{1},\chi_{2},\ldots$, and $u_{1},u_{2},\ldots,v_{1},v_{2},\ldots,w_{1},w_{2},\ldots$ as associated eigenvalues. Let $R$ be associated with the quantity $\mathfrak{a}$.
                   
                   We now construct a wall with a great many holes, but each hole is closed with a door at which we place an apparatus which accomplishes the following: every time a molecule of our gas approaches it, it catches the
                   molecule, measures the value of the quantity $\mathfrak{a}$ in the system $\mathfrak{S}$ in the molecule, and then lets it go on its way (with the original momentum). In addition, if $\mathfrak{a}$ has a value $u_{\mu}$ (resp.\ $v_{\nu}$)
                   the door is opened (resp.\ closed).\footnoteA{These capabilities remind one in a striking fashion of those of the ``Maxwell demon.'' They are nevertheless harmless. L. Szil\'ard has shown, in a work presently in proof (``On the
                   reduction of entropy in a thermodynamics system due to the intervention of intelligent beings''), which are the properties of the ``Maxwell demon''  that enable it to perform entropy-reducing actions, and how these
                   are to be compensated thermodynamically. It turns out, as is presented there, that it derives primarily from the gift of ``memory'', and our arrangement does not belong to this troublesome type.}\footnoteB{Leo Szilard had already addressed
                   the paradox posed by Maxwell's demon (a putative intelligent entity capable of selectively passing or blocking particles at a partition in such  way as to induce an uncompensated reduction of entropy) in his doctoral dissertation (Berlin, 1922)
                   on thermodynamic fluctuation phenomena. He undoubtedly discussed this problem in person with von Neumann in the mid-1920s; the article mentioned here (in 1927) by von Neumann was only published in 1929, as a writeup of
                   his habilitation lecture \autocite{Szilard:1929}. As von Neumann points out, the ```quantum semipermeable wall'' imagined in his argument does not involve any ``record keeping'' (or ``memory'') of the measurement 
                   apparatus, so avoids the paradox, as Szilard showed. It can no more violate the 2nd law of thermodynamics than would, say, a semipermeable membrane allowing oxygen molecules, but not nitrogen molecules to pass through.}
                   
                   This wall is semipermeable in the sense that molecules whose system $\mathfrak{S}$ is in a state $\varphi_{\mu}$ (resp.\ $\psi_{\nu}$) are smoothly passed through (resp.\ reflected back). Indeed, the $\varphi_{\mu}$ (resp.\ $\psi_{\nu}$
                   are eigenfunctions of the operator [$R$] of $\mathfrak{a}$, and must therefore, on measurement, give with certainty the value $u_{\mu}$ (resp.\ $v_{\nu}$), and therefore are left intact, being either let through [the door], or reflected.
                   Other states are possibly changed by the measurement, and have an uncertain fate. 
                   
                   During these measurements, moreover, the measuring apparatus---and also our semipermeable wall---do not remain quite unchanged. The measurement of a quantity $\mathfrak{a}$ certainly forces $\mathfrak{S}$ to
                   become an eigenfunction of the operator of $\mathfrak{a}$, hence, in certain cases, changes the system $\mathfrak{S}$, and also the expectation value of its energy (if $\mathfrak{a}$ is not simultaneously measurable
                   with the energy). Consequently this energy difference must be compensated in the measurement apparatus.\footnoteA{Here it is not a question of a quantum mechanical law, but---with the
                   validity of quantum mechanics presumed--of the first law of phenomenological thermodynamics. (Admittedly, in its statistical formulation, as given by L.\ Szil\'ard, \emph{Zeitschrift f\"ur Physik}, Vol.\ 32, pp.\ 753--758, 1925 [Szilard \autocite*{Szilard:1925}])} 
                   We will however assume, that this is the only side-effect: i.e., that the change in energy which accompanies the change of state of $\mathfrak{S}$ corresponds perhaps to the tensing or relaxation of a spring.
                   
                   In a gas, in which only such molecules occur whose system $\mathfrak{S}$ is in a state $\varphi_{\mu}$ or $\psi_{\nu}$, our semipermeable wall operates as its name suggests: it does not react with the molecules, remains itself
                   unaltered (as no energy changes [in the wall] take place), and lets the $\varphi_{\mu}$ through while reflecting the $\psi_{\nu}$.
                   
                   \section{[Reversible transformation of pure states]}
                   
                   We want to indicate a procedure that reversibly transforms a pure ensemble $\mathfrak{G}_{1}$ into [another] pure ensemble $\mathfrak{G}_{2}$. $\mathfrak{G}_{1}$ (resp.\ $\mathfrak{G}_{2}$)
                   have the operator $P_{\varphi}$ (resp.\ $P_{\psi}$), i.e., the states $\varphi$ (resp.\ $\psi$) of the system $\mathfrak{S}$. We will show that the reversible transformation can be accomplished, with the
                   only side-effect being the [transfer of the] energy difference of $\varphi$ and $\psi$ into the corresponding tensing or relaxation of a spring. More precisely, it will actually be shown that this can proceed
                   in an arbitrarily good approximation (as is the case mostly in thermodynamic considerations).
                   
                   One should not try to directly measure, in the ensemble $\mathfrak{G}_{1}$, a quantity of which $\psi$ is an eigenfunction, as this would then only produce the desired transition from $\varphi$ to $\psi$
                   in a fraction $|Q(\varphi,\psi)|^{2}$ of the elements of $\mathfrak{G}_{1}$ (in the rest $\varphi$ would be transformed into other eigenfunctions of this quantity). For orthogonal $\varphi,\psi$ therefore, not a
                   single time. We now wish to examine this worst case scenario: it is nonetheless sufficiently general as, for any two $\varphi,\psi$, there is always a $\chi$ orthogonal to both, so that we only then need to transform
                   $\varphi$ to $\chi$ and then $\chi$ to $\psi$.
                     
                     Let therefore $\varphi$ and $\psi$ be orthogonal; further, let $p=1,2,\ldots$---later we shall let $p\rightarrow\infty$. Set
                     \begin{equation}
                     \label{stateseq}
                      \varphi_{r} = \cos{(\frac{\pi r}{2p})}\cdot\varphi + \sin{(\frac{\pi r}{2p})}\cdot\psi,\quad (r=0,,,1,\ldots,p),
                      \end{equation}
                 then $\varphi_{0}=\varphi, \varphi_{p}= \psi$. We complete each $\varphi_{r}$ to a complete orthonormal system $\omega_{r,1}=\varphi_{r}, \omega_{r,2}, \omega_{r,3},\ldots$, and choose an absolutely discrete
                 non-degenerate operator $R_{r}$  (cf.\ W.A.Q., end of section 6)
                 with these eigenfunctions. $R_{r}$ corresponds to the quantity $\mathfrak{a}_{r}$. We now perform the following operations on $\mathfrak{G}_{1}$.
                 
                 We measure the quantity $\mathfrak{a}_{1}$ on each system $\mathfrak{S}$ in $\mathfrak{H}_{0} \equiv \mathfrak{G}_{1}$, generating thereby the ensemble $\mathfrak{H}_{1}$. We measure on every $\mathfrak{S}$ 
                 in $\mathfrak{H}_{1}$ the quantity $\mathfrak{a}_{2}$, generating the ensemble $\mathfrak{H}_{2}$, and so on.\footnoteA{The idea of the procedure is as follows: transform $\varphi=\varphi_{0}$
                 into $\psi=\varphi_{p}$ via the states $\varphi_{1},\varphi_{2},\ldots,\varphi_{p-1}$.} We thereby produce the sequence $\mathfrak{H}_{0},,,\mathfrak{H}_{1},\ldots,\mathfrak{H}_{p}$. It will
                 now be shown that for sufficiently large $p$, $\mathfrak{H}_{p}$ differs as little as we wish from $\mathfrak{G}_{2}$. Given that the measurements incur no other effects than the required tensing or relaxing of springs
                 corresponding to energy changes, we have the following result: $\mathfrak{G}_{1}$ can be brought arbitrarily close to $\mathfrak{G}_{2}$, provided only that the energy changes which occur be compensated. If
                 $\mathfrak{G}_{2}$ is changed back to $\mathfrak{G}_{1}$ by an analogous procedure [interchange $\varphi$ and $\psi$ on right hand side of Eq.\ (\ref{stateseq})], the compensations are exactly removed, i.e., the
                 process is reversible. We have therefore achieved our goal, once the above-said assertion is proved.
                 
                 Let the [density] operators of $\mathfrak{H}_{0},\mathfrak{H}_{1},\ldots,\mathfrak{H}_{p}$ be $U_{0},U_{1},,,,\ldots,U_{p}$; then, by definition,
                 \begin{eqnarray}
                 U_{0} &=& U, \\
                 U_{r} &=& \sum_{\mu} Q(\omega_{r,\mu},U_{r-1}\omega_{r,\mu})P_{\omega_{r,\mu}},\quad (r=1,2,\ldots,p),
                 \end{eqnarray}
                 and it is to be shown that $U_{p}$ tends to $P_{\psi}$ (for $p\rightarrow\infty$).\footnoteA{If the quantity $\mathfrak{a}$ with eigenfunctions $\omega_{1},\omega_{2},\ldots$
                 and eigenvalues $u_{1},u_{2},\ldots$ is measured on the ensemble $\mathfrak{G}$
                 with [density] operator $U$, then the following occurs: it [$\mathfrak{a}$] must assume the values $u_{1},u_{2},\ldots$, $u_{\mu}$ has the probability $Q(\omega_{\mu},U\omega_{\mu})$ (cf.\ W.A.Q.\ section 7, note 36),
                 and when that value is found, the system is in the state $\omega_{\mu}$. Accordingly, $\mathfrak{G}$ is transformed into a mixture of pure ensembles $P_{\omega_{\mu}}$ with the corresponding 
                 weights $Q(\omega_{\mu},U\omega_{\mu})$: and its [density] operator is transformed to $\sum_{\mu}Q(\omega_{\mu},U\omega_{\mu})P_{\omega_{\mu}}$.}\footnoteB{To summarize von Neumann's argument here, we note
                        that for large $p$, $\cos^{2}{(\pi/2p)}\approx 1 -\frac{\pi^{2}}{4p^{2}}$, so a fraction $\frac{\pi^{2}}{4p^{2}}$ of the states  $\varphi_{r}$ are ``lost''---i.e., are not converted to $\varphi_{r+1}$---at each of the $p$ stages of the
                        rotation of the state through an angle $\pi/2$. The net loss is therefore of order $\pi^{2}/4p$, which can be made as small as we please by making $p$ large enough, as is implicit in the execution of a truly quasistatic
                        transformation. Von Neumann (uncharacteristically) makes heavy work of this basically simple idea, as he admits in footnote 23.}
 
                 The second equation has the consequence
                 that $\mathrm{Tr}\; U_{r} = \mathrm{Tr} \;U_{r-1}$, whence $\mathrm{Tr} \;U_{p} = \mathrm{Tr}\; U_{0} =1$, and further (from the non-negative definiteness of $U_{r-1}$ and all $P_{\omega_{r,\mu}}$),
                 \begin{eqnarray}
                   Q(\varphi_{r+1}, U_{r}\varphi_{r+1}) &\geq& Q(\varphi_{r},U_{r-1}\varphi_{r})\cdot Q(\varphi_{r+1},P_{\varphi_{r}}\varphi_{r+1}) \\
                   &=& Q(\varphi_{r}, U_{r-1}\varphi_{r})\cdot |Q(\varphi_{r},\varphi_{r+1})|^{2}.
                   \end{eqnarray}
                  However, as we obviously have
                  \begin{equation}
                   Q(\varphi_{r},\varphi_{r+1}) = \cos{(\frac{\pi}{2p})},
                   \end{equation}
                   it follows that
                   \begin{eqnarray}
                   Q(\psi, U_{p}\psi) &=& Q(\varphi_{p},U_{p}\varphi_{p}) = Q(\varphi_{p},U_{p-1}\varphi_{p}) \\
                   &\geq& (\cos{(\frac{\pi}{2p})})^{2p-2}Q(\varphi_{1},U_{0}\varphi_{1}) \\
                   &=& (\cos{(\frac{\pi}{2p})})^{2p-2}Q(\varphi_{1},P_{\varphi_{0}}\varphi_{1}) \\
                   &=& (\cos{(\frac{\pi}{2p})})^{2p-2}\;|Q(\varphi_{0},\varphi_{1})|^{2} = (\cos{(\frac{\pi}{2p})})^{2p}.
                   \end{eqnarray}
                   To summarize, we always have $\mathrm{Tr} \;U_{p} = 1$ and
                   \begin{equation}
                    Q(\psi,U_{p}\psi) \geq  (\cos{(\frac{\pi}{2p})})^{2p}.
                    \end{equation}
                    As $\cos{(\frac{\pi}{2p})}^{2p}\rightarrow 1$ for $p\rightarrow\infty$, it follows that $U_{p}\rightarrow P_{\psi}$,\footnoteA{This can be shown as follows. As $U_{p}$ is [non-negative] definite,
                    \begin{equation}
                     (\cos{\frac{\pi}{2p}})^{2p} \leq Q(\psi, U_{p}\psi) \leq \mathrm{Tr}\;U_{p} = 1,
                     \end{equation}
                     so that for $p\rightarrow\infty$, $Q(\psi, U_{p}\psi)\rightarrow 1$. If $\chi$ is orthogonal to $\psi$ and of unit norm, then
                     \begin{eqnarray}
                     Q(\psi, U_{p}\psi) + Q(\chi, U_{p}\chi) &\leq& \mathrm{Tr}\; U_{p} = 1,\\
                     Q(\psi, U_{p}\psi) + Q(\chi, U_{p}\chi) &\geq& Q(\psi, U_{p}\psi),
                     \end{eqnarray}
                     so that both the entire left hand side, as well as its first term, will tend to 1 as $p\rightarrow\infty$. Therefore $Q(\chi, U_{p}\chi)$ will tend to 0. Therefore for all $f$ orthogonal to $\psi$, $Q(f, U_{p}f)\rightarrow 0$.  
                     
                     Further (see note 13 for the inequality)
                     \begin{equation}
                       |Q(\psi,U_{p}f)| \leq \sqrt{Q(\psi,U_{p}\psi)Q(f,U_{p}f)} \rightarrow 0,
                       \end{equation}
                       which implies
                       \begin{equation}
                       Q(\psi, U_{p}f) \rightarrow 0,\quad Q(f,U_{p}\psi) \rightarrow 0.
                       \end{equation}
                       If we now take an arbitrary $g$, it can be written $g=c\psi +f$, with $f$ orthogonal to $\psi$, and from the preceding,
                       \begin{equation}
                          Q(g, U_{p}g) \rightarrow |c|^{2}.
                          \end{equation}
                        However, as $c = Q(g,\psi)$, this means
                        \begin{eqnarray}
                          Q(g,U_{p}g) &\rightarrow& |Q(g,\psi)|^{2} = Q(g,P_{\psi}g),\\
                          Q(g,[U_{p}-P_{\psi}]g) &\rightarrow& 0.
                          \end{eqnarray}
                            By replacing $g$ in the last equation successively with $\frac{1}{2}(f+g),\frac{1}{2}(f-g),\frac{1}{2}(f+ig),\frac{1}{2}(f-ig)$ (with $f,g$ arbitrary), and subtracting the second from the first (and the fourth from the third), one
                            finds, for all $f,g$,
                            \begin{equation}
                            Q(f,[U_{p}-P_{\psi}]g) \rightarrow 0.
                            \end{equation}
                            Therefore, in any matrix representation of $U_{p}-P_{\psi}$ all matrix elements tend to zero [as $p\rightarrow\infty$], which completes the proof.} 
                            which completes the proof.\footnoteA{The underlying idea of this somewhat formal proof is the following: on measurement of $\mathfrak{a}_{r}$ a fraction $|Q(\varphi_{r-1},\varphi_{r})|^{2}=\cos{\frac{\pi}{2p}}|^{2}$ of all systems
                        transition from the state $\varphi_{r-1}$  to the state $\varphi_{r}$. Accordingly, there will be at least a fraction $(\cos{\frac{\pi}{2p}})^{2p}$ of such systems that go from the state $\varphi=\varphi_{0}$, via
                        $\varphi_{1},\varphi_{2},\ldots,\varphi_{p-1}$, over to state $\varphi=\varphi_{p}$. And exactly this fraction will tend to 1 as $p\rightarrow\infty$.}                      
  
  \section{Transmission of quantum systems through semipermeable walls}
  
    Finally, we demonstrate a sort of converse of Section 5. Namely, we show: if a semipermeable wall (however constructed) exists, that does not react with molecules of a system $\mathfrak{S}$ which is in state $\varphi$ (resp.\ $\psi$),
    but smoothly transmits (resp.\ reflects) the molecules, then $\varphi, \psi$ must be orthogonal. For the proof of this fact we will need to use the result mentioned in Section 2, according to which the ensemble $\mathfrak{G}$
    with the [density] operator $U$ ($\mathrm{Tr}\; U = 1$) has entropy
    \begin{equation}
    -Nk\mathrm{Tr}\; (U\ln{U}),
    \end{equation}
    although this result will first be proven in Section 9. This may be permitted, given that the proof there will not make any use of this section.
    
    Let $\mathfrak{G}_{1}$ be an ensemble with operator $P_{\varphi}$, i.e., in state $\varphi$. We now consider the corresponding gas, which is contained in a volume $V$. We introduce a wall in the middle of the gas so that one half
   of the gas is on the right and one half on the left of the wall. On the right half we transform the gas reversibly into the pure ensemble $P_{\psi}$ (state $\psi$, cf.\ Section 6). Now we can allow both halves to merge without
   performing work with the use of the previously mentioned semipermeable wall. In this way an ensemble $\mathfrak{G}_{2}$ arises with the [density] operator
   \begin{equation}
      U = \frac{1}{2}P_{\varphi}+\frac{1}{2}P_{\psi},
      \end{equation}
      but with half the volume. Finally, we let this expand reversibly and isothermally to volume $V$.
      
      During all these steps an entropy alteration of the heat reservoirs only occurs in the last one: namely, a reduction by $Nk\ln{2}$. On the other hand, the entropy of the gas undergoes a change of $-Nk\mathrm{Tr}\;(U\ln{U})$ in the
      transition from $\mathfrak{G}_{1}$ to $\mathfrak{G}_{2}$. Therefore, we must have
      \begin{equation}
       -Nk\mathrm{Tr}\;(U\ln{U}) \geq Nk\ln{2}.
       \end{equation}
       
   
 One may easily verify that the operator $U$ has the eigenvalues $\frac{1+x}{2}$, $\frac{1-x}{2}$, $0$ (with $x=Q(\varphi,\psi)$), where the first two are simple, the third infinitely degenerate.\footnoteB{The real
 variable $x$ here is actually the absolute value of the inner product $|\langle\psi|\varphi\rangle|$, in von Neumann's notation $x=|Q(\varphi,\psi)|$. The entropy is best calculated by evaluating the trace relative to the orthonormal basis ${|\varphi\rangle,|\chi\rangle}$ where the unit vector $|\chi\rangle \equiv C(|\psi\rangle-|\varphi\rangle\langle\varphi|\psi\rangle)$ (with $C=1/\sqrt{1-x^{2}}$ is constructed to be orthogonal to $|\varphi\rangle$. In the ${|\varphi\rangle,|\chi\rangle}$ basis $U$ has the matrix
            \begin{equation}
 \mathbf{U}= \left(
\begin{array}{cc}
\frac{1}{2}(1+x^{2}) & \frac{1}{2}\sqrt{1-x^{2}}\langle\varphi|\psi\rangle \\[.2cm]
 \frac{1}{2}\sqrt{1-x^2}\langle\psi|\varphi\rangle & \frac{1}{2}(1-x^{2})
\end{array}
\right)
\label{Umatrix}
\end{equation}
which clearly satisfies $\mathrm{Tr}\; U = 1$, as required. The eigenvalues of $U$ are easily found from the secular equation $\lambda^{2}-\lambda+\frac{1-x^{2}}{4}=0$: the solutions are $\lambda = \frac{1\pm x}{2}$, from which von Neumann's Eq.\ (\ref{Uentr}) follows immediately.} 
We can calculate $\mathrm{Tr}(U\ln{U})$ from this,
       and one obtains the condition
       \begin{equation}
       \label{Uentr}
          -\frac{1+x}{2}\ln{\frac{1+x}{2}} -\frac{1-x}{2}\ln{\frac{1-x}{2}} \geq \ln{2}.
          \end{equation}
          This will occur however, as one can easily calculate, only for $x=0$, $Q(\varphi,\psi)=0$, i.e., if $\varphi, \psi$ are orthogonal.

      \section{Determination of entropy changes}
      
      We are now in a position to calculate how large an entropy increase occurs when we transform a pure ensemble $\mathfrak{G}_{1}$ of the state $\varphi$ into a mixture $\mathfrak{G}_{2}$
      of pure ensembles of the states $\psi_{1},\psi_{2},\ldots$ (which constitute a complete orthonormal system) with the corresponding weights (mixture composition)\footnoteB{A positive hermitian statistical 
      operator $U$ with $\mathrm{Tr}(U)=1$ necessarily has a discrete spectrum only, with eigenvalues $w_{\mu}$ summing to unity (see footnote 13). Thus, for an arbitrary such  $U$, there exists a complete orthonormal
      set $\psi_{\mu}$ with $U=\sum_{\mu}w_{\mu}P_{\psi_{\mu}}$. The ensemble $\mathfrak{G}_{2}$ with 
      statistical operator $U=\sum_{\mu}w_{\mu}P_{\psi_{\mu}}$ is statistically indistinguishable from an ensemble where a fraction $w_{1}$ of the systems are in the pure state $\psi_{1}$, $w_{2}$ are in state $\psi_{2}$,
      and so on. Cf.\ note $ac$ of annotated \autocite{VonNeumann:1927b}. Henceforth, the given gas ensemble will be regarded as so constituted.}
      \begin{equation}
        w_{1},w_{2},\ldots\quad(w_{1}\geq 0,\;w_{2}\geq 0,\ldots, w_{1}+w_{2}+\ldots=1).
        \end{equation}
      A reversible transformation must be found which in particular can be executed on the gases of these ensembles (cf.\ Section 3).\footnoteB{We outline here the strategy of von Neumann's remarkable (from a modern viewpoint)
      argument leading to his famous entropy formula---see following footnote for a detailed dissection. It seems at first sight amazing that von Neumann was able to invent a direct line leading from the \emph{classical formula for the 
      entropy of an ideal gas} to the \emph{general expression for entropy for any ensemble of states of an arbitrary quantum system}!  We hope the annotations presented in this and the following note will clarify how this works.
      
      The ``gas ensemble'' referred to here is obtained by taking the ensemble of $N$ quantum systems (for which we desire a formula for the entropy) and enclosing each system in a massive box,
      and then placing all these boxes (now called by von Neumann the ``molecules'' of the gas ensemble, which are so massive they  behave classically) 
      in a much larger container (of volume $V$), wherein they bounce around, reaching \emph{classical} thermal equilibrium at some temperature $T$. The interaction of these boxes---called by
      von Neumann ``molecules'', cf.\ Sec 2---with each other
      and the walls of the large box is presumed to have no effect whatsoever on the internal quantum dynamics of the systems of the ensemble, which are physically isolated from the external classical gas behavior.  The total entropy of this setup
      then amounts to the sum of the entropy $S_{\mathrm{quant}}$ of the ensemble of quantum systems, plus the classical ideal gas entropy of  $N$ particles in a volume $V$ and at temperature $T$, namely (as in \autocite[p.\ 821]{Einstein:1914b}, cf.\ footnote $a$)
       $S_{\mathrm{class}}= Nk(c\ln{T}+\ln{V})$, where $c$ is the specific heat at constant volume per molecule (e.g., $c=3/2$ if the boxes are allowed purely translational motion). The strategy of the argument is to calculate the change in
       entropy if a pure ensemble $\mathfrak{G}_{1}$
       (where $S_{\mathrm{quant}}=0$), with the gas of boxes (all containing a quantum system in the same state $\varphi$)
       at temperature $T$, is transformed into a mixed ensemble $\mathfrak{G}_{2}$, with a nonzero $S_{\mathrm{quant}}$, at the same temperature $T$. As the classical
       gas contribution is the same at the start and end, the change of entropy gives just the desired $S_{\mathrm{quant}}$ of the final mixed ensemble.}
          
           We introduce into the gas ensemble $\mathfrak{G}_{1}$ a series of walls in such a way that they divide the full volume $V$ into sub-volumes  $w_{1}V$, $w_{2}V\ldots$. Each of these compartments contains only molecules in
      the state $\varphi$---these should, following Section 6, be transformed reversibly into the state $\psi_{1}$ (resp.\ $\psi_{2}$, \ldots)
        The gases of pure ensembles with states $\psi_{1},\psi_{2},\ldots$ (with the numbers $w_{1}N, w_{2}N,\ldots$)
      we then expand isothermally and reversibly from their [original] volumes $w_{1}V, w_{2}V,\ldots$ to volume $V$. At this point we enclose the first gas in a box with semipermeable walls, which allow $\psi_{2},\psi_{3}, \ldots$ to pass through, but
      reflect $\psi_{1}$; the second in a box with semipermeable walls, which allow $\psi_{1},\psi_{3}, \ldots$ to pass through, but reflect $\psi_{2}$; etc. etc. We now superimpose \footnoteA{This ``superimposing'', which already plays a role
      in Section 6, one finds described, for example, in Planck, \emph{Thermodynamik}, sections 235, 236 (7th edition, 1922).} all these boxes, and obtain thereby the desired gas mixture $\mathfrak{G}_{2}$ (with the unchanged volume $V$).\footnoteB{Here,
      we analyze in more detail von Neumann's very short, but crucial argument, whereby the transition from the original pure (quantum) ensemble 
      $\mathfrak{G}_{1}$ (quantum entropy zero, classical ``box gas'' entropy equal to that of a classical ideal gas at temperature $T$, volume $V$)
      to the final ensemble $\mathfrak{G}_{2}$ with quantum statistical operator $U=\sum_{\mu}w_{\mu}P_{\varphi_{\mu}}$ (quantum entropy $-Nk\mathrm{Tr}(U\ln{U})$, and unchanged classical ``box gas'' entropy, as initially, for temperature $T$, volume $V$),
       is accomplished in four steps, each of which is reversible:
      \begin{enumerate}
      \item The first step involves inserting walls randomly so as to produce boxes with fractional volumes $w_{1}V$, number of systems (all in state $\varphi$) $w_{1}N$ ``molecules'', another with sub-volume 
      $w_{2}V$, number of systems (all in state $\varphi$) $w_{2}N$ ``molecules'',
      etc.  The random choice of sub-volumes $w_{\mu}V$ will of course produce the desired system (``molecule'') numbers $w_{\mu}N$ only in the limit of very large $N$. There is clearly no change in total entropy in this step: it remains zero.
      \item Within each of the boxes obtained in the first step, the (uniform) state of all the ``molecules'' is isentropically (cf.\ section 6) rotated to the appropriate new state, $\psi_{1}$ in the first box (of volume $w_{1}V$), $\psi_{2}$ in the second box, etc.;
      thus to a pure ensemble of state $\psi_{\mu}$ in the box of volume $w_{\mu}V$.
      \item Each of the boxes obtained in the preceding step is now isothermally expanded (from volume $w_{\mu}V$, for the $\mu$'th box) to the original volume $V$. 
      In this procedure we must imagine the ``molecules'' containing our quantum states
       bouncing around as a classical ideal gas (while preserving the internal quantum state
      $\psi_{\mu}$ of all the molecules in the $\mu$'th box) as the walls are moved. There is a positive change of entropy in this step, \emph{deriving entirely from the classical entropy change} $kNw_{\mu}\ln{\frac{V}{w_{\mu}V}}=-Nkw_{\mu}\ln{w_{\mu}}$
       for an isothermal expansion of an ideal gas of
      $N w_{\mu}$ molecules, from volume $w_{\mu}V$ to volume $V$ (the appropriate entropy formula can be found in \autocite{Einstein:1914b}, where it appears as the second formula on p.\ 821: the only entropy change arises from the second term in
      the first sum in this formula). 
      Summing over all the boxes (indexed by $\mu$) we obtain the expressions Eqs.\ (\ref{isoexp},\ref{entropychange}) in the text. Note that the entropy change is positive, as $w_{\mu}\leq 1$.
      Physically, heat must be supplied to each box as it expands to maintain the temperature at a constant value as the ``molecules'' do work expanding the walls.
      \item In the final step, all the boxes (now of the original volume $V$) are merged sequentially using semipermeable membranes. Planck had shown \autocite[pp.\ 219-220]{Planck:1922} that two boxes of equal volume containing distinguishable gas molecules
      (thus, molecules of type 1 in box 1, and molecules of type 2 in box 2) can be merged via a reversible isentropic process (actually, Planck demonstrates the opposite, separation, process---but as it is reversible we can imagine a merging process
      just as well). He thus demonstrated that ``the entropy of a mixture is equal to the sum of the entropies of each gas considered separately, if each gas were to occupy the whole volume of the mixture at the given temperature by itself.''  Here,
      von Neumann's ``molecules'' are themselves boxes containing a quantum system, and are distinguishable as different boxes are in orthogonal quantum states (one of the $\psi_{\mu}$). At the conclusion of this sequence of operations, we have converted the 
      initial pure ensemble (all systems in state $\varphi$, with the $N$ ``molecules'' containing them forming an ideal \emph{classical} gas at temperature $T$ in volume $V$) to a mixed ensemble with weight $w_{\mu}$ for the state $\psi_{\mu}$.
      The only change in entropy in the overall process occurs in step 3, whence the desired 
      result Eq.\ (\ref{entropychange}). As the classical entropy component due to the ``molecules'' is the same at the end as at the beginning, namely, the classical ideal gas entropy for $N$ particles at temperature $T$ and volume $V$, this entropy
      change is associated solely with the ensemble of quantum systems contained in the ``molecules.'' The final compact expression Eq.\ (\ref{finalentropy}) follows immediately, as $U$ diagonalizes to an infinite matrix with eigenvalues $w_{1},w_{2},\ldots,w_{\mu}, \ldots$
       Effectively, we have modeled an arbitrary quantum ensemble as a \emph{classical} gas of  (arbitrary) \emph{quantum} systems---which could be a quantum gas itself, for example!
      \end{enumerate}
      }
      
      Only in the penultimate step above (the isothermal reversible expansion) are entropy changes incurred in heat reservoirs: indeed, it results in a reduction [in the entropy of the heat reservoirs] by
      \begin{equation}
      \label{isoexp}
          \sum_{\mu} -w\;\;Nk\ln{w} = -Nk\sum_{\mu}w_{\mu}\ln{w_{\mu}}.
          \end{equation}
          As the process is reversible, $\mathfrak{G}_{2}$ has an entropy greater than $\mathfrak{G}_{1}$ by
          \begin{equation}
          \label{entropychange}
          \mathbf{S} = -Nk\sum_{\mu}w_{\mu}\ln{w_{\mu}}.
          \end{equation}
          
          The ensemble $\mathfrak{G}_{2}$ has, by its construction, the [density] operator
          \begin{equation}
          U = \sum_{\mu}w_{\mu}P_{\psi_{\mu}}.
          \end{equation}
          Consequently $U$ has eigenfunctions $\psi_{1},\psi_{2},\ldots$ with eigenvalues $w_{1},w_{2},\ldots$; and $U\ln{U}$ has the same eigenfunctions with the eigenvalues $w_{1}\ln{w_{1}}, w_{2}\ln{w_{2}},\ldots$. Therefore,
          \begin{equation}
             \mathrm{Tr} (U) = \sum_{\mu} w_{\mu} =1,
             \end{equation}
             and
             \begin{equation}
             \label{finalentropy}
               \mathbf{S} = -Nk\sum_{\mu} w_{\mu}\ln{w_{\mu}} = -Nk\mathrm{Tr}(U\ln{U}).
               \end{equation}
               
               \section{Determination of entropy changes (contd.)}
         
      Now let $\mathfrak{G}_{1}$ be a pure ensemble, and $\mathfrak{G}_{2}$ an arbitrary ensemble with the [density] operator $U$, normalized to $\mathrm{Tr} U = 1$ according to Section 2. Following note 13
      the matrix of $U$ assumes diagonal form in a suitable complete orthonormal system [basis]---let this [basis] be $\psi_{1},\psi_{2},\ldots$ and the diagonal elements (eigenvalues) of $U$ be $w_{1},w_{2},\dots$.
      Then
      \begin{equation}
       U = \sum_{\mu} w_{\mu}P_{\psi_{\mu}},
       \end{equation}
       and on account of the [non-negative] definite property of $U$ we have $w_{1}\geq 0, w_{2}\geq 0,\ldots$, and from $\mathrm{Tr} U = 1$ it follows that $w_{1}+w_{2}=\dots = 1$. We can therefore apply the final
       formula from Section 8: $\mathfrak{G}_{2}$ has, relative to $\mathfrak{G}_{1}$, an entropy surplus of
       \begin{equation}
       \mathbf{S} = -Nk\mathrm{Tr}(U\ln{U}).
       \end{equation}
       Note that this expression is quite independent of $\mathfrak{G}_{1}$, as all pure ensembles have the same entropy (this follows already from Section 6), which we shall normalize to zero. So $\mathfrak{G}_{2}$ has
       the entropy $\mathbf{S}$. And as $0\leq w_{\mu}\leq 1$, for all $\mu$ we have $w_{\mu}\ln{w_{\mu}}\leq 0$, vanishing only if $w_{\mu}= $0 or 1. Thus, this entropy is always positive, and can vanish only when
       all $w_{\mu}$ except for one vanish: in other words, when $U=P_{\psi_{\mu}}$, whence $\mathfrak{G}_{2}$ is a pure ensemble.
       
       With this we have established all the results announced at the end of Section 2.
       
       \section{[Determination of the equilibrium (maximum entropy) ensemble]}
       
       In accordance with the program announced in Section 3, we now find, among all the ensembles with a given expectation value of the energy $\mathbf{E}$, the stable one, i.e., the ensemble with greatest entropy.
       If $H$ is the energy operator, we must therefore (cf.\ Section 3), among all [non-negative] definite $U$ with
       \begin{equation}
       \label{Uconstraints}
           \mathrm{Tr} (U) = 1,\quad \mathrm{Tr}(UH) = \mathbf{E},
           \end{equation}
           find that one with the maximal value for
           \begin{equation}
            \mathbf{S} = -Nk\mathrm{Tr}(U\ln{U}),
            \end{equation}
            i.e., with minimal
            \begin{equation}
            \mathrm{Tr}(U\ln{U}).
            \end{equation}
       
       One sees at once: at the minimum, it must be the case for all (symmetric) operators $V$ for which $U+\varepsilon V$ satisfies the associated constraints, i.e.
       \begin{equation}
       \mathrm{Tr}(V) = 0,\quad \mathrm{Tr}(VH) = 0,
       \end{equation}
       that also\footnoteA{Actually, we also need to demand that $U+\varepsilon V$ (for sufficiently small $\varepsilon$) also be [non-negative] definite, for example by requiring that $U+V, U-V$ be [non-negative] definite. This 
       last requires
       \begin{equation}
          |Q(f, Vf)| \leq Q(f, Uf).
          \end{equation}
          Writing $U$ as a diagonal matrix, with diagonal elements $w_1,w_2,\ldots,$ and let the matrix of $V$ be $v_{\mu\nu}$; the condition above becomes in this case
          \begin{equation}
            |\sum_{\mu,\nu}v_{\mu\nu}x_{\mu}x_{\nu}^{*}| \leq \sum_{\mu}w_{\mu}|x_{\mu}|^{2},
            \end{equation}
            (for all $x_{1},x_{2},\ldots$. By the definiteness of $U$, all $w_{\mu}\geq 0$.
            
            If all $w_{\mu}>0$ it is not difficult to find a positive-definite system [i.e., matrix] $v^{\prime}_{\mu\nu}$ such that $|v_{\mu\nu}|\leq v^{\prime}_{\mu\nu}$ implies the previous inequality. In that case the
            condition derived in the text for $U$ remains valid, as the condition for $V$ do not conflict with the purely algebraic processes used there.
            
            On the other hand, if there exists a $w_{\mu}=0$, these arguments fail, but one can show directly that, for such $U$, $\mathrm{Tr}(U\ln{U})$ can still be decreased while keeping $\mathrm{Tr}(U)$ and
            $\mathrm{Tr}(U\ln{U})$ constant. Here we shall not enter into any further discussion of this case, which presents absolutely no difficulties.}
       \begin{equation}
       \frac{d}{d\varepsilon}\mathrm{Tr}\left\{(U+\varepsilon V)\ln{(U+\varepsilon V)}\right\}_{\varepsilon=0} = 0.
       \end{equation}
       
       Now, for all analytic functions $f(x)$ we have the formula:\footnoteA{Note that this is only self-evident if $U$ commutes with $V$! More generally, the proof goes as follows. It suffices to prove the asserted [formula]
       for polynomial $f(x)$---for analytic $f(x)$ it then follows by a limiting argument. It follows for polynomials, provided it holds for all $f(x)=x^n$, i.e., if
       \begin{equation}
         \frac{d}{d\varepsilon}\mathrm{Tr}(U+\varepsilon V)^{n}_{\varepsilon=0} = n\mathrm{Tr}(VU^{n-1}).
         \end{equation}
         Now clearly
      \begin{equation}
      \label{derivU}
        \frac{d}{d\varepsilon}\mathrm{Tr}(U+\varepsilon V)^{n}_{\varepsilon=0} = \mathrm{Tr}(U^{n-1}V)+\mathrm{Tr}(U^{n-2}VU)+\cdots+\mathrm{Tr}(VU^{n-1});
        \end{equation}
        and, as the following general [cyclic] identity follows from the definition of the trace,
        \begin{equation}
        \mathrm{Tr}(ABC) = \mathrm{Tr}(BCA),
        \end{equation}
        all $n$ factors on the right side [of Eq.\ (\ref{derivU})] are equal to the final one. This concludes the proof.}
         
       \begin{equation}
       \frac{d}{d\varepsilon}\mathrm{Tr}\;f(U+\varepsilon V)_{\varepsilon=0} = \mathrm{Tr}(Vf^{\prime}(U)),
       \end{equation}
       whence we require
       \begin{equation}
       \label{lnUorth}
         \mathrm{Tr}\left\{V(\ln{U}+1)\right\} = 0.
         \end{equation}
         In order that this should hold for all $V$ that satisfy the constraints
        \begin{equation}
        \label{Vconstr}
         \mathrm{Tr}\; V = 0,\quad\mathrm{Tr}(VH) = 0,
         \end{equation}
         there must exist two numbers $\alpha,\beta$, so that\footnoteB{The reasoning here may be clarified if we think of the trace operation, $\mathrm{Tr}(XY) \equiv \sum_{\mu,\nu}X_{\mu\nu}Y_{\nu\mu}$, as an inner product of two (infinite dimensional)
         vectors with components given by the matrix elements of $X$ and $Y$ ordered linearly. Note that for $X, Y$ hermitian (as here), $\mathrm{Tr}(XY) \equiv \sum_{\mu,\nu}X_{\mu\nu}Y^{*}_{\mu\nu}$, so the inner product takes the usual form for
         a complex space. Eqs.\ (\ref{Vconstr}) can thus be rephrased ``the vector $V$ is orthogonal to the vectors corresponding to the identity matrix $1$ and the Hamiltonian matrix $H$.''  Eq.\ (\ref{lnUorth}) says ``the vector corresponding to 
         $\ln{U}+1$ is orthogonal to all such vectors $V$. It follows that $\ln{U}+1$ lies in the subspace spanned by the identity vector $1$ and the vector $H$, which is just Eq.\ (\ref{lnUsol}).}
         \begin{eqnarray}
         \label{lnUsol}
          \ln{U} + 1 &\!\!=\!\!& \alpha\cdot 1 + \beta H, \\
           U &\!\!=\!\!& e^{\alpha -1}\exp{\beta H}.
           \end{eqnarray}
           Or, if we write $\alpha$ in place of $e^{\alpha -1}$,
           \begin{equation}
            U = \alpha\exp{\beta H}.
            \end{equation}
            
            It still remains to determine $\alpha, \beta$ from the constraints [Eq.\ (\ref{Uconstraints})]. One determines $\alpha$ from $\mathrm{Tr} (U) = 1$,
            \begin{equation}
              \alpha = \frac{1}{\mathrm{Tr}(\exp{\beta H})},
              \end{equation}
              while for $\beta$ the condition $\mathrm{Tr}(UH)=\mathbf{E}$ produces the less manageable result
              \begin{equation}
              \frac{\mathrm{Tr}(H\exp{\beta H})}{\mathrm{Tr}(\exp{\beta H})} = \mathbf{E}.
              \end{equation}

       \section{The temperature}
        
        With the tools developed in Sections 9, 10 we can complete the discussion of equilibrium ensembles $\mathfrak{G}$ in the usual manner. To this end we now introduce the concept of temperature.
        
        With the aid of the final formulas of Section 10 we can calculate the entropy of the equilibrium state. We set\footnoteA{If $H$ has the eigenvalues $w_{1},w_{2},\ldots$, then $Z(\beta)=\sum_{\mu}e^{\beta w_{\mu}}$, i.e., the well-known ``partition sum.''}
        \begin{equation}
         \mathrm{Tr} (\exp{\beta H}) = Z(\beta),
         \end{equation}
       whence
       \begin{equation}
        \mathbf{E} = \frac{Z^{\prime}(\beta)}{Z(\beta)},\quad U = \frac{1}{Z(\beta)}\exp{\beta H},
        \end{equation}
        and therefore also
        \begin{equation}
        \mathbf{S} = -Nk\mathrm{Tr}(U\ln{U}) = -Nk\beta\frac{Z^{\prime}(\beta)}{Z(\beta)}+Nk\ln{Z(\beta)}.
        \end{equation}
        Any given one of the three quantities $\beta,\mathbf{E},\mathbf{S}$ determines the two others; in particular, we calculate $N {\displaystyle \frac{d\mathbf{E}}{d\mathbf{S}}}$. 
        This is:
        \begin{eqnarray}
           \frac{d\mathbf{S}}{d\beta} &=& -Nk\beta \left(\frac{Z^{\prime}(\beta)}{Z(\beta)}\right)^{\prime} = -Nk\beta \frac{d\mathbf{E}}{d\beta},\\
           N\frac{d\mathbf{E}}{d\mathbf{S}} &=& N\frac{d\mathbf{E}}{d\beta}/\frac{d\mathbf{S}}{d\beta} = -\frac{1}{k\beta}.
           \end{eqnarray}
        $N {\displaystyle \frac{d\mathbf{E}}{d\mathbf{S}}}$ 
        is now, by the usual definition,\footnoteA{Because, as one easily shows, $\mathfrak{G}$ [in contact] with every
        reservoir of lower temperature gains total entropy by loss of energy, and with every one of higher temperature in the same way absorbs energy;
        thus can only be in equilibrium with this  [i.e., with a reservoir of this temperature].} the temperature $\mathbf{T}$ of $\mathfrak{G}$, from which it follows that
        \begin{equation}
          \beta = -\frac{1}{k\mathbf{T}}.
          \end{equation}
          Now we can immediately express $\mathbf{E}$ and $\mathbf{S}$ as functions of $\mathbf{T}$:
          \begin{eqnarray}
             \mathbf{E} &=& \frac{Z^{\prime}(-1/k\mathbf{T})}{Z(-1/k\mathbf{T})},\\
             \mathbf{S} &=& \frac{N}{\mathbf{T}}\frac{Z^{\prime}(-1/k\mathbf{T})}{Z(-1/k\mathbf{T})}+Nk\ln{Z(-1/k\mathbf{T})}.
             \end{eqnarray}
        And for $U$ one arrives at
        \begin{equation}
         U = \frac{1}{Z(-1/k\mathbf{T})}\exp{-\frac{H}{k\mathbf{T}}}.
         \end{equation}
              This accomplishes the derivation of the formulas of Section 3.
              
             In concluding, one more remark. Our derivations are only possible to the extent that $Z(-1/k\mathbf{T}), Z^{\prime}(-1/k\mathbf{T})$ are meaningful, i.e., (in the terminology of Section 3),
             to the extent that the series
             \begin{equation}
             \sum_{\mu}\exp{(-\frac{w_{\mu}}{k\mathbf{T}})},\quad  \sum_{\mu}w_{\mu}\exp{(-\frac{w_{\mu}}{k\mathbf{T}})}
             \end{equation}
             converge. Thus $H$ in any case cannot have a continuous spectrum, and its eigenvalues $w_1,w_2,\ldots$ must tend to $+\infty$.  This is clearly the formalization of the fact already employed in Section 2, that
             the system $\mathfrak{S}$ can be enclosed in a box.\footnoteA{If, as for example with the hydrogen atom, a continuous spectrum is present, the system $\mathfrak{S}$ must certainly be  spatially indefinitely
             extended.} The requirement is even more striking in the circumstance that the series above begins to diverge above a certain temperature: then one can indeed place the system $\mathfrak{S}$ at 
             sufficiently low temperature in a box, but at the specified temperature a sort of ``explosion'' takes place.

\printbibliography 
\end{document}